\documentclass[
11pt, 
english,
singlespacing, 
headsepline, 
]{book} 

\usepackage[numbers,sort&compress]{natbib}
\input{packages.tex}

\definecolor{VertexColor}{RGB}{163,53,101}
\definecolor{vertexcolor}{RGB}{163,53,101}

\definecolor{SEcolor}{RGB}{255,110,182}
\definecolor{SEColor}{RGB}{255,110,182}

\definecolor{GluonColor}{RGB}{0,163,180}
\definecolor{gluoncolor}{RGB}{0,163,180}

\definecolor{phisix}{RGB}{255,161,122}

\newif\ifstartcompletesineup
\newif\ifendcompletesineup
\pgfkeys{
    /pgf/decoration/.cd,
    start up/.is if=startcompletesineup,
    start up=true,
    start up/.default=true,
    start down/.style={/pgf/decoration/start up=false},
    end up/.is if=endcompletesineup,
    end up=true,
    end up/.default=true,
    end down/.style={/pgf/decoration/end up=false}
}
\pgfdeclaredecoration{complete sines}{initial}
{
    \state{initial}[
        width=+0pt,
        next state=upsine,
        persistent precomputation={
            \ifstartcompletesineup
                \pgfkeys{/pgf/decoration automaton/next state=upsine}
                \ifendcompletesineup
                    \pgfmathsetmacro\matchinglength{
                        0.5*\pgfdecoratedinputsegmentlength / (ceil(0.5* \pgfdecoratedinputsegmentlength / \pgfdecorationsegmentlength) )
                    }
                \else
                    \pgfmathsetmacro\matchinglength{
                        0.5 * \pgfdecoratedinputsegmentlength / (ceil(0.5 * \pgfdecoratedinputsegmentlength / \pgfdecorationsegmentlength ) - 0.499)
                    }
                \fi
            \else
                \pgfkeys{/pgf/decoration automaton/next state=downsine}
                \ifendcompletesineup
                    \pgfmathsetmacro\matchinglength{
                        0.5* \pgfdecoratedinputsegmentlength / (ceil(0.5 * \pgfdecoratedinputsegmentlength / \pgfdecorationsegmentlength ) - 0.4999)
                    }
                \else
                    \pgfmathsetmacro\matchinglength{
                        0.5 * \pgfdecoratedinputsegmentlength / (ceil(0.5 * \pgfdecoratedinputsegmentlength / \pgfdecorationsegmentlength ) )
                    }
                \fi
            \fi
            \setlength{\pgfdecorationsegmentlength}{\matchinglength pt}
        }] {}
    \state{downsine}[width=\pgfdecorationsegmentlength,next state=upsine]{
        \pgfpathsine{\pgfpoint{0.5\pgfdecorationsegmentlength}{0.5\pgfdecorationsegmentamplitude}}
        \pgfpathcosine{\pgfpoint{0.5\pgfdecorationsegmentlength}{-0.5\pgfdecorationsegmentamplitude}}
    }
    \state{upsine}[width=\pgfdecorationsegmentlength,next state=downsine]{
        \pgfpathsine{\pgfpoint{0.5\pgfdecorationsegmentlength}{-0.5\pgfdecorationsegmentamplitude}}
        \pgfpathcosine{\pgfpoint{0.5\pgfdecorationsegmentlength}{0.5\pgfdecorationsegmentamplitude}}
}
    \state{final}{}
}

\tikzset{
corner/.style={line width=1pt,dashed,draw=black,dash pattern=on 6pt off 4pt},
scalar/.style={line width=1pt,draw=black},
gluon/.style={line width=1pt,decorate, draw=GluonColor,
    decoration={complete sines,aspect=0,amplitude=1.25mm,segment length=1.5mm,start up,end up}},
gluontwo/.style={line width=1pt,decorate, draw=GluonColor,
    decoration={complete sines,aspect=0,amplitude=.7mm,segment length=1mm,start up,end up}},
ghost/.style={line width=1pt,loosely dotted,draw=black},
wilson/.style={line width=2pt,draw=black},
 }
\NewDocumentCommand\semiloop{O{black}mmmO{}O{above}}
{%
\draw[#1] let \p1 = ($(#3)-(#2)$) in (#3) arc (#4:({#4+180}):({0.5*veclen(\x1,\y1)})node[midway, #6] {#5};)
}

\newif\ifstartcompletesineup
\newif\ifendcompletesineup
\pgfkeys{
    /pgf/decoration/.cd,
    start up/.is if=startcompletesineup,
    start up=true,
    start up/.default=true,
    start down/.style={/pgf/decoration/start up=false},
    end up/.is if=endcompletesineup,
    end up=true,
    end up/.default=true,
    end down/.style={/pgf/decoration/end up=false}
}
\pgfdeclaredecoration{complete sines}{initial}
{
    \state{initial}[
        width=+0pt,
        next state=upsine,
        persistent precomputation={
            \ifstartcompletesineup
                \pgfkeys{/pgf/decoration automaton/next state=upsine}
                \ifendcompletesineup
                    \pgfmathsetmacro\matchinglength{
                        0.5*\pgfdecoratedinputsegmentlength / (ceil(0.5* \pgfdecoratedinputsegmentlength / \pgfdecorationsegmentlength) )
                    }
                \else
                    \pgfmathsetmacro\matchinglength{
                        0.5 * \pgfdecoratedinputsegmentlength / (ceil(0.5 * \pgfdecoratedinputsegmentlength / \pgfdecorationsegmentlength ) - 0.499)
                    }
                \fi
            \else
                \pgfkeys{/pgf/decoration automaton/next state=downsine}
                \ifendcompletesineup
                    \pgfmathsetmacro\matchinglength{
                        0.5* \pgfdecoratedinputsegmentlength / (ceil(0.5 * \pgfdecoratedinputsegmentlength / \pgfdecorationsegmentlength ) - 0.4999)
                    }
                \else
                    \pgfmathsetmacro\matchinglength{
                        0.5 * \pgfdecoratedinputsegmentlength / (ceil(0.5 * \pgfdecoratedinputsegmentlength / \pgfdecorationsegmentlength ) )
                    }
                \fi
            \fi
            \setlength{\pgfdecorationsegmentlength}{\matchinglength pt}
        }] {}
    \state{downsine}[width=\pgfdecorationsegmentlength,next state=upsine]{
        \pgfpathsine{\pgfpoint{0.5\pgfdecorationsegmentlength}{0.5\pgfdecorationsegmentamplitude}}
        \pgfpathcosine{\pgfpoint{0.5\pgfdecorationsegmentlength}{-0.5\pgfdecorationsegmentamplitude}}
    }
    \state{upsine}[width=\pgfdecorationsegmentlength,next state=downsine]{
        \pgfpathsine{\pgfpoint{0.5\pgfdecorationsegmentlength}{-0.5\pgfdecorationsegmentamplitude}}
        \pgfpathcosine{\pgfpoint{0.5\pgfdecorationsegmentlength}{0.5\pgfdecorationsegmentamplitude}}
}
    \state{final}{}
}

\newcommand{\tr}{\text{tr}\ }

\newcommand{\Op}{\mathcal{O}}

\newcommand{\id}{\mathds{1}}
\newcommand{\Wl}{\mathcal{W}_\ell}

\newcommand{\vvev}[1]{\langle\!\langle\, #1 \, \rangle\!\rangle}
\newcommand{\vev}[1]{\langle\, #1 \, \rangle}

\def\Dds{{\mathbb{D}}}
\def\Fds{{\mathbb{F}}}
\def\Ids{{\mathbb{I}}}
\def\Am{{\mathcal{A}}}
\def\Bm{{\mathcal{B}}}

\def\Gm{{\mathcal{G}}}
\def\Fds{{\mathds{F}}}
\def\Km{{\mathcal{K}}}
\def\Lm{{\mathcal{L}}}
\def\Nm{{\mathcal{N}}}
\def\Om{{\mathcal{O}}}
\def\a{{\alpha}}

\def\c{{\gamma}}

\usepackage{bbm}

\def\veps{\varepsilon}

\newcommand\g{{\gamma}}

\newcommand{\bal}{\begin{equation}\begin{aligned}}
\newcommand{\eal}{\end{aligned} \end{equation}}
\def\id{\protect{{1 \kern-.28em {\rm l}}}}

\newcommand{\be}{\begin{equation}}
\newcommand{\ee}{\end{equation}}

\newcommand{\mc}{\mathcal}

\renewcommand{\a}{\alpha}

\renewcommand{\c}{\chi}

\newcommand{\pd}{\partial}
\newcommand{\spd}{\slashed{\partial}}
\newcommand{\sx}{\slashed{x}}

\newcommand{\D}{\Delta}
\newcommand{\e}{\epsilon}

\newcommand{\m}{\mu}

\newcommand{\n}{\nu}

\newcommand{\x}{\chi}

\newcommand{\p}{\pi}

\newcommand{\y}{\upsilon}

\usepackage{color}

\definecolor{mypink1}{rgb}{0.958, 0.188, 0.478}

\newcommand{\ba}{\begin{eqnarray}}
\newcommand{\ea}{\end{eqnarray}}
\def\ov{\over}
\def \del{\partial}

 \def\n{\nu}

\hypersetup{}

\tikzset{Witten diagram/.style={execute at begin picture={%
\draw[blue ,fill=blue!05] circle[radius=\pgfkeysvalueof{/tikz/Witten/radius}];
\path node (X){\phantom{X}};
},baseline={(X.base)}},vertex/.style={circle,fill,inner sep=1.414pt,node
contents={}},
Witten/.cd,radius/.initial=1.414cm}

\RequirePackage{geometry}
\input{./Figures/images.tex}

\begin{document}

\pagestyle{plain} 

\vspace{1cm}

\vspace*{.005\textheight}

\begin{center}
\noindent \textcolor{Emerald}{\rule{14.5cm}{.5mm}}
\hfill
\vspace{0.4cm}
\begin{spacing}{1.2}
	\huge\bfseries  \textcolor{PineGreen}{Correlators on the Wilson Line \\ Defect CFT} 
\end{spacing}
\noindent\textcolor{Emerald}{\rule{14.5cm}{.5mm}}

\vspace{1cm}
 
\large \textit{Dissertation  zur Erlangung des akademischen Grades} \par
\vspace{0.5cm}
\textsc{doctor rerum naturalium (Dr. rer. nat.)} \par
\vspace{0.7cm}
\textit{im Fach:} \textsc{Physik}  \par
\textit{Spezialisierung:} \textsc{Theoretische Physik}  \par
\vspace{0.2cm}
\textit{eingereicht an der} \par
\vspace{0.7cm}
\large \textit{Mathematisch-Naturwissenschaftlichen Fakultät}  \par
\large\textit{der Humboldt-Universität zu Berlin}  \par
\vspace{1cm}
\textit{von} \par
\vspace{0.5cm}
\textcolor{PineGreen}\Large\textsc{\textcolor{PineGreen}{Giulia Peveri}}\par
\vspace{0.5cm}

\end{center}
\vspace{2cm}

Präsidentin der Humboldt-Universität zu Berlin: \par
\vspace{0.1cm}
Prof. Dr.  Julia von Blumenthal

\vspace{0.5cm}

Dekanin der Mathematisch-Naturwissenschaftlichen Fakultät: \par
\vspace{0.1cm}
Prof. Dr. Caren Tischendorf

\vspace{0.5cm}

Gutachter/innen: \par
\vspace{0.1cm}
1. Prof. Dr.  Jan Plefka \par
2. Prof. Dr.  Valentina Forini \par
3. Prof. Dr.  Andrea Cavaglià\par

\vspace{1cm}
Tag der mündlichen Prüfung: \,22.09.2023

\thispagestyle{empty}

\vfill

\newpage
\thispagestyle{empty}
\mbox{}


\newpage

\begin{center}
\huge \textbf{Eidesstattliche Erkl\"arung} \par
\end{center}
\normalsize \noindent Ich erkläre, dass ich die Dissertation selbständig und nur unter Verwendung der von mir gemäß $\S 7$ Abs. 3 der Promotionsordnung der Mathematisch-Naturwissenschaftlichen Fakultät, veröffentlicht im Amtlichen Mitteilungsblatt der Humboldt-Universität zu Berlin Nr. 42/2018 am 11.07.2018 angegebenen Hilfsmittel angefertigt habe.
\vspace{1cm}

\noindent Ort,\,Datum:\\
\rule[0.5em]{25em}{0.5pt}
 
\noindent Unterschrift:\\
\rule[0.5em]{25em}{0.5pt} 

\thispagestyle{empty}

\newpage
\thispagestyle{empty}
\mbox{}


\newpage
\vspace*{.1\textheight}

\begin{flushright}
\textit{Alla mia famiglia, che rimarrà sempre \\ il mio punto di riferimento.}
\end{flushright}

\thispagestyle{empty}

\newpage
\thispagestyle{empty}
\mbox{}


\newpage

\begin{center}
\huge \textbf{Abstract} \par
\end{center}
\normalsize \noindent Conformal field theory (CFT) plays a key role in modern theoretical physics, with applications in string theory,  condensed matter, statistical mechanics, and pure mathematics as well.  Through CFT we describe real physical systems at criticality and fixed points of the renormalization group flow.  It is also central in the study of quantum gravity, thanks to the AdS/CFT correspondence. This thesis originates in the context of the $\mathcal{N}=4$ supersymmetric Yang-Mills (SYM) theory, which represents the CFT side of this correspondence.  

$\mathcal{N}=4$ SYM has been widely studied in the last years, being the simplest interacting CFT in four dimensions, thanks to its rich pool of symmetries.  Interesting extended operators live in $\mathcal{N}=4$ SYM. This work mainly revolves around the supersymmetric Wilson line and its interpretation as a conformal defect.  Particularly, we focus on excitations localized on the defect called insertions, whose correlators are described by a one-dimensional CFT.  This 1$d$ theory is an exceptional holographic laboratory to study conformal correlators due to the ample interplay of different techniques to evaluate them, such as bootstrap, supersymmetric localization, and integrability.

After introducing the technical background, we dive into the first main result: an efficient algorithm for computing multipoint correlation functions of scalar insertions on the Wilson line.  This algorithm consists of recursion relations encoding the possible interactions between operators of elementary scalar fields up to next-to-leading order at weak coupling.  We show various computations of such four-, five- and six-point correlators, and discuss their properties.  Moreover, we use the four-point function case to illustrate the power of the Ward identities, which are crucial in deriving a next-to-next-to-leading order result.

The four-point Ward identities also set the stage for a conjecture.  Thanks to the cornucopia of perturbative correlators obtained through the recursive formulae, we find a family of differential operators annihilating our correlation functions, which we hypothesize to be a multipoint extension of the Ward identities satisfied by the four-point functions. These non-perturbative constraints are shown to be fundamental ingredients in the bootstrap of a five-point function at strong coupling.

We conclude this thesis with a wider look at 1$d$ CFTs in general.  We explore another tool to represent 1$d$ correlators: the Mellin representation.  In the higher-dimensional case, the Mellin representation of conformal correlators has proven to be an excellent tool for the study of holographic CFTs.  Therefore, we define an inherently one-dimensional Mellin amplitude, which can be defined at the non-perturbative level with appropriate subtractions and analytical continuations. This definition allows us to derive an infinite set of non-perturbative sum rules whose characteristics are discussed in detail, and applications are sketched.  The efficiency of the 1$d$ Mellin formalism is manifest at the perturbative level.  We find a closed-form expression for the Mellin transform of leading order contact interactions and use it to extract CFT data.

\thispagestyle{empty}

\newpage
\thispagestyle{empty}
\mbox{}


\newpage
\begin{center}
\huge \textbf{Zusammenfassung} \par
\end{center}

\normalsize \noindent Konforme Feldtheorien (CFT) spielen eine Schlüsselrolle in der modernen theoretischen Physik, mit Anwendungen in der Stringtheorie, der kondensierten Materie, der statistischen Mechanik und auch in der reinen Mathematik.  Mit CFT beschreibt man reale physikalische Systeme bei Kritikalität und Fixpunkte des Renormierungsgruppenflusses.  Dank der AdS/CFT-Korrespondenz spielt sie auch bei der Untersuchung der Quantengravitation eine zentrale Rolle. Diese Arbeit hat ihren Ursprung im Kontext der $\mathcal{N}=4$ supersymmetrischen Yang-Mills-Theorie (SYM), welche die CFT-Seite dieser Korrespondenz darstellt.  

$\mathcal{N}=4$ SYM wurde in den letzten Jahren ausgiebig untersucht, da sie, dank ihrer vielen Symmetrien, die einfachste wechselwirkende CFT in vier Dimensionen ist.  In $\mathcal{N}=4$ SYM existieren interessante nicht lokale Operatoren. Diese Arbeit dreht sich hauptsächlich um die supersymmetrische Wilson-Linie und ihre Interpretation als konformer Defekt.  Insbesondere konzentrieren wir uns auf Anregungen, die auf dem Defekt lokalisiert sind, sogenannte Einfügungen, deren Korrelatoren durch eine eindimensionale CFT beschrieben werden.  Diese 1$d$-Theorie ist ein außergewöhnliches holographisches Laboratorium, um konforme Korrelatoren zu studieren, da sie ein reichhaltiges Zusammenspiel verschiedener Techniken zu ihrer Auswertung bietet, wie Bootstrap, supersymmetrische Lokalisierung und Integrabilität.

Nach einer Einführung in den technischen Hintergrund wird das erste Hauptergebnis präsentiert: Ein effizienter Algorithmus zur Berechnung von Mehrpunkt Korrelationsfunktionen von Skalareinfügungen auf der Wilson-Linie.  Dieser Algorithmus besteht aus Rekursionsbeziehungen, welche die möglichen Wechselwirkungen zwischen Operatoren elementarer Skalarfelder bis zur nächsten Ordnung bei schwacher Kopplung kodieren.  Es werden verschiedene Berechnungen solcher Vier-, Fünf- und Sechspunkt-Korrelatoren gezeigt und ihre Eigenschaften diskutiert.  Darüber hinaus wird am Beispiel der Vierpunkt-Funktion die Leistungsfähigkeit der Ward-Identitäten veranschaulicht, die für die Ableitung eines Ergebnisses nächster, vorletzter und führender Ordnung entscheidend sind.

Die Vierpunkt-Ward-Identitäten bilden auch die Grundlage für eine Vermutung.  Dank der Fülle von Störungskorrelatoren, die durch die Rekursionsformeln erhalten wurden, findet man eine Familie von Differentialoperatoren. Diese vernichten die Korrelationsfunktionen und es wird angenommen, dass sie eine Mehrpunkt-Erweiterung der Ward-Identitäten sind, die durch die Vierpunkt-Funktion erfüllt wird. Diese nichtperturbativen Beschränkungen erweisen sich als fundamentale Bestandteile des Bootstraps einer Fünfpunkt-Funktion bei starker Kopplung.

Abschließend wird der Blick auf 1$d$-CFTs im Allgemeinen gerichtet.  Untersucht wird ein weiteres Werkzeug zur Darstellung von 1$d$-Korrelatoren: die Mellin-Darstellung.  Im höherdimensionalen Fall hat sich die Mellin-Darstellung von konformen Korrelatoren als hervorragendes Werkzeug für die Untersuchung von holographischen CFTs erwiesen.  Daher definiert man eine inhärent eindimensionale Mellin-Amplitude, die auf der nichtperturbativen Ebene mit geeigneten Subtraktionen und analytischen Fortsetzungen definiert werden kann. Diese Definition erlaubt es, einen unendlichen Satz von nichtperturbativen Summenregeln abzuleiten, deren Eigenschaften im Detail diskutiert und deren Anwendungen skizziert werden.  Die Effizienz des 1$d$-Mellin-Formalismus zeigt sich auf der perturbativen Ebene.  Man findet einen Ausdruck in geschlossener Form für die Mellin-Transformation von Kontaktwechselwirkungen führender Ordnung, den man verwendet, um CFT-Daten zu extrahieren.

\thispagestyle{empty}

\newpage
\thispagestyle{empty}
\mbox{}


\newpage

\begin{center}
\huge \textbf{Statement of Originality} \par
\vspace{0.3cm}
\end{center}
\normalsize \noindent The second part of this thesis is based on original research in collaboration with various researchers whose work we gratefully acknowledge. In the following, we list the papers this part is based on, some still under completion.

\bibliographystyle{./JHEP}
\nobibliography*{}
\begin{itemize}
\vspace{0.3cm}
\item \bibentry{Bianchi:2021piu} 
\vspace{0.3cm}
\item \bibentry{Barrat:2021tpn}
\vspace{0.3cm}
\item \bibentry{Barrat:2022eim}
\vspace{0.05cm}
\item \bibentry{Artico:2023abp}
\vspace{0.3cm}
\item \bibentry{Barrat:2023pev}
\end{itemize}

\thispagestyle{empty}

\newpage
\thispagestyle{empty}
\mbox{}


\newpage

\begin{center}
\huge \textbf{Acknowledgments} \par
\end{center}
\small \noindent 
This PhD has been a real journey, from moving to a new country to a world pandemic, from grief to deep joy. It was a hard but illuminating experience and I am extremely grateful to all the people that I met along the way that enriched it.

First and foremost, I want to express my gratitude to my supervisors. Valentina Forini, I appreciate the trust and freedom you granted me to explore various research directions independently.  Your encouragement to consistently strive for excellence and step beyond my comfort zone has been very much appreciated. Jan Plefka, I'm grateful for your guidance during a tough phase of my PhD and for your valuable advice. Your support has played a significant role in my progress. Thank you both also for taking the time to review my thesis and provide feedback. I would also like to thank Andrea Cavaglià for carefully reading through this work and offering insightful suggestions.

Throughout my PhD, I was fortunate to collaborate with exceptional colleagues. Daniele Artico,  thanks for your continuous support and for the countless brainstorming sessions spanning from physics to climate. Your energy is without a doubt contagious, as well as your proactive spirit. Julien Barrat, our work together was a lot of fun! I am appreciative of the profound discussions we shared regarding both physics and life. Your dedication to physics has been inspiring. Lorenzo Bianchi, I value your guidance during the earliest phase of my PhD, and your genuine excitement towards physics has set a motivating example. Gabriel Bliard, I appreciate the opportunity to truly get to know you and to share remarkable moments both within and outside of work. Our conversations about the PhD journey and our future aspirations have proven to be really insightful.  I would like to extend my gratitude to Pedro Liendo for consistently being available to provide support, explanations, and invaluable advice. Your pragmatic approach and concrete counsel are truly valued. I would also like to extend my thanks to Davide Bonomi, Pietro Ferrero, Francesco Galvagno, Carlo Meneghelli, and Aaditya Salgarkar for the valuable exchange of ideas and discussions related to our common research topics.

Undoubtedly, one of the most remarkable experiences of my PhD journey was the outreach project ``Non-Standard Models'', a collaborative effort involving Allison, Claudio, Daniele, Ilaria, Julien, and Michele. Building this project from scratch and sharing the challenges, successes, and excitement with all of you was truly amazing. 

I was fortunate to be part of the RTG ``Rethinking Quantum Field Theory'', an experience that transformed my PhD journey. Interacting with a diverse group of PhD students from various nationalities and backgrounds has been incredibly valuable, allowing me to gain insights from different perspectives and explore other research fields. Thanks to all the students, especially to Alessandro, Davide, Felipe, Moritz, Tim, Tomàs.  I am also grateful to Andrea and Luke, who accompanied me during the initial years of my PhD.

Throughout these years, I could also count on the support of incredible friends with whom I shared the ups and downs of this experience. I am thinking of Churci, Christian, Luca, Marti, Miguel, and Sbuzzi.  A special thank goes to Remus, without whom I could not have navigated the challenges of the lockdowns and who provided essential support during the delicate beginning of this PhD.  This year wouldn't have been as exciting without you, Ciup. Thank you for always being there, for my complaints, for my crazy moments, and for amazing adventures together.

Occasioni come queste, la fine di un lungo percorso, invitano sempre a fermarsi e a guardarsi indietro. Ho iniziato questo dottorato in uno dei momenti più difficili. Mi ricordo quello che ho scritto per i ringraziamenti della tesi magistrale, che rimane ancora molto attuale purtroppo. Perché casa manca e mancherà sempre, ma devo a questa esperienza la possibilità di spingermi molto oltre i miei confini e di conoscermi più in profondità. Senza di voi, sempre presenti nonostante la distanza, questo sarebbe stato impossibile. Grazie mamma, papà, zia e Gi. Vi amo immensamente.  Ma devo tantissimo anche a chi non c'è più e la cui assenza rimane incomprensibile e lacerante. Vorrei poteste essere qui anche voi a festeggiare questo traguardo.  

\thispagestyle{empty}

\newpage
\thispagestyle{empty}
\mbox{}


\newpage

\pagenumbering{gobble}

\frontmatter 

\hypersetup{linkcolor=black}
\tableofcontents 

\newpage
\thispagestyle{empty}
\mbox{}

\newpage

\pagestyle{fancy} 
\fancyhf{}
\fancyhead[LE]{\thepage}
\fancyhead[RE]{\fontsize{9}{12}\selectfont\itshape\nouppercase{\leftmark}}
\fancyhead[RO]{\thepage}
\fancyhead[LO]{\fontsize{9}{12}\selectfont\itshape\nouppercase{\rightmark}}
\renewcommand{\chaptermark}[1]{\markboth{\chaptername \ \thechapter.\ \ #1}{}}  
\renewcommand{\sectionmark}[1]{\markright{\thesection.\ \ #1}{}} 


\part{Introduction}

\hypersetup{linkcolor=mypinegreen}
\pagenumbering{arabic}
\mainmatter 

\chapter{Motivation and Main Results}

\section{Why Do We Care About Conformal Field Theory?}

The modern theoretical framework to describe our physical world is called \textit{Quantum Field Theory} (QFT), and it developed as a fusion of three different precursors: classical field theory, special relativity, and quantum mechanics. In this paradigm, particles are intended as excitations of more fundamental structures,  namely quantum fields. The interactions between particles are encoded in a particular functional called Lagrangian, where these fields appear.  Throughout the past century, this fruitful paradigm has given deeper insights into the physical world, and many different branches of research streamed from it. 

In particular, there is one that is quite central in this paradigm: \textit{Conformal Field Theory} (CFT).  With CFT, we mean a quantum theory invariant under conformal transformations, which are maps preserving angles, such as rotations, translations, etc.  Conformal field theory is ubiquitous in modern theoretical physics, and for this reason, it has been one of the most active areas of research in the last decades.  In particular, it has relevant applications in string theory, condensed matter, and statistical mechanics, and it has been inspirational for some development in pure mathematics as well. 

To understand its importance, we first need to point out that conformal field theory plays a central role in describing \textit{phase transitions}. In general, we can say that when a system changes its properties, often discontinuously,  as a result of a modification of the external thermodynamical quantities, we have a phase transition.  We can also classify them based on their characteristics. When the temperature remains constant even if heat is added to the system, we call them first-order or discontinuous phase transitions. On the other hand, there are second-order or \textit{continuous} phase transitions, where the free energy is continuous across the transition.  Examples of this type are the ferromagnetic, superconducting, and superfluid transitions.  

Alongside the concept of phase transition is necessary to consider also the notion of \textit{critical point}, which is the endpoint of a phase equilibrium curve.  Most notably, the example of the liquid-vapor critical point,  the end point of the pressure-temperature curve where liquid and vapor can coexist.  At these points, we can study \textit{critical phenomena}, namely phenomena associated with continuous phase transitions.  Interestingly,  we can characterize them with parameters known as \textit{critical exponents}.

One can finally ask what this has to do with CFT.  It turns out that CFT can be used to study the behavior of systems at the critical point. In some cases, this formalism even allows one to solve some models exactly. The most prominent example is the \textit{Ising model}, the simplest theoretical description of ferromagnetism. This model undergoes a phase transition, and at the critical point, it is described by a two-dimensional conformal field theory, which has been exactly solved \cite{Onsager:1943jn}. That is not true for the three-dimensional case, even if it is well-understood at a very high numerical precision. Both instances point to a remarkable phenomenon called \textit{universality,} namely when completely different systems can be described by the same critical exponents, e.g. the critical exponents of the 3$d$ Ising model coincide in value with the ones found for a liquid/gas transition. This suggests that the behavior of a system near a critical point depends on the dimension and the symmetries rather than on the underlying dynamics.

Another related aspect that motivates the study of CFT is the \textit{renormalization group flow}.  The renormalization group allows us to systematically investigate the changes in a physical system at different distances or energy/momentum scales.  The running of the parameters, like the couplings, of a certain model induced by a change of scale leads to what is called renormalization group flow. This tool is powerful and useful in understanding how the system behaves at very long (IR) and very short (UV) distances.  In addition, we know that the theories flow to a so-called \textit{fixed point}, where they are scale-invariant, and their couplings become dimensionless. Excluding rare examples, a quantum field theory that is scale-invariant is also conformally invariant.  Therefore, every local quantum field theory approaches a CFT in the large- and small-distance limits. 

Last but not least, CFT plays a key role in the study of quantum gravity through the \textit{AdS/CFT correspondence},  a duality stating the equivalence of type IIB superstring theory in AdS$_5 \times$ S$^5$ background and four-dimensional $\mathcal{N}=4$ supersymmetric Yang-Mills (SYM) theory \cite{Maldacena:1997re,Witten:1998qj,Gubser:1998bc}.  The great success of this conjecture is addressing, in a unified framework,  two of the most relevant open problems in theoretical physics: quantum gravity and strongly-coupled QFTs.  In quantum field theory, one typically computes the probabilities of various physical events using the technique of perturbation theory, an expansion of the solution as a power series of a small parameter.  The regime in which we can apply this technique is called weak coupling because this expansion is performed around small values of the coupling of a theory.  Opposed to this, we have a strong coupling regime. The AdS/CFT correspondence bridges these two realms, allowing to use perturbative methods to probe strong coupling regimes.  Performing perturbative computations in a CFT allows ideally to develop a non-perturbative formulation of string theory and vice versa.

\section{Realm of Research and Main Results} 

The research work presented in this thesis originates in the context of the AdS/CFT correspondence,  with a particular focus on the CFT side of this duality.  $\mathcal{N}=4$ SYM is considered to be the simplest interacting non-abelian gauge field theory in four dimensions,  especially thanks to its rich pool of symmetries. It is conformal, maximally supersymmetric, and it is believed to be integrable in the planar limit. Therefore, it has been at the crossroad of various techniques in the past decades. 

In this context, alongside local operators, i.e.  local product of fields, their spacetime derivatives, and their correlation functions,  we can study \textit{extended operators}, which are non-local. An important non-local observable is the Wilson loop,  describing the coupling between a heavy (probe) particle and the gauge field of the theory.
In $\mathcal{N}=4$ SYM, one may consider supersymmetric extensions of this operator, known as Maldacena-Wilson loop operators. In particular, one can also study the configuration of an infinite straight line \cite{Maldacena:1997re,sjrey}:
\begin{equation}
\mathcal{W}_{\ell} := \frac{1}{N} \,\text{tr}\,  \mathcal{P}\,  \text{exp}\, \int^{+\infty}_{-\infty} \, d\tau\left(i\dot{x^{\mu}}(\tau) A_{\mu}(x) + |\dot{x}(\tau)| \phi^6(x) \right)\,,\notag
\end{equation}
with $A_{\mu}(x)$ the gauge fields, $\phi^i(x)$ the adjoint scalar fields $\mathcal{N}=4$ SYM and $N$ the rank of the gauge group.  In recent years, there has been a renewal interest in this operator and its circular brother from the point of view of \textit{conformal defects} due to a revival of the conformal bootstrap program. We refer generically to defects as extended $p$-dimensional operators placed in a $d$ dimensional Euclidean space. Defects break the original symmetries of a CFT. In particular, for the conformal group $SO(d+1,1)$ we have that $SO(d+1,1) \rightarrow SO(p+1,1) \times SO(q)$, with $p+q=d$ being $q$ the \textit{co-dimension} of the defect. Defects are an interesting frontier in the study of CFTs since they enrich a system's dynamic by probing new physics.  Moreover,  they bring us closer to the study of concrete setups, as every real physical system contains impurities.

There exist several interesting configurations that can be studied in this setup. In this work, we focus on excitations localized on the defect itself, for which the correlators are described by a $1d$ (non-local) CFT.  Such defect makes the theory simpler to study compared to its parent theory $\mathcal{N}=4$ SYM, but not trivial since the interactions among these defect fields happen in the bulk of this theory, which is four-dimensional.  Moreover, the defect theory inherits some properties and symmetries from its parent theory $\mathcal{N}=4$ SYM, which allows to study it using the same rich interplay of various techniques,  from supersymmetric localization \cite{Giombi:2018qox,Giombi:2018hsx,Correa:2012at} and integrability \cite{Drukker:2012de, Correa:2012hh, Gromov:2015dfa, Kiryu:2018phb, Grabner:2020nis, Cavaglia:2021bnz,Cavaglia:2022qpg,Cavaglia:2022yvv}, to the numerical/analytical bootstrap \cite{Liendo:2018ukf,Ferrero:2021bsb} and the large charge limit \cite{Miwa:2006vd,Giombi:2021zfb,Giombi:2022anm}. This also makes it a perfect laboratory to test novel techniques, with the ultimate goal to transfer the know-how to higher-dimensional theories. 

Interestingly this setup, on the strong coupling side, corresponds to a particular gauge fixing of the non-linear sigma model describing the motion of a string in AdS$_5 \times $ S$^5$ \cite{Drukker:2006xg,Giombi:2017cqn,Cooke:2017qgm}.  However, in this context, the gauge fixing provides a worldsheet effective field theory in AdS$_2$ background.  This AdS$_2$/CFT$_1$ correspondence of the open string/Wilson line can be viewed as an example of ``non-gravitational'' or ``rigid'' holography \cite{Aharony:2015zea}.  At zero string coupling and in the limit of large string tension, the worldsheet decouples from closed string modes in the bulk, and its fluctuations are suppressed.  If one works in static gauge,  the worldsheet theory does not contain a dynamical metric and shares many similarities with QFTs in non-dynamical AdS$_2$.

\subsubsection{NNLO Four-Point Function}

Even if two- and three-point functions carry non-trivial information about a CFT, namely the CFT data, their kinematics is trivial, as they are fixed by conformal symmetry up to normalization. That is not the case for four-point functions. For this reason,  four-point functions in this $1d$ CFT have been studied extensively both at weak \cite{Kiryu:2018phb,Cavaglia:2022qpg} and strong \cite{Giombi:2017cqn,Liendo:2018ukf,Ferrero:2021bsb} coupling.  There is also an exact topological limit that has been studied using localization \cite{Giombi:2009ds,Giombi:2012ep,Beccaria:2020ykg}.  In addition, numerical results have been obtained for arbitrary coupling using a mix of integrability and bootstrap techniques \cite{Cavaglia:2022qpg}.  

We also contribute to this line of work by computing the next-to-next-to-leading order of the simplest four-point correlator at weak coupling, where the 't Hooft coupling $\lambda:= g^2 N$ is the parameter of the perturbative expansion.  Our calculation is made possible by the application of the Ward identity annihilating this correlator. This differential operator constraints the correlator to such an extent that we can trade the computation of a huge number of complicated Feynman diagrams with a handful of simple ones. This ultimately allows us to retrieve the full expression of this correlator.  This result has also been obtained in \cite{Cavaglia:2022qpg} with a novel technique, combining bootstrap and integrability. Our derivation offers then an important check of this new method too.

\subsubsection{Multipoint Correlators and Ward Identities}

The next interesting objects to study after four-point functions are obviously higher-point correlators.  However, there has not yet been a broad study of \textit{multipoint correlators} on the Wilson line defect CFT, which in general is also lacking in $\mathcal{N}=4$ SYM, with the exception of some remarkable work done in \cite{Drukker:2008pi,Drukker:2009sf}.
The study of multipoint correlators is essential if we want to embark on the quest to solve exactly the theory living on the Wilson line defect (and similar CFTs). 

More generally, higher-point functions have been one of the long-term goals of the bootstrap program for CFTs. They could work as a testing ground for multipoint bootstrap techniques. However, they are also interesting in their own right, as they contain an infinite amount of CFT data, which includes both protected operators, i.e. operators whose scaling dimension does not acquire quantum corrections, and unprotected ones. Moreover,  higher-point correlators of simple operators also hold information about lower-point functions of very complicated operators.

For this reason, in this thesis, we explore multipoint correlators in the large $N$ limit and in the weak coupling regime. In particular, we derive explicit recursion relations that encode next-to-leading order correlators with an arbitrary number of fundamental scalar fields inserted on the Wilson line.  By pinching fundamental scalars together, we also build operators of higher length\footnote{Throughout this thesis, we use length as a synonym for scaling dimension.}.  This algorithmic procedure is not limited to protected operators as it includes unprotected ones. Our result then encompasses arbitrary $n$-point correlators of arbitrary operators made out of fundamental scalar fields. 

Remarkably, we observe that a special class of differential operators annihilates all the correlators of protected operators. We conjecture these constraints to be valid non-perturbatively and to be extensions of the superconformal Ward identities satisfied by the four-point functions.  We also expand correlators of protected and unprotected operators in conformal blocks to perform consistency checks and extract new CFT data. Additionally, we pave the way to the bootstrap of the simplest five-point function of protected operators at strong coupling by deriving the superconformal blocks using the power of the conjectured Ward identities.

\subsubsection{Mellin Transform} \label{sec:Mellinintro}

We could say that the results sketched up to this point are, in one way or another, contributing or, more generally, connecting to the research area of conformal bootstrap. However, as we mentioned, 1$d$ CFTs, as the Wilson line we have just looked at in $\mathcal{N}=4$ SYM but also in $\mathcal{N}=6$ super Chern-Simons theory with matter (ABJM) \cite{Aharony:2008ug}, are extremely interesting from the point of view of integrability too, as they are believed to be integrable. A lot of remarkable work has been done exploiting this property, for example, to derive the spectrum of the defect theory exactly \cite{Correa:2012hh,Grabner:2020nis,Cavaglia:2021bnz}. However, it is still an interesting open question how the power of integrability can be exploited in this setting. 

More generally, the study of integrable field theories in curved backgrounds is an active and largely unexplored research subject, which has recently witnessed some interesting developments \cite{Beccaria:2019stp,Beccaria:2019mev,Beccaria:2019dju,Beccaria:2020qtk}. It has been pointed out that, e.g. \cite{Giombi:2017cqn, Komatsu:2020sag}, a crucial ingredient for our understanding of integrability in curved space would be the analog of flat space $S$-matrix factorization, and we believe Mellin space may provide the correct setting to look for such a feature.  

More in general, the Mellin representation of conformal correlators~\cite{Mack:2009mi,Penedones:2010ue} has proven to be an excellent tool, especially for the study of holographic CFTs~\cite{Fitzpatrick:2011ia,Paulos:2011ie,Rastelli:2017udc} in the higher-dimensional case.  The number of independent cross-ratios for a $n$-point correlation function of local operators in a $d$-dimensional CFT is identical to that of independent variables for a $d+1$-dimensional scattering amplitude.  The Mellin representation, or Mellin amplitude, makes this correspondence manifest, expressing the correlators in a form that is the natural AdS counterpart of flat-space scattering amplitudes. This form has several nice features too.  First, the Mellin amplitude has simple poles located at the values of the twist of exchanged operators (there are, however, infinitely many accumulation points of such poles). Secondly, the crossing symmetry of the correlator maps to the amplitude crossing symmetry. Finally, the language of Mellin amplitudes is particularly suitable for large $N$ gauge theories, where perturbation theory is described in terms of Witten diagrams. 

In this thesis, we use these properties as guiding principles for the definition of an inherently one-dimensional Mellin transform for 1$d$ four-point correlators. We also discuss the finite number of subtractions and the analytic continuations necessary to get a fully \textit{non-perturbative} definition of this transform. This tool is then used to derive an infinite set of non-perturbative sum rules for CFT data of exchanged operators, which is tested on known examples. This formalism is finally applied to a perturbative setup with quadratic interactions and an arbitrary number of derivatives in a bulk AdS$_2$ field theory to obtain a closed-form expression for the Mellin transform of leading order contact interactions and for the first correction to the scaling dimension of the ``two-particles'' operators exchanged in the generalized free field theory correlator.

\section{Outline}

This thesis is divided into three parts. In the first part we briefly introduce the technical building blocks necessary to clearly present the research results. These are gathered in the second part, while all the appendices are collected in the last part.

More in detail, the first part is structured as follows.  Chapter~\ref{ch:CFT} includes the basics of conformal symmetry,  including the representation of the conformal algebra, the definition of correlation functions, the operator product expansion, and a short introduction to the philosophy behind conformal bootstrap.  Chapter~\ref{ch:susyen=4} introduces supersymmetry and promotes the conformal algebra to a superconformal algebra. It also features the $\mathcal{N}=4$ SYM theory with details on its action, the planar limit, and the Feynman rules. The main character of Chapter~\ref{ch:wilsonline} is the Maldacena-Wilson line. Its nature of conformal defect is there explained, as for the 1$d$ theory defined on the line with its correlators and insertion rules. Chapter~\ref{ch:Mellin} concludes the first part with an introduction to the higher-dimensional Mellin formalism and its role in holographic CFTs.

The second part contains all the novel research results, starting from the recursion relations for both protected and unprotected operators in Chapter~\ref{sec:recursionrelations}, moving to the applications in the following chapters. Chapter~\ref{sec:23point} focuses on the study of two- and three-point functions of protected, unprotected, and composite operators, which give important checks and are useful to set conventions for the rest of the work.  In Chapter~\ref{ch:4point}, we explore four-point functions, not only using the recursion relations but also a Ward identity, which allows us to greatly reduce the computational complexity of the next-to-next-to-leading order of the simplest four-point correlator. In Chapter~\ref{ch:higherpoint}, all this technology is finally applied to higher-point correlators. There an extension of the Ward identity for four-point correlators is conjectured for multipoint correlation functions. The ingredients for a bootstrap of a five-point function are also set, and the bootstrap algorithm is put into action.  The last chapter of this part, Chapter~\ref{ch:Mellinres}, explores instead another representation of correlators in generic 1$d$ CFTs through the development of an inherently one-dimensional Mellin formalism.  Finally, Chapter~\ref{sec:conclusion} presents a summary of the results and prospects.

Part III gathers all the relevant appendices.

\chapter{Conformal Field Theory} \label{ch:CFT}

In this chapter, we review the basics of Conformal Field Theory, which is one of the main ingredients of this thesis.  The content is mostly based on \cite{DiFrancesco:1997nk,Simmons-Duffin:2016gjk,Osborn:2019lec}.

In particular, we first introduce conformal symmetry, discussing in detail the generators and the representations of the conformal algebra. We then move to correlation functions, operator product expansion and conformal blocks, which combined allow us to access the CFT data of a theory.  In the final section, we briefly explain the philosophy behind conformal bootstrap.

\section{Conformal Symmetry} \label{sec:conformalsymmetry}

A conformal transformation is an invertible mapping $x \rightarrow x^{\prime}$ which leaves the metric tensor $g_{\mu \nu}, \mu,\nu=1,\dots, d$ invariant up to a scale $\Omega(x)$:
\begin{equation}
g_{\mu \nu} (x^{\prime}) = \Omega(x)\, g_{\mu \nu} (x) \,.
\end{equation}
A conformal transformation preserves angles but not distances. The set of conformal transformations forms a group, which has the Poincaré group as a subgroup ($\Omega(x)=1$).

\begin{figure}[h]
\centering
\includegraphics[scale=0.28]{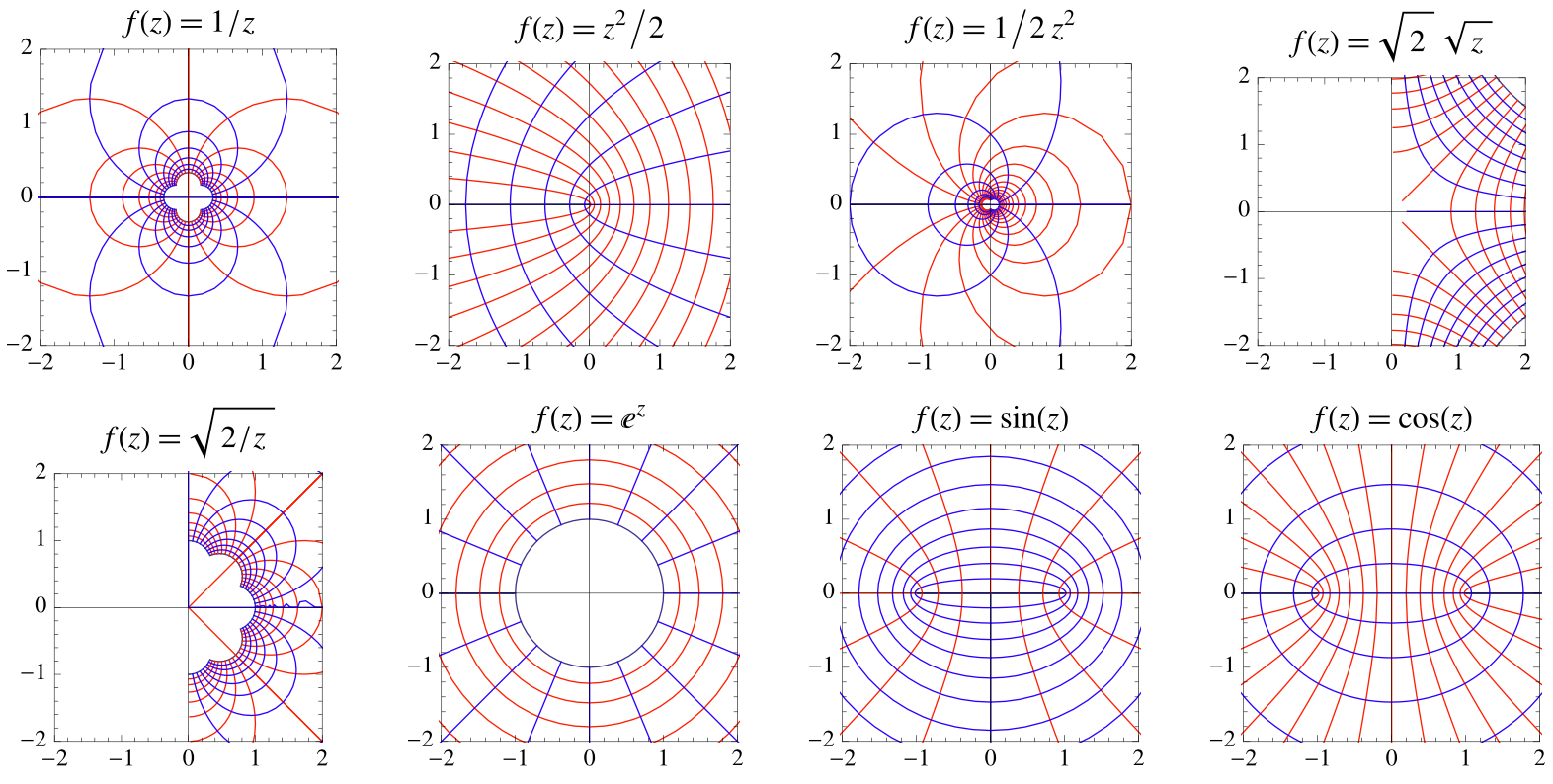}
\caption{\textit{Various conformal transformations of the regular grid \cite{Churchill:1990brw}.}}
\end{figure}

To classify conformal transformations, we consider the infinitesimal transformation $x_\mu^{\prime}=x_\mu+\epsilon_\mu(x)$. Requiring that the transformation is conformal, we obtain
\begin{equation} \label{eq:conformalkillingeq}
\partial_\mu \epsilon_\nu+\partial_\nu \epsilon_\mu=c(x) \g_{\mu \nu}\,,
\end{equation}
where $c(x)$ is a scalar function. Contracting both sides with $g^{\mu \nu}$ gives 
\begin{equation}
c(x) = \frac{2}{d}\, \partial_d\, \epsilon^d\,,
\end{equation}
for generic $d$ spacetime dimensions. Equation (\ref{eq:conformalkillingeq}) is the \textit{conformal} Killing equation and it is possible to classify all its solutions (Killing vectors). The most generic Killing vector reads:
\begin{equation}
\epsilon^\mu(x)=\underbrace{a_\mu}_{\text {translation }}+\underbrace{\omega_{\mu \nu} x^\nu}_{\text {rotation }}+\underbrace{\lambda x_\mu}_{\text {dilatation }}+\underbrace{b_\mu x^2-2 x_\mu b_\nu x^\nu}_{\text {special conformal}},\qquad \omega_{\mu \nu}=-\omega_{\nu \mu} \,,
\end{equation}
where a \textit{special conformal transformation} (SCT) is nothing but a translation, preceded and followed by an inversion $ x^\mu \rightarrow x^\mu / x^2$.
The total number of parameters defining conformal transformations is $\frac{1}{2} (d+2)(d+1)$, so long as $d \neq 2$.\\
The infinitesimal conformal transformations can be exponentiated to give finite transformations. In particular, for every Killing vector, we can define a generator $Q_{\epsilon}$ associated with it:
\begin{subequations} \label{eq:conformalgenerators}
\begin{align}
\text {translation:} \qquad \qquad & P_\mu=-i \partial_\mu\,, \\
\text {dilation:} \qquad \qquad & D=-i x^\mu \partial_\mu\,, \\
\text {rotation:} \qquad \qquad & L_{\mu \nu}=i\left(x_\mu \partial_\nu-x_\nu \partial_\mu\right)\,, \\
\mathrm{SCT}: \qquad \qquad & K_\mu=-i\left(2 x_\mu x^\nu \partial_\nu-x^2 \partial_\mu\right)\,.
\end{align}
\end{subequations}
We can now derive the commutation relations of the generators from those of the Killing vectors:
\begin{equation} \label{eq:Killinggen}
\left[Q_{\epsilon_1}, Q_{\epsilon_2}\right]=Q_{-\left[\epsilon_1, \epsilon_2\right]}\,,
\end{equation}
where $\left[\epsilon_1,\epsilon_2\right]$ is a commutator of vector fields\footnote{The minus sign in (\ref{eq:Killinggen}) comes from the fact that when the generators $Q_i$ are represented by differential operators $\mathcal{D}_i$, repeated action reverses the order $\left[Q_1,\left[Q_2, \mathcal{O}\right]\right]=\mathcal{D}_2 \mathcal{D}_1 \mathcal{O}$.}.
We then obtain combining (\ref{eq:Killinggen}) with (\ref{eq:conformalgenerators}):
\begin{subequations}
\begin{align}
{\left[D, P_\mu\right] } & =i P_\mu\,, \\
{\left[D, K_\mu\right] } & =-i K_\mu\,, \\
{\left[K_\mu, P_v\right] } & =2 i\left(\eta_{\mu \nu} D-L_{\mu \nu}\right)\,, \\
{\left[K_\rho, L_{\mu \nu}\right] } & =i\left(\eta_{\rho \mu} K_v-\eta_{\rho v} K_\mu\right) \,,\\
{\left[P_\rho, L_{\mu \nu}\right] } & =i\left(\eta_{\rho \mu} P_v-\eta_{\rho v} P_\mu\right)\,, \\
{\left[L_{\mu v}, L_{\rho \sigma}\right] } & =i\left(\eta_{v \rho} L_{\mu \sigma}+\eta_{\mu \sigma} L_{v \rho}-\eta_{\mu \rho} L_{v \sigma}-\eta_{\nu \sigma} L_{\mu \rho}\right)\,.
\end{align}
\end{subequations}
We can rewrite these relations in a more elegant form by redefining the generators:
\begin{equation} \label{eq:diff1}
J_{d+1,\mu}= \frac{P_{\mu}-K_{\mu}}{2} \, , \quad J_{d+2,\mu} = \frac{P_{\mu}+K_{\mu}}{2}\,, \quad J_{\mu \nu} = L_{\mu \nu}\,, \quad J_{d+1,d+2}=D\,.
\end{equation}
If we now introduce the metric $\eta_{ab}=\text{diag}(-1,1,1\dots,1)$ in $\mathbb{R}^{d+1,1}$, then these new generators obey $SO(d+1,1)$ commutation relations:
\begin{equation}
[J_{ab},J_{cd}]= i(\eta_{ad}J_{bc}+\eta_{bc}J_{ad}-\eta_{ac}J_{bd}-\eta_{bd}J_{ac})\,.
\end{equation}
This is how we observe the isomorphism between the conformal group in $d$ dimensions and the group $SO(d+1,1)$ with $\frac{1}{2} (d+2)(d+1)$ parameters. It is also the starting point of the \textit{embedding space formalism}, which rethink the action in terms of $\mathbb{R}^{d+1,1}$ instead of $\mathbb{R}^d$,  providing a simple and powerful way to understand the constraints of conformal invariance \cite{Dirac:1936fq,Mack:1969rr,Boulware:1970ty,Ferrara:1973eg,Weinberg:2010fx,Costa:2011mg,Costa:2011dw,Costa:2014rya,Iliesiu:2015qra,Cuomo:2017wme}.\par
To conclude this section, we construct conformal invariants or \textit{conformal cross-ratios}, i.e. functions of $n$ points that are unchanged by all types of conformal transformations. Translations and rotations imply that cross-ratios can depend only on the distances between a pair of distinct points:
\begin{equation}
x_{ij}= |x_i-x_j|\,.
\end{equation}
Scale invariance implies that only ratios of such distances appear. Finally, applying a special conformal transformation, we observe that it is impossible to construct an invariant with only 2 or 3 points. The simplest cross-ratios are the following functions of four-points:
\begin{equation} \label{eq:crossratios4pt}
u=\frac{x_{12}^2\,x_{34}^2}{x_{13}^2\,x_{24}^2}\,,\qquad v=\frac{x_{23}^2\,x_{14}^2}{x_{13}^2\,x_{24}^2} \,.
\end{equation}

The reason behind having exactly two independent cross-ratios with four-points can be understood by following these steps:
\begin{itemize}
\itemsep0em 
\item using translations, we move the point $x_1$ to zero,
\item using special conformal transformations, we move the point $x_4$ to infinity,
\item using rotations and dilatations, we move $x_3$ to $(1,0,\dots,0)$,
\item using again rotations, we can move $x_2$ to $(x,y,0,\dots,0)$.
\end{itemize}

\begin{figure}[h]
\centering 
\includegraphics[scale=0.33]{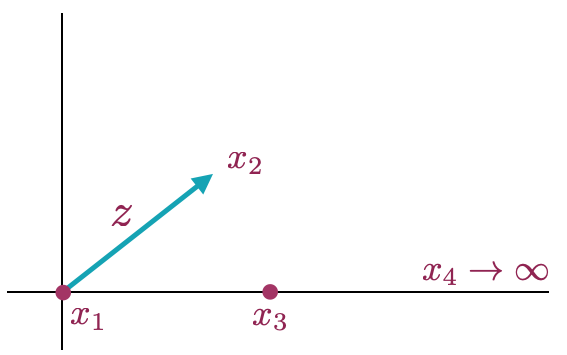}
\caption{\textit{We can use conformal transformations to move four points on a plane in the above configuration. }}
\end{figure}
At the end of this procedure, which we refer to as taking the \textit{conformal limit}, only two quantities $x,y$ are undetermined, precisely the two independent conformal invariants. We can then evaluate $u$ and $v$ in the conformal limit, obtaining
\begin{equation}
u=z\,\bar{z} \,, \qquad v=(1-z)(1-\bar{z})\,,
\end{equation}
with $z=x+iy$.
The number of independent cross-ratios in a $d$-dimensional spacetime is given by
\begin{equation}
\begin{split}
& n\, <\, d+1 : \,\,\frac{1}{2} \, n(n-3),\\
& n\, \geq\, d+1 : \,\, nd-\frac{1}{2} (d+1)(d+2)\,.
\end{split}
\end{equation}

\section{Representations of the Conformal Algebra} \label{sec:conformalalgebra}

In conformal field theory, as in any quantum field theory,  physical operators transform in representations of the global symmetries, which in this case is the conformal symmetry. To understand how fields are affected by conformal transformations, we can use the same trick as for the smaller Poincaré algebra, where the subgroup of Poincaré that leaves the point $x=0$ invariant, the Lorentz group, is studied. In the case of the conformal group, this subgroup is represented by the rotations, the dilatations and the special conformal transformations. Therefore the operators at the origin must transform as irreducible representations of $L_{\mu \nu}$, $D$ and $K_{\mu}$. We assume that they transform as
\begin{equation}
[D, \mathcal{O}(0)]= \Delta \mathcal{O}(0)\,, \qquad [L_{\mu \nu}, \mathcal{O}(0)]=(S_{\mu \nu})_b^{\,a} \,\mathcal{O}^b(0)\,,
\end{equation}
where $\Delta$ is the eigenvalue of the dilatation operator, and it is referred to as \textit{scaling dimension} of $\mathcal{O}$.  Regarding rotations, the operators transform in a representation of $SO(d)$. Note that in the following, we suppress the indices $a,b$, appearing in the second equation,  to simplify the notation. Acting now on the operator $\mathcal{O}(0)$ with $P_{\mu}/K_{\mu}$ leads to a higher/lower scaling dimension, allowing us to form the \textit{conformal multiplet}, our irreducible representation.  

In physical interesting theories, the spectrum of the dilatation operator is real and bounded from below, the so-called \textit{unitarity bound}\footnote{In our case of interest, namely for scalars, the unitarity bound is satisfied for $\Delta\geq \frac{d}{2} -1$ or $\Delta=0$.}, so that the conformal multiplet must contain an operator of the lowest dimension. We take then, without loss of generality,  $\mathcal{O}(0)$ to be the operator with the lowest scaling dimension in its conformal multiplet:
\begin{equation} \label{eq:conformalprimop}
[K_{\mu}, \mathcal{O}(0)]=0\,.
\end{equation}
Such operators are called \textit{primary operators}. They are characterized by their scaling dimension $\Delta$, and they are the highest weight representation of the conformal group.  All the other operators in the multiplet, the \textit{descendants},  are obtained from the primaries by acting with $P_{\mu}$, meaning they are simply their derivatives. In general, we refer to these multiplets as \textit{long multiplets}. However, when the unitarity bound is saturated, the conformal representation becomes reducible and contains a sub-representation of states with zero norm, known as the \textit{null states}.  Quotienting out these states in a consistent way leads us to a final irreducible representation containing fewer states than the initial multiplet. Hence, we refer to this as \textit{short multiplet}. The most notable short multiplets are
\begin{subequations}
\begin{align}
\text {free scalar multiplet:} \qquad \qquad &  {[P^2, \mathcal{O}(0)]}  =0 \,, \\
\text {conserved current multiplet:} \qquad \qquad & {[P_\mu, J^\mu(0)]} =0 \,,\\
\text {stress-tensor multiplet:} \qquad \qquad & {[P_\mu, T^{\mu \nu}(0)]}=0 \,.
\end{align}
\end{subequations}
We come back to representation theory in the context of supersymmetry in Section \ref{sec:superconformalalgebra}.

We end this section by deriving the action of the conformal generators on an operator at an arbitrary location. Therefore, we need to translate the operator inserted at the origin with $P_{\mu}$:
\begin{equation}
\mathcal{O}(x)= e^{x^{\mu}P_{\mu}} \, \mathcal{O}(0)\, e^{-x^{\mu}P_{\mu}} \,,
\end{equation}
which we then explicitly calculate using the Hausdorff formula:
\begin{equation}
e^{-A} \, B \, e^A= B+ [B,A]+\frac{1}{2!}[[B,A],A]+\frac{1}{3!} [[[B,A],A],A]+\dots\,,
\end{equation}
to get
\begin{subequations} \label{eq:diff2}
\begin{align}
[P_{\mu},\mathcal{O}(x)]&=\partial_{\mu}\mathcal{O}(x)\,,\\
[D,\mathcal{O}(x)]&=(x^{\mu} \partial_{\mu}+\Delta) \,\mathcal{O}(x)\,,\\
[L_{\mu \nu},\mathcal{O}(x)]&=(x_{\nu} \partial_{\mu}-x_{\mu} \partial_{\nu} +S_{\mu \nu}) \,\mathcal{O}(x)\,,\\
[K_{\mu},\mathcal{O}(x)]&=(2x_{\mu}(x\cdot \partial)-x^2 \partial_{\mu} + 2\Delta x_{\mu} - 2x^{\nu} S_{\mu \nu})\, \mathcal{O}(x)\,.
\end{align}
\end{subequations}

\section{Correlation Functions and the CFT Data} \label{sec:corrfunctionsandCFTdata}

After having discussed the conformal group with its generators and representations, we can now introduce the main observables in conformal field theories: correlation functions of local operators.  We then enter into detail about what we intend for CFT data, how we define the OPE for a conformal field theory and how we can use conformal symmetry to constrain our correlation functions.

\subsection{Correlation Functions} \label{sec:correlationfunctions}

We start by exploring correlators of scalar operators since the whole thesis revolves around them.  As mentioned above, the conformal operators transform in rank-$s$ symmetric-traceless representations of the rotation group $SO(d)$. Scalar operators are then the zero-rank case. To read about more general representations, we advise \cite{Costa:2011mg,Costa:2011dw,Costa:2014rya,Iliesiu:2015qra}.

Let us analyze then the constraints that conformal symmetry imposes on the scalar correlators of primary operators, starting from the simplest case of one-point functions. By Lorentz invariance, only scalars can acquire one-point functions. Translational invariance implies that they have to be constant, while scale invariance fixes this constant to zero unless the operator has $\Delta=0$, meaning it is the identity.  It is convenient to normalize the path integral such that $\langle \mathbb{1} \rangle =1$.

Moving to two- and three-point functions, we observe that conformal symmetry is so constraining that they are completely fixed up to a multiplicative constant.  In particular,
\begin{equation} \label{eq:conformaltwopoint}
\langle \mathcal{O}(x_1)\mathcal{O}(x_2) \rangle = 
\left\{ \begin{aligned} 
\frac{c_{\Om_1\Om_2}}{x_{12}^{\,\,2\Delta}},  \qquad & \text{if}\,\Delta_1=\Delta_2 \equiv \Delta,\\
0,\,\,\,\,\qquad & \text{otherwise} \,.
\end{aligned} \right.
\end{equation}
where normally the constant $c_{\Om_1\Om_2}$ is reabsorbed in the definition of the field in theories without defects.\\
Similarly, three-point functions are constrained by conformal symmetry to take the form:
\begin{equation}\label{eq:conformalthreepoint}
\langle \mathcal{O}(x_1) \mathcal{O}(x_2) \mathcal{O}(x_3) \rangle = \frac{c_{\Om_1\Om_2\Om_3}}{(x_{12}^2)^{\Delta_{123}}(x_{23}^2)^{\Delta_{231}} (x_{13}^2)^{\Delta_{132}}}\,,
\end{equation}
with $\Delta_{ijk}=\Delta_i+\Delta_j-\Delta_k$. The constant $c_{\Om_1\Om_2\Om_3}$ is often called \textit{three-point coupling}, \textit{structure constant} of the operator algebra or \textit{OPE coefficient}, since it has a fundamental role in the operator product expansion, as we observe in the next section. 

Conformal symmetry is remarkably powerful, as it constrains the form of two- and three- point functions, but it is insufficient to fix the form of four-point functions. Nevertheless, we can still infer the following form for identical scalars of scaling dimension $\Delta$:
\begin{equation} \label{eq:genericfourpoint}
\langle \mathcal{O}(x_1) \mathcal{O}(x_2) \mathcal{O}(x_3) \mathcal{O}(x_4) \rangle = \frac{1}{x_{12}^{2\Delta}\, x_{34}^{2\Delta}}\, \mathcal{A}(u,v)\,,
\end{equation}
where $u$ and $v$ are the cross-ratios defined in (\ref{eq:crossratios4pt}).

We could go on and consider correlation functions with a higher number of points. This topic is actually central to this thesis, and we study it in the next part.  Though, it is worth noticing here that in principle, higher-point functions can be reconstructed from the three-point functions and the scaling dimensions, what it is commonly referred to as \textit{CFT data}.  This implies that the full knowledge of the CFT data is sufficient to determine completely the theory. Unfortunately the set of CFT data is infinite. It is true that there are some consistency checks that we could use as input to determine the CFT data, as we see in the following. However, it is quite hard to solve these constraints. That is why we study higher-point correlators. As mentioned in the introduction,  they contain information about an infinite number of lower-point functions by means of the OPE, that we now introduce.

\subsection{Operator Product Expansion} \label{sec:OPE}

The \textit{operator product expansion} or \textit{OPE} is a general tool of quantum field theory, which consists of approximating the product of two local operators with an infinite sum of local operators:
\begin{equation} \label{eq:OPE1}
\mathcal{O}(x_1) \mathcal{O}(x_2) \overset{x_1\rightarrow x_2}{=} \sum_{\Om_k} f_{12k} (x_{12}) \, \mathcal{O}_k\,,
\end{equation}
where the sum runs over all the conformal primaries $\mathcal{O}_k$.  Since all operators can be constructed from primaries, we can rewrite this expression in the following way:
\begin{equation} \label{eq:OPE2}
\mathcal{O}(x_1) \mathcal{O}(x_2)  \overset{x_1\rightarrow x_2}{=}  \sum_{\Om_k} c_{\Om_1\Om_2\Om_k} C_{12k} (x_{12},\partial_2) \, \mathcal{O}_k\,,
\end{equation}
where $c_{\Om_1\Om_2\Om_k}$ are the coefficient of the three-point functions (\ref{eq:conformalthreepoint}) and the operator $C_{12k} (x_{12},\partial_2)$ encodes the contributions from the descendants and it is completely fixed by conformal symmetry in terms of the operators scaling dimensions $\Delta_k$. For this reason and considering that the OPE has a finite radius of convergence \cite{Dolan:2003hv, Pappadopulo:2012jk},  it is more powerful in CFT than in regular QFT,  being there just an asymptotic expansion.

Coming back to higher-point functions, they can be then expressed as infinite sums of products of lower-point functions by repetitively applying the OPE. In this sense, no new data is contained in higher-point functions: all dynamical information is encompassed by three-point functions, which determine, together with scaling dimensions, higher-point correlation functions.

\begin{figure}[h]
\centering
\includegraphics[scale=0.35]{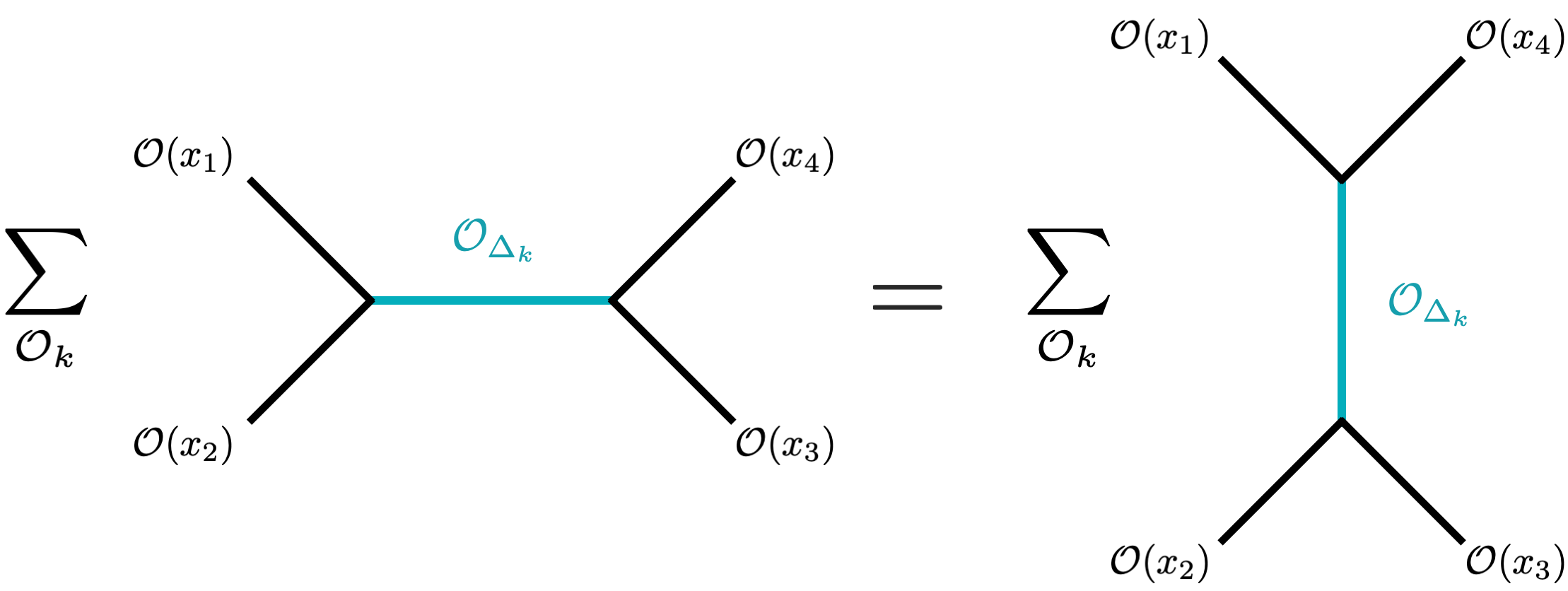}
\caption{\textit{Graphical representation of the OPE associativity spelled out in (\ref{eq:OPEassociativity}).}}
\label{fig:OPEassociativity}
\end{figure}

If we now think about applying the OPE to a correlator, we realize that there are many different options since it can affect pairs of operators in different orders. However, every combination should be equivalent, meaning that the correlator cannot depend on the order the OPE is taken.  Let us consider the canonical example of a four-point function. It can be decomposed in the following ways, which must be equal by associativity:
\begin{equation} \label{eq:OPEassociativity}
\langle \wick{ \mathcal{O} \c1{(x_1)} \mathcal{O} \c1{(x_2)} \mathcal{O} \c2{(x_3)} \mathcal{O} \c2{(x_4)} } \rangle = \langle \wick{ \mathcal{O} \c1{(x_1)} \mathcal{O} \c2{(x_2)} \mathcal{O} \c2{(x_3)} \mathcal{O} \c1{(x_4)} } \rangle \,,
\end{equation}
where the square brackets stand for taking the OPE between those operators. We refer to the OPE limit on the LHS as \textit{s-channel} ($\chi \rightarrow 0$) and on the RHS as \textit{t-channel} ($\chi \rightarrow 1$). This relation is represented graphically in Figure \ref{fig:OPEassociativity}.  

This equation is an important consistency check.  It points to a fundamental property known as \textit{crossing symmetry}, the starting point of the so-called \textit{conformal bootstrap}. We enter into more detail in Section \ref{sec:bootstrap}. 

\subsection{Conformal Blocks} \label{sec:conformalblocks}

We can now effectively take the OPE between operators in a correlation function.  In particular, we consider again a generic four-point function, and we perform the OPE as in the LHS of (\ref{eq:OPEassociativity}). This result in
\begin{equation}
\langle \mathcal{O}(x_1)\mathcal{O}(x_2)\mathcal{O}(x_3) \mathcal{O}(x_4) \rangle = \sum_{\mathcal{O}_k} c_{\Om_1\Om_2\Om_k} c_{\mathcal{O}_3 \mathcal{O}_4 \mathcal{O}_k} C_{12k} (x_{12},\partial_2) C_{34k} (x_{34},\partial_4) \,  \langle \mathcal{O}_k (x_2) \mathcal{O}_k (x_4) \rangle \,.
\end{equation}
If we compare this equation to the expression of the four-point (\ref{eq:genericfourpoint}), we find
\begin{equation} \label{eq:conformalblocks4pt}
\mathcal{A}(u,v)= \sum_{\mathcal{O}_k}  c_{\Om_1\Om_2\Om_k} c_{\mathcal{O}_3 \mathcal{O}_4 \mathcal{O}_k}  \,g^{12,34}_{\Delta_k} (u,v)\,,
\end{equation}
where the functions $g_{\Delta_k}(u,v)$ are called \textit{conformal blocks} and allow us to obtain a representation of the otherwise arbitrary function $\mathcal{A}(u,v)$ in terms of the CFT data. For $d>1$ conformal blocks also feature an implicit sum over the spins and depend on the spin representation.

These functions have been widely studied by Dolan and Osborn \cite{Dolan:2000ut,Dolan:2003hv,Dolan:2011dv}, which observed that acting with the Casimir\footnote{The Casimir of a Lie algebra $\mathfrak{g}$ is the operator $\mathcal{C}$ that commutes with all the elements of the algebra: $[\mathcal{C},x]=0\,,\quad \forall x \in \mathfrak{g}$.} operator on a four-point function allows to select the contribution from a single conformal family to the OPE. This implies that conformal blocks are solutions to a differential equation of the form
\begin{equation}
\left(	\mathcal{C}_2^{(1+2)}-c_2 \right)\, \langle \mathcal{O}(x_1)\mathcal{O}(x_2)\mathcal{O}(x_3) \mathcal{O}(x_4) \rangle = 0\,,
\end{equation}
where $\mathcal{C}_2^{(1+2)}$ is the \textit{quadratic Casimir} acting of the points $x_1$ and $x_2$ defined by
\begin{equation}
\mathcal{C}_2^{(1+2)}=-\frac{1}{2}\sum_{A,B}\left(L_{AB}^{(1+2)}\right)^2\,, \qquad \qquad L_{AB}^{(1+2)}=L^{(1)}_{AB}+L^{(2)}_{AB}\,,
\end{equation}
with the differential operators being the generators introduced in (\ref{eq:diff1}) and $c_2$ being the eigenvalue of the Casimir operator when acting on an operator with scaling dimension $\Delta$ in $d$ dimensions, i.e. $c_2=\Delta(\Delta-d)$.

Conformal blocks are known in closed-form expressions in even dimensions (up to six dimensions),  where they can be written as hypergeometric functions. However, they are much more complicated in odd dimensions, where a closed form has yet to be found.  We discuss them more in detail for our cases of interest in Section \ref{sec:conformalblockexpansion}.

\subsection{Conformal Bootstrap} \label{sec:bootstrap}

To conclude this section on conformal field theory,  we combine the ingredients we have just introduced to explain the philosophy behind conformal bootstrap.  \textit{Conformal bootstrap} is a set of non-perturbative techniques to characterize, constrain and solve CFTs using physical consistency conditions like symmetry, unitarity and causality. 

In the previous section, we showed that we can expand a four-point function in conformal blocks:
\begin{equation} \label{eq:conformalblocks4pt}
\mathcal{A}(u,v)= \sum_{\mathcal{O}_k}  c_{\Om_1\Om_2\Om_k} c_{\Om_3\Om_4\Om_k}  \,g_{\Delta_k} (u,v)\,.
\end{equation}
This equation is quite powerful for two reasons. If we know the correlator explicitly, then we can use the expansion to extract infinitely many OPE coefficients $c_{\Om_i\Om_j\Om_k}$.  Conversely, if we know the OPE coefficients, then the expansion provides a way to reconstruct the correlation function.

In this thesis, we mostly encounter computations of the first type, meaning we compute some correlators using perturbation theory and extract some CFT data from those. The bootstrap method approaches the problem of solving a CFT somewhat differently.  Even when the four-point function $\mathcal{A}(u,v)$ and the CFT data are not known, equation (\ref{eq:conformalblocks4pt}) can lead to powerful constraints on the CFT data, allowing even to reconstruct the correlator in specific cases. This is possible thanks to crossing symmetry, which we introduced in Section \ref{sec:OPE}.  

As a matter of fact, when we combine the conformal blocks decomposition with crossing symmetry, we obtain these \textit{crossing equations}:
\begin{equation}
\sum_{\Om_k} \left( c_{\Om_1 \Om_2 \Om_k} c_{\Om_3\Om_4\Om_k} \, v^{\frac{\Delta_2+\Delta_3}{2}} \, g_{\Delta_k}^{12,34}(u,v) \right) = \sum_{\Om_k} \left(c_{\Om_2\Om_3\Om_k} c_{\Om_1\Om_4\Om_k} \, u^{\frac{\Delta_1+\Delta_2}{2}} \, g_{\Delta_k}^{23,14}(v,u) \right) \,,
\end{equation}
which are highly non-trivial consistency conditions that any set of CFT data must satisfy.  The goal of the bootstrap program is to solve these equations, but unfortunately, this is everything but trivial.  First of all, because there are infinitely many equations to solve to fully determine the infinite set of CFT data, and moreover, the conformal dimensions are real numbers. An additional complication is that one block on the LHS is not directly mapped to a block on the RHS; quite the opposite. It corresponds indeed to infinitely many blocks on the RHS.

\begin{figure}[h]
\centering
\includegraphics[scale=0.45]{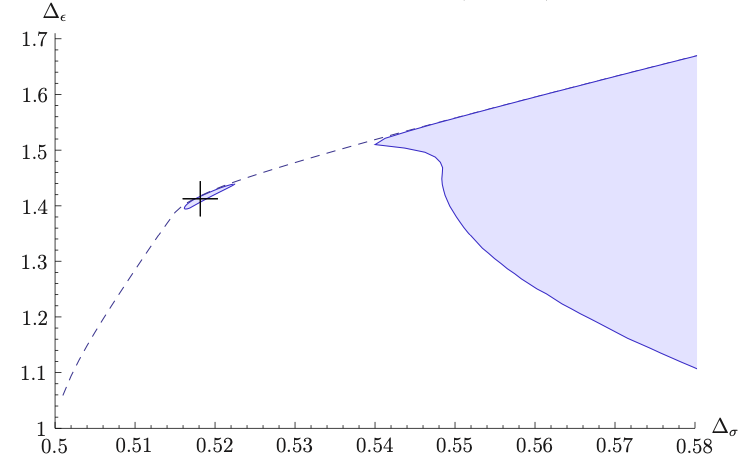}
\caption{\textit{This picture is taken from \cite{Kos:2014bka}, where the 3$d$ Ising model was studied. It depicts the allowed regions for the scaling dimensions ($\Delta_\sigma,\Delta_\epsilon$) of the two operators $\sigma$ and $\epsilon$ in the space of possible CFT data.  These bounds are obtained using the crossing symmetry and unitarity of three different correlators made up of the two operators. These constraints yield a small region in the CFT data space compatible with the known values in the 3$d$ Ising CFT that are indicated with black crosshairs.}}
\label{bootstrapisland}
\end{figure}

However,  some progress has been made in solving these equations numerically,  which allowed to at least carve out the space of legit CFT data from the space of all possible parameters (see Figure \ref{bootstrapisland}). The ultimate hope is that solving these equations for a sufficiently high number of correlators shrinks the space of parameters to a point, allowing the theory to be solved.  For further information, especially recent progress, see e.g.  \cite{Poland:2016chs,Poland:2018epd,Poland:2022qrs}.

Finally, there has been a comprehensive exploration of analytical methods to solve these constraints. For an overview of the advancement in this topic, check \cite{Hartman:2022zik}.

\chapter{Superconformal Field Theory} \label{ch:susyen=4}

In this chapter, we review another important symmetry which plays a central role in modern quantum field theory: \textit{supersymmetry} or SUSY.  In particular, we first consider supersymmetry alone, with a particular focus on the four-dimensional case,  and then combine it with conformal symmetry. In the context of superconformal symmetry, we discuss the famous case of $\mathcal{N}=4$ super Yang-Mills, the simplest interacting superconformal QFT in 4$d$. We then explore various facets of this theory: the action, the planar limit,  and the Feynman rules.

\section{Supersymmetry} \label{sec:SUSY}

Supersymmetry relates the two fundamental classes of particles: bosons (integer spin) and fermions (half-integer spin).  In this theory, each particle in one class has an associated \textit{superpartner} in the other class.
For a detailed introduction to this topic, see e.g. \cite{Weinberg:2000cr,Terning:2006bq}.

Supersymmetry represents a way to evade the no-go theorem introduced by Coleman and Mandula in 1967 \cite{Coleman:1967ad}, which states that it is only possible to combine spacetime and internal symmetries trivially. This theorem is valid under a list of reasonable assumptions, and supersymmetry relaxes one of these assumptions, allowing the presence of fermionic generators together with bosonic ones.

Thanks to these additional elements, this extension of the Poincaré algebra is promoted to a \textit{graded-Lie algebra}\footnote{Note that, despite the name, this is not a Lie algebra because the fermionic generators break the antisymmetry property.} or a \textit{superalgebra}.   Therefore, in addition to the usual Poincaré generators (translations, boosts, and rotations), the SUSY generators include complex, anticommuting spinors $Q$ and their conjugates $\bar{Q}$:
\begin{equation} \label{eq:anticomQ}
\{Q_{\alpha},Q_{\beta}\}=\{\bar{Q}_{\dot{\alpha}},\bar{Q}_{\dot{\beta}}\}=0\,,
\end{equation}
where $\alpha,\beta =1,2$ are the indices usually used for the spinors and $\dot{\alpha}, \dot{\beta}=1,2$ for the conjugates.
The non-trivial extension of the Poincaré symmetry arises because the anticommutator of $Q$ and $\bar{Q}$ gives a translation generator $P_{\mu}$:
\begin{equation}
\{Q_{\alpha},\bar{Q}^{\dot{\alpha}}\}=2 \sigma^{\mu}_{\alpha\dot{\alpha}} P_{\mu}\,,
\end{equation}
with
\begin{equation}
\sigma_{\alpha \dot{\alpha}}^\mu=\left(\mathbb{1}, \sigma^i\right), \quad \bar{\sigma}^{\mu \dot{\alpha} \alpha}=\left(\mathbb{1},-\sigma^i\right)
\end{equation}
and $\sigma^i$ being the usual three-dimensional Pauli matrices. Moreover, we have
\begin{subequations}
\begin{align}
[Q_\alpha, L^{\mu \nu}] & = \left(\sigma^{\mu \nu}\right)_{\alpha}^{\,\beta} Q_{\beta} \,,\\
[\bar{Q}_{\dot{\alpha}}, L^{\mu \nu}] & = \epsilon_{\dot{\alpha} \dot{\beta}} \left(\bar{\sigma}^{\mu \nu} \right)^{\dot{\beta}}_{\,\dot{\gamma}} \bar{Q}^{\dot{\gamma}} \,,
\end{align}
\end{subequations}
with the generators of Lorentz transformations, $\sigma^{\mu \nu}$ and $\bar{\sigma}^{\mu \nu}$, defined by
\begin{subequations}
\begin{align}
\left(\sigma^{\mu \nu}\right)_\alpha{ }^\beta & =\frac{i}{4}\left(\sigma^\mu{ }_{\alpha \dot{\alpha}} \bar{\sigma}^{v \dot{\alpha} \beta}-\sigma^v{ }_{\alpha \dot{\alpha}} \bar{\sigma}^{\mu \dot{\alpha} \beta}\right), \\
\left(\bar{\sigma}^{\mu \nu}\right)^{\dot{\alpha}}{ }_{\dot{\beta}} & =\frac{i}{4}\left(\bar{\sigma}^{\mu \dot{\alpha} \alpha} \sigma^v{ }_{\alpha \dot{\beta}}-\bar{\sigma}^{v \dot{\alpha} \alpha} \sigma^\mu{ }_{\alpha \dot{\beta}}\right) .
\end{align}
\end{subequations}
All the other (anti)commutation relations vanish, meaning, for example, that the SUSY generators commute with translations:
\begin{equation}
[P_{\mu},Q_{\alpha}]=[P_{\mu},\bar{Q}_{\dot{\alpha}}]=0\,.
\end{equation} 

There is an additional global symmetry called \textit{R-symmetry} which makes $Q_\alpha$ invariant under multiplication by a phase:
\begin{equation}
Q_\alpha \rightarrow e^{i p} Q_\alpha\,, \quad \bar{Q}_{\dot{\alpha}} \rightarrow e^{-i p} \bar{Q}_{\dot{\alpha}}\,,
\end{equation}
with a global parameter $p$, so that in general there is one linear combination of $U(1)$ charges, the \textit{R-charges}, that does not commute with $Q$ and $\bar{Q}$:
\begin{equation} \label{eq:U1Rsymm}
[Q_\alpha,R]=Q_\alpha\,, \qquad [\bar{Q}_{\dot{\alpha}},R]=-\bar{Q}_{\dot{\alpha}}\,.
\end{equation}
The corresponding $R$-symmetry group is called $U(1)_R$.

\subsection{Extended Supersymmetry}

So far, we have considered the case where there is only one spinor \textit{supercharge},  but the superalgebra can be easily extended to accommodate $\mathcal{N}$ supercharges. After having introduced a new index to keep track of this, our extended superalgebra then becomes:
\begin{subequations}
\begin{align}
\{Q^A_\alpha, \bar{Q}^{\dot{\alpha}\,B}\} & = 2\sigma^\mu_{\alpha \dot{\alpha}} P_\mu \delta^A_B \,,\\
[Q^A_\alpha, L^{\mu \nu}] & = \left(\sigma^{\mu \nu}\right)_{\alpha}^{\,\beta} Q^A_{\beta} \,,\\
[\bar{Q}^A_{\dot{\alpha}}, L^{\mu \nu}] & = \epsilon_{\dot{\alpha} \dot{\beta}} \left(\bar{\sigma}^{\mu \nu} \right)^{\dot{\beta}}_{\,\dot{\gamma}} \bar{Q}^{A\,\dot{\gamma}} \,,
\end{align}
\end{subequations}
where $A,B=1,\dots,\mathcal{N}$.  This allows for a generalization of the symmetry of equation (\ref{eq:U1Rsymm}) to a $U(\mathcal{N})_R$ symmetry, which rotates the supercharges.

The superalgebra can be further extended to include \textit{central charges} $Z^{AB}$ and $\bar{Z}^{AB}$, such that equations (\ref{eq:anticomQ}) turn into
\begin{equation}
\{Q^A_\alpha, Q^B_\beta\} = \epsilon_{\alpha \beta} Z^{AB}\,,\qquad \{\bar{Q}_{\dot{\alpha}\,A},\bar{Q}_{\dot{\beta}\,B}\}  = \epsilon_{\dot{\alpha} \dot{\beta}}\bar{Z}_{AB}\,,
\end{equation}
where the central charges are antisymmetric with respect to $A$ and $B$.

\subsection{SUSY Representations} \label{sec:SUSYrep}

In general, single particles states fall into irreducible representations of the SUSY algebra called \textit{supermultiplets}.
Since $Q$ is a spinor, it produces a fermionic state when it acts on a bosonic state. The supermultiplets contain then both fermions and bosons. Actually, it can be shown that each supermultiplet with non-zero energy contains the same number of bosonic and fermionic degrees of freedom. 

Since $P_{\mu} P^{\mu}$ commutes with $Q$ and $\bar{Q}$, all the particles in the same supermultiplet have the same mass, and for an analogous argument, they also have to be in the same representation of the gauge group.

We now focus on massless representations of the SUSY algebra, since massive ones are not relevant for this work. Massless states are labeled by their momentum $p_\mu$ and helicity $\lambda$. If we choose the frame where $p_\mu = (E, 0,0,E)$, then the SUSY algebra reduces to
\begin{subequations}
\begin{align}
\{Q_1, \bar{Q}_1 \} & = 4 E\,, \\
\{Q_2, \bar{Q}_2\} & = 0 \,, \\
\{Q_\alpha, Q_\beta\} & =0\,, \\
\{\bar{Q}_{\dot{\alpha}}, \bar{Q}_{\dot{\beta}}\} & =0\,.
\end{align}
\end{subequations}
Therefore we have an algebra with only one raising operator that we use to define creation and annihilation operators as follows
\begin{equation} \label{eq:cr/ann_op}
a^A \equiv \frac{Q^A_1}{2\sqrt E}\,, \qquad a^\dag_A \equiv \frac{Q_{A\, \dot{1}}}{2 \sqrt E} \,, \qquad \text{with}\,\, \{a^A, a^\dag_B\}=\delta^A_{\,B}\,.
\end{equation}
To construct a multiplet, we, therefore, first pick a vacuum of a fixed helicity, which satisfies
\begin{equation}
Q_1^A|\Omega\rangle= 0\,,
\end{equation}
and we then act with the creation/annihilation operators. $Q_1^A$ lowers the helicity by $1 / 2$, i.e. $Q_1^A|p, \lambda\rangle$ has helicity $\lambda-1 / 2$, while $Q_{A \dot{1}}$ raises the helicity by $1 / 2$.

\section{$\mathcal{N}=4$ supersymmetric Yang-Mills} \label{sec:N=4SYM}

Now that we are acquainted with the concepts of supersymmetry, superalgebra and their extensions, we can introduce one of the main players of this work: $\mathcal{N}=4$ supersymmetric Yang-Mills theory (SYM). This theory is a realization of \textit{maximally extended} supersymmetry, as we explore in the next section. 

Historically the interest in this theory arose due to its finiteness property: its $\beta$ function has been proven to vanish in perturbation theory, and the same is supposed to be true at the non-perturbative level.  The theory, in fact, contains only massless particles, and conformal symmetry is preserved at the quantum level.

There was also a renewed interest due to the AdS/CFT correspondence proposal by Maldacena \cite{Maldacena:1997re}, which relates type IIB supergravity in $d+1$ dimensional Anti-de-Sitter space and $d$ dimensional (super)conformal theories. $\mathcal{N}=4$ SYM is also believed to be integrable in the planar limit ($N \rightarrow \infty$), which we discuss in Section \ref{sec:planarlimit}.  Finally, the theory possesses exact electric-magnetic duality due to its invariance under $S$-duality group $SL(2,\mathbb{Z})$ \cite{Montonen:1977sn}. 

Thanks to these numerous symmetries and interesting features, this theory is used as a laboratory to develop and improve methods to approach more complex theories and, therefore, it has been studied with a variety of different techniques, from perturbation theory to conformal bootstrap, from localization to integrability.

\subsection{Maximally Extended SUSY} \label{sec:maxextsusy}

As we have just briefly mentioned, $\mathcal{N}=4$ SYM has the maximal amount of supersymmetry possible for a supersymmetric theory in four dimensions. A greater number of supercharges would require fields of spin larger than one and, therefore, the inclusion of gravity.  

Using the $\mathcal{N}=4$ supersymmetry,  we can transform the gauge bosons into $\mathcal{N}=4$ different supersymmetric fermionic partners and vice versa. By applying the creation/annihilation operator (\ref{eq:cr/ann_op}) with $A=1,\dots,4$, we can explicitly construct the supermultiplet that contains one vector field $A_\mu$ (spin-1 gauge boson), four spinor fields $\psi_\alpha^a(x),\,a \in\{1,2,3,4\}$ (spin-1/2 Weyl fermions) and six scalar fields $\phi^I(x),\, I \in\{1,\ldots, 6\})$ (spin-0 bosons), all in the adjoint representation of the gauge group. $R$-symmetry allows the four spinors to transform in the fundamental of $SU(4)_R$, the six real scalars in the rank 2 antisymmetric representation, which is the fundamental of $SO(6)_R$.

\subsection{Superconformal Algebra} \label{sec:superconformalalgebra}

The superconformal algebra is an extension of the conformal algebra we reviewed in Sections~\ref{sec:conformalsymmetry} and \ref{sec:conformalalgebra}.   The superconformal algebra is unique given an $\mathcal{N}$-extended supersymmetry algebra and a $d$-dimensional conformal algebra. In our case of interest, i.e. $d=4$, we have 
\begin{equation}
\mathfrak{su}(2,2|\mathcal{N}) \supset \mathfrak{so}(5,1) \times \mathfrak{su}(\mathcal{N})_R \times \mathfrak{u}(1)_R\,.
\end{equation}
In particular, for $\mathcal{N}=4$ SYM, $\mathfrak{u}(1)_R$ commutes with the rest of the generators and can be quotiented out, leading to $\mathfrak{psu}(2,2|4)$.

To build the superalgebra, we could naively think that it is sufficient to add the supercharges $Q,\bar{Q}$ to the generators of the conformal group. However, to ensure the closure of the algebra, we need to add some fermionic supercharges $S_\alpha^A, \bar{S}_{A\,\alpha}$, leading to:
\begin{subequations}
\begin{align}
\{S^A_\alpha, \bar{S}_{\dot{\beta}\,B}\} & = 2 (\sigma^\mu)_{\alpha \dot{\beta}} K_{\mu} \delta^A_{\,B}\,,\\
\{Q^A_\alpha, S_{\beta\,B}\}  & = \epsilon_{\alpha \beta}\,(\delta^A_{\,B} D+ R^A_{\,B}) + \frac{1}{2}\delta^A_{\,B} L_{\mu \nu}(\sigma^{\mu \nu})_{\alpha \beta}\,.
\end{align}
\end{subequations}
The other numerous (anti)commutation relations are gathered in Appendix \ref{app:4dsuperconformalalgebra}.

As we did for the conformal group, we want to explore the representations of the superconformal algebra. We focus on local and gauge-invariant operators $\mathcal{O}(x)$, that are labeled by their scaling dimension $\Delta$ and by their spin $[l,\bar{l}]$. To construct a supermultiplet, we start from a \textit{superconformal primary} operator. Besides being a conformal primary (\ref{eq:conformalprimop}), it is annihilated by $S, \bar{S}$:
\begin{equation} \label{eq:superconfprimop}
[S^A_\alpha,\mathcal{O}(0)]=[\bar{S}_{A\,\dot{\alpha}}, \mathcal{O}(0)]=0\,, \qquad \forall A,\alpha\,,
\end{equation}
and forms a representation of the $R$-symmetry group. To produce the rest of the multiplet, the \textit{superdescendants} $\mathcal{O}^\prime$,  we just apply any product of generators of the superconformal algebra to the primaries:
\begin{equation}
\mathcal{O}^{\prime}: \qquad [Q,\mathcal{O}(0)]\,,\qquad [Q,[Q,\Op(0)]]\,, \qquad [Q,[\bar{Q}, \Op(0)]]\,,  \qquad [P,\Op(0)]\,,\quad \dots \,\,\,.
\end{equation}
Because the Poincaré supercharges are anticommuting spinors, a superconformal multiplet contains bosons and fermions, which are related to each other by supersymmetry. Furthermore, $Q$ and $\bar{Q}$ act at most $2^{N_Q+N_{\bar{Q}}}$ times, with $N_Q + N_{\bar{Q}}$ being the total number of supercharges. Finally, we choose a basis where the superdescendants are conformal primaries on their own. In other words, they form an irreducible representation of the conformal group.

It is fundamental to mention an important class of superconformal primaries, the \textit{chiral primaries}. These operators fulfill an additional condition besides (\ref{eq:superconfprimop}):
\begin{equation} \label{eq:BPScondition}
[Q^A_\alpha,\Op]=0\,,
\end{equation}
meaning they are annihilated by at least one of the supercharges. This is what we call \textit{BPS-condition} \cite{Bogomolny:1975de,Prasad:1975kr}. Consequently, these operators are protected, meaning their scaling dimension does not receive quantum corrections \cite{Intriligator:1999ff,Petkou:1999fv}.

As for the conformal case, we can distinguish again \textit{short} from \textit{long} supermultiplets. The difference between the two is analogous to what we have already outlined in Section~\ref{sec:conformalalgebra}.  Long supermultiplets are generated from superprimaries with generic scaling dimension above the unitarity bound, while short multiplets are generated from superprimaries that satisfy a shortening condition, namely, they are killed by combinations of Poincaré supercharges.

In this work, we focus on a particular kind of operators, since they are leading in the large $N$ limit, which we discuss in Section \ref{sec:planarlimit}. They are called \textit{single-trace} operators of scalar primaries $\phi$:
\begin{equation}
\Op_n(x)\equiv \tr \phi^{I_1} \dots \phi^{I_n}\,,
\end{equation}
where the trace acts on gauge group indices, and the index $I$ keeps track of the $SO(6)_R$ $R$-symmetry.  A subclass of these operators are \textit{$1/2$-BPS operators}, since they satisfy (\ref{eq:BPScondition}) for half of the supercharges. For this reason, these operators are protected, as their scaling dimension $\Delta=n$ does not receive quantum corrections.  They saturate the unitarity bound and belong, therefore, to a short multiplet. Finally, these operators are characterized by their quantum numbers $\Delta,\,s$ (spin) and the $R$-symmetry Dynkin labels $[0,n,0]$. 

It is worth noticing that there are also unprotected single-trace operators such as the \textit{Konishi operator}, which for $\mathcal{N}=4$ SYM reads
\begin{equation}
K=\tr(\phi^I \phi^I)\,.
\end{equation}
This operator does not satisfy the BPS condition.

Lastly, it is also possible to construct \textit{multi-trace operators}, which are products of single-trace operators. $1/4$- and $1/8$-BPS are realized by such operators, but we are not going to deal with these operators. For further details, see e.g. \cite{Ryzhov:2001bp,Gava:2006pu}.

\subsection{$\mathcal{N}=4$ SYM Action} \label{sec:N=4action}

Since this thesis is basically built upon perturbation theory, we need to introduce the starting point of this method: the action of the theory. There are two ways of deriving it. On the one hand, we can use the fact that $\mathcal{N}=4$ supersymmetric field theory is by default also $\mathcal{N}=1$ supersymmetric. We can then use the $\mathcal{N}=1$  \textit{superspace formalism}\footnote{The superspace formulation consists in adding some ``anticommuting'' dimensions alongside ordinary space dimensions. The latter parametrizes the bosonic degrees of freedom, while the anticommuting dimensions the fermionic ones.}, provided that the coupling constants and the superpotential of the $\mathcal{N}=1$ theory preserve certain constraints. On the other hand, we can obtain the $\mathcal{N}=4$ SYM Lagrangian from dimensional reduction of the $\mathcal{N}=1$ SYM theory in 10 dimensions. We do not enter into further detail, but both derivations can be found in e.g. \cite{Ammon:2015wua}.

Coming back to the action,  the quantization can be done using the conventional Faddeev-Popov method, which is standard knowledge and can be found explicitly in many QFT textbooks, e.g. \cite{Peskin:1995ev}. This procedure introduces ghost fields $c$ and leads us to the following action (after gauge fixing):
\begin{align} \label{eq:N=4action}
S &= \frac{1}{g^2} \int d^4 x\ \text{Tr} \left\lbrace \frac{1}{2} \tensor{F}{_{\mu\nu}} \tensor{F}{^{\mu\nu}} + \tensor{D}{_\mu} \tensor{\phi}{_I} \tensor{D}{^\mu} \tensor{\phi}{^I} - \frac{1}{2} [ \tensor{\phi}{_I} , \tensor{\phi}{_J} ] [ \tensor{\phi}{^I} , \tensor{\phi}{^J} ] \right. \notag \\
& \left. \qquad \qquad \qquad \qquad \qquad \qquad +\, i \bar{\psi} \tensor{\gamma}{^\mu} \tensor{D}{_\mu} \psi + \bar{\psi} \tensor{\Gamma}{^I} [ \tensor{\phi}{_I} , \psi ] + \tensor{\partial}{_\mu} \bar{c} \tensor{D}{^\mu} c + \xi \left( \tensor{\partial}{_\mu} \tensor{A}{^\mu} \right)^2 \right\rbrace\,,
\end{align}
with the covariant derivative $D_\mu =\partial_\mu - i\left[A_\mu, \,\,\right]$. The various fields were introduced in Section \ref{sec:maxextsusy} and the coupling constant $g$ is a dimensionless quantity. In particular, to define the action, we used the \textit{covariant gauge fixing condition} $\partial^\mu A_\mu=0$, and we set the gauge parameter $\xi=1$ to work in the \textit{Feynman gauge}.  Appendix \ref{sec:convidentities} collects other conventions and useful identities. \\

\subsection{Planar Limit} \label{sec:planarlimit}

Another fundamental ingredient we need to apply the perturbative method to this theory is the expansion in the large $N$ limit.  It was introduced by 't Hooft in \cite{tHooft:1973alw}, and it consists of taking the limit $N \rightarrow \infty$, where $N$ is the number of color indices of the gauge group $U(N)$.  To properly discuss this regime, we need to introduce the so-called \textit{'t Hooft coupling}: 
\begin{equation}
\lambda\equiv g^2 N\,,
\end{equation}
where $g$ is the coupling constant of the $\Nm=4$ SYM theory that we encountered in the action (\ref{eq:N=4action}). If we keep $\lambda$ fixed and we take $N\rightarrow\infty$ , we get a divergent factor that is, however, compensated by the divergence of the number of components $N^2$ in the fields, leaving the action finite \cite{tHooft:1973alw}. The radius of convergence of a large $N$ expansion is known to be non-zero \cite{tHooft:2002ufq}, although renormalization effects can spoil it.

We connect now to Feynman diagrams. There the dependence from $N$ is encoded in the color factors. A propagator contributes with a factor $1/N$, while a vertex with $N$. The Wilson line (\ref{eq:wilsonline}) that we describe in detail later brings a factor $1/N$ and a color contraction factor $N$. These contributions are related to the \textit{topologies} of Feynman diagrams, which can be distinguished by their \textit{Euler's characteristics}:
\begin{equation}
\chi\equiv V-E+F=2-2G\,,
\end{equation}
where $V$ is the number of vertices, $E$ the number of edges (propagators), $F$ the number of faces (index contractions) and $G$ the \textit{genus} of the diagram.
The leading diagrams in the large $N$ limit are the so-called \textit{planar diagrams}, and they have the lowest genus so they can be drawn in two dimensions without crossings.

\begin{figure}[h] 
\centering
\includegraphics[scale=0.27]{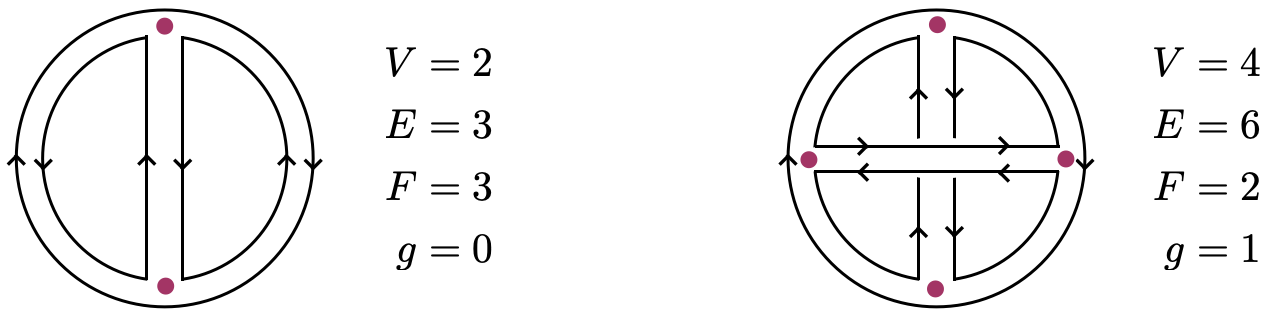}
\caption{\textit{Vacuum diagrams in the double line notation. Interaction vertices are marked with a small \textbf{\textcolor{vertexcolor}{dot}}. The left diagram is planar, while the diagram on the right has the topology of a torus (genus 1 surface).}}
\end{figure}

\subsection{Feynman Rules} \label{sec:N=4Feynmanrules}

We now move to deriving the Feynman rules in position space that we use throughout this thesis. First of all, we define the bosonic propagators:
\begin{equation} \label{eq:I12}
I_{12}\equiv\frac{1}{(2 \pi)^2\,x_{12}^2}\,,\qquad \qquad x_{ij}\equiv x_i-x_j\,,
\end{equation}
which are the Green's functions of the d'Alembert operator $\Box=\partial^{\mu} \partial_\mu$.

Similarly, we can derive the fermionic propagator, that we write down together with the other propagators and their graphical representation:
\begin{subequations}
\begin{align}
\text{Scalars:} \qquad 
& \propagatorS = g^2 \tensor{\delta}{_{IJ}} \tensor{\delta}{^{ab}} I_{12}\;, \label{subeq:propagatorS} \\
\text{Gluons:} \qquad 
& \propagatorG = g^2 \tensor{\delta}{_{\mu\nu}} \tensor{\delta}{^{ab}} I_{12}\;, \label{subeq:propagatorG} \\
\text{Gluinos:} \qquad 
& \propagatorF = i g^2 \tensor{\delta}{^{ab}} \slashed{\partial}_{\Delta} I_{12}\;, \label{subeq:propagatorF} \\
\text{Ghosts:} \qquad 
& \propagatorGh = g^2 \tensor{\delta}{^{ab}} I_{12}\;, \label{subeq:propagatorGh}
\end{align}
\label{eq:propagators}
\end{subequations}
with $a,b$ being $SU(N)$ group indices and where
\begin{equation*}
\slashed{\partial}_{\Delta} := \gamma \cdot \frac{\partial}{\partial \Delta}\;, \qquad \qquad \Delta^{\mu} := x_{12}^{\mu}\;,
\end{equation*}
with $\gamma_\mu$ the Dirac matrices.

We now draw all the vertices of the $\mathcal{N}=4$ theory:
\begin{equation} \label{fig:n=4vertices}
\VertexScalarScalarGluon \qquad \VertexFermionFermionScalar \qquad \VertexFermionFermionGluon \qquad \VertexGhostGhostGluon \qquad \VertexGluonGluonGluon \qquad \VertexFourScalars \qquad \VertexScalarScalarGluonGluon \qquad \VertexGluonGluonGluonGluon \,,
\end{equation}
which can be directly read from the action (\ref{eq:N=4action}). Not all of them are relevant to this work. We focus in detail only on the useful ones in the next sections.

In the computation of the Feynman diagrams at next-to-leading order, we encounter three-, four- and five-point massless Feynman integrals, which we define as follows:
\begin{subequations}
\begin{gather}
Y_{123} =\Yintegral := \int d^4 x_4\ I_{14} I_{24} I_{34}\,, \label{subeq:Y123} \\
X_{1234}=\Xintegral := \int d^4 x_5\ I_{15} I_{25} I_{35} I_{45}\,, \label{subeq:X1234} \\
H_{12,34} =\Hintegral := \int d^4 x_5 d^4 x_6\ I_{15} I_{25} I_{36} I_{46} I_{56}\,, \label{subeq:H1324}
\intertext{with $I_{ij}$ the propagator function defined in (\ref{eq:I12}). The letter assigned to each integral is evocative of the drawing of the propagators. Another expression that is encountered is the following:}
F_{12,34} := \frac{\left( \partial_1 - \partial_2 \right) \cdot \left(\partial_3 - \partial_4 \right)}{I_{12}I_{34}} H_{12,34}\,. \label{subeq:F1324}
\end{gather}
\end{subequations}
The notation presented above is standard and has already been used in e.g. \cite{Beisert:2002bb, Drukker:2008pi}. The three- and four-point massless integrals in Euclidean space are conformal and have been solved analytically (see e.g. \cite{tHooft:1978jhc, Usyukina:1992wz} and \cite{Drukker:2008pi,Kiryu:2018phb} for the modern notation).  The $Y$- and $X$-integrals have been solved analytically and can be found in Appendix \ref{subsec:bulkintegrals}. The $H$-integral seems to have no known closed form so far, but the F-integral (\ref{subeq:F1324}) can, fortunately, be reduced to a sum of $Y$- and $X$-integrals which we also show in Appendix \ref{subsec:bulkintegrals}.

In the computation of the Feynman diagrams at next-to-next-to-leading order, we encounter new types of integrals. In particular,
\begin{subequations}
\begin{gather}
K_{ij} := \int d^4 x_5\, I_{15} I_{25} I_{35}\, Y_{ij5}\,,  \label{eq:Kij} \\
A_k := \int d^4 x_5\, I_{15} I_{25} I_{35} \log x_{k5}^2\,. \label{eq:Ak}
\end{gather}
\end{subequations}
More details, such as explicit expressions and identities for these integrals, can be found in Appendix \ref{subsec:bulkintegrals}.

\subsubsection{Insertions Rules} \label{sec:insertionsbulk}

We now introduce the insertion rules that we use to compute Feynman diagrams. Keep in mind that we gather some useful conventions and identities in Appendix \ref{sec:convidentities}. In particular, here we recall that with $SU(N)$ as the gauge group and working in the large $N$ limit, the generators obey the following commutation relation:
\begin{equation}
[ T^a\,, T^b ] = i f^{abc}\, T^c,
\end{equation}
in which $f^{abc}$ are the structure constants of the $\mathfrak{su}(N)$ Lie algebra. The generators are normalized as
\begin{equation}
\tr T^a \tensor{T}{^b} = \frac{\delta^{ab}}{2}.
\end{equation}
Note that $f^{ab0} = 0$ and $\tr \tensor{T}{^a} = 0$. The (contracted) product of structure constants gives $f^{abc} f^{abc} = N (N^2 - 1) \sim N^3$, where the second equality holds in the large $N$ limit.

We can now move to the three- and four-point insertions. Keeping in mind that the $\mathcal{N}=4$ SYM theory contains eight vertices, as shown in (\ref{fig:n=4vertices}), there are two interesting three-point vertices for us: two scalar fields with one gauge field and the Yukawa interaction.  The insertion rule for the first vertex can be easily obtained from the action (\ref{eq:N=4action})\footnote{Note that for compactness of the expression,  we write $\phi_1$ to intend $\phi(x_1)$.}:
\begin{equation} \label{eq:vertex2scalars1gluon}
\begin{aligned}
\vertexSSG\ & =\frac{2 i}{g^2} \operatorname{Tr} T^d\left[T^e, T^f\right] \int d^4 x_4\left\langle\phi_1^{I, a} \phi_2^{J, b} A_{3, \mu}^c \partial^v \phi_{4, d}^K \phi_{4, e, K} A_{4, v, f}\right\rangle \\[-1.5em]
& =-g^4 f^{a b c} \delta^{IJ}\left(\partial_1-\partial_2\right)_\mu Y_{123} \,,
\end{aligned}
\end{equation}
where $Y_{123}$ has been defined in (\ref{subeq:Y123}) and its analytical expression in $1d$, the interesting case for us, as we will see, can be found in (\ref{eq:Y123}). We do not use an insertion rule for the Yukawa term, as it arises only in one Feynman diagram among the ones we considered.

Moving to four-point insertions, from the action we can derive a Feynman rule for the four-scalar vertex:
\begin{equation} \label{eq:vertexSSSS}
\begin{aligned}
\vertexSSSS\ & =-g^6\left\{f^{a b e} f^{c d e}\left(\delta_{IK} \delta_{JL}-\delta_{IL} \delta_{JK}\right)+f^{a c e} f^{b d e}\left(\delta_{IJ} \delta_{KL}-\delta_{IL} \delta_{JK}\right)\right.\\[-1.5em]
& \left.\qquad \qquad \qquad +f^{a d e} f^{b c e}\left(\delta_{IJ} \delta_{KL}-\delta_{IK} \delta_{JL}\right)\right\} X_{1234} \,,
\end{aligned}
\end{equation}
where $X_{1234}$ as an integral can be found in (\ref{subeq:X1234}).

Finally using the vertex (\ref{eq:vertex2scalars1gluon}), we can obtain this four-point insertion:
\begin{equation} \label{eq:Hinsertone}
\Hinsertone\ =g^6\left\{\delta_{IK} \delta_{JL} f^{a c e} f^{b d e} I_{13} I_{24} F_{13,24}+\delta_{IL} \delta_{JK} f^{a d e} f^{b c e} I_{14} I_{23} F_{14,23}\right\}\,,
\end{equation}
with $I_{IJ}$ the propagator function defined in (\ref{eq:I12}) and $F_{IJ,KL}$ as defined in (\ref{subeq:F1324}). An analytical expression for $F_{IJ,KL}$ in terms of $X$- and $Y$-integrals is given in (\ref{eq:FXYidentity}).

\subsubsection{Scalar Self-Energy}

To conclude, we move our attention to the one-loop correction to the scalar-propagator,  the only two-point insertion we need.  It consists of the following diagrams:

\begin{equation} \label{eq:selfenergydiagrams}
\propagatorSSEnotext\ =\ \SSEone\ +\ \SSEtwo\ +\ \SSEthree\ +\ \SSEfour \,.
\end{equation}

All these diagrams can be computed easily. The first one gives a contribution of
\begin{equation}
\begin{aligned}
\SSEone\ & =(-i)^2 \frac{2}{g^4} \int d^4 x_3 d^4 x_4\left\langle\phi_{1, I}^a \phi_{2, J}^b \operatorname{Tr} \partial_\mu \phi_3^K A_3^\mu \phi_{3, K} \operatorname{Tr} \partial_v \phi_4^L A_4^v \phi_{4, L}\right\rangle \\
& =2 g^4 N \delta^{a b} \delta_{IJ} Y_{112}+g^4 N \delta^{a b} \delta_{IJ} \frac{1}{(2 \pi)^2 \varepsilon^2} \int d^4 x_3 I_{13} I_{23} \,.
\end{aligned}
\end{equation}
Note that the $Y$-integral of the first term is given explicitly in (\ref{eq:Y112}) and contains a logarithmic divergence, while the second term contains a quadratic divergence encoded by the $1 / \varepsilon^2$. This factor arises when defining:
\begin{equation}
I_{33} \equiv \frac{1}{(2 \pi)^2 \varepsilon^2}\,,
\end{equation}
which consists of a regularization method called \textit{point-splitting regularization}, where the zero is replaced by an infinitesimal distance. Effectively,  we split two coincident points by inserting a small distance $\epsilon$.

The second diagram reads:
\begin{equation}
\SSEtwo\ =-4 g^4 N \delta^{a b} \delta_{IJ}\left\{Y_{112}-\frac{2}{(2 \pi)^2 \varepsilon^2} \int d^4 x_3 I_{13} I_{23}\right\}\,,
\end{equation}
where we recognize the same types of divergences as in the first diagram.

The last two diagrams differ only by a multiplicative factor and give:
\begin{subequations}
\begin{align}
\SSEthree\ &=-5 g^4 N \delta^{a b} \delta_{IJ} \frac{1}{(2 \pi)^2 \varepsilon^2} \int d^4 x_3 I_{13} I_{23}\,,\\
\SSEfour\ &=-4 g^4 N \delta^{a b} \delta_{IJ} \frac{1}{(2 \pi)^2 \varepsilon^2} \int d^4 x_3 I_{13} I_{23} \,.
\end{align}
\end{subequations}
Those diagrams contain only quadratic divergences that vanish, as they cancel each other when we sum up all the contributions. We are left with an expression containing only one log divergence:
\begin{equation} \label{eq:selfenergy}
\propagatorSSEnotext =-2 g^4 N \delta^{a b} \delta_{IJ} Y_{112}\,.
\end{equation}
This expression is well-known and is the same as the one given in e.g. \cite{Erickson:2000af,Plefka:2001bu,Drukker:2008pi}. 

We are not going to need the expression for the gluon self-energy. In any case, at next-to-leading order, it is very similar to the one of the scalar and it can also be found in \cite{Erickson:2000af,Plefka:2001bu}.

\chapter{The Wilson Line Defect CFT} \label{ch:wilsonline}

We can now introduce another key player in this work: the \textit{Maldacena-Wilson line} operator.  In particular, we first present the concept of \textit{conformal defect} as an extended operator inserted in a generic $d$-dimensional theory,  which partially breaks the symmetries of the original theory.  In fact, the Wilson line can be interpreted as a point-like defect, an impurity, which breaks the symmetries of the host theory: $\mathcal{N}=4$ SYM.  

We especially explore the one-dimensional CFT living on this line defect in detail, focusing on its correlators and their various interesting properties. To conclude the chapter, we present the Feynman rules needed to compute correlators of operators living on the Wilson line.

\section{Conformal Defects} \label{sec:conformaldefects}

Let us introduce conformal defects by first considering a $d$-dimensional CFT on a flat Euclidean space $\mathbb{R}^d$ with coordinates 
\begin{equation}
x^\mu =(x^a, x^i)\,,\qquad a=1,\dots,p \quad \text{and} \quad i=p+1,\dots,d\,.
\end{equation}
We insert then an infinite flat defect extending in the $x^a$ directions and located at $x^i=0$, which splits the space between parallel ($x^a$) and orthogonal ($x^i$) directions. The defect is then identified by $p$, and $q \equiv d-p$ is known as the \textit{co-dimension}. 

As anticipated above, the insertion of a defect breaks the original conformal symmetry we discussed in Section \ref{sec:conformalsymmetry} to a subgroup of it:
\begin{equation}
SO(d+1,1)\rightarrow SO(p+1,1) \times SO(q)\,, \qquad p+q=d\,,
\end{equation}
namely conformal transformations along the defect and rotations around it. 

Moreover,  the original rotation group $SO(d)$ (see Section \ref{sec:conformalalgebra}) is broken to $SO(p) \times SO(q)$. We can consider $SO(q)$ as an ``internal symmetry'' group of the defect. From a defect point of view, we have then a $p$-dimensional conformal field theory (CFT$_p$) with a $SO(q)$ flavour group.  However, we have to make an important remark: in such CFTs, there exists, in general, a stress-tensor, while no such stress-tensor is part of the spectrum of defect operators.

Lastly,  one can analyze these defect conformal theories as usual CFTs, meaning computing correlations functions and extracting CFT data. Particularly there are two possibilities: either to consider correlators of \textit{insertions}, namely operators living on the defect, or a mixture of \textit{bulk} local operators and defect operators.  In this work, we focus on the first kind of correlation functions. Before looking at them closely, let us meet our defect operator: the Maldacena-Wilson line.

\section{Defining the Wilson Line} \label{sec:wilsonloopandline}

Before diving into the details of the defect operator we consider, let us briefly introduce the historical importance of it. The \textit{Wilson loop} is a gauge invariant operator that was introduced by Wilson in 1974 \cite{Wilson:1974sk}, and it is related to many important observables \cite{Giles:1981ej,Migdal:1983qrz}.

We know that there is confinement in QCD at low energies; namely, color-charged particles cannot be isolated. In particular, quarks appear only in pairs quark-antiquark $q\bar{q}$. To keep their distance fixed in time, we consider external heavy quarks which are no more dynamical. In this setting, the Wilson loop defines and measures the interaction potential between pairs of quark-antiquark $V_{q \bar{q}}$.  It is also the traced holonomy of the gauge connection $A_\mu$, meaning the phase factor picked up by an external quark moving along a closed path $C$, parametrized by the vector $x^\mu$:
\begin{equation} \label{eq:wilsonloop}
\mathcal{W}_C=\frac{1}{\text{dim}_R} \text{tr}_R\, \mathcal{P}  \exp \left(i\oint_C dx^\mu\,  A_\mu\right)\,,
\end{equation}
where dim$_R$ is the dimension of the representation $R$ and $\mathcal{P}$ is the path-ordering exponential.

The Wilson loop is also central in the context of the AdS/CFT correspondence \cite{Maldacena:1997re,Berenstein:1998ij}: a duality between type IIB string theory compactified on AdS$_5\times$S$^5$ and a four-dimensional $\Nm=4$ theory. In this context, it is possible to define a generalization of the Wilson loop, the so-called \textit{Maldacena-Wilson} loop \cite{Maldacena:1997re}, which in the fundamental representation is 
\begin{equation} \label{eq:susywilsonloop}
\mathcal{W}_C=\frac{1}{N} \text{tr} \, \mathcal{P}  \exp \oint_C d\tau  \, \left( i \dot{x}^\mu A_\mu(\tau) + |\dot{x}|\, \theta_I \phi^I \right)\,,
\end{equation}
where again $ x^\mu$ parametrizes the loop and $\theta_I$ is an $SO(6)$ vector satisfying $\theta^2=1$.  Moreover, $A_\mu \equiv T^a A^a_\mu \equiv T^a \phi^I_a$, where $T^a$ is a generator of the gauge group of the SYM theory, and $I$ is a $\mathfrak{so}(6)_R$ index. Interestingly, this operator has been studied for multiple geometries of the path $C$, as a circle or as a line \cite{Erickson:2000af} but also with cusps \cite{Korchemsky:1987wg,Drukker:2012de}. 

In general (\ref{eq:susywilsonloop}) is only locally supersymmetric \cite{Drukker:1999zq}, however depending on the geometry and on the choice of the coupling, also global supersymmetry can be preserved. The exact amount can be computed following the analysis carried out in \cite{Zarembo:2002an,Drukker:2006zk,Drukker:2007dw,Drukker:2007qr}. There are two cases in which the highest amount of supersymmetry is preserved, i.e. \textit{1/2 BPS loops}: the straight line and the circular loop. One can move from one to the other via a conformal transformation (an inversion).  We focus on the first object, and we finally introduce one of the main characters of this work: the Maldacena-Wilson line in $\Nm=4$ SYM theory.

To define it, we just take the path $C$ in (\ref{eq:susywilsonloop}) to be an infinite straight line, obtaining then:
\begin{equation} \label{eq:wilsonline}
\Wl := \frac{1}{N} \tr \mathcal{P} \exp \int_{-\infty}^{\infty} d\tau\, \left( i \dot{x}^\mu A_\mu(\tau) + |\dot{x}|\, \phi^6(\tau) \right)\,.
\end{equation}
Here we have chosen the scalar $\phi^6$ to be the one coupling to the line by setting $\theta = (0,0,0,0,0,1)$ in (\ref{eq:susywilsonloop}). Note that we have Wick-rotated to Euclidean space and defined the path such that the line extends in the Euclidean time direction, i.e. $\dot{x}_\mu = (0,0,0,1)$ and $|\dot{x}|=1$.  As just stated above, this operator preserves the maximal amount of supersymmetry, i.e. half of the supercharges (16 out of 32), hence it is 1/2-BPS. From the discussion in Section \ref{sec:superconformalalgebra} follows that it is a protected operator; namely it does not receive anomalous contributions to its conformal dimension. In addition, it was shown perturbatively in \cite{Erickson:2000af} that the expectation value of the Wilson line is simply
\begin{equation} \label{eq:vevwilsonline}
\vev{\Wl} = 1\,.
\end{equation}

Before exploring this operator closely, it is interesting to notice that at strong coupling, Wilson loops are related by duality to open string minimal surfaces in AdS$_5$ ending on the contour defining the loop operator at the boundary. In our case of interest, the 1/2-BPS Wilson line is dual to an AdS$_2$ minimal surface which is embedded in AdS$_5$ and sits at a point on S$^5$.

\begin{figure}[h] 
\centering
\includegraphics[scale=0.35]{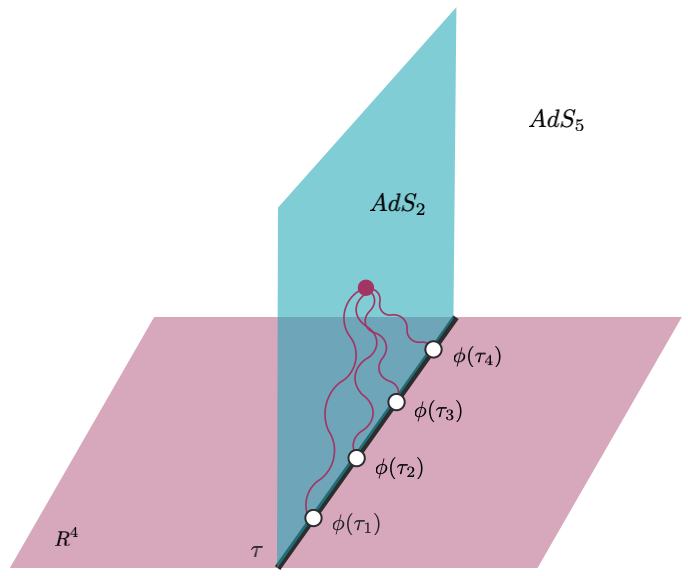}
\caption{\textit{Four-point function of local operators inserted on the Wilson line from a Witten diagram on the AdS$_2$ worldsheet.}}
\end{figure}

\section{The 1$d$ Defect CFT} \label{sec:1ddefectCFT}

Now that we have introduced this extended operator, it is time to explain its defect nature.  In Minkowski space, it corresponds to a point-like impurity in the $3d$ space, which evolves in time.  As a consequence, if we consider the defect to be part of the vacuum, the conformal symmetry of $\Nm = 4$ SYM is broken:
\begin{equation}
SO(5,1) \rightarrow SO(2,1) \times SO(3)\,.
\end{equation}

If we restrict ourselves to operators inserted \textit{on} the line, what we introduced in Section~\ref{sec:conformaldefects} under the name of insertions, then the symmetry group $SO(1,2)$ corresponds to the $1d$ conformal group and its generators correspond to the dilatations, translations and special conformal transformations along the line.  On the other hand, the subgroup $SO(3)$ refers to rotations orthogonal to the defect.

Because of the presence of the scalar field $\phi^6$ in \eqref{eq:wilsonline}, the defect also breaks the $R$-symmetry from $SO(6)_R$ to $SO(5)_R$. This choice entails that $SO(5)_R$ rotates the five scalars $\phi^i\, (i =1, \ldots, 5)$ that do not couple to the line and are therefore protected operators. In this setup, the full superconformal algebra $\mathfrak{psu}(2,2|4)$ of $\Nm=4$ SYM breaks into the $\Nm=8$ superconformal quantum mechanics algebra $\mathfrak{osp}(4^*|4)$.

The leftover scalar field $\phi^6$ is the only scalar of length $L=1$ that couples to the Wilson line, due to our choice of $\theta = (0,0,0,0,0,1)$. This operator can be seen as the $1d$ analog of the famous Konishi operator in $4d$ $\Nm=4$ SYM: it is the lowest-dimensional unprotected operator at weak coupling, and it is not degenerate. Perhaps the main difference is that it flows to a ``two-particle'' state at strong coupling (i.e. $\Delta = 2$), whereas the Konishi operator decouples in this limit. As a result, the conformal dimension of $\phi^6$ has been determined up to five loops at weak coupling \cite{Grabner:2020nis} and up to four loops at strong coupling \cite{Ferrero:2021bsb}.

In the next section, we enter into detail about the correlation functions of insertions on the Wilson line. For now, let us underline the fact that they are also constrained by the $1d$ conformal symmetry in a way analogous to higher-dimensional CFTs, and they can be interpreted as characterizing a defect CFT$_1$ living on the Wilson line \cite{Drukker:2006xg,Cooke:2017qgm}. This CFT$_1$ should then be fully determined by the spectrum of scaling dimensions and OPE coefficients.

The states of this CFT$_1$ live in unitary representations of $OSp(4^*|4)$, and they can be organized in superconformal multiplets labeled by four quantum numbers: the scaling dimension $\Delta$ associated with the $1d$ conformal group, the spin $s$ associated with the internal $SO(3)$ symmetry and finally the $SO(5)$ Dynkin labels $[a,b]$ associated with the $R$-symmetry. These representations were classified in \cite{Gunaydin:1990ag,Liendo:2016ymz}.

This CFT$_1$ particularly hosts 1/2-BPS multiplets denoted by $\mathcal{B}_k$, whose superconformal primaries have the quantum numbers 
\begin{equation}
\left\{\Delta, s, [a,b]\right\}=\left\{k,k,[0,k]\right\}\,,\qquad k \in \mathbb{Z}\,.
\end{equation}
These operators are protected, and they are relevant for this work. 

The first multiplet we discuss in detail is $\mathcal{B}_1$, which contains the simplest superconformal primaries of the theory: the scalars $\phi^i$ (with $\Delta=1$). This multiplet also contains the components of the field-strength $\mathbb{F}_{t i} \equiv i F_{t i}+D_i \phi^6$ (with $\Delta=2$) along the directions $i=1,2,3$ transverse to the line.  $\mathbb{F}_{t i}$ is also known as the \textit{displacement operator}, which can be defined for any defect in any CFT, and it measures generically the changes of the Wilson line/loop under deformations orthogonal to the contour.

The other important multiplet is $\mathcal{B}_2$ which contains the superconformal primary operators
\begin{equation}
\mathcal{O}^{ij}_S := \phi^i\phi^j-\phi^j\phi^i-\frac{2}{5}\delta^{ij}\phi^k\phi^k
\end{equation}
of protected $\Delta=2$.

As introduced in Section \ref{sec:conformalalgebra}, alongside short multiplets, there are long ones, which in principle, do not preserve any supercharges.  They are denoted by $\mathcal{L}_{s,[a, b]}^{\Delta}$, with the indices corresponding to the quantum numbers of the primary. For long operators, $\Delta$ is a non-trivial function of the coupling $g$.

In \cite{Liendo:2016ymz} several selection rules for the OPE were deduced. In particular, the followings OPEs are relevant to us:
\begin{subequations} \label{eq:OPEB} 
\begin{align} 
&\mathcal{B}_1 \times \mathcal{B}_1 \rightarrow \mathbb{1}+\mathcal{B}_2+\sum_{\Delta>1} \mathcal{L}_{0,[0,0]}^{\Delta} \,,\label{eq:OPEB1B1} \\ 
&\mathcal{B}_1 \times \mathcal{B}_2 \rightarrow \mathcal{B}_1+\mathcal{B}_3+\sum_{\Delta} \mathcal{L}_{0,[0,1]}^{\Delta}\,, \label{eq:OPEB1B2}
\end{align}
\end{subequations}
where $\mathcal{L}_{0,[0,0]}^{\Delta}$ and $\mathcal{L}_{0,[0,1]}^{\Delta}$ are the unprotected multiplets. The first transforms as a singlet under the global $\mathrm{SO}(5) \times \mathrm{SO}(3)$ symmetry and the second is a singlet under $SO(3)$ but transforms in the $[0,1]$ representation of $SO(5)$. There are infinitely many multiplets with such quantum numbers, all with unprotected scaling dimensions. As shown in (\ref{eq:OPEB}), the whole infinite set of such multiplets can appear in the fusion of two operators in the displacement multiplet.

Before moving to correlation functions,  we conclude this section by coming back to the dual Wilson line picture introduced in Section \ref{sec:wilsonloopandline}. The fundamental open string stretching in AdS preserves the same $OSp(4^*|4)$ as the Wilson line \cite{Gomis:2006sb}. In particular, the $1d$ conformal group is realized as the isometry of AdS$_2$. 

Expanding the string action in static gauge around the minimal surface solution, one finds \cite{Drukker:2000ep} that the AdS$_2$ fluctuations transverse to the string include five massless scalars $y_a$ corresponding to the $S^5$ directions, three massive scalars $x^i$ with $m^2=2$ corresponding to the AdS$_5$ fluctuations and finally eight fermionic modes. 

It is, therefore, natural to identify these excitations, which can be thought of as fields living in AdS$_2$, with the CFT$_1$ insertions that we have just introduced \cite{Sakaguchi:2007ba, Faraggi:2011bb}. Particularly,  the massless $y^a$ fields should be dual to the scalars $\phi^i$, while the three AdS$_5$ fluctuations $x^i$ should be dual to the field strength operators $\mathbb{F}_{t i}$, according to the standard relation $m^2=\Delta(\Delta-d)$ between AdS$_{d+1}$ scalar masses and the corresponding CFT$_d$ operator dimensions. More information about this CFT$_1$ dual can be found in \cite{Giombi:2017cqn}, where four-point functions of bosonic excitations have been computed at strong coupling using Witten diagrams.

\section{Correlation Functions on the Defect} \label{sec:wilsonlinecorrelators}

As anticipated, in this thesis, we consider correlation functions of operators in the scalar sector, which involve only the six fundamental scalar fields $\phi^I (\tau)$ ($I=1, \ldots, 6$) of the bulk theory. Operators are constructed by effectively inserting them inside the trace of the Wilson line, and we can now understand why we refer to them as insertions.  Moreover, we consider only single-trace representations of the algebra, which we introduced in Section~\ref{sec:superconformalalgebra}.

\begin{figure}[h]
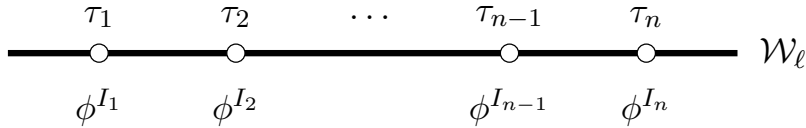

\centering
\scalebox{1.2}{\wilsonline}
\caption{\textit{Representation of the correlation functions \eqref{eq:correlators} in the $1d$ defect CFT, defined by inserting operators on the Maldacena-Wilson line. At the points $\tau_1\,, \ldots\,, \tau_n$, scalar fields are inserted \textit{inside} the trace of the Wilson line operator.}}
\label{fig:insertions}
\end{figure}

A generic $n$-point correlation function of the defect single-trace operators is to be understood in the following way\footnote{Generally, one normalizes the correlators by the expectation value of the Wilson loop without insertions. However, the Wilson line has a trivial expectation value (\ref{eq:vevwilsonline}).}:
\begin{equation} \label{eq:correlators}
\vev{\phi^{I_1}\, \ldots \phi^{I_n}}_{1d} := \frac{1}{N} \vev{\tr \mathcal{P} \left[ \phi^{I_1}\, \ldots \phi^{I_n} \exp \int_{-\infty}^{\infty} d\tau \bigl( i \tensor{\dot{x}}{^{\smash{\mu}}} \tensor{A}{_{\smash{\mu}}} + | \dot{x} |\, \tensor{\phi}{^6} \bigr) \right]}_{4d}\,,
\end{equation}
where we suppress the dependency on $\tau_1, \ldots, \tau_n$ (for the local insertions) and on $\tau$ (for the Wilson line itself) for compactness. Without loss of generality we consider the $\tau$'s to be ordered, i.e. $\tau_1 < \tau_2 < \ldots < \tau_n$.
This type of correlators is illustrated in Figure \ref{fig:insertions}.
The brackets on the left-hand side refer to correlators in the $1d$ defect theory, while the ones on the right-hand side correspond to correlators in the $4d$ bulk theory. From now on $\vev{\ldots}$ always refers to $1d$ correlators, hence we drop the subscript.

These correlation functions correspond to single-trace operators\footnote{In principle, \textit{multi-trace} operators with the same quantum numbers can also be constructed. See footnote 4 in \cite{Barrat:2021tpn} for more detail.}, in the sense that there is only one overall color trace in \eqref{eq:correlators}. This differs from the bulk theory case, where each operator carries its own trace. This property is specific to the defect theory, and it is crucial to construct correlators involving operators of higher $R$-charge. This can be done by bringing two operators close to each other. We refer to this limit as \textit{pinching}, and we explain it in more detail at the end of this section.

\begin{figure}
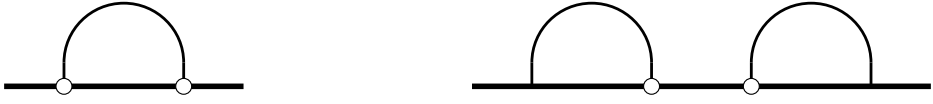

\centering
\begin{subfigure}{.4\textwidth}
  \centering
  \scalebox{1.5}{\twopointpr}
\end{subfigure}%
\begin{subfigure}{.6\textwidth}
  \centering
  \scalebox{1.5}{\twopointnpr}
\end{subfigure}
\caption{\textit{Examples of diagrams contributing to the two-point functions of fundamental scalars $\phi^I$. The diagram on the left contributes to both $\vev{\phi^i \phi^j}$ and $\vev{\phi^6 \phi^6}$, while the one on the right contributes only to $\vev{\phi^6 \phi^6}$. 
We refer to this type of diagram as \textbf{U-diagram} and they are described in detail in Section \ref{sec:mainU}. }}
\label{fig:Lambda}
\end{figure}

In Part II, where we present our perturbative results, we consider in particular \textit{unit-normalized} correlation functions, which are defined in the following way:
\begin{equation}
\vvev{\phi^{I_1}\, \ldots \phi^{I_n}} := \frac{ \vev{\phi^{I_1}\, \ldots \phi^{I_n}} }{\sqrt{n_{I_1} \ldots n_{I_n}}}\,,
\label{eq:unitnormalized}
\end{equation}
with $n_I$, the normalization constants related to two-point functions. Indeed this definition is chosen such that
\begin{equation}
\vvev{\phi^{I} (\tau_1) \phi^{J} (\tau_2)} = \frac{\delta^{IJ}}{\tau_{12}^2}\,,
\label{eq:twopt}
\end{equation}
with $\tau_{ij} := \tau_i - \tau_j$. Note that the (classical) scaling dimension of the fundamental scalar fields $\phi^I$ is $\Delta = 1$ due to their origin from a 4$d$ bulk theory, which explains the form of the propagator in (\ref{eq:twoptni}),  in spite of the theory being one-dimensional.  The derivation of normalization constants and their explicit expressions are covered in Section \ref{subsec:twopoint}, where we also discuss in detail the difference between two-point functions of protected $\phi^i$ and of the unprotected scalar $\phi^6$, which can be already partially appreciated in Figure \ref{fig:Lambda}.
We also assume normal ordering in the correlators, such that all operators have vanishing one-point functions, as required by conformal symmetry.

Three-point functions of the scalars $\phi^I$ are also kinematically fixed by conformal symmetry and read
\begin{equation}
\vvev{\phi^I (\tau_1) \phi^J (\tau_2) \phi^K (\tau_3)} = \frac{c_{\phi^I \phi^J \phi^K}}{\tau_{12}^{\Delta_{IJK}} \tau_{23}^{\Delta_{JKI}} \tau_{31}^{\Delta_{KIJ}}}\,,
\label{eq:threept}
\end{equation}
with $\Delta_{IJK} := \Delta_I + \Delta_J - \Delta_K$. For protected operators, an appropriate tensor with $R$-symmetry indices must be inserted in \eqref{eq:threept}, and it follows directly that three-point functions with an odd number of fundamental fields $\phi^i$ vanish since all $R$-symmetry indices must be contracted. We discuss this tensor in detail in the following for the case of four-point functions.

For higher $n$-point functions, conformal symmetry is not strong enough to fix the kinematical form of the correlators.  Nevertheless, it constrains them to be functions of $n-3$ cross-ratios $\chi_i$. For convenience, we restrict ourselves to correlators of the operators $\phi^I$, and we use the following factorized form: 
\begin{equation}
\vvev{\phi^{I_1}\, \ldots \phi^{I_n}}= \mathcal{K} (\tau_1, \Delta_{\phi^{I_1}}; \dots\,; \tau_n, \Delta_{\phi^{I_n}}) \,\mathcal{A}^{I_1 \dots I_n}(\chi_1\,, \ldots\,, \chi_{n-3})\, ,
\label{eq:factorisedcorrunprotected}
\end{equation}
where we refer to $\mathcal{A}^{I_1 \ldots I_n}$ as the \textit{reduced correlator} and $\chi_i$ are the spacetime cross-ratios, define such that the following limit holds:
\begin{equation}
(\chi_1, \chi_2, \ldots, \chi_{n-3}) \overset{(\tau_1, \tau_{n-1}, \tau_n) \rightarrow (0,1,\infty)}{\longrightarrow} (\tau_2, \ldots, \tau_{n-2})\,,
\end{equation}
resulting in the following expressions:
\begin{equation} \label{eq:crossratios}
\chi_1 = \frac{\tau_{12} \tau_{(n-1)n}}{\tau_{1(n-1)} \tau_{2n}}\,, \quad \chi_2 = \frac{\tau_{13} \tau_{(n-1)n}}{\tau_{1(n-1)} \tau_{3n}}\,, \ldots,
\quad \chi_i = \frac{\tau_{1(i+1)} \tau_{(n-1)n}}{\tau_{1(n-1)} \tau_{(i+1)n}}\, .
\end{equation}
Notice that the $1-\chi_i$ (which would be independent cross-ratios in a higher-dimensional CFT) are given by
\begin{equation}
1-\chi_1 = \frac{\tau_{1n} \tau_{2(n-1)}}{\tau_{1(n-1)} \tau_{2n}}\,, \quad 1-\chi_2 = \frac{\tau_{1n} \tau_{3(n-1)}}{\tau_{1(n-1)} \tau_{3n}}\,, \ldots,
\quad 1-\chi_i = \frac{\tau_{1n} \tau_{(i+1)(n-1)}}{\tau_{1(n-1)} \tau_{(i+1)n}}\, .
\end{equation}

Expression (\ref{eq:factorisedcorrunprotected}) is adopted whenever we deal with unprotected operators, which basically means whenever $\phi^6$ appears, alone or alongside other (protected) operators. In these cases, the $R$-symmetry indices, whenever present, are kept open. 

However, anytime we deal \textit{only} with protected operators, in order to have a more transparent and useful expression, the reduced correlator depends explicitly not only on the spacetime cross-ratios but also on $n(n-3)/2$ $R$-symmetry cross-ratios $r_i\,, s_i\,, t_{ij}$, to be defined shortly:
\begin{equation} \label{eq:factorisedcorrprotected}
\vvev{\phi^{i_1}\, \ldots \phi^{i_n}}= \mathcal{K} (\tau_1, \Delta_{\phi^{i_1}}, u_1; \dots\,; \tau_n, \Delta_{\phi^{i_n}},u_n)\, \mathcal{A}_{\Delta_1\dots\Delta_n}(\chi_i;\,r_i\,, s_i\,, t_{ij})\,,
\end{equation}
where $u_i$ are complex vectors accounting for the $R$-symmetry indices of the protected operators: 
\begin{equation} \label{eq:uvector}
\phi^i \equiv u_i \cdot \phi\,,
\end{equation}
and they satisfy $u^2=0$ and $u\cdot \theta=0$, with $\theta$ introduced below (\ref{eq:wilsonline}). Again the prefactor $\Km$ is chosen such that the reduced correlator depends only on these cross-ratios. 
 
Moving then to the definition of the $R$-symmetry cross-ratios, in analogy with the spacetime cross-ratios,  one can start to define the $R$-symmetry cross-ratios $r_i$ such that the indices have a one-to-one correspondence with the spacetime cross-ratios, i.e.
\begin{equation} \label{eq:rRsymmecross}
r_1 = \frac{(u_1 \cdot u_2) (u_{n-1} \cdot u_n)}{(u_1 \cdot u_{n-1}) (u_2 \cdot u_n)}\,, \,r_2 = \frac{(u_1 \cdot u_3) (u_{n-1} \cdot u_n)}{(u_1 \cdot u_{n-1}) (u_3 \cdot u_n)}\,, \ldots \, ,\,
r_i = \frac{(u_1 \cdot u_{i+1}) (u_{n-1} \cdot u_n)}{(u_1 \cdot u_{n-1}) (u_{i+1} \cdot u_n)}\, .
\end{equation}
This correspondence implies that we have $n-3$ $r_i$.

Then the $s_i$ are defined in analogy to $1-\chi_i$. This gives the following $n-3$ cross-ratios:
\begin{equation} \label{eq:sRsymmecross}
s_1 = \frac{(u_1 \cdot u_n) (u_2 \cdot u_{n-1})}{(u_1 \cdot u_{n-1}) (u_2 \cdot u_n)}\,,\, s_2 = \frac{(u_1 \cdot u_n) (u_3 \cdot u_{n-1})}{(u_1 \cdot u_{n-1}) (u_3 \cdot u_n)}\,, \ldots\, , \,
s_i = \frac{(u_1 \cdot u_n) (u_{i+1} \cdot u_{n-1})}{(u_1 \cdot u_{n-1}) (u_{i+1} \cdot u_n)}\,.
\end{equation}

This convention for the prefactor and the spacetime cross-ratios is prevalent throughout the paper. However,  please note that in Sections \ref{subsubsec:66666}, \ref{subsubsec:sixpt-buildingblocks}, and \ref{subsubsec:666666} we adopt a different convention for the correlators presented there. In particular, the prefactor reads
\begin{equation}
\tilde{\mathcal{K}} (\tau_1, \Delta_{\phi^{I_1}}; \dots\,; \tau_n, \Delta_{\phi^{I_n}}) = \left(\frac{\tau_{32}}{\tau_{21} \tau_{31}}\right)^{\Delta_{1}}\left(\frac{\tau_{n-1, n-2}}{\tau_{n, n-2} \tau_{n, n-1}}\right)^{\Delta_{n}} \prod_{i=1}^{n-2}\left(\frac{\tau_{i+2, i}}{\tau_{i+1, i} \tau_{i+2, i+1}}\right)^{\Delta_{i+1}}\,,
\label{eq:prefactor}
\end{equation}
with $\tau_{ij}:=\tau_i-\tau_j$ as usual, while the spacetime cross-ratios are defined as
\begin{equation}
\tilde{\chi}_i:= \frac{\tau_{i,i+1}\tau_{i+2,i+3}}{\tau_{i,i+2}\tau_{i+1,i+3}}\,,
\label{eq:crossratio}
\end{equation}
and they are positive-definite. The prefactor as well as the cross-ratios are adopted from \cite{Rosenhaus:2018zqn}\footnote{Note that in \cite{Rosenhaus:2018zqn} the points are ordered as $\tau_1 > \tau_2 > \ldots > \tau_n$.}, where they emerge naturally in the derivation of the conformal blocks in a particular channel, called comb channel.  This choice is particularly useful for extracting bosonic CFT data at weak coupling from the correlators we present in the next part.  However, this is not always a convenient choice, and that is why we do not adopt it throughout the thesis.

\subsection{Topological Limit} \label{sec:topologicallimit}

It is well-known that the correlators defined in \eqref{eq:factorisedcorrprotected} are \textit{topological} when we set $u_i \cdot u_j = \tau_{ij}^2$, i.e.~the functions $\Am_{\Delta_1 \ldots \Delta_n}$ are constant in this limit, in the sense that they do not depend on the variables $u$ and $\tau$ \cite{Drukker:2009sf}\footnote{They can however still depend non-trivially on the 't Hooft coupling $\lambda$ and the number of colors $N$.}:
\begin{equation} \label{eq:topologicallimit}
\Am(\chi;r,s,t) \overset{\text{topological lim}}{\longrightarrow} \text{const}\,.
\end{equation}
Therefore, we find it useful to define the remaining $(n-3)(n-4)/2$ $R$-symmetry cross-ratios $t_{ij}$ in such a way that they reduce in the topological sector to the analogs of the following spacetime cross-ratios:
\begin{equation}
t_{ij} \to (\chi_i - \chi_j)^2\,,
\end{equation}
namely
\begin{equation} \label{eq:tRsymmecross}
t_{ij}= \frac{(u_{i+1} \cdot u_{j+1})(u_1 \cdot u_n)(u_{n-1} \cdot u_n)}{(u_1 \cdot u_{n-1})(u_{i+1} \cdot u_n)(u_{j+1} \cdot u_n)}\,,
\end{equation}
with $i=[1,n-4]\, , j=[i+1, n-3]$ and $i<j$.

\subsection{Pinching Technique} \label{sec:pinching}

To conclude this section, we introduce an important technique.  Since the operators are inserted inside the trace of the Wilson line, the \textit{pinching} of two operators or more produces again single-trace operators but with a higher length. For example,
\begin{equation}
\vvev{\phi^{I_1} (\tau_1)\, \ldots\, \underbrace{\phi^{I_{n-1}} (\tau_{n-1}) \phi^{I_n} (\tau_n)}_{\text{two operators of length $1$}}} \overset{\tau_n \to \tau_{n-1}}{\longrightarrow} \vvev{\phi^{I_1} (\tau_1)\, \ldots\, \underbrace{\phi^{I_{n-1}} (\tau_{n-1}) \phi^{I_n} (\tau_{n-1})}_{\text{one operator of length $2$}}}\,.
\end{equation}
This pinching technique allows to construct \textit{any} single-trace scalar operator made of fundamental scalar fields from correlation functions involving operators of length $L=1$.
Note that this is \textit{not} the case in the bulk theory, where the pinching of two single-trace operators produces a double-trace operator, since each operator carries its own trace.  Then,  interestingly the \textit{information} required to solve the scalar sector of this theory is very much reduced compared to e.g. its bulk counterpart.

We might have concerns about divergences that occur when we bring the operators close together. However, in the specific context of protected operators, as addressed in this study, it's worth noting that all these divergences arising from individual contributions to these correlators actually cancel out in the final result, as we expect.

\subsection{Expansion in Conformal Blocks} \label{sec:conformalblockexpansion}

As we discussed in Section \ref{sec:conformalblocks}, expanding a correlation function in terms of conformal blocks is a useful tool, as e.g.  it allows to extract CFT data.  In our context, the block expansion is relevant for two reasons. On one hand, it serves as an important consistency check of the results we derive at weak coupling; on the other hand, it is central to the bootstrap of a five-point function at strong coupling.

Expanding a four-point correlator, as we illustrated in Section \ref{sec:conformalblocks}, is quite established.  However, an analogous expansion for multipoint correlators is a realm that nowadays remains largely unexplored.  In this thesis, we move a step further in this exploration.

In this section, we introduce the blocks topologies that we will encounter and use: the \textit{comb} and the \textit{snowflake} OPE channels. 

\subsubsection{Comb Channel} 

Let us start by reviewing the comb channel.  Four-point blocks in $d=1$ have been known for a long time \cite{Ferrara:1974ny}, but only recently this work was extended to higher-point functions \cite{Rosenhaus:2018zqn}. Five-point point blocks were also derived for generic dimension $d$ in \cite{Goncalves:2019znr}.

\begin{figure}[h]
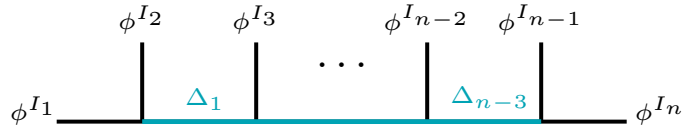

\centering
\combblocks
\caption{\textit{Representation of the comb channel for $n$-point correlation functions. The vertices correspond to bosonic OPE coefficients, which can be interpreted as three-point functions in the bosonic theory. For $n$ external operators, there are $n-3$ operators being exchanged.}}
\label{fig:combch}
\end{figure}

This channel consists of taking one by one the OPE of an external operator with an internal operator, as represented in Figure \ref{fig:combch}\footnote{The exception being, of course, the two extremities, where we have to take the OPE of two external operators.}.

We focus particularly on the case where all the external operators are identical scalar fields of length $L=1$.

For a given \textit{$R$-symmetry channel}\footnote{An $R$-symmetry channel corresponds to a fixed choice of indices $I_1, \ldots, I_6$.},  the reduced correlator of such $n$-point functions can be expanded in blocks in the following way:
\begin{align}
\mathcal{A}^{I_1 \dots I_n}
=& \sum_{\Op_1,\dots,\Op_n} c_{\phi^{I_1} \phi^{I_2} \Op_1} c_{\Op_1 \phi^{I_3} \Op_2} \dots c_{\Op_{n-4} \phi^{I_{n-2}} \Op_{n-3}} c_{\Op_{n-3} \smash{\phi^{I_{n-1}}} \smash{\phi^{I_n}}} 
g_{\Delta_1,\dots,\Delta_{n-3}} (\chi_1, \dots, \chi_{n-3})\,,
\label{eq:expansionblocks}
\end{align}
where $\Delta_k$ refer to the scaling dimension of the exchanged operators, and $c_{\Op_1 \Op_2 \Op_3}$ are the three-point coefficients defined in \eqref{eq:threept}. In the case where all the scalar fields are protected, we consider the highest-weight $R$-symmetry channel $F_0$. If the operators are all unprotected, then there is only one $R$-symmetry channel, which is labeled $\Am^{6 \ldots 6}$.

The functions $g_{\Delta_1, \ldots, \Delta_{n-3}}$ correspond to the comb conformal blocks derived in \cite{Rosenhaus:2018zqn} and for identical external operators $\phi$, they are defined as
\begin{align}
&g_{\Delta_1,\dots,\Delta_{n-3}} (\chi_1, \dots, \chi_{n-3}) := \prod_{k=1}^{n-3}\, \chi_k^{\Delta_k} \notag \\
& \qquad \times F_{K}\left[\begin{array}{c}
\left.  \Delta_1, \Delta_1+\Delta_2-\Delta_{\phi}, \ldots, \Delta_{n-4}+\Delta_{n-3}-\Delta_{\phi},\Delta_{n-3}\right.\\
 2 \Delta_1, \ldots, 2 \Delta_{n-3}\end{array} ; \chi_{1}, \ldots, \chi_{n-3} \right]\,,
\label{eq:combblocks}
\end{align} 
where the function $F_K$ is a multivariable hypergeometric function defined by the following expansion:
\small
\begin{align}
& F_{K} \Bigg[\begin{array}{c}a_{1}, b_{1}, \ldots, b_{k-1}, a_{2} \\ c_{1}, \ldots, c_{k}\end{array} ; \, x_1, \ldots, x_{k}\Bigg]  \notag \\
& \qquad \qquad = \sum_{n_{1}, \ldots, n_{k}=0}^{\infty} \frac{\left(a_{1}\right)_{n_{1}}\left(b_{1}\right)_{n_{1}+n_{2}}\left(b_{2}\right)_{n_{2}+n_{3}} \cdots\left(b_{k-1}\right)_{n_{k-1}+n_{k}}\left(a_{2}\right)_{n_{k}}}{\left(c_{1}\right)_{n_{1}} \cdots\left(c_{k}\right)_{n_{k}}} \frac{x_{1}^{n_{1}}}{n_{1} !} \cdots \frac{x_{k}^{n_{k}}}{n_{k} !}\,.
\end{align}
\normalsize
Here $(a)_{n}=\Gamma(a+n) / \Gamma(a)$ refers to the Pochhammer symbol.

\subsubsection{Snowflake Channel} \label{sec:snowflake}

\begin{figure}[h]
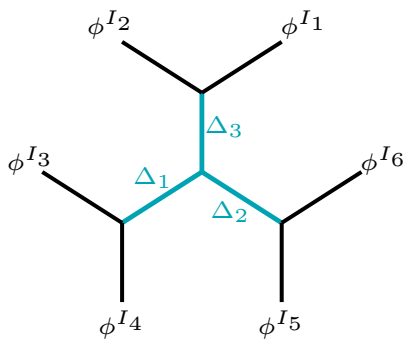

\centering
\snowblocks
\caption{\textit{Representation of six-point functions in the snowflake channel. Here the OPEs are taken pairwise between external operators and lead to the OPE coefficient consisting of products of four three-point functions, represented by the vertices. }}
\label{fig:snowch}
\end{figure}

We now move our attention to the other topology appearing for multipoint functions with $n = 6$,  the so-called \textit{snowflake} channel,  represented diagrammatically in Figure \ref{fig:snowch}. 

In this case, the OPE limits consist of bringing two neighboring external operators close to each other pairwise, and this has for consequence that the OPE coefficient in the middle can consist of operators that are all different from the external ones, as opposed to the comb channel of the previous section, where at least one external operator is present in the three-point functions. As above, we specialize our analysis to the case where all the external operators are identical and are of length $L=1$, i.e. correlation functions that involve either the protected fundamental scalars $\phi^i$ or the unprotected one $\phi^6$.

Explicit expressions for these blocks have already appeared in the literature \cite{Fortin:2020zxw}, however here we use different cross-ratios that make the blocks symmetric in all its arguments.\footnote{We thank Lorenzo Quintavalle for sharing these formulae with us.} Here is the new set of cross-ratios that we use:
\begin{equation}
z_1=\frac{\tau_{12}\tau_{46}}{\tau_{16}\tau_{24}}\,, \qquad z_2=\frac{\tau_{26}\tau_{34}}{\tau_{23}\tau_{46}}\,, \qquad z_3=\frac{\tau_{24}\tau_{56}}{\tau_{26}\tau_{45}}\,.
\end{equation}
As usual, six-point functions can then be decomposed into conformal prefactor and reduced correlator:
\begin{equation}
\vvev{\phi^{I_1} (\tau_1) \ldots \phi^{I_6} (\tau_6)} = \Km(\tau_1, \Delta_{\smash{\phi^{I_1}}}; \dots\,;\tau_6, \Delta_{\smash{\phi^{I_6}}})\, \Am^{I_1 \ldots I_6  } (z_1\,, z_2\,, z_3)\,.
\end{equation}
For the choice of the conformal prefactor, we also adopt the convention of \cite{Fortin:2022grf}, that we specialize to identical operators:
\begin{equation}
\Km(\tau_1, \Delta_{\phi}; \dots\,;\tau_6, \Delta_{\phi}) = \frac{1}{\tau_{12}^{2 \smash{\Delta_{\phi}}} \tau_{34}^{2 \smash{\Delta_{\phi}}} \tau_{56}^{2 \smash{\Delta_{\phi}}}}\,.
\end{equation}

For a given $R$-symmetry channel,  correlators can be expanded in the following way:
\begin{align}
\Am^{I_1 \ldots I_6} (z_1\,, z_2\,, z_3) =& \sum_{\Op_1\,, \Op_2\,, \Op_3} c_{\phi^{I_1} \phi^{I_2} \Op_1} c_{\phi^{I_3} \phi^{I_4} \Op_2} c_{\phi^{I_5} \phi^{I_6} \Op_3} c_{\Op_1 \Op_2 \Op_3} g_{\Delta_1, \Delta_2, \Delta_3} ( z_1, z_2,z_3 )\,,
\end{align}
where now the function $g_{\Delta_1, \Delta_2, \Delta_3}$ corresponds to the snowflake conformal blocks. For our purposes, we write a series expansion of the form
\begin{equation}
\,g_{\Delta_{1}, \Delta_2, \Delta_3}\left(z_1\,, z_2\,,z_3 \right) = z_1 ^{\Delta_1} z_2^{\Delta_2} z_3^{\Delta_3} \sum_{n_1, n_2, n_3} \bar{c}_{n_1, n_2, n_3} z_1^{n_1} z_2^{n_2} z_3^{n_3} \,,
\end{equation}
where we only need the coefficients $\bar{c}_{n_1, n_2, n_3}$ for low values of $n_1$, $n_2$, $n_3$\footnote{In \cite{Fortin:2020zxw} a different Taylor expansion is used with a closed-form expression for the corresponding coefficients $\bar{c}_{n_1, n_2, n_3}$.}. It is easy to determine the coefficients up to an overall normalization by applying the Casimir equations on the blocks order by order (see Appendix \ref{app:casimir}), and this results in the following expansion of the full correlator:
\begin{equation}
\Am^{I_1 \ldots I_6} (z_1\,, z_2\,, z_3) = 1 + c_{\phi^{I_1} \phi^{I_2} \Op_{\Delta=1}} c_{\phi^{I_3} \phi^{I_4} \Op_{\Delta=1}} c_{\phi^{I_5} \phi^{I_6} \mathds{1}} c_{\Op_{\Delta=1} \Op_{\Delta=1} \mathds{1}} z_1 z_2 + \ldots\,,
\label{eq:snowexp}
\end{equation}
where we have used the fact that terms with two $\Delta$'s set to zero vanish since one-point functions vanish. We note that $c_{\phi^{I_5} \phi^{I_6} \mathds{1}} c_{\Op_{\Delta=1} \Op_{\Delta=1} \mathds{1}} = 1$ because of the unit-normalization of two-point functions. We have labeled the exchanged operator with (bare) scaling dimension $\Delta=1$ as $\Op_{\Delta=1}$, but we see in Section \ref{sec:blockexpansion6pt} that in our two cases of interest, this operator always turns out to be $\phi^6$.

\section{Insertion Rules} \label{sec:insertionrulesdefect}

In Section \ref{sec:N=4Feynmanrules}, we discussed the Feynman diagrams and the insertions rules for the \textit{bulk} diagrams and their respective integrals.  However, in the presence of the line, new types of integral arise, in addition to the bulk ones.

In particular, we have to deal with two types of boundary integrals that we name \textit{$T$-integrals} and \textit{$U$-integrals}.

\subsection{$T$-Integrals} \label{sec:mainT}

We denote the first type of integrals that we encounter by $T_{ij;kl}$\footnote{This class of integrals also appears in \cite{Kiryu:2018phb}, where they are defined slightly differently and labeled as $B_{ij;kl}$.}, and define them to be
\begin{equation} \label{eq:Tdef}
T_{ij;kl} := \partial_{ij} \,\int_{\tau_k}^{\tau_l} d\tau_m\, \epsilon(ijm) \, Y_{ijm}\,,
\end{equation}
where $\epsilon(ijk)$ encodes the change of sign due to the path ordering, formally defined as
\begin{equation} \label{eq:epsdef}
\epsilon(ijk) := \text{sgn}\, \tau_{ij}\, \text{sgn}\, \tau_{ik}\, \text{sgn}\, \tau_{jk}\,.
\end{equation}

When the range of integration is the entire line, the integral is easy to perform and results in
\begin{equation} \label{eq:T1}
T_{ij;(-\infty)(+\infty)} = - \frac{I_{ij}}{12}\,.
\end{equation}
In the case where $(i,j) = (k,l)$ it gives
\begin{equation} \label{eq:T2}
T_{ij;ij} = \frac{I_{ij}}{12} \,.
\end{equation}

We enter into further detail about this class of integrals in Appendix \ref{subsubsec:Tintegrals}.

\subsection{$U$-Integrals} \label{sec:mainU}

Since the scalar $\phi^6$ couples directly to the Wilson line,  there is another class of integrals that we have to consider. We denote these integrals by $U_{a;ij}$ and they are defined as
\begin{equation} \label{eq:Uint}
U_{a;ij} := \int_{\tau_i}^{\tau_j} d\tau_n\,  I_{an}\,,
\end{equation}
where $a$ is the insertion point of the scalar $\phi^6$ on the Wilson line and $ij$ indicate the range of integration. 
These integrals can be easily performed explicitly, and their expression can be found in (\ref{eq:defU}).

Given that we are interested in next-to-leading order computations,  we have to consider also integrals of the type (\ref{eq:Uint}),  that arise when two scalars $\phi^6$ couple to the Wilson line.  We refer to these integrals as $U^{(2)}_{ab;ij}$ and they are defined as
\begin{equation} \label{eq:Utwo} 
U^{(2)}_{ab;ij}:=\int^{\tau_j}_{\tau_i}\, d\tau_n \,I_{an} U_{b;nj}= \int^{\tau_j}_{\tau_i}\, d\tau_n I_{an} \int^{\tau_j}_{\tau_n} d\tau_{m} I_{bm}  \,.
\end{equation}

In Appendix \ref{subsubsec:Uintegrals}, the explicit expressions of these integrals in three different configurations can be found (see equations (\ref{eq:Utwoexpl})). 

\begin{figure}[h]
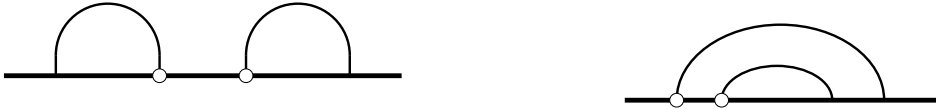

\centering
\begin{subfigure}{.5\textwidth}
  \centering
  \scalebox{1.3}{\twopointnpr}
\end{subfigure}%
\begin{subfigure}{.5\textwidth}
  \centering
  \scalebox{1.3}{\Utwo}
\end{subfigure}
\caption{\textit{Illustration of the two types of $U$-integrals that one can encounter: on the left, there are two $U$-integrals of the first type, while on the right, we have one of the second type. The main difference between them lies in the integration limits.}}
\label{fig:Udiagrams}
\end{figure}

\chapter{Mellin Representation Formalism} \label{ch:Mellin}

In this chapter, we introduce the last character of this thesis: the Mellin formalism for conformal theories.  In particular, we outline its main features for a generic higher-dimensional theory. The following quick review is mostly based on \cite{Penedones:2016voo}.

As we mentioned in the introduction Section \ref{sec:Mellinintro}, the Mellin representation has proven to be an excellent tool for finding simpler representations of correlation functions of local operators in CFTs. These are rather complicated functions of the cross-ratios, of crucial importance in the AdS/CFT correspondence. For this reason, we would like to apply this useful representation to the correlators we introduced in the previous chapter, the insertions on the Wilson line, and their holographic counterparts. Therefore, we set the stage for an inherently one-dimensional Mellin transform, which is the primary goal of this work \cite{Bianchi:2021piu}, and thus will be presented in detail in Chapter \ref{ch:Mellinres}.

The Mellin transform has been used in various contexts, for example, in computer science for the analysis of algorithms due to its scale-invariant properties. Although, it is only in 2009 that Mellin amplitudes were properly introduced by Mack \cite{Mack:2009mi,Mack:2009gy} in the context of CFT and further developed by Penedones \cite{Penedones:2010ue}. Throughout the last ten years, they turned out to be particularly useful for the study of holographic CFTs \cite{Fitzpatrick:2011ia,Paulos:2011ie,Rastelli:2017udc}, thanks to their similarity with scattering amplitudes of dual resonance models. In particular, they are crossing symmetric and meromorphic functions, manifesting a correspondence between their poles and the OPE coefficients.  We can take this analogy further, and in the case of holographic CFTs, we can get bulk flat space scattering amplitudes as a limit of dual CFT Mellin amplitudes.

\section{Definition}

Let us consider the $n$-point function of scalar primary operators and let us write this generic correlator in terms of the Mellin amplitude $M(\gamma_{ij})$\footnote{We use the shorthand notation $M(\gamma_{ij})$ to denote a function $M(\gamma_{12}, \gamma_{13},\dots)$ that depend on all the Mellin variables $\gamma_{ij}$.}:
\begin{equation} \label{eq:genericMellin}
\vev{\phi_1(x_1) \dots \phi_n(x_n)} =\int [d\gamma_{ij}]\, M(\gamma_{ij}) \prod^{n}_{1 \leq i < j \leq n} \frac{\Gamma(\gamma_{ij})}{|x_i-x_j|^{2\gamma_{ij}}}\,,
\end{equation}
where $\gamma_{ij}$ are the \textit{Mellin variables}. These need to satisfy some constraints imposed by conformal symmetry:
\begin{equation} \label{eq:Mellinconstraints}
\sum^n_{j=1} \gamma_{ij}=0\,, \qquad \gamma_{ij}=\gamma_{ji}\,, \qquad \gamma_{ii}=-\Delta_i\,,
\end{equation}
being $\Delta_i$ the scaling dimension of the operators. We can also define a \textit{reduced} Mellin amplitude $\hat{M}(\gamma_{ij})=M(\gamma_{ij}) \prod^n_{i<j} \Gamma(\gamma_{ij})$ for convenience\footnote{The choice of the $\Gamma$-functions to factorize out is important to compensate an essential singularity of the Mellin amplitude. For further details, check out Section \ref{sec:reggeandbound}.}.
We should notice that in the case of $n=2,3$, the Mellin variables are entirely fixed by the constraints (\ref{eq:Mellinconstraints}), therefore the Mellin representation just gives back the known form of the conformal two- (\ref{eq:conformaltwopoint}) and three-point functions (\ref{eq:conformalthreepoint}). The integration measure $[d\gamma_{ij}]$ runs over the independent Mellin variables, which depends on the number of spacetime dimensions $d$:
\begin{equation}
\begin{split}
n<d+1: & \qquad \frac{1}{2}\,n(n-3)\,, \\
n \geqslant d+1: & \qquad n d-\frac{1}{2}(d+1)(d+2)\,,
\end{split}
\end{equation}
where we include a factor of $\frac{1}{2\pi i}$ for each variable.
Finally, the integration contour runs typically parallel to the imaginary axis. However, the precise contour in the complex plane depends on the poles of the integrand. Normally, the contour must lie on the left/right of the semi-infinite sequences of poles that run to the left/right. Although, in some cases, to ensure the convergence of the integral, it is necessary to perform some ``regularizations'' in the form of subtractions. This procedure is explained in \cite{Penedones:2019tng} for the higher-dimensional case. My collaborators and I specialized it to the one-dimensional case as well in \cite{Bianchi:2021piu}. We explain this regularization in Section \ref{sec:convergenceandsubtractions}.

Coming back to the constraints (\ref{eq:Mellinconstraints}), we can actually solve them by introducing some fictitious ``momentum'' variables $p_i$, living in a $D$-dimensional space:
\begin{equation}
\gamma_{ij}=p_i \cdot p_j \,.
\end{equation}
These variables obey two conditions:
\begin{subequations}
\begin{align}
\text {``momentum conservation'':} \qquad \qquad &  \sum^n_{i=1}p_i=0 \,, \\
\text {``on-shell'' condition:} \qquad \qquad & p_i^2 =-\Delta_i \,.
\end{align}
\end{subequations}
Counting now these ``Mandelstam variables'':
\begin{equation}
\begin{split}
n<D: & \qquad \frac{1}{2} n(n-3) \,,\\
n \geqslant D: & \qquad n(D-1)-\frac{1}{2} D(D+1) \,,
\end{split}
\end{equation}
we conclude that the number of independent Mandelstam variables in $D$ dimensions coincides precisely with the number of independent conformal cross-ratios in $d$ dimensions if we set $D=d+1$, and of course with the number of integration variables.

\begin{figure}[h]
\centering
\includegraphics[scale=0.35]{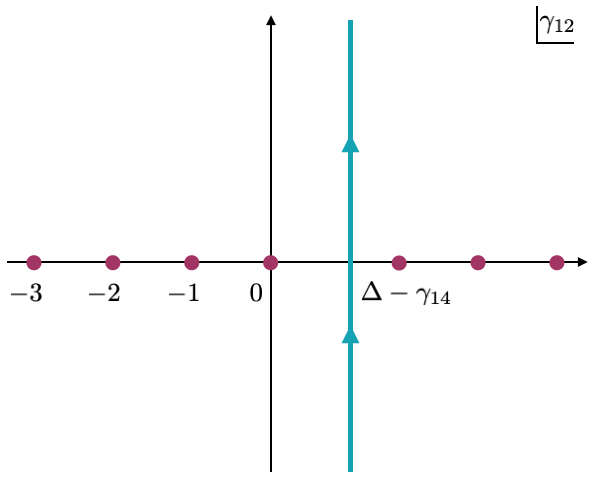}
\caption{\textit{Integration contour for the Mellin variable $\gamma_{12}$. The \textbf{\textcolor{vertexcolor}{dots}} represent the (double) poles of the $\Gamma$-functions given by (\ref{eq:polesgamma12}). In general, the Mellin amplitude has several semi-infinite sequences of poles, and each sequence should stay entirely on one side of the \textbf{\textcolor{gluoncolor}{contour}}.}}
\label{mellingamma12}
\end{figure}

We can consider the more specific case of a four-point function of scalar operators of dimension $\Delta$. In this case, there are two independent Mellin variables, which we choose to be $\gamma_{12}$ and $\gamma_{14}$, leading to
\begin{equation}
\vev{\phi_{\Delta}(x_1) \dots \phi_{\Delta}(x_4)} =\frac{1}{\left(x_{13} x_{24}\right)^{2 \Delta}} \int_{-i \infty}^{+i \infty} \frac{d \gamma_{12} \gamma_{14}}{(2 \pi i)^2} \hat{M}\left(\gamma_{12}, \gamma_{14}\right) u^{-\gamma_{12}} v^{-\gamma_{14}}\,,
\end{equation}
where $u$ and $v$ are the usual cross-ratios defined in (\ref{eq:crossratios4pt}) and 
\begin{equation}
\hat{M}\left(\gamma_{12}, \gamma_{14}\right)=M\left(\gamma_{12}, \gamma_{14}\right) \Gamma^2\left(\gamma_{12}\right) \Gamma^2\left(\gamma_{14}\right) \Gamma^2\left(\Delta-\gamma_{12}-\gamma_{14}\right) \,.
\end{equation}
To understand where the contour lies, we need to consider that the $\Gamma$-functions give rise to semi-infinite sequences of (double) poles at
\begin{equation} \label{eq:polesgamma12}
\begin{aligned}
& \gamma_{12}=0,-1,-2, \ldots \,,\\
& \gamma_{12}=\Delta-\gamma_{14}, \Delta-\gamma_{14}+1, \Delta-\gamma_{14}+2, \ldots\,,
\end{aligned}
\end{equation}
and the Mellin amplitude $M\left(\gamma_{i j}\right)$ also has the same type of semi-infinite sequences of poles, as we explore in the next section. The integration contour should then pass in the middle of these sequences of poles, as shown in Figure \ref{mellingamma12}. The invariance of the four-point function, under permutation of the insertion, points $x_i$, leads to the crossing symmetry of the Mellin amplitude:
\begin{equation}
M\left(\gamma_{12}, \gamma_{13}, \gamma_{14}\right)=M\left(\gamma_{13}, \gamma_{12}, \gamma_{14}\right)=M\left(\gamma_{14}, \gamma_{13}, \gamma_{12}\right) \,,
\end{equation}
where the three variables obey a single constraint: $\gamma_{12}+\gamma_{13}+\gamma_{14}=\Delta$. This is reminiscent of crossing symmetry for scattering amplitudes, written in terms of Mandelstam invariants.

\section{Connection with the OPE}

A characteristic that makes the Mellin representation so interesting and useful is that it encodes the OPE into simple analytical properties for $M(\gamma_{ij})$.  Let us write again the OPE of two operators (\ref{eq:OPE2}) that we discussed in Section \ref{sec:OPE}, this time making explicit the first contribution to the sum:
\begin{equation} 
\phi_i(x_i) \phi_j(x_j)  \overset{x_1\rightarrow x_2}{=}  \sum_{\phi_k} c_{\phi_1\phi_2\phi_k} \left( \left(x_{i j}^2\right)^{-\frac{\Delta_i+\Delta_j-\Delta_k}{2}} \phi_k\left(x_k\right)+\text {descendants} \right)\,,
\end{equation}
where for simplicity, $\phi_k$ is taken to be a scalar operator. To reproduce the leading behavior as $x_{i j}^2 \rightarrow 0$,  $M$ must have a pole at $\gamma_{i j}=\frac{\Delta_i+\Delta_j-\Delta_k}{2}$, as it can be seen by closing the $\gamma_{i j}$ integration contour to the left of the complex plane. More generally, the location of the leading pole is controlled by the twist $\tau$ of the exchanged operator\footnote{The twist $\tau$ of an operator is given by the conformal dimension minus the spin: $\tau \equiv \Delta-s$.}. For any primary operator $\phi_k$ of twist $\tau_k$ that contributes to the OPE, $\hat{M}(\gamma_{i j})$ has poles at
\begin{equation}
\gamma_{i j}=\frac{\Delta_i+\Delta_j-\tau_k-2 n}{2}, \quad n=0,1,2 \ldots\,,
\end{equation}
where the poles with $n>0$ correspond to the contributions from the descendants since they contribute with an infinite sequence of satellite poles. If the CFT has a discrete spectrum of scaling dimensions,  then its Mellin amplitudes are analytic functions with single poles at its singularities (meromorphic functions).  It is also clear that the residues of these poles are proportional to the product of the OPE coefficient $c_{\phi_1\phi_2\phi_k}$ and the Mellin amplitude of the lower point correlator $\vev{\phi_k\dots}$. For the precise formulae, check \cite{Mack:2009mi,Goncalves:2014rfa}.

We can point out another reason why it is useful to distinguish between the Mellin $M$ and the reduced Mellin $\hat{M}$.  In fact, $M$ has simpler factorization properties. In particular, we come back to the four-point function, and for convenience, we write the Mellin amplitude in terms of the ``Mandelstam invariants'':
\begin{equation}
\begin{aligned}
& \mathtt{s}=-\left(p_1+p_2\right)^2=\Delta_1+\Delta_2-2 \gamma_{12}\,, \\
& \mathtt{t}=-\left(p_1+p_3\right)^2=\Delta_1+\Delta_3-2 \gamma_{13}\,.
\end{aligned}
\end{equation}
Now the s-channel\footnote{This corresponds to take the OPE as illustrated in the LHS of (\ref{eq:OPEassociativity}).} OPE implies that the Mellin amplitude $M(\mathtt{s}, \mathtt{t})$ has poles in $\mathtt{s}$ with residues that are \textit{polynomials} of $\mathtt{t}$:
\begin{equation}
M(\mathtt{s}, \mathtt{t}) \approx c_{\phi_1\phi_2 \phi_k} c_{\phi_3\phi_4 \phi_k} \frac{Q_{s_k, n}(\mathtt{t})}{\mathtt{s}-\Delta_k+s_k-2 n}, \quad n=0,1,2, \ldots\,,
\end{equation}
where $Q_{s, n}(\mathtt{t})$ are kinematical polynomials of degree $s$ in the variable $\mathtt{t}$, called \textit{Mack polynomials}.
This strengthens the analogy with scattering amplitudes. Each operator of spin $s$ in the OPE $\phi_i \times \phi_j$ gives rise to poles in the Mellin amplitude that are very similar to the poles in the scattering amplitude associated with the exchange of a particle of the same spin.

\section{Planar Correlators}

We now focus our attention on correlators that are relevant in our case, which is in the planar limit. We discussed the details of this regime in Section \ref{sec:planarlimit}.

The Mellin formalism is quite powerful when applied to planar correlators. While in a general CFT the analytic structure of Mellin amplitudes is quite complicated, it gets much simpler at large $N$.  To appreciate this characteristic, we can analyze two different perspectives.

On the one hand, we can note that in the Mellin representation the $\Gamma$-functions have themselves poles at fixed positions.  However, in a generic CFT, there are no operators with these scaling dimensions. Therefore the Mellin amplitude must have zeros at these values to cancel these unwanted OPE contributions.  Nevertheless, we expect precisely this type of contributions in correlation functions of single-trace operators in large $N$ CFTs. 

On the other hand,  a remarkable theorem \cite{Fitzpatrick:2012yx,Komargodski:2012ek} about the spectrum of CFTs in dimension $d>2$ states that for any two primary operators $\phi_i$ and $\phi_k$ of twists $\tau_i$ and $\tau_j$, and for each non-negative integer $k$, the CFT must contain an infinite family of the so-called ``double-twist'' operators with increasing spin $s$ and twist approaching $\tau_1+\tau_2+2 k$ as $s \rightarrow \infty$. This implies that the Mellin amplitude has infinite sequences of poles accumulating at these asymptotic values of the twist, so it is not a meromorphic function. 

Merging together these two perspectives, we can note where a key simplification occurs in large $N$ CFTs: the double-twist operators are recognized as the usual double-trace operators \cite{Penedones:2010ue}. Therefore we witness a factorization: the poles corresponding to the exchanged double-trace operators are precisely captured by the product of the gamma functions $\prod^n_{i<j} \Gamma(\gamma_{ij})$, while the Mellin $M$ has only poles associated with single-trace operators.

\section{Holographic CFTs} \label{sec:holographiccfts}

We conclude this introductory chapter on Mellin amplitudes by showing how this formalism can be extremely effective in simplifying the expressions of holographic correlators.  This representation could be considered analogous to the Fourier transform for these correlators.

\begin{figure}[h]
\centering
\includegraphics[scale=0.3]{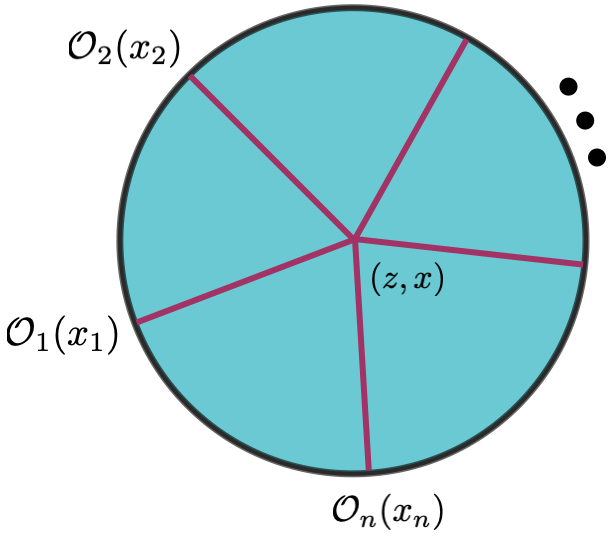}
\caption{\textit{Witten diagram for an $n$-point contact interaction in AdS. The interior of the disk represents the bulk of AdS, and the circumference represents its conformal boundary. The \textbf{\textcolor{vertexcolor}{lines}} connecting the boundary points $x^\prime_i$ to the bulk point $(z,x)$ represent bulk-to-boundary propagators.}}
\label{contactdiagram}
\end{figure}

The best way to illustrate this property is by considering a calculation of a Witten diagram. In particular, we consider a contact Witten diagram as in Figure \ref{contactdiagram}, which corresponds to an interaction vertex $\lambda \phi_1\dots \phi_n$ in the bulk lagrangian, what it is commonly called a \textit{D-function}.  

For external dimensions $\Delta_i$ and in the general case of AdS$_{d+1}$, it can be defined as
\begin{equation}
\vev{\Om_1(x_1)\dots \Om_n(x_n)}=\int \frac{d z d^d x}{z^{d+1}} \tilde{K}_{\Delta_i}\left(z, x ; x^\prime_1\right)\dots \tilde{K}_{\Delta_i}\left(z, x ; x^\prime_n\right)
\end{equation}
via the bulk-to-boundary propagator in $d$ dimensions
\begin{equation}
K_{\Delta_i}\left(z, x ; x^{\prime}\right)=\mathcal{C}_{\Delta_i}\left[\frac{z}{z^2+\left(x-x^{\prime}\right)^2}\right]^{\Delta_i} \equiv \mathcal{C}_{\Delta_i} \tilde{K}_{\Delta_i}\left(z, x ; x^{\prime}\right), \quad \mathcal{C}_{\Delta_i}=\frac{\Gamma\left(\Delta_i\right)}{2 \sqrt{\pi} \Gamma\left(\Delta_i+\frac{1}{2}\right)} \,.
\end{equation}
Surprisingly the Mellin amplitude of this diagram is just a constant \cite{Penedones:2010ue}:
\begin{equation}
M=\int [d\gamma_{ij}] \left(\frac{\pi^{d / 2} \Gamma\left[\frac{\sum \Delta_i-d}{2}\right]}{\prod \Gamma\left[\Delta_i\right]}\right) \times \prod_{i<j} \Gamma\left[\delta_{i j}\right]\left(x_{i j}^2\right)^{-\delta_{i j}} \,.
\end{equation}
This can also be generalized to interaction vertices with derivatives \cite{Penedones:2010ue}. 

\begin{figure}[h]
\centering
\includegraphics[scale=0.3]{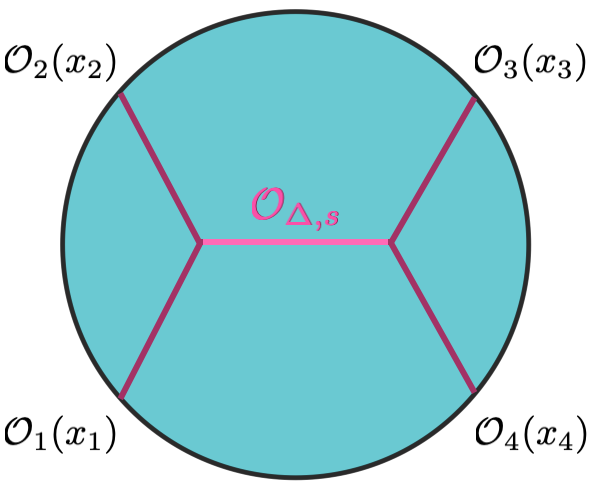}
\caption{\textit{Witten diagram depicting the exchange of a bulk field, dual to an operator of dimension $\Delta$ and spin $s$.}}
\label{exchangediagram}
\end{figure}

Moreover, one can consider more complicated Witten diagrams, which are rather involved functions of the cross-ratios but are again much simpler in Mellin space. They describe the exchange of a bulk field dual to a single-trace boundary operator $\mathcal{O}_\Delta$ of spin $s$, as depicted in Figure \ref{exchangediagram}.

The conformal block decomposition of this diagram in the s-channel contains the single-trace operator $\mathcal{O}$ plus double-trace operators, schematically of the form $\mathcal{O}_1\left(\partial^2\right)^n \partial_{\mu_1 \ldots \mu_j} \mathcal{O}_2$ and $\mathcal{O}_3\left(\partial^2\right)^n \partial_{\mu_1 \ldots \mu_j} \mathcal{O}_4$. Moreover, the OPE in the crossed channels only contains double-trace operators. This means that the Mellin amplitude is of the form
\begin{equation}
M(\mathtt{s},\mathtt{t})=c_{\Om_1\Om_2 \mathcal{O}} c_{\Om_3\Om_4 \mathcal{O}} \sum_{n=0}^{\infty} \frac{Q_{s, n}(\mathtt{t})}{\mathtt{s}-\Delta+s-2 n}+R(\mathtt{s}, \mathtt{t})\,,\quad n=0,1,2,\dots\,,
\end{equation}
where the OPE coefficients $c_{\Om_1\Om_2 \mathcal{O}}$ and $c_{\Om_3\Om_4 \mathcal{O}}$ are proportional to the bulk cubic couplings and $R(\mathtt{s}, \mathtt{t})$ is an analytic function. The residues are proportional to degree $s$ Mack polynomials $Q_{s,n}(\mathtt{t})$, which are entirely fixed by conformal symmetry.

To conclude, the Mellin amplitude was also determined for general tree-level scalar Witten diagrams \cite{Fitzpatrick:2011ia,Paulos:2011ie,Fitzpatrick:2011hu}.


\part{Research}

\interfootnotelinepenalty=10000 

\chapter{Recursion Relations} \label{sec:recursionrelations}

After having introduced all the necessary technical background, it is time to dive into research. The first stop is dedicated to the development of the techniques and the tools and it is based on \cite{Barrat:2021tpn,Barrat:2022eim}. 

The goal for this chapter is to derive a set of recursion relations that allow us to compute arbitrary correlators of insertions of fundamental scalar fields $\phi^I$ on the Maldacena-Wilson line.

Particularly, we derive recursive formulae at weak coupling for both leading and next-to-leading orders, in the case where an even number of $\phi^6$ scalars are inserted, while we obtain only the leading order for the odd case.

We restrict this analysis to the large $N$ limit, introduced in Section \ref{sec:planarlimit}, and use the 't Hooft coupling $\lambda := g^2 N$ as the parameter of the perturbative expansion.

The plan is to first focus on correlators of the protected operators $\phi^i$. Once we master those relations, we include the (unprotected) scalar $\phi^6$ and study which complications brings to the recursive formulae.

\section{Recursion for Protected Operators} \label{sec:recursionprotected}

As a first step, we only consider operator of length $L=1$, not only because it is, of course, the simplest case but also because, thanks to the pinching property mentioned in Section~\ref{sec:wilsonlinecorrelators}, it is enough to compute all the higher-point correlators of protected scalars.

For compactness, we define the following shorthand notation:
\begin{equation}  \label{eq:Arecursionprotected}
A_n(1,\ldots , n) := \vev{\phi^{i_1}_1 (u_1,\tau_1) \ldots \phi^{i_n}_1 (u_n, \tau_n)}\,,
\end{equation}
which we shall be studying at leading order (LO) and next-to-leading (NLO) order precision.  Notice that for odd $n$ the correlators $A_n$ vanish due to $R$-symmetry. Moreover, we consider correlation functions that are \textit{not} unit-normalized, since this is the natural normalization to work with when doing perturbative computations. However, the results presented in the subsequent chapters are unit-normalized.

\subsection{Leading Order} \label{subsec:tree}

We start by deriving a leading-order formula. For operators of scaling dimension $\Delta = 1$, it is easy to find a recursive expression for $n$-point functions at leading order.  In fact, this problem is related to a more mathematical one concerning meanders and arch statistics, which was already solved in \cite{DiFrancesco:1995cb}. Adapting (3.1) of that paper to our case of interest, we obtain the recursion
\begin{equation} \label{eq:recursiontree}
A_n^\text{\tiny{LO}}(1,\ldots , n)  = \sum_{j=0}^{\frac{n}{2}-1} A_2^\text{\tiny{LO}}(1,2j+2) A_{2j}^\text{\tiny{LO}}(2,\ldots, 2j+1) A_{n-2-2j}^\text{\tiny{LO}}(2j+3,\ldots, n)\,,
\end{equation}
which can be represented diagrammatically as
\begin{equation} \label{eq:recursiontreediagrams}
A_n^\text{\tiny{LO}}(1,\ldots , n) = \sum_{j=0}^{\frac{n}{2}-1} \recursiontree\,,
\end{equation}
where \treelevelblob \, stands for the leading-order correlation function of appropriate length.

In the expression above, the starting values for the recursion are given by the vacuum expectation value and by the two-point functions:
\begin{equation} \label{eq:startingvalstree}
A_0^\text{\tiny{LO}} = 1\,, \qquad A_2^\text{\tiny{LO}}(i,j) = \frac{\lambda}{8 \pi^2} (ij)\,,
\end{equation}
with $(ij)$ being
\begin{equation}
(i j):=\frac{\left(u_i \cdot u_j\right)}{\tau_{i j}^2}\,,\qquad \qquad \tau_{ij}:= \tau_i-\tau_j\,.
\end{equation}
The two-point $A_2^{\text{LO}}$ is discussed more in detail in Section \ref{subsubsec:twoptone}.

As mentioned above, only correlators with an even number of operators are non-vanishing, and two- and four-point functions can be compared with the results of \cite{Kiryu:2018phb}, with which they agree perfectly.

\subsection{Next-to-Leading Order} \label{subsec:oneloop}

At next-to-leading order, the situation becomes more intricate, not only because of the appearance of $4d$ vertices, but also because some of them couple to the Wilson line. Nevertheless, we can write a recursive diagrammatic formula, which produces all the relevant Feynman diagrams for an arbitrary $n$-point function of $\phi^i_1$ operators:
\begin{align} \label{eq:recursionloop}
A_n^\text{\tiny{NLO}}(1,\ldots , n) =& \sum_{i=1}^{n-3} \sum_{j=i+1}^{n-2} \sum_{k=j+1}^{n-1} \sum_{l=k+1}^{n} \recursionFour + \sum_{i=1}^{n-1} \sum_{j=i+1}^{n} \recursionSE\notag \\
& + \sum_{i=1}^{n-1} \sum_{j=i+1}^{n} \Biggl( \sum_{k=0}^{i-1} \recursionYNone  + \sum_{k=i}^{j-1} \recursionYNtwo  \notag \\
&+ \sum_{k=j}^{n+1} \recursionYNthree \Biggr)  + \sum_{i=1}^{n-3} \sum_{j=i+3}^{n} \recursionbridge\,.  
\end{align}

Let us analyze this expression. First of all,  its recursiveness is encoded in two different aspects: on the one hand in the dependency on the leading-order correlators, which obey the relation given in \eqref{eq:recursiontree},  on the other hand on the NLO correlators, as one can note from the last term of the sum, where one should insert the NLO correlation function with $n=j-i-1$\footnote{Note that it is recursive with respect to correlators on a \textit{finite} line and not extending to $\pm \infty$. This is explained in more detail below \eqref{eq:formulaOneloop}.}. 

Let us now describe in detail each term that appears in \eqref{eq:recursionloop}.

\subsubsection{Bulk Diagrams}

We start by looking at the diagrams associated with the bulk interactions, which are displayed on the first line of (\ref{eq:recursionloop}). In particular, the first one corresponds to the possible NLO insertions involving four scalar lines, namely
\begin{equation}
\vertexFourno\ := \vertexSSSSno\ + \Hinsertoneno\ + \Hinserttwono\ ,
\end{equation}
and the associated diagram reads
\begin{align} \label{eq:contribX}
\recursionFour &= A_4^4 (i,j,k,l) A_{i-1}^\text{\tiny{LO}} (1, \ldots, i-1) A_{j-i-1}^\text{\tiny{LO}} (i+1, \ldots, j-1) \notag \\
& \qquad\qquad \raisebox{2ex}{$\times A_{k-j-1}^\text{\tiny{LO}} (j+1, \ldots, k-1) A_{l-k-1}^\text{\tiny{LO}} (k+1, \ldots, l-1)$} \notag \\
& \qquad\qquad\qquad \times A_{n-l}^\text{\tiny{LO}} (l+1, \ldots, n)\,.
\end{align}
Using the insertion rules \eqref{eq:vertexSSSS} and \eqref{eq:Hinsertone}, we define the following four-point correlator as a NLO building block:
\begin{align}
A_4^4 (i,j,k,l) :=& \frac{\lambda^3}{8} \left[\left(2 (u_i \cdot u_k) (u_j \cdot u_l) -(u_i \cdot u_l)(u_j \cdot u_k)-(u_i \cdot u_j)(u_k \cdot u_l)\right) X_{ijkl} \right.  \notag \\
& \left. + (u_i \cdot u_l)(u_j \cdot u_k) I_{il} I_{jk} F_{il;jk}-(u_i \cdot u_j)(u_k \cdot u_l) I_{ij} I_{kl} F_{ij;kl} \right]\,.
\end{align}
The results for the integrals $X_{ijkl}$ and $F_{ij;kl}$ associated with the insertion rules can be found in Appendix \ref{subsec:bulkintegrals}. Note that all the diagrams encompassed by this term are perfectly finite as long as the external points are distinct. 

The second and third lines in \eqref{eq:recursionloop}, corresponding to self-energy and $Y$-diagrams, also add up to a finite result since divergences coming from the self-energy cancel with the ones arising in the $Y$-diagrams.  To be precise, the self-energy diagrams read
\begin{equation}
\recursionSE = A_2^\text{\tiny{SE}} (i,j) A_{i-1}^\text{\tiny{LO}} (1, \ldots, i-1) A_{j-i-1}^\text{\tiny{LO}} (i+1,  \ldots, j-1) A_{n-j}^\text{\tiny{LO}} (j+1, \ldots, n)\,,
\label{eq:formulaSE}
\end{equation}
where we use as a building block the well-known expression for the scalar propagator at NLO (see \eqref{eq:selfenergy})
\begin{equation} 
A_2^\text{\tiny{SE}} (i,j) := - \lambda^2\, (u_i \cdot u_j)\, Y_{iij}\,.
\label{eq:A2SE}
\end{equation}

\subsubsection{Boundary Diagrams}

We turn now our attention to the boundary contributions, which are encoded in the diagrams on the second line and in the first diagram of the third line of (\ref{eq:recursionloop}). In particular, we start with the $Y$-diagrams. Note that inserting a gluon field on the Wilson line,  which we indicate with \textcolor{gluoncolor}{light blue} dots, corresponds to expanding the exponential of \eqref{eq:correlators} up to the first order, and thus it results in a one-dimensional integral between the points before and after the insertion. For example,
\begin{equation}
\recursionYexplainone := \int_{\tau_k}^{\tau_{k+1}} d\tau_{\alpha}\ \recursionYexplaintwo\,.
\end{equation}

In order to show that the divergences cancel with the ones coming from the self-energy graphs, it is convenient to express the $Y$-diagrams differently:
\begin{align}
\sum_{k=0}^{i-1} \recursionYNone =&\ \recursionYLidone \notag \\
& - \sum_{\alpha = 1}^{i-2} \sum_{\beta = \alpha+1}^{i-1} \recursionYLtwo\,,
\label{eq:reformulateY}
\end{align}
where the red dots indicate the places where the gluon line should be connected. The sum over $i, j$ is implied here. 
It should be clarified that the diagrams on the right-hand side should \textit{not} be considered non-planar when the gluon line is crossing a scalar line. Similarly, such a crossing does \textit{not} generate an additional 4$d$ vertex. Here the dots are intended to only indicate the range of integration.
It is easy to check that the relation \eqref{eq:reformulateY} holds by rewriting the integration limits of the left-hand side as $\int_a^b = \int_{-\infty}^{\infty} - \int_{-\infty}^a - \int_b^{\infty}$.

The same can be performed for the other two terms with $Y$-vertices, and we are left with the following diagrams to compute:
\begin{gather*}
\recursionYone - \sum_{\alpha = 1}^{i-2} \sum_{\beta = \alpha+1}^{i-1} \recursionYLtwo \\ - \sum_{\alpha = i+1}^{j-2} \sum_{\beta = \alpha+1}^{j-1} \recursionYCtwodue -  \sum_{\alpha = i+1}^{n-1} \sum_{\beta = \alpha+1}^{n} \recursionYRtwo\,,
\end{gather*}
which of course also have to be summed over $i$ and $j$. By doing so, we have isolated the divergences inside the first term since the integration ranges of the other terms do not include the points $\tau_i$ and $\tau_j$. Moreover, since the limits of integration of the first term are $-\infty$ and $+\infty$, we can perform the integral analytically and extract the divergences. We then find
\begin{equation}
\recursionYone = A_2^\text{\tiny{Y,div}} (i,j) A_{i-1}^\text{\tiny{LO}} (1, \ldots, i-1) A_{j-i-1}^\text{\tiny{LO}} (i+1,  \ldots, j-1) A_{n-j}^\text{\tiny{LO}} (j+1, \ldots, n)\,,
\label{eq:formulaYdiv}
\end{equation}
with
\begin{equation}
A_2^\text{\tiny{Y,div}} (i,j) := \lambda^2 (u_i \cdot u_j)\, Y_{iij} + \frac{\lambda^2}{4} (u_i \cdot u_j)\, T_{ij;0(n+1)}\,,
\label{eq:A2Ydiv}
\end{equation}
where the integral $T_{ij;kl}$ is defined in \eqref{eq:Tdef}, with the subscripts $0$ and $n+1$ referring respectively to $-\infty$ and $+\infty$. The case $(0,n+1) = (-\infty,+\infty)$ can be found in \eqref{eq:T1}. Since the recursive structures of \eqref{eq:formulaSE} and \eqref{eq:formulaYdiv} are identical, it is clear that the divergences of \eqref{eq:A2SE} and \eqref{eq:A2Ydiv} cancel perfectly, and since the remaining $T$-integrals are finite, we are left with a finite expression.

The remaining term on the RHS of \eqref{eq:reformulateY} is also finite and reads
\begin{align}
\recursionYLtwo =& A_4^\text{\tiny{Y}} (i,j,\alpha,\beta) A^\text{\tiny{LO}}_{\alpha-1} (1, \ldots, \alpha-1) \notag \\
& \raisebox{2ex}{$\times A^\text{\tiny{LO}}_{\beta-\alpha-1} (\alpha+1, \ldots, \beta-1) A^\text{\tiny{LO}}_{i-\beta-1} (\beta+1, \ldots, i-1)$} \notag \\
& \times A^\text{\tiny{LO}}_{j-i-1} (i+1, \ldots, j-1) A^\text{\tiny{LO}}_{n-j} (j+1, \ldots, n)\,,
\label{eq:Ynodiv1}
\end{align}
where the starting point evaluates to
\begin{equation}
A_4^\text{\tiny{Y}} (i,j,\alpha,\beta) := \frac{\lambda^3}{32 \pi^2 \tau_{\alpha\beta}^2} (u_\alpha \cdot u_\beta) (u_i \cdot u_j)\, T_{ij;\alpha\beta}\,.
\label{eq:A4Y}
\end{equation}

The two other terms (center and right) can be implemented in the very same way. For the center term, we have
\begin{align}
\recursionYCtwodue =& A_4^\text{\tiny{Y}} (i,j,\alpha,\beta) A^\text{\tiny{LO}}_{i-1} (1, \ldots, i-1) \notag \\
& \raisebox{2ex}{$\times A^\text{\tiny{LO}}_{\alpha-i-1} (i+1, \ldots, \alpha-1) A^\text{\tiny{LO}}_{\beta-\alpha-1} (\alpha+1, \ldots, \beta-1)$} \notag \\
& \times A^\text{\tiny{LO}}_{j-\beta-1} (\beta+1, \ldots, j-1) A^\text{\tiny{LO}}_{n-j} (j+1, \ldots, n)\,,
\label{eq:Ynodiv2}
\end{align}
while for the right one, we obtain
\begin{align}
\recursionYRtwo =& A_4^\text{\tiny{Y}} (i,j,\alpha,\beta) A^\text{\tiny{LO}}_{i-1} (1, \ldots, i-1) \notag \\
& \raisebox{2ex}{$\times A^\text{\tiny{LO}}_{j-i-1} (i+1, \ldots, j-1) A^\text{\tiny{LO}}_{\alpha-j-1} (j+1, \ldots, \alpha-1)$} \notag \\
& \times A^\text{\tiny{LO}}_{\beta-\alpha-1} (\alpha+1, \ldots, \beta-1) A^\text{\tiny{LO}}_{n-\beta} (\beta+1, \ldots, n)\,.
\label{eq:Ynodiv3}
\end{align}
The integrals in \eqref{eq:A4Y} give different results depending on the ordering of the variables $\tau_i, \tau_j, \tau_\alpha, \tau_\beta$. The results have been collected in \eqref{eq:Tint}.

\subsubsection{Recursive Diagram}

The last term of the formula given in \eqref{eq:recursionloop} encodes the recursiveness with respect to the full next-to-leading order formula:
\begin{equation}
\recursionbridge = A_{j-i-1}^\text{\tiny{NLO}} (i+1,\ldots,j-1) A_{i-1}^\text{\tiny{LO}} (1,\ldots,i-1) A_2^\text{\tiny{LO}}(i,j) A_{n-j}^\text{\tiny{LO}} (j+1,\ldots,n)\,.
\label{eq:formulaOneloop}
\end{equation}
However, we must be careful here, because the limits of integration for the inserted NLO expression are \textit{not} $\pm \infty$ but $(0, n+1) = (i,j)$ for the $Y$-diagrams in \eqref{eq:A2Ydiv}.

The recursion relation \eqref{eq:recursionloop} and the associated expressions are enough to determine $n$-point functions of protected operators of length $\Delta=1$.  We meet correlators computed with this recursive formula in the following chapters. Anyway,  many correlation functions have been obtained using this recursion and can be found in \cite{Barrat:2021tpn} and in the ancillary \textsc{Mathematica} notebook.

\section{Recursion Including Unprotected Operators} \label{sec:recursiongeneral}
\begingroup
\allowdisplaybreaks

It is now time to add the remaining field $\phi^6$ in the recursion relation.  In particular,  we have to distinguish between two cases: the formulae depend on whether an \textit{even} or \textit{odd} number of $\phi^6$ operators is inserted on the Wilson line.

Similarly to (\ref{eq:Arecursionprotected}), in this section, we use for compactness the following shorthand notation for the correlators:
\begin{equation} \label{eq:defA}
A^{I_1 \ldots I_n} := \vev{\phi^{I_1} (\tau_1) \ldots \phi^{I_n} (\tau_n)}\,.
\end{equation}
Note that this differs from (\ref{eq:Arecursionprotected}) by the fact that we keep the $R$-symmetry indices open. 

Of course, it is still valid that a correlator containing an odd number of protected scalars $\phi^i$ vanishes, because of the $R$-symmetry indices. Therefore, in the following, we consider the number of $\phi^i$ to always be even.

\subsection{Even Case} \label{subsec:evencase}

We start our analysis by studying the case where an \textit{even} number of unprotected scalars $\phi^6$ is included in the correlator. This provides a generalization of the equations (\ref{eq:recursiontreediagrams}) and (\ref{eq:recursionloop}).

\subsubsection{Leading Order} \label{subsubsec:evenLO}

We begin with a formula for the leading order. In this case, computing correlation functions with an even number of $\phi^6$ operators is the same as computing correlation functions of only protected operators $\phi^i$,  and thus the recursion relation is the same as equation (\ref{eq:recursiontree}),  with the difference that we now keep the $R$-symmetry indices open:
\begin{equation}
A_\text{\tiny{LO}}^{I_1 \ldots I_n}  = \sum_{j=0}^{\frac{n}{2}-1} A_\text{\tiny{LO}}^{I_1 I_{2j+2}} A_\text{\tiny{LO}}^{I_2 \ldots I_{2j+1}} A_\text{\tiny{LO}}^{I_{2j+3} \ldots I_n}\,.
\label{eq:recursioneventree}
\end{equation}
The diagrammatical representation is identical to (\ref{eq:recursiontreediagrams}).

Arbitrary correlation functions of scalar fields can then be obtained by selecting $R$-symmetry indices, as long as the number of $\phi^6$ is kept even.

In the expression above, the starting values for the recursion are given by the vacuum expectation value and by the two-point functions:
\begin{equation} 
A_\text{\tiny{LO}} = 1\,, \qquad \quad A_\text{\tiny{LO}}^{I_1 I_2} = \frac{\lambda}{8 \pi^2} \frac{\delta^{I_1 I_2}}{\tau_{12}^2}\,.
\label{eq:evenstart}
\end{equation}

\subsubsection{Next-to-Leading Order} \label{subsubsec:evenNLO}

We turn our attention to the next-to-leading order, where the recursion relation becomes more involved.
For arbitrary operators (still with an even number of $\phi^6$), it consists of the diagrams appearing in $\vev{\phi^{i_1} \ldots \phi^{i_n}}$, which must be complemented with the $U$-diagrams, already introduced in Section \ref{sec:mainU}, that account for the coupling of $\phi^6$ to the Wilson line. An example of such diagrams can be found e.g.  on the right in Figure \ref{fig:Lambda}. Explicitly, we have
\begin{equation}
A_\text{\tiny{NLO}}^{I_1 \ldots I_n} = \left. A_\text{\tiny{NLO}}^{I_1 \ldots I_n} \right|_\text{old} + \left. A_\text{\tiny{NLO}}^{I_1 \ldots I_n} \right|_\text{new}\,,
\label{eq:oldnew}
\end{equation}
where $\left. A_\text{\tiny{NLO}}^{I_1 \ldots I_n} \right|_\text{old}$ refers to (\ref{eq:recursionloop})\footnote{Note again that when we deal with unprotected operators we keep the indices open. This change is easy to implement in (\ref{eq:recursionloop}) by removing the null-vectors $u_i$ and keeping the $R$-symmetry indices of the fundamental fields open.}, while the second term can be determined by considering all the possible $U$-contractions:
\begin{align}
\left. A_\text{\tiny{NLO}}^{I_1 \ldots I_n} \right|_\text{new} =& \sum_{j=1}^{n-1} \sum_{k=j+1}^n \Biggl( \sum_{l=k}^{n-2} \sum_{m=l+2}^n \recursionphione + \sum_{l=k}^n \sum_{m=j}^{k-1} \recursionphitwo \notag \\
& \qquad\qquad\quad + \sum_{l=0}^{j-3} \sum_{m=l+2}^{j-1} \recursionphithree + \sum_{l=j}^{k-1} \sum_{m=0}^{j-1} \recursionphifour  \notag \\
& \qquad\qquad\quad + \sum_{l=j}^{k-1} \sum_{m=k}^n \recursionphifive  + \sum_{l=j}^{k-3} \sum_{m=l+2}^{k-1} \recursionphisix \notag \\
& \qquad\qquad\quad + \sum_{l=0}^{j-1} \sum_{m=k}^n \recursionphiseven + \sum_{l=0}^{j-1} \sum_{m=j}^{k-1} \recursionphieight
\Biggr) \notag \\
& + \sum_{j=1}^{n-3} \sum_{k=j+2}^{n-1} \Biggl( \sum_{l=k+1}^n \sum_{m=l}^n \recursionphinine + \sum_{l=k+1}^n \sum_{m=k}^{l-1} \recursionphiten \Biggr) \notag \\
& + \sum_{j=2}^{n-2} \sum_{k=j+2}^n \Biggl( \sum_{l=1}^{j-1} \sum_{m=l}^{j-1} \recursionphieleven + \sum_{l=1}^{j-1} \sum_{m=0}^{l-1} \recursionphitwelve \Biggr) \notag \\
& + \sum_{j=1}^{n-3} \sum_{k=j+2}^{n-1} \sum_{l=k+1}^n \sum_{m=0}^{j-1} \recursionphithirteen +  \sum_{j=3}^{n-2} \sum_{k=j+2}^{n} \sum_{l=1}^{j-1} \sum_{m=k}^{n} \recursionphifourteen \notag \\
& + \sum_{j=1}^{n-5} \sum_{k=j+2}^{n-3} \sum_{l=k+1}^{n-2} \sum_{m=l+2}^n \recursionphififthteen\,,
\label{eq:recursionevenNLO}
\end{align}
where every sum should be considered as going in steps of $2$. In the recursion, we find \treelevelblobodd\, to indicate that a leading-order contribution of appropriate \textit{odd} length has to be inserted there. These contributions are derived in Section \ref{subsec:oddcase} and are given in equation \eqref{eq:recursionoddtree}.

This recursive expression is lengthy but easy to understand: it corresponds to summing over all the possible $U$-diagrams. When propagators end on the Wilson line without a dot, it means that the integration limit of the $U$-integral goes from the previous propagator to the next. More concretely:
\begin{equation}
\Uoneint\, :=  \int_{\tau_i}^{\tau_j} d\tau_n\, I_{an}\,,
\label{eq:Uoneint}
\end{equation}
where we have not included the leading-order insertions \treelevelblob\, on the right-hand side for the sake of clarity.

The explicit form of this diagrammatic expression is particularly long, therefore we give it in Appendix \ref{sec:recursionapp}. It is important to note that two types of $U$-integrals appear in that expression, the ones defined in \eqref{eq:Uint} and \eqref{eq:Utwo}.

This formula has been implemented in the ancillary notebook of \cite{Barrat:2022eim} and can readily be used for producing arbitrary correlators composed of fundamental scalar fields $\phi^I$ ($I=1\,, \ldots\,, 6$). The technical details related to the recursion relation and the $U$-integrals can be found in Appendices \ref{subsubsec:Uintegrals} and \ref{sec:recursionapp}.

In later sections, we refer to the terms $\left. A_\text{\tiny{NLO}}^{I_1 \ldots I_n} \right|_\text{old}$ appearing in equation \eqref{eq:oldnew} as \textit{building blocks} since these terms appear in all the correlation functions involving fundamental scalar fields. On the other hand, the second term $\left. A_\text{\tiny{NLO}}^{I_1 \ldots I_n} \right|_\text{new}$ is only relevant when some of the $R$-symmetry indices are set to $I_k = 6$.

The recursive nature of this relation is again encoded in (\ref{eq:formulaOneloop}), where now the \textit{full} expression $A_\text{\tiny{NLO}}^{I_1 \ldots I_n}$ on the left-hand side of \eqref{eq:oldnew} should be used as input for this recursive term.

\subsection{Odd Case} \label{subsec:oddcase}

We finally consider the case where an \textit{odd} number of $\phi^6$ appears in the correlators, while the number of protected scalars $\phi^i$ is still kept even. We restrict our analysis to the leading order since a coupling to the Wilson line already appears here and hence it corresponds to the interacting theory.

Diagrammatically the formula reads
\begin{align}
A_{\text{\tiny{LO}}}^{I_1 \ldots I_n} (1,\ldots , n) =& \sum^{n}_{i=1} \left( \sum_{j=0}^{\frac{i-1}{2}}  \recursionAodd + \sum^{\frac{n}{2}}_{j=\frac{i}{2}} \recursionBodd \right) \notag \\
&+\sum^{n-1}_{i=1} 	\sum_{j=i+2}^{n} \recursionCodd \,,
\label{eq:recursionoddtree}
\end{align}
where the sum in the second line goes in steps of $2$, and where \treelevelblobodd\, are the leading-order correlation functions with an \textit{odd} number of points of the appropriate length.
Again it is fairly easy to understand the formula: these three terms ensure that all the possible $U$-diagrams are represented, either when the propagator of equation \eqref{eq:Uoneint} closes over leading-order contractions (the first two terms) or when the $U$-integral is contained \textit{inside} a leading-order propagator (the third one).

The diagrammatic expression given above can be expressed formally as
\begin{align}
(\ref{eq:recursionoddtree})&= \sum_{i=1}^n \Biggl( \sum_{j=0}^{\frac{i-1}{2}} \frac{\lambda}{8\pi^2} \delta^{i6} U_{i;2j(2j+1)} A^{I_1,\dots,I_{2j}}_{\text{\tiny{LO}}} A^{I_{2j+1},\dots,I_{i-1}}_{\text{\tiny{LO}}} A^{I_{i+1},\dots,I_{n}}_{\text{\tiny{LO}}} \notag \\
&\quad\quad\,\,+\sum_{j=\frac{i}{2}}^{\frac{n}{2}} \frac{\lambda}{8\pi^2} \delta^{i6} U_{i;2j(2j+1)} A^{I_1,\dots,I_{i-1}}_{\text{\tiny{LO}}} A^{I_{i+1},\dots,I_{2j}}_{\text{\tiny{LO}}} A^{I_{2j+1},\dots,I_{n}}_{\text{\tiny{LO}}} \Biggr) \notag \\
&\,\,+ \sum_{i=1}^{n-1} \sum^n_{j=i+2} A^{I_i I_j}_{\text{\tiny{LO}}} A^{I_{1},\dots,I_{i-1}}_{\text{\tiny{LO}}} A^{I_{j+1},\dots,I_{n}}_{\text{\tiny{LO}}} A^{I_{i+1},\dots,I_{j-1}}_{\text{\tiny{LO}}}\,,
\end{align}
\label{eq:recursionoddtreeexp}
again with the sum in the last line going in steps of $2$. The starting values of the recursion are the same as in \eqref{eq:evenstart}. 

To conclude, in this chapter, we presented recursion relations that allow us to compute arbitrary correlators of fundamental scalar fields $\phi^I$, both at leading and next-to-leading orders for the even case and at leading order for the odd case.  We underline once more that these recursive formulae are implemented in the ancillary \textsc{Mathematica} notebook of \cite{Barrat:2022eim} and they are ready to be used.

In the following chapters, we consider concrete examples of correlators that can be computed using these expressions.

\endgroup

\chapter{Two- and Three-Point Functions}\label{sec:23point}

Now that the main techniques and tools have been introduced, we can start to explore their applications. In particular, as already delineated in the preliminaries, we focus on insertions of scalar operators on the Maldacena-Wilson line.  This chapter is based on \cite{Barrat:2021tpn, Barrat:2022eim}.

The starting point when looking at conformal correlators is the study of two- and three-point functions\footnote{In general, in CFTs one-point functions are zero. It is also the case in this setup. Though, notice that if we had considered the correlator of a bulk operator in the presence of a defect (the Wilson line in our case), we would have obtained a non-vanishing one-point function.}.We have already introduced their general features in Section \ref{sec:correlationfunctions}, but we now specialize their analysis to protected and unprotected insertions on the Wilson line.

In particular, for operators of lengths $L=1,2$, we obtain the normalization constants and scaling dimensions up to next-to-leading order, and compare the results to the literature when possible.  The combination of the recursion relations and the pinching technique allow us to compute correlation functions of some composite operators too.

\begingroup
\allowdisplaybreaks

\section{Two-Point Functions and Anomalous Dimensions} \label{subsec:twopoint}

We start by computing two-point functions both for protected and unprotected operators of lengths $L=1,2$. We obtain normalization constants as well as anomalous dimensions, which for the latter can be compared to the existing literature, while to the best of our knowledge, the normalization constants are new results. The method presented here can be extended straightforwardly to operators of higher length consisting of the elementary scalar fields $\phi^I$.

\subsection{Operators of Length $L=1$} \label{subsubsec:twoptone}

As explained in Section \ref{sec:1ddefectCFT}, there are two distinct operators of length $L=1$, which are the half-BPS operators $\phi^i$ and the unprotected scalar field $\phi^6$.  Computing their two-point correlators is not only useful to extract the normalization constants but it is also a sanity check.

Starting from the protected operators $\phi^i$,  which have exact conformal dimensions $\Delta = 1$, conformal symmetry fixes their two-point functions to be of the form
\begin{equation}
\vev{\phi^i(\tau_1) \phi^j(\tau_2)} = \twopttree = n_i \frac{u_i\cdot u_j}{\tau_{12}^2}\,,
\label{eq:twoptni}
\end{equation}
where the normalization constant $n_i$  is known to be \cite{Drukker:2011za,Correa:2012at}
\begin{equation} \label{eq:normalization1}
n_i=2\,\mathds{B}(\lambda)= \frac{\sqrt{\lambda}}{2\pi^2}\frac{I_2(\sqrt{\lambda})}{I_1(\sqrt{\lambda})}\,,
\end{equation}
with $I_i$ the modified Bessel function of the first kind,  taking the form
\begin{equation}
I_i(x)=\sum^\infty_{m=0} \frac{(-1)^m}{m!\,\Gamma(m+\alpha+1)} \left(\frac{x}{2}\right)^{2m+i}\,.
\end{equation}
The function $\mathds{B}(\lambda)$ in (\ref{eq:normalization1}) is the Bremsstrahlung function with the leading weak-coupling terms being explicitly 
\begin{equation}
\mathds{B}(\lambda)= \frac{\lambda}{16 \pi^2}-\frac{\lambda^2}{384 \pi^2} +O(\lambda^3)\,.
\end{equation}

At leading order we have explicitly
\begin{equation}
\vev{\phi^i(\tau_1) \phi^j(\tau_2)}_{\text{LO}}= \twopttree = \frac{\lambda}{8\pi^2}\frac{u_i\cdot u_j}{\tau_{12}^2}\,,
\label{eq:twoptidiagLO}
\end{equation}
which is trivially produced by inputting the starting value in \eqref{eq:evenstart} into the recursion relation \eqref{eq:recursioneventree}.
At next-to-leading order, we use equation \eqref{eq:oldnew} in order to generate the following diagrams:
\begin{equation}
\vev{\phi^i(\tau_1) \phi^j(\tau_2)}_{\text{NLO}} = \twoptSE\, + \twoptY\,,
\label{eq:twoptidiags}
\end{equation}
where the \textcolor{gluoncolor}{light blue} dots in the second diagram indicate the places where the gluon line should be connected. This again corresponds to integrals along the Wilson line, similarly to the case of the $U$-diagrams and as explained for the boundary diagrams in Section \ref{subsec:oneloop}. The two diagrams are individually divergent and refer to the starting values given in equations (\ref{eq:A2SE}) and (\ref{eq:A2Ydiv}). The divergences cancel and the two-point function at next-to-leading order reads
\begin{equation}
\vev{\phi^i(\tau_1) \phi^j(\tau_2)}_{\text{NLO}} = - \frac{\lambda^2}{192 \pi^2} \frac{u_i \cdot u_j}{\tau_{12}^2}\,,
\label{eq:twoptiNLO}
\end{equation}
in perfect agreement with \eqref{eq:normalization1}.

One can easily obtain the closed form for two-point functions by taking correlators with an even number of $\phi_1$ and pinching each half together. This results in
\begin{equation} \label{eq:twopt}
n_\Delta = \frac{\lambda^\Delta}{2^{3\Delta} \pi^{2\Delta}} \left( 1 - \frac{\lambda}{24} + O(\lambda^2) \right)\,,
\end{equation}
in perfect agreement with \cite{Kiryu:2018phb}.

For the unprotected operator $\phi^6$, the non-normalized two-point function reads
\begin{equation}
\vev{\phi^6 (\tau_1) \phi^6 (\tau_2)} = \frac{n_6}{\tau_{12}^{2 \Delta_6}}\,.
\label{eq:2ptphi6}
\end{equation}
Here the normalization constant takes the form:
\begin{equation}
n_6 = 2\,\mathbb{B}(\lambda) +\Lambda(\lambda)\,,
\label{eq:normalization6}
\end{equation}
which can be understood from Feynman diagrams in the following way \cite{Alday:2007he,Correa:2018fgz}: the first term corresponds to diagrams that are common to both $\phi^i$ and $\phi^6$, while the term $\Lambda(\lambda)$ refers to the diagrams unique to $\phi^6$, i.e. the diagrams where the scalar field couples directly to the Wilson line (see Figure \ref{fig:Lambda} for an example). 

At leading order, however, we find that it coincides with $\vev{\phi^i \phi^j}$:
\begin{align}
\vev{\phi^6(\tau_1) \phi^6(\tau_2)}_{\text{LO}} = \left. \vev{\phi^i(\tau_1) \phi^j(\tau_2)}_{\text{LO}} \right|_{i=j} =\, \twopttree\, = \frac{1}{\tau_{12}^2} \frac{\lambda}{8 \pi^2}\,,
\end{align}
and so it is clear that the function $\Lambda(\lambda)$ defined in \eqref{eq:normalization6} satisfies $\Lambda(0) = 0$.

However, at next-to-leading order, we observe that new diagrams contribute:
\begin{align}
\vev{\phi^6(\tau_1) \phi^6(\tau_2)}_{\text{NLO}} =& \left. \vev{\phi^i(\tau_1) \phi^j(\tau_2)}_{\text{NLO}} \right|_{i=j} \notag \\
& +\, \twoptUone\, +\, \twoptUtwo\, +\, \twoptUthree \notag \\
& +\, \twoptUfour\, +\, \twoptUfive\, +\, \twoptUsix \notag \\
& +\, \twoptUseven\, +\, \twoptUeight\, + O(\lambda^3)\,.
\end{align}
The new diagrams are $U$-integrals, which are the ones contributing to the function $\Lambda(\lambda)$. 

To compute the NLO is necessary to make a digression on the scaling dimension of $\phi^6$, which receives quantum corrections. It takes the following form:
\begin{equation}
\Delta_{\phi^6} = 1 + \sum_{n=1}^\infty \lambda^n \gamma_{\phi^6}^{(n)}\,,
\label{eq:scalingdim6}
\end{equation}
where the anomalous dimensions $\gamma_{\phi^6}^{(n)}$ are known up to order $n=5$ \cite{Grabner:2020nis}. The first correction reads
\begin{equation}
\gamma_{\phi^6}^{(1)} = \frac{1}{4\pi^2}\,.
\label{eq:gamma6}
\end{equation}
As a consequence, the operator $\phi^6$ requires a renormalization procedure. Expanding the two-point function at $\lambda \sim 0$, we have
\begin{equation} \label{eq:renormalize1}
\vev{\phi^6(\tau_1)\phi^6(\tau_2)}= \frac{1}{\tau^2_{12}} \frac{\lambda}{8 \pi^2}\left\lbrace 1+\lambda\gamma_{\phi^6}^{(1)}\, \log\frac{\epsilon^2}{\tau^2_{12}}+\dots \right\rbrace \,,
\end{equation}
with $\epsilon \to 0$. In order to cancel the divergence, we promote $\phi^6$ to its renormalized version:
\begin{equation}
\phi_R^6(\tau):=\phi^6(\tau) \left\lbrace 1-\lambda\gamma_{\phi^6}^{(1)}\, \log\frac{\epsilon^2}{\mu^2}+\dots \right\rbrace \,,
\end{equation}
where $\mu$ corresponds to some choice of scale. This results in a finite two-point function:
\begin{equation}
\vev{\phi^6_R(\tau_1)\phi^6_R(\tau_2)}= \frac{1}{\tau^2_{12}} \frac{\lambda}{8 \pi^2} \left\lbrace 1+\lambda\gamma_{\phi^6}^{(1)}\, \log\frac{\mu^2}{\tau^2_{12}}+\dots \right\rbrace\,.
\end{equation}
This correlation function is still conformal upon renormalization of the dilatation operator. In the rest of this thesis, we drop the subscript since we always refer to renormalized operators. 

Now that we defined the renormalization procedure, using the integrals given in Appendix \ref{subsubsec:Uintegrals}, we find the following result for the leading and next-to-leading orders combined:
\begin{equation}
\vev{\phi^6(\tau_1) \phi^6(\tau_2)} = \frac{1}{\tau_{12}^2} \frac{\lambda}{8 \pi^2} \left(1 - \frac{\lambda}{24} \frac{6+\pi^2}{\pi^2} + O(\lambda^2) \right) \left( 1 - \frac{\lambda}{4\pi^2} \log \tau_{12}^2  + O(\lambda^2) \right)\,.
\end{equation}
This factorized form is useful for reading off the normalization coefficient as well as the anomalous dimension since it can be compared to \eqref{eq:2ptphi6}. The first-order correction to the scaling dimension agrees with (\ref{eq:scalingdim6}), while the normalization constant is
\begin{equation}
n_6 = \frac{\lambda}{8 \pi^2} \left(1 - \frac{\lambda}{24}  \frac{6+\pi^2}{\pi^2} + O(\lambda^2) \right)\,,
\label{eq:norm6}
\end{equation}
which to the best of our knowledge has not been given explicitly in the literature yet.
Comparing this result to \eqref{eq:normalization6}, we determine
\begin{equation}
\Lambda(\lambda) = - \frac{\lambda^2}{32 \pi^4} + O(\lambda^3)\,.
\end{equation}

\subsection{Operators of Length $L=2$} \label{subsubsec:twopttwo}

We now move our attention to operators of length $L=2$. Orthogonal eigenstates of the dilatation operator at next-to-leading order have been constructed in \cite{Correa:2018fgz}:
\begin{subequations} \label{eq:opL2}
\begin{align} 
O^{ij}_S :=&\ \phi^i \phi^j + \phi^j \phi^i - \frac{2}{5} \delta^{ij} \phi^k \phi^k\,, \label{subeq:OS} \\
O^{ij}_A :=&\, \phi^i \phi^j  - \phi^j  \phi^i \,, \\
O^{i}_A :=&\, \phi^6 \phi^i - \phi^i \phi^6 \,, \\
O^{i}_S :=&\, \phi^6 \phi^i + \phi^i \phi^6 \,, \\
O_{\pm} :=&\, \phi^i \phi^i \pm \sqrt{5}\, \phi^6 \phi^6\,.
\end{align}
\end{subequations}
In this case, the operator $O^{ij}_S$ is protected, while the other ones are not.  

Let us see how to compute the two-point function of the protected operator $O^{ij}_S$.  Inserting the definition given in \eqref{subeq:OS} results in
\begin{align}
\vev{O^{ij}_S (\tau_1) O^{kl}_S (\tau_2)} =&\, \vev{\phi^i_1 \phi^j_1 \phi^k_2 \phi^l_2} + \vev{\phi^i_1 \phi^j_1 \phi^l_2 \phi^k_2} - \frac{2}{5} \delta^{kl} \vev{\phi^i_1 \phi^j_1 \phi^m_2 \phi^m_2} \notag \\
&+ \vev{\phi^j_1 \phi^i_1 \phi^k_2 \phi^l_2} + \vev{\phi^j_1 \phi^i_1 \phi^l_2 \phi^k_2} - \frac{2}{5} \delta^{kl} \vev{\phi^j_1 \phi^i_1 \phi^m_2 \phi^m_2} \notag \\
&- \frac{2}{5} \delta^{ij} \vev{\phi^m_1 \phi^m_1 \phi^k_2 \phi^l_2} - \frac{2}{5} \delta^{ij} \vev{\phi^m_1 \phi^m_1 \phi^l_2 \phi^k_2} + \frac{4}{25} \delta^{ij} \delta^{kl} \vev{\phi^m_1 \phi^m_1 \phi^n_2 \phi^n_2}\,,
\label{eq:fullexpOS}
\end{align}
where we defined $\phi^i_1 := \phi^i (\tau_1)$ for compactness. Each term can be seen as the \textit{pinching limit} of a four-point function of the fundamental protected scalars $\phi^i$, e.g.
\begin{equation}
\vev{\phi^i_1 \phi^j_1 \phi^k_3 \phi^l_3} = \lim\limits_{(2,4) \to (1,3)} \vev{\phi^i_1 \phi^j_2 \phi^k_3 \phi^l_4}\,.
\end{equation}
The recursion relations \eqref{eq:recursiontreediagrams} and \eqref{eq:recursionloop} can be used to efficiently compute each of these terms up to next-to-leading order.

We now illustrate with an example at leading order how the pinching of the recursion relation works. In the planar limit, the four-point function consists of the following two diagrams:
\begin{equation}
\vev{\phi^i_1 \phi^j_2 \phi^k_3 \phi^l_4}_\text{LO} = \fourpttreeone\, +\, \fourpttreetwo\,.
\label{eq:fourptLO}
\end{equation}
In order to generate the first term of \eqref{eq:fullexpOS}, $\vev{\phi^i_1 \phi^j_1 \phi^k_2 \phi^l_2}$, we need to pinch $(\tau_2, \tau_4) \to (\tau_1, \tau_3)$ and then relabel $\tau_3$ to $\tau_2$. Only the second diagram in \eqref{eq:fourptLO} survives this pinching, since the first one results in self-contractions\footnote{The $u$ vectors introduced in (\ref{eq:uvector}) ensure that the self-contractions of these protected operators are evaluated to zero.}, and we have
\begin{equation}
\vev{\phi^i_1 \phi^j_1 \phi^k_2 \phi^l_2}  =\, \twoptOijLO\,  = \delta^{il} \delta^{jk} \frac{\lambda^2}{64 \pi^4 \tau_{12}^4}\,.
\end{equation}
We can repeat the same procedure for the other terms at leading order and for the next-to-leading order\footnote{See Section \ref{subsubsec:fourpt-buildingblocks} and in particular equation \eqref{eq:buildingblocksNLO} for more detail on the four-point function at next-to-leading order.}. We finally obtain
\begin{equation}
\vev{O^{ij}_S (\tau_1) O^{kl}_S (\tau_2)} = 2 \left( \delta^{ik} \delta^{jl} + \delta^{il} \delta^{jk} - \frac{2}{5} \delta^{ij} \delta^{kl} \right) \frac{\lambda^2}{64 \pi^4 \tau_{12}^4} \left(1 - \frac{\lambda}{24} + O(\lambda^2) \right)\,.
\end{equation}
In this case, there is no factor corresponding to the correction to the scaling dimension since this operator is half-BPS and hence protected ($\Delta = 2$). The normalization constant is
\begin{equation}
n_{O^{ij}_S} = \frac{\lambda^2}{64 \pi^4} \left(1 - \frac{\lambda}{24} + O(\lambda^2) \right)\,,
\end{equation}
which agrees with (\ref{eq:twopt}) after identifying $(u_1 \cdot u_2)^2 \to 2 \left( \delta^{ik} \delta^{jl} + \delta^{il} \delta^{jk} - \frac{2}{5} \delta^{ij} \delta^{kl} \right)$.\

One can proceed similarly for the other unprotected operators in order to read their normalization constants as well as the anomalous dimensions. For example, the two-point function of $\Op_A^{ij}$ can be obtained in the following way:
\begin{equation}
\vev{O_A^{ij} (\tau_1) O_A^{kl} (\tau_2)} =\, \vev{\phi^i_1 \phi^j_1 \phi^k_2 \phi^l_2} - \vev{\phi^i_1 \phi^j_1 \phi^l_2 \phi^k_2} - \vev{\phi^j_1 \phi^i_1 \phi^k_2 \phi^l_2} + \vev{\phi^j_1 \phi^i_1 \phi^l_2 \phi^k_2}\,. 
\end{equation}
Note that there are only correlators of protected operators of length $L=1$ on the right-hand side, but that the pinching operation generates logarithmic divergences that can be related to the anomalous dimension of the operator, as explained in \eqref{eq:renormalize1} and below. We find that the normalization constant is
\begin{equation}
n_{O_A^{ij}} = - \frac{\lambda^2}{32 \pi^4} \left(1 - \frac{\lambda}{24} + O(\lambda^2) \right)\,,
\end{equation}
while the anomalous dimension turns out to be
\begin{equation}
\gamma^{(1)}_{O_A^{ij}} = \frac{1}{4\pi^2}\,,
\end{equation}
in perfect agreement with \cite{Correa:2018fgz}.

All the other operators can be treated the same way, even when they involve $\phi^6$. We find their normalization constants to be
\begin{subequations}
\begin{align}
n_{O^i_A} &= - \frac{\lambda^2}{32 \pi^4} \left(1 - \frac{\lambda}{24} \frac{6 + \pi^2}{\pi^2} + O(\lambda^2) \right)\,,\\
n_{O^i_S} &= \frac{\lambda^2}{32 \pi^4} \left(1 - \frac{\lambda}{24} + O(\lambda^2) \right)\,,\\
n_{O_{\pm}} &= \frac{5\lambda^2}{32 \pi^4} \left(1 - \frac{\lambda}{24 \pi^2} \left( \pi^2 - \frac{9}{2} (1 \pm \sqrt{5}) \right) + O(\lambda^2) \right)\,.
\end{align}
\end{subequations}
while their anomalous dimensions read
\begin{subequations}
\begin{align}
\gamma^{(1)}_{O^i_A} &= \frac{3}{8\pi^2}\,,\\ \label{eq:anomalousOA}
\gamma^{(1)}_{O^i_S} &= \frac{1}{8\pi^2}\,,\\
\gamma^{(1)}_{O_{\pm}} &= \frac{5\pm\sqrt{5}}{16\pi^2}\,.
\end{align}
\end{subequations}

All the anomalous dimensions listed above perfectly match the results of \cite{Correa:2018fgz} for the supersymmetric case $\zeta = 1$.

\section{Three-Point Functions} \label{subsec:threepoint}

We now move to the computation of selected three-point functions using the recursion relations given in Chapter \ref{sec:recursionrelations}. We focus our attention on correlators involving the two operators of length $L=1$: $\phi^i$ and $\phi^6$.

Note that from now (if not explicitly stated), we consider unit-normalized correlation functions, following the definition given in \eqref{eq:unitnormalized} and using the normalization constants just derived in Section \ref{subsubsec:twoptone}.

Considering first protected operators, we can obtain a closed formula in $\Delta$ for the LO and NLO of three-point functions, as in (\ref{eq:twopt}):
\begin{equation}
c_{\phi^{i} \phi^{j} \phi^{k}}  = \left( \frac{\sqrt{\lambda}}{2\sqrt{2} \pi} \right)^{\Delta_i+\Delta_j+\Delta_k} \left( 1 - \frac{\lambda}{24} (\delta_{\Delta_i,\Delta_j+\Delta_k} + \delta_{\Delta_j,\Delta_k+\Delta_i} + \delta_{\Delta_k,\Delta_i+\Delta_j}) \right)\,,
\label{eq:threept}
\end{equation}
which matches again \cite{Kiryu:2018phb}. For all the cases we looked at, we observed perfect agreement between (84) in \cite{Kiryu:2018phb} and the pinching of our NLO formula. 

We can then compute three-point functions of ``mixed'' operators, i.e. involving two protected operators $\phi^i$ together with the only unprotected operator of length $L=1$ $\phi^6$.
Using the recursion relation for an odd number of $\phi^6$ operators given in \eqref{eq:recursionoddtree}, we find the following result:
\begin{equation}
\vvev{\phi^i \phi^j \phi^6} = \frac{\vev{\phi^i \phi^j \phi^6}}{n_i \sqrt{n_6}} = \frac{\delta^{ij}}{\tau_{12} \tau_{23} \tau_{31}} \left( - \frac{\sqrt{\lambda}}{2\sqrt{2} \pi} + \ldots \right)\,,
\end{equation}
which yields, by comparison to \eqref{eq:threept}, the OPE coefficient
\begin{equation}
c_{\phi^i \phi^j \phi^6} = - \frac{\sqrt{\lambda}}{2\sqrt{2} \pi} + O(\lambda^{3/2})\,.
\label{eq:ij6}
\end{equation}

The same computation can easily be performed for three unprotected operators $\phi^6$. In this case, we obtain
\begin{equation}
c_{\phi^6 \phi^6 \phi^6} = - \frac{3\sqrt{\lambda}}{2\sqrt{2} \pi} + O(\lambda^{3/2})\,.
\label{eq:3ptphi6}
\end{equation}
These results are used as consistency checks for the correlation functions that we expand in conformal blocks in Sections \ref{sec:blockexpansion4pt}, \ref{sec:blockexpansion5pt} and \ref{sec:blockexpansion6pt}.

In the ancillary \textsc{Mathematica} notebook coming along with \cite{Barrat:2022eim}, there are examples of three-point functions involving unprotected operators of length $L=2$ as well.

\chapter{Four-Point Functions} \label{ch:4point}

After having explored two- and three-point functions, we delve into the study of the first objects populating a CFT, not completely fixed by the symmetries: four-point functions.  In particular, we start by looking closely at the simplest four-point function $\vvev{\phi_1, \phi_1, \phi_1, \phi_1}$\footnote{This is the simplest four-point correlator because all operators are protected and have length 1.}, which is retrievable almost by definition from the recursion relations we introduced in Chapter \ref{sec:recursionrelations}.  

Interestingly, this correlator is annihilated by a particular differential operator,  giving rise to a powerful Ward identity that we use to greatly simplify the computation of this correlator at NNLO. We then include in this study unprotected operators and expand these correlators in blocks to perform consistency checks and extract new CFT data.

The results presented in this chapter are based on \cite{Barrat:2021tpn, Barrat:2022eim}, and on \cite{Artico:2023abp} which contains unpublished work.

\section{The Simplest Four-Point Correlator} \label{subsubsec:fourpt-buildingblocks}
 
We introduce the first correlator we analyze in detail, which is the simplest four-point function:
\begin{equation}
\vvev{\phi_1(\tau_1) \phi_1(\tau_2) \phi_1(\tau_3) \phi_1(\tau_4)} = \Km_{1111}(u,\tau) \Am_{1111} (\chi; r,s)\,,
\end{equation}
where 
\begin{equation}
\vvev{\phi_1 \phi_1 \phi_1 \phi_1} \equiv \vvev{\phi_1^i \phi_1^j \phi_1^k \phi_1^l}\,,\quad i,j,k,l = 1\,, \ldots\,, 5\,,
\end{equation}
namely, we are considering only protected operators.  Notice that in the following we also omit the insertion points of the operators on the Wilson line, assuming them to be always distinct.

As explained in Section \ref{sec:wilsonlinecorrelators}, we use the $u$ vectors to keep track of the $R$-symmetry indices. Therefore the prefactor and the cross-ratios read
\begin{equation}
\Km_{1111} := \frac{(u_1 \cdot u_2)(u_3 \cdot u_4)}{\tau_{12}^2 \tau_{34}^2}\,,
\end{equation}
and
\begin{equation}
\chi := \frac{\tau_{12} \tau_{34}}{\tau_{13} \tau_{24}}\,, \quad r := \frac{(u_1 \cdot u_2)(u_3 \cdot u_4)}{(u_1 \cdot u_3)(u_2 \cdot u_4)}\,, \quad s := \frac{(u_1 \cdot u_4)(u_2 \cdot u_3)}{(u_1 \cdot u_3)(u_2 \cdot u_4)}\,,
\end{equation}
where $\chi$ satisfies $0 < \chi < 1$, with the ordering of the spacetime points $\tau_1 < \tau_2 < \tau_3 < \tau_4$.

The reduced correlator can be expanded into three $R$-symmetry channels:
\begin{equation} \label{eq:splittingR}
\Am_{1111} := F_0 (\chi) + \frac{\chi^2}{r} F_1 (\chi) + \frac{s}{r} \frac{\chi^2}{(1-\chi)^2} F_2 (\chi)\,,
\end{equation}
where the prefactors have been chosen such that they satisfy on their own the superconformal Ward identities, which we discuss in detail later.

This correlator is important because it is the starting point to build correlators of protected and unprotected operators too. On the one hand, it is the input for next-to-leading-order recursion relations that we explored in Chapter \ref{sec:recursionrelations}, which allow us to compute correlators of protected operators, on the other hand, it provides the \textit{building blocks} for other correlators involving an even number of unprotected operators.

This correlator has been computed perturbatively up to the next-to-leading order in \cite{Kiryu:2018phb}.  We repeat this computation being it instructive not only to compute the next-to-next-to-leading order (NNLO) but as well for all the other computations that follow.

In particular, the $R$-symmetry channels, with a unit-normalized correlator, have the following perturbative expansion:
\begin{equation} \label{eq:perturbative}
F_j (\chi) = \sum_{k=0}^{\infty} \lambda^{k} F^{(k)}_j (\chi)\,.
\end{equation}

At leading order, equation (\ref{eq:recursiontree}) becomes
\begin{align}
\vev{\phi_1 \phi_1 \phi_1 \phi_1}_\text{LO} = & A_2^\text{LO} (1,2) A_2^\text{LO} (3,4) + A_2^\text{LO} (1,4) A_2^\text{LO} (2,3)\,,
\end{align}
which can be represented diagrammatically using (\ref{eq:recursiontreediagrams}):
\begin{equation}
\vev{\phi_1 \phi_1 \phi_1 \phi_1}_\text{LO}  =\, \fourpttreeone\, + \fourpttreetwo\,.
\end{equation}
We first input the starting values given in \eqref{eq:startingvalstree} and the decomposition into $R$-symmetry channels of \eqref{eq:splittingR}. We then unit-normalize the correlator following \eqref{eq:unitnormalized} and using the normalization constant computed in \eqref{eq:normalization1} resulting in the following channels:
\begin{equation} \label{eq:buildingblocksLO}
F_0^{(0)} (\chi) = F_2^{(0)} (\chi) = 1 \,, \quad F_1^{(0)} (\chi) = 0\,.
\end{equation}

At next-to-leading order, the recursion relation given in \eqref{eq:recursionloop} produces the following diagrams:
\begin{align}
\vev{\phi_1 \phi_1 \phi_1 \phi_1}_\text{NLO} = & \phantom{+}\,\,\, \,\fourptX\, +\, \fourptHone\, +\, \fourptHtwo \notag \\
& +\, \fourptSEone\, +\, \fourptSEtwo\, +\, \fourptSEthree\, \notag \\
& +\, \fourptYone\,+\, \fourptYtwo\, +\, \fourptYthree \notag \\
& +\, \fourptSEfour\, +\, \fourptYfour\,.
\end{align}
The first line comes from the term \eqref{eq:contribX}, while the self-energy diagrams in the second line are generated by \eqref{eq:formulaSE} and the $Y$-diagrams in the third line come from \eqref{eq:formulaYdiv}, \eqref{eq:Ynodiv1}, \eqref{eq:Ynodiv2} and \eqref{eq:Ynodiv3}. Finally, the last line is obtained from the recursive term \eqref{eq:formulaOneloop}. Using the integrals of Appendix \ref{sec:integralsapp}, we find the following expressions for the unit-normalized $R$-symmetry channels at NLO:
\begin{subequations} \label{eq:buildingblocksNLO}
\begin{align}
F_0^{(1)} (\chi) =& \frac{1}{8\pi^2} \left( 2 L_R (\chi) + \frac{\ell (\chi,1)}{1-\chi} \right) \,, \\
F_1^{(1)} (\chi) =& - \frac{1}{8 \pi^2} \frac{\ell (\chi,1)}{\chi(1-\chi)} \,, \\
F_2^{(1)} (\chi) =& - \frac{1}{8\pi^2} \left( 2 L_R (\chi) - \frac{\ell (\chi,1)}{\chi} - \frac{\pi^2}{3}  \right) \,.
\end{align}
\end{subequations}
This computation was first performed in \cite{Kiryu:2018phb}, and then repeated in \cite{Barrat:2021tpn} with the use of the recursion relation. 

Note that we have used the Rogers dilogarithm, defined as
\begin{equation} \label{eq:Rogers}
L_R(\chi) := \text{Li}_2 (\chi) + \frac{1}{2} \log (\chi) \log(1-\chi)\,,
\end{equation}
and satisfying the following properties:
\begin{subequations}
\begin{align}
&L_R (x) + L_R (1-x) = \frac{\pi^2}{6}\,, \\
&L_R (x) + L_R(y) = L_R (xy) + L_R \left( \frac{x(1-y)}{1-xy} \right) + L_R \left( \frac{y(1-x)}{1-xy} \right)\,.
\end{align}
\end{subequations}
We also use the following two-variable function introduced in \cite{Barrat:2021tpn}:
\begin{equation} \label{eq:ell}
\ell(\chi_1, \chi_2) := \chi_1 \log \chi_1 - \chi_2 \log \chi_2 + (\chi_2 - \chi_1) \log (\chi_2 - \chi_1)\,.
\end{equation}
Note that the function $\ell(\chi,1)$ is manifestly crossing-symmetric, i.e.
\begin{equation}
\ell(\chi,1) = \ell(1-\chi,1)\,,
\end{equation}
and it is related to a special limit of the Bloch-Wigner function $D(\chi, \bar{\chi})$ in the following sense:
\begin{equation}
\ell(\chi, 1) = \chi(1-\chi) \lim\limits_{\bar{\chi} \to \chi} \frac{D(\chi, \bar{\chi})}{2(\bar{\chi} - \chi)}\,,
\end{equation}
with
\begin{equation}
D(\chi, \bar{\chi}) = 2 \text{Li}_2 (\chi) - 2 \text{Li}_2 (\bar{\chi}) + \log \chi \bar{\chi} \log \frac{1 - \chi}{1 - \bar{\chi}}\,.
\end{equation}
The function $\ell$, which appears in higher-point functions as well, satisfies the following identities:
\begin{subequations}
\begin{align}
&\ell(\chi_1, \chi_2) + \ell(\chi_2, \chi_1) = i \pi (\chi_1 - \chi_2)\,, \\
&\ell(\chi_1, \chi_2) = \chi_1\, \chi_2\, \ell(\chi_2^{-1}, \chi_1^{-1}) \quad \text{for } 0 < \chi_1 < \chi_2 < 1\,, \\
&\ell(\chi_1, \chi_2) + \ell(1-\chi_2, 1-\chi_1) = \ell(\chi_1,1) - \ell(\chi_2,1)\,.
\end{align}
\end{subequations}

To conclude, there are some easy checks that one can perform on these results. In particular, we can use the pinching technique introduce in Section \ref{sec:pinching} to get 
\begin{equation} 
\vvev{\phi_1(\tau_1) \phi_1(\tau_2) \phi_1(\tau_3) \phi_1(\tau_4)} \rightarrow \vvev{\phi_2(\tau_1) \phi_2(\tau_2)} \,,
\end{equation}
matching the result given in equation \eqref{eq:twopt} for the case $\Delta = 2$. In this case, the limit one has to perform is
\begin{equation}
(u_2, \tau_2) \to (u_1, \tau_1)\,, \quad (u_4, \tau_4) \to (u_3, \tau_3)\,.
\end{equation}

\section{Ward Identity} \label{sec:wardidentity4pt}

As briefly mentioned, when expressed in terms of spacetime and $R$-symmetry cross-ratios, this correlator and more generally four-point functions of arbitrary half-BPS operators satisfy some elegant superconformal Ward identities. In general, Ward identities are a quantum version of Noether's theorem and encode the symmetries of the theory. In our case, they translate superconformal symmetry into powerful constraints that take the following form \cite{Liendo:2016ymz}:
\begin{equation} \label{WIpedro}
\left. \left( \frac{1}{2} \partial_\chi + \partial_{\zeta_1} \right) \Am_{\Delta_1 \Delta_2 \Delta_3 \Delta_4} \right|_{\zeta_1 = \chi} = 0\,, \qquad \left.  \left( \frac{1}{2} \partial_\chi + \partial_{\zeta_2} \right) \Am_{\Delta_1 \Delta_2 \Delta_3 \Delta_4} \right|_{\zeta_2 = \chi} = 0\,.
\end{equation}
Equivalently, these two independent equations can be rewritten into one single equation using our definition of $R$-symmetry cross-ratios, (see (\ref{eq:rRsymmecross}) and (\ref{eq:sRsymmecross})), which in terms of $\zeta_1$ and $\zeta_2$ are respectively
\begin{equation}
r = \zeta_1 \zeta_2\,, \qquad s = (1-\zeta_1)(1-\zeta_2)\,,
\end{equation}
resulting in
\begin{equation} \label{eq:WI4}
\left( \frac{1}{2} \partial_{\chi} + \alpha \partial_r - (1-\alpha) \partial_s \right) \,  \Am_{\Delta_1 \Delta_2 \Delta_3 \Delta_4} \raisebox{-1ex}{$\biggr |$}_{\raisebox{.75ex}{$\begin{subarray}{l} r = \alpha \chi\\
s = (1-\alpha)(1-\chi) \end{subarray}$}}=0\,,
\end{equation}
which is valid for any $\alpha$ real.  In Section \ref{subsec:conjectureWI}, we conjecture a multipoint extension of this Ward identity based on our perturbative results.

This differential equation encodes the constraints of superconformal symmetry on the correlators and turned out to be essential for bootstrapping the four-point function $\vev{\phi_1 \phi_1 \phi_1 \phi_1}$ at strong coupling \cite{Liendo:2018ukf,Ferrero:2021bsb}.  

Interestingly, if we now consider (\ref{WIpedro}) where we now exchange the spacetime and the $R$-symmetry cross-ratios,  we still annihilate the four-point functions. The analytic continuation between the two different types of cross-ratios was first discovered in \cite{Liendo:2016ymz} and leads to a curious web of correspondences between different defect setups. We briefly discuss this peculiar fact in the outlook Section \ref{sec:morewardidentities}.

We explore further this powerful constraint since it turns out to be quite fundamental to compute the four-point function $\vvev{\phi_1 \phi_1 \phi_1 \phi_1}$ as NNLO, as we do in the next section.  In particular, in our case where $\Delta_i = 1$ for all the operators,  the $R$-symmetry channels are related algebraically, and remarkably the correlator depends only on \textit{one} function of the kinematical variable $\chi$, $f(\chi)$, and on one constant $\Fds$ derived using supersymmetric localization, which are sufficient to completely fix it. 

To see this, we can explicitly solve the Ward identity (\ref{eq:WI4}), and the solution for the reduced correlator defined in (\ref{eq:splittingR}) reads \cite{Liendo:2018ukf}:
\begin{equation} \label{eq:SolutionWI4Point}
\Am (\chi; r,s) = \mathds{D} f(\chi) + \mathds{F}\, \frac{\chi^2}{r}\,,
\end{equation}
where we have introduced the differential operator
\begin{equation} \label{eq:SolutionWI4Point}
\mathds{D} := \left( \frac{2}{\chi} - 1 + \frac{1-s}{r} \right) - \left( 1-\chi + \frac{\chi}{r} (1- \chi - s) \right) \pd_\chi\,.
\end{equation}
$\mathds{F}$ refers to the \textit{topological sector} (\ref{eq:topologicallimit}) discussed in Section \ref{sec:topologicallimit}, which is constant (in the sense that it does not depend on kinematic variables\footnote{It could seem surprising that summing all the channels together gives a constant term, since they depend on $\chi$, but we can also understand it using the Ward identities. In fact $\partial_\chi \sum^5_{i=1} F_i (\chi) = 0$.}) and known to all orders in the coupling $\lambda$ \cite{Drukker:2009sf}:
\begin{align} \label{eq:TopologicalSector}
\Fds &= F_0(\chi) + F_1(\chi)  + F_2(\chi)  \notag \\
&= \frac{3}{\pi^2 \mathds{B}^2} \left( \frac{\lambda}{16 \pi^2} - \mathds{B}\right) = \frac{3 I_1 (\sqrt{\lambda}) I_3 (\sqrt{\lambda})}{I_2^2 (\sqrt{\lambda})}\,,
\end{align}
with $\mathds{B}$ the Bremsstrahlung function defined in (\ref{eq:normalization1}). 
At weak coupling, the topological sector can be expanded as
\begin{equation}
\Fds = 2 - \frac{\lambda}{24} - \frac{\lambda^2}{480} + O(\lambda^3)\,.
\label{eq:TopologicalSectorAtWeakCoupling}
\end{equation}

The $R$-symmetry channels introduced in (\ref{eq:splittingR}) can be related to $f(\chi)$ and $\mathds{F}$ in the following way:
\begin{subequations}
\begin{align}
F_0 (\chi) &= \left( \frac{2}{\chi} - 1 \right) f(\chi) - ( 1 - \chi ) f'(\chi)\,, \label{eq:F0FromLittlef} \\
F_1 (\chi) &= \mathds{F} - \frac{f(\chi)}{\chi^2} - \frac{1-\chi}{\chi} f'(\chi)\,, \label{eq:F1FromLittlef} \\
F_2 (\chi) &= \frac{(1-\chi)^2}{\chi^2} \left( f(\chi) - \chi f'(\chi) \right)\,.\label{eq:F2FromLittlef}
\end{align}
\end{subequations}
Therefore, it is sufficient to compute \textit{only} one channel in order to obtain the full correlator. In the following, we use this property for computing the next order of the four-point function.

\section{NNLO Correlator} \label{sec:2loop}

In Section \ref{subsubsec:fourpt-buildingblocks} we showed how to compute $\vvev{\phi_1 \phi_1 \phi_1 \phi_1}$ at LO and NLO in the planar limit. We now want to push this computation further, meaning we aim at the NNLO. The result has been already derived in \cite{Cavaglia:2022qpg} using a combination of bootstrap and integrability. We reproduce this result using perturbative techniques.

To reach our goal,  we use a major simplification that the Ward identity offers us in combination with a constraint that crossing symmetry\footnote{We discussed crossing symmetry more in detail in Section \ref{sec:OPE}. } imposes on $f(\chi$).
The full correlator is invariant under the exchange of spacetime and $R$-symmetry variables $(u_1, \tau_1) \leftrightarrow (u_3, \tau_3)$\footnote{Or, equivalently, $(u_2, \tau_2) \leftrightarrow (u_4, \tau_4)$.}, from which it is easy to deduce that the reduced correlator satisfies
\begin{equation}
\Am (\chi; r,s) = \frac{s}{r} \frac{\chi^2}{(1-\chi)^2} \Am (1-\chi; r,s)\,.
\label{eq:CrossingSymmetryOfTheeducedCorrelator}
\end{equation}
As a consequence, $f(\chi)$ satisfies
\begin{equation}
\chi^2 f(1- \chi) + (1-\chi)^2 f(\chi) = 0\,,
\label{eq:CrossingSymmetryOfLittlef}
\end{equation}
which channel-wise translates to
\begin{subequations}
\begin{align}
F_0 (\chi) &= F_2 (1-\chi)\,, \label{eq:CrossingSymmetryOfF0AndF2} \\
F_1 (\chi) &= F_1 (1-\chi)\,.\label{eq:CrossingSymmetryOfF1}
\end{align}
\end{subequations}

It is convenient to follow the conventions of \cite{Cavaglia:2022qpg} and define
\begin{equation} \label{eq:hDefinition}
f(\chi) := \frac{\chi}{1-\chi} h (\chi)\,,
\end{equation}
such that the crossing equation becomes
\begin{equation}
h(\chi) + h(1-\chi) = 0\,.
\label{eq:CrossingSymmetryOfLittleh}
\end{equation}

In the weak coupling regime, it is easy to derive the expression for $h(\chi)$ at leading and next-to-leading orders:
\begin{subequations}
\begin{align}
h^{(0)} (\chi) &= 1 - 2\chi\,, \label{eq:LittlehLO} \\
h^{(1)} (\chi) &= - \frac{2\pi^2}{3} \chi - 2 (H_{1,0} - H_{0,1})\,,\label{eq:LittlehNLO}
\end{align}
\end{subequations}
with $H_{\vec{a}} := H_{\vec{a}} (\chi)$ the Harmonic Polylogarithms (HPL) defined in Appendix \ref{app:HPL}. In this case, it is convenient to introduce the HPL's not only to make contact with \cite{Cavaglia:2022qpg} but also to write the NNLO more compactly and to have a glance at its transcendentality properties.

Note here that the transcendentality weight $w(\ell)$ of $h^{(\ell)} (\chi)$ is homogeneous (since $\pi^2$ has weight $2$) and $w(\ell) = 2 \ell$, being $\ell$ the number of loops.
Moreover, $h^{(\ell)} (\chi)$ appears to be composed of one rational function and some transcendental functions \textit{without rational prefactors}.
This fact was used in \cite{Cavaglia:2022qpg} for bootstrapping the correlator, and we use it here as well for determining some of the contributions numerically.

Then, if one channel is enough to retrieve the full correlator,  we have to understand what is the simplest channel to compute.  It is quite clear that $F_1$ is the best candidate.  This consideration is based on the lower orders (\ref{eq:buildingblocksLO}) and (\ref{eq:buildingblocksNLO}) and on considering and counting the possible diagrams in all three channels.  It turns out that instead of computing many (more than fifty) diagrams, namely all the diagrams in all the three channels (this includes self-energy diagrams at the next order and complicated couplings to the line defect), we only need to compute fifteen diagrams\footnote{Of course, we have to agree on how we identify \textit{one} Feynman diagram. Here each self-energy insertion gives rise to an individual Feynman diagram. A similar reasoning is valid for the XY-diagrams, even if, in principle, each gluon line insertion could give rise to a new diagram.}. One may be tempted to think that this is a consequence of the large $N$ limit. However, although this regime greatly simplifies the computation, this feature appears to be present at finite $N$ too.  In any case, without the help of the Ward identity, this computation would be much more challenging.
\begin{table}[h]
\centering
\caption{\textit{The relevant Feynman diagrams for the computation of $F_1(\chi)$ at next-to-next-to-leading order.
The double line separates \textbf{bulk} from \textbf{boundary} diagrams.
For the latters, the \textbf{\textcolor{gluoncolor}{dots}} placed on the Wilson line indicate where the gluon can possibly connect.}}
\begin{tabular}{lc}
\hline
Self-energy & \DefectSSSSTwoLoopsSelfEnergyOneS\ \DefectSSSSTwoLoopsSelfEnergyTwoS\ \DefectSSSSTwoLoopsSelfEnergyThreeS\ \DefectSSSSTwoLoopsSelfEnergyFourS\ \\[3ex]
\hline
$XX$ & \DefectSSSSTwoLoopsXXOneS\ \DefectSSSSTwoLoopsXXTwoS\ \\[3ex]
\hline
$XH$ & \DefectSSSSTwoLoopsXHOneS\ \DefectSSSSTwoLoopsXHTwoS\ \DefectSSSSTwoLoopsXHThreeS\ \DefectSSSSTwoLoopsXHFourS\ \\[3ex]
\hline
Spider & \DefectSSSSTwoLoopsSpiderS\ \\[3ex]
\hline \hline
$XY$ & \DefectSSSSTwoLoopsXYOneS\ \DefectSSSSTwoLoopsXYTwoS\ \DefectSSSSTwoLoopsXYThreeS\ \DefectSSSSTwoLoopsXYFourS\ \\[3ex]
\hline
\end{tabular}
\label{table:DiagramsAtNNLO}
\end{table}

The relevant planar diagrams are listed in Table \ref{table:DiagramsAtNNLO}. As for the NLO computation,  we separate the contributing diagrams into two groups: the \textit{bulk} and \textit{boundary} ones.

In the next section, we explore in detail how to compute all the diagrams in this channel.  Note that we can solve analytically all the bulk integrals in the limit $\tau_4 \rightarrow \infty$ or in the full conformal frame, i.e. $(\tau_1, \tau_2, \tau_3, \tau_4) \to (0,\chi,1,\infty)$.  Unfortunately, this limit is not enough to do the same for the boundary integrals, which have been solved only numerically.

Moreover, the results for Feynman diagrams are \textit{not} unit-normalized. This is the natural framework when dealing with Wick contractions, otherwise, there would be a mixing of diagrams belonging to different orders. We, therefore, unit-normalize only \textit{after} the computations.

\subsection{Bulk Diagrams} \label{subsec:BulkDiagrams}

We first focus on computing the diagrams where the interactions happen in the bulk.

\subsubsection{Self-Energy Diagrams}

We can start with the self-energy diagrams, which are obtained considering one quartic scalar vertex (\ref{eq:vertexSSSS}) and a self-energy insertion (\ref{eq:selfenergy}). We then perform the Wick contractions to find that the first diagram evaluates to
\begin{equation} \label{eq:TwoLoopsSelfEnergyDiagramOne}
\DefectSSSSTwoLoopsSelfEnergyOne\ = - \frac{\lambda^4}{2} \int d^4 x_5\,  I_{25} I_{35} I_{45}\,Y_{115}\,,
\end{equation}
where $Y_{115}$ is given in (\ref{eq:Y112}) and log-divergent.
It is easier to evaluate this integral in the conformal frame $\tau_4 \rightarrow \infty$.
We can then extract the propagator $I_{45}$ from the integral, by doing e.g. the replacement $I_{45} \to I_{24}$, and the diagram becomes
\begin{equation} \label{eq:TwoLoopsSelfEnergyDiagramOneb}
\DefectSSSSTwoLoopsSelfEnergyOne\ = - \frac{\lambda^4}{2} I_{24} H_{11,23} + O\left(\frac{1}{\tau_4^3}\right)\,,
\end{equation}
with the $H$-integral defined in (\ref{subeq:H1324}). We do not insert the explicit expression of this integral here as these terms drop at the end of the computation. This is not unexpected, as the same happens in the NLO case, where the self-energy diagrams are canceled by some other diagrams.

The other self-energy diagrams can be evaluated in the same way and their sum gives
\begin{align} \label{eq:SigmaSelfEnergy}
F^{\text{SE}}_1 &:= \DefectSSSSTwoLoopsSelfEnergyOne\ + \DefectSSSSTwoLoopsSelfEnergyTwo\ +\DefectSSSSTwoLoopsSelfEnergyThree\ + \DefectSSSSTwoLoopsSelfEnergyFour \notag \\
& = - \frac{\lambda^4}{32\pi^2} \left( \frac{1}{I_{13}}( H_{11,23} + H_{22,13} + H_{33,12} ) - \frac{K_{44}}{I_{13}} \right)\,,
\end{align}
where $K_{44}$ is a special case of $K_{ij}$ defined in (\ref{eq:Kij}), that we write explicitly in (\ref{eq:K44}).

If we write explicitly all these integrals, we see contributions of the type $\log \tau_4^2$ and $\log \veps^2$. However for the correlator to be conformal, the $\log \tau_4^2$ terms must drop at the end of the computation. This actually provides a strong check of our computations. On top of that the $\log \veps^2$ terms must also cancel each other for the correlator to be finite.

\subsubsection{$XX$-Diagrams}

In this channel, i.e. $F_1$,  we can only obtain two diagrams using the $X$-vertex (\ref{eq:vertexSSSS}), which we call $XX$-diagrams and that are also easy to compute.
The first one reads
\begin{equation}
\DefectSSSSTwoLoopsXXOne\ = - \frac{\lambda^4}{4} \int d^4 x_5\, X_{1255}\,  I_{35} I_{45}\,,
\label{eq:TwoLoopsXXDiagramOne}
\end{equation}
which it is convenient to symmetrize by applying
\begin{equation}
\int d^4 x_5\, X_{1255}\,  I_{35} I_{45} = \frac{1}{2} \left( \int d^4 x_5\, X_{1255}\,  I_{35} I_{45}
+ \int d^4 x_5\, X_{3455}\,  I_{15} I_{25} \right)\,.
\label{eq:SymmetrizationOfX}
\end{equation}
This integral can be computed in the conformal frame, where it can be shown to give
\begin{align}
\DefectSSSSTwoLoopsXXOne\ =\ & 
- \frac{\lambda^4}{16} \left( I_{34} ( H_{11,23} + H_{22,13} + H_{33,12} ) + K_{44} \right) \notag \\
& + \frac{\lambda^4}{128 \pi^2} ( \log \tau_{12}^2 + \log \tau_4^2 ) I_{24} Y_{123}\notag \\
&- \frac{\lambda^4}{256 \pi^2} I_{24} \left( A_1 + A_2 + A_3 \right) + O\left(\frac{1}{\tau_4^3}\right)\,,
\label{eq:TwoLoopsXXDiagramOneb}
\end{align}
where the $A$-integrals are defined in (\ref{eq:Ak}). These integrals drop if we use the identity (\ref{eq:AkIdentity}).
We can then analogously compute the other $XX$-diagram, obtaining this final expression for the sum of the two diagrams:
\begin{align} \label{eq:SigmaXX}
F^{XX}_1 &:= \DefectSSSSTwoLoopsXXOne\ + \DefectSSSSTwoLoopsXXTwo\ \notag \\
&\,\,= - \frac{\lambda^4}{8} \left( I_{24} ( H_{11,23} + H_{22,13} + H_{33,12} ) + K_{44} \right) \notag \\
&\,\,\phantom{=\ } + \frac{\lambda^4}{128 \pi^2} \log \tau_4^2\, I_{24} Y_{123} + O\left(\frac{1}{\tau_4^3}\right)\,.
\end{align}
This sum is invariant under the interchange of variables $\tau_1 \leftrightarrow \tau_3$. In any case, note again that all these terms should drop from the final expression since they are either divergent or not conformal.

\begingroup
\allowdisplaybreaks

\subsubsection{$XH$-Diagrams}

The diagrams with two $Y$-vertices (\ref{eq:vertex2scalars1gluon}) and one $X$-vertex (\ref{eq:vertexSSSS}) are more involved, but they can also be solved in the conformal frame.
The first one gives
\begin{align}
\DefectSSSSTwoLoopsXHOne\ &= \frac{\lambda^4}{8} \int d^4x_5\, F_{15,25}\, I_{15} I_{25} I_{35} I_{45} \notag \\
&= \frac{\lambda^4}{32} \left( I_{24} ( 3 H_{11,23} + 3 H_{22,13} - H_{33,12} ) + K_{44} \right) \notag \\
&\phantom{=\ } + \frac{\lambda^4}{8} I_{24} \left( H_{12,13} + H_{12,23} - \frac{2K_{12}}{I_{12}} \right) \notag \\
&\phantom{=\ } + \frac{\lambda^4}{512 \pi^2} \left( \log \frac{\tau_{12}^2}{\tau_{23}^2} + \log \tau_4^2 \right) I_{24} Y_{123} + O\left(\frac{1}{\tau_4^3}\right)\,.
\label{eq:TwoLoopsXHDiagramOne}
\end{align}
This can be obtained using (\ref{eq:AkIdentity}), followed by the symmetrization \eqref{eq:SymmetrizationOfX}. The integral $K_{12}$ is defined in (\ref{eq:Kij}) and further details can be found in Appendix \ref{subsec:bulkintegrals}. To simplify this expression we can also apply some interesting identities relating $K$-integrals to $H$-integrals (\ref{eq:KHidentity}) but also $H$-integrals only (\ref{eq:Hidentity}).

The other three $XH$-diagrams can be computed analogously and their sum reads
\begin{align}
F^{XH}_1 :&= \DefectSSSSTwoLoopsXHOne\ + \DefectSSSSTwoLoopsXHTwo\ + \DefectSSSSTwoLoopsXHThree\ + \DefectSSSSTwoLoopsXHFour \notag \\
&= \frac{\lambda^4}{128\pi^2} \frac{1}{I_{13}} \left( ( H_{11,23} + H_{22,13} + H_{33,12} ) + K_{44} \right) \notag \\
&\phantom{=\ } + \frac{\lambda^4}{128\pi^2} \frac{1}{I_{13}} ( H_{12,13} + H_{13,23} + 2 H_{12,23} ) \notag \\
&\phantom{=\ } - \frac{\lambda^4}{128\pi^2} \frac{1}{I_{13}} \left(-\left( \frac{K_{12}}{I_{12}} + \frac{K_{23}}{I_{23}} \right) + K_{14}+ K_{34} \right) \notag \\
&\phantom{=\ } + \frac{\lambda^4}{1638 \pi^4} \log \tau_4^2 \frac{Y_{123}}{I_{13}} + O\left(\frac{1}{\tau_4^3}\right)\,,
\label{eq:SigmaXH}
\end{align}
where we discarded the terms suppressed in the limit $\tau_4 \to \infty$, such as
\begin{equation} \label{eq:VanishingH}
I_{24} H_{12,34} = I_{24} H_{14,23} = O\left(\frac{1}{\tau_4^3}\right)\,.
\end{equation}

\endgroup

\subsubsection{Spider Diagram}

We compute now the fermionic loop, which we refer to as the \textit{spider} diagram.
The Wick contractions yield the following sixteen-dimensional integral:
\begin{align}
\DefectSSSSTwoLoopsSpider\ =\ & \frac{\lambda^4}{4} \int d^4 x_5 \int d^4 x_6 \int d^4 x_7 \int d^4 x_8\, I_{15} I_{26} I_{37} I_{48} \notag \\
& \times \tr \spd_6 I_{56} \spd_6 I_{67} \spd_8 I_{78} \spd_5 I_{58}\,,
\label{eq:TwoLoopsSpiderDiagram}
\end{align}
where the trace is acting on the $\gamma$-matrices. Here we have already extracted the $\Gamma$-matrices with $R$-symmetry indices by applying the identity (\ref{eq:TwoLoopsTraceOfGammas}) and selecting the $R$-symmetry channel corresponding to $\delta^{ik} \delta^{jl}$.
We can use the fermionic star-triangle identity (\ref{eq:startriangle}) twice in order to lift two integrals:
\begin{align}
\DefectSSSSTwoLoopsSpider\ =\ & - 4 \pi^4 \lambda^4 \int d^4 x_5\, I_{15} I_{25} I_{45} \int d^4 x_7\,  I_{27} I_{37} I_{47} I_{57}^2 \notag \\
& \times \tr \sx_{25} \sx_{27} \sx_{47} \sx_{45}\,.
\label{eq:TwoLoopsSpiderDiagramb}
\end{align}
The trace is easy to perform with the $\gamma$ identity (\ref{eq:TwoLoopsTraceOfGammasb}) and after using the algebraic identity (\ref{eq:TwoLoopsAlgebraicIdentity}), the diagram reduces to
\begin{align} \label{eq:TwoLoopsSpiderDiagramc}
\DefectSSSSTwoLoopsSpider\ =\ & - 2 \lambda^4 \frac{1}{I_{24}} \int d^4 x_5\, X_{2345}\, I_{15} I_{25} I_{45} \notag \\
& + 2 \lambda^4 \int d^4 x_5\,  \left( X_{2355}\, I_{15} I_{45} + X_{1455}\, I_{25} I_{35} \right)\,.
\end{align}
The first line corresponds to the \textit{kite} integral defined in \eqref{eq:KiteIntegralDefinition}, where its solution is also given in terms of HPL's.
The second line is identical to the $XX$-diagrams and can be found in \eqref{eq:SigmaXX} up to an overall prefactor.

All in all the spider diagram reads
\begin{align} \label{eq:TwoLoopsSpiderDiagramd}
F^{\text{Spider}}_1 & =\ \DefectSSSSTwoLoopsSpider\ \notag \\
& = \frac{\lambda^4}{64\pi^2} \frac{1}{I_{13}} \left( ( H_{11,23} + H_{22,13} + H_{33,11} ) + K_{44} \right)
- \frac{\lambda^4}{32\pi^2} \frac{1}{I_{13}} H_{12,23} \notag \\
& \phantom{=\ } - \frac{\lambda^4}{8192 \pi^4} \log \tau_4^2  \frac{Y_{123}}{I_{13}} + O\left(\frac{1}{\tau_4^3}\right)\,.
\end{align}
This concludes our computation of the bulk diagrams, for which we can obtain an analytical expression in the conformal limit.

\subsection{Boundary Diagrams} \label{subsec:BoundaryDiagrams}

\begingroup
\allowdisplaybreaks

We move now our attention to the boundary integrals.
The first diagram in the last line of Table \ref{table:DiagramsAtNNLO} reads
\begin{equation}
\DefectSSSSTwoLoopsXYOne\ = \frac{\lambda^4}{8} \left( \int_{-\infty}^{\tau_2} + \int_{\tau_4}^{\infty} \right) d\tau_6\ \veps (\tau_1\, \tau_3\, \tau_6) \int d^4 x_5\, \pd_{15} Y_{156}\, I_{25} I_{35} I_{45}\,.
\label{eq:TwoLoopsXYDiagramOne}
\end{equation}
This expression is straightforward to obtain from the Wick contractions.
The function $\veps (\tau_i\, \tau_j\, \tau_k)$ is defined in (\ref{eq:epsilon}).
This integral can be shown to give
\begin{align}
\DefectSSSSTwoLoopsXYOne\ =\ & \frac{\lambda^4}{16 \pi^2} \int d^4 x_5\, \frac{1}{|x_5^\perp|} I_{15}\, \pd_5 X_{2345} \notag \\
& \times \left\lbrace \tan^{-1} \Bigl( \frac{\tau_{45}}{|x_5^\perp|} \Bigr) + \tan^{-1} \Bigl( \frac{\tau_{25}}{|x_5^\perp|} \Bigr) - 2 \tan^{-1} \Bigl( \frac{\tau_{15}}{|x_5^\perp|} \Bigr) \right\rbrace \notag \\
& - \frac{\lambda^4}{8} \int d^4 x_5\, I_{25} I_{35} I_{45} \left( Y_{125} + Y_{145} \right) \notag \\
& + \frac{\lambda^4}{4} \int d^4 x_5\, Y_{115}\, I_{25} I_{35} I_{45}\,.
\label{eq:TwoLoopsXYDiagramOneb}
\end{align}
In order to obtain this result, we use the elementary identity
\begin{equation}
\int d^4 x_5\, \pd_{5} Y_{ij5}\, I_{k5} I_{l5} I_{m5} = - \int d^4 x_5\, \pd_5 X_{klm5}\, I_{i5} I_{j5}\,,
\label{eq:IdentityYXToXY}
\end{equation}
as well as integration by parts and the identity $(\partial_i-\partial_j) Y_{ijk}=-(2\partial_j+\partial_k)Y_{ijk}$.
In the limit $\tau_4 \to \infty$, this turns into
\begin{align}
\DefectSSSSTwoLoopsXYOne\ =\ & \frac{\lambda^4}{16 \pi^2} I_{24} \int d^4 x_5\, \frac{1}{|x_5^\perp|} I_{15} \pd_5 Y_{235} \notag \\
& \times \left\lbrace \tan^{-1} \Bigl( \frac{\tau_{25}}{|x_5^\perp|} \Bigr) - 2 \tan^{-1} \Bigl( \frac{\tau_{15}}{|x_5^\perp|} \Bigr) - \frac{\pi}{2} \right\rbrace \notag \\
& - \frac{\lambda^4}{8} I_{24} \left( H_{12,23} - 2 H_{11,23} \right) + O\left(\frac{1}{\tau_4^3}\right)\,.
\label{eq:TwoLoopsXYDiagramOnec}
\end{align}
Similarly, we have
\begin{align}
\DefectSSSSTwoLoopsXYTwo\ =\ & \frac{\lambda^4}{16 \pi^2} I_{24} \int d^4 x_5\, \frac{1}{|x_5^\perp|} I_{25} \pd_5 Y_{135} \notag \\ 
& \times \left\lbrace \tan^{-1} \Bigl( \frac{\tau_{15}}{|x_5^\perp|} \Bigr) + \tan^{-1} \Bigl( \frac{\tau_{35}}{|x_5^\perp|} \Bigr) - 2 \tan^{-1} \Bigl( \frac{\tau_{25}}{|x_5^\perp|} \Bigr) \right\rbrace \notag \\
& - \frac{\lambda^4}{8} I_{24} \left( H_{12,13} + H_{13,23} - 2 H_{22,13} \right) + O\left(\frac{1}{\tau_4^3}\right)\,, \label{eq:TwoLoopsXYDiagramTwo} \\
\DefectSSSSTwoLoopsXYThree\ =\ & \frac{\lambda^4}{16 \pi^2} I_{24} \int d^4 x_5\, \frac{1}{|x_5^\perp|} I_{35} \pd_5 Y_{125} \notag \\
& \times \left\lbrace \tan^{-1} \Bigl( \frac{\tau_{25}}{|x_5^\perp|} \Bigr) - 2 \tan^{-1} \Bigl( \frac{\tau_{35}}{|x_5^\perp|} \Bigr) + \frac{\pi}{2} \right\rbrace \notag \\
& - \frac{\lambda^4}{8} I_{24} \left( H_{12,23} - 2 H_{33,12} \right) + O\left(\frac{1}{\tau_4^3}\right)\,, \label{eq:TwoLoopsXYDiagramThree} \\
\DefectSSSSTwoLoopsXYFour\ =\ & \frac{\lambda^4}{16 \pi^2} \int d^4 x_5\, \frac{1}{|x_5^\perp|} I_{45} \pd_5 X_{1235} \notag \\
& \times \left\lbrace \tan^{-1} \Bigl( \frac{\tau_{15}}{|x_5^\perp|} \Bigr) + \tan^{-1} \Bigl( \frac{\tau_{35}}{|x_5^\perp|} \Bigr) - 2 \tan^{-1} \Bigl( \frac{\tau_{45}}{|x_5^\perp|} \Bigr) + \pi \right\rbrace \notag \\
& - \frac{\lambda^4}{8} \left( K_{14} + K_{34} \right) + \frac{\lambda^4}{4} K_{44} + O\left(\frac{1}{\tau_4^3}\right)\,. \label{eq:TwoLoopsXYDiagramFour}
\end{align}
Unfortunately, this is as far as we can go analytically because we are not able to solve further the first line of each diagram  \eqref{eq:TwoLoopsXYDiagramOneb}-\eqref{eq:TwoLoopsXYDiagramFour}, what we call $F^{XY,\mathit{n}}_1$.
The other terms luckily have been all encountered before, and by gathering them together we have
\begin{align}
F_1^{XY,\mathit{a}} =\ & 
\frac{\lambda^4}{64\pi^2}\frac{1}{I_{13}} \left( ( H_{11,23} + H_{22,13} + H_{33,11} ) + K_{44} \right) \notag \\
&- \frac{\lambda^4}{128\pi^2}\frac{1}{I_{13}} ( H_{12,13} + H_{13,23} + 2 H_{12,23} + K_{14} + K_{34} ) \notag \\
&+ \frac{\lambda^4}{8192\pi^4 \pi^2} \log \tau_4^2\, I_{34} \frac{Y_{123}}{I_{13}} + O\left(\frac{1}{\tau_4^3}\right)\,.
\label{eq:SigmaXYAnalytical}
\end{align}
Note that the last line was extracted from the first line of \eqref{eq:TwoLoopsXYDiagramFour}, such that in the following we consider expressions that are finite in the limit $\tau_4 \to \infty$.

\endgroup

We now solve the remaining integrals numerically. Before this last step, we gather all the analytical expressions encounter so far and we simplify them further. The expression shortens quite dramatically:
\begin{align}
F_1^{\mathit{a}} &:= F^{\text{SE}}_1+ F^{XX}_1 + F^{XH}_1 + F^{\text{Spider}}_1 + F^{XY,\mathit{a}}_1  = \notag \\
& \phantom{=\ }- \frac{\lambda^4}{64\pi^2} \frac{1}{I_{13}} \left(H_{12,13} + H_{13,23} + 2 H_{12,23} \right) \notag \\
& \phantom{=\ } + \frac{\lambda^4}{16384 \pi^4} \left( \log \tau_{12}^2 + \log \tau_{23}^2 - 8 \right) \frac{Y_{123}}{I_{13}}  + O\left(\frac{1}{\tau_4^3}\right)\,.
\label{eq:SigmaAnalytical}
\end{align}
We note in particular that all the divergent terms canceled each other. Moreover, the $\log \tau_4^2$ terms of $K_{14}$ and $K_{34}$ cancel the last line of \eqref{eq:SigmaXYAnalytical} as expected.

We can also rewrite (\ref{eq:SigmaAnalytical}) in terms of HPL's:
\begin{align}
F_1^{\mathit{a}} =\ &
\frac{1}{8192 \pi^8 \chi (1 - \chi)}
\Bigl\lbrace-4 H_1+H_{1,0} + H_{0,1} -2 H_{1,1} \notag \\
& + 3 ( H_{0,0,1} + H_{0,1,0} - 2 H_{1,0,0} )- 2 ( H_{1,1,0} + H_{1,0,1} - 2 H_{0,0,1} ) \notag \\
& + \chi\left(- 3 \zeta_3- 4 ( H_0 + H_1 )- 2 ( H_{0,0} - H_{1,1} ) \right. \notag \\
& \left. - H_{0,0,1} + H_{0,1,0} - 2 H_{1,0,0} - H_{1,0,1} + H_{1,1,0} - 2 H_{0,1,1}\right)\Bigr\rbrace \,.
\label{eq:SigmaAllAnalytical}
\end{align}

\subsection{Numerical Integration} \label{subsec:NumericalIntegration}

We now proceed with the numerical integration of the remaining terms. In order to be efficient, we first formulate an ansatz, based on the first two orders of the correlator and the form of the other terms. This ansatz is a complete basis of HPL's\footnote{We remind that the relevant details about HPL's can be found in Appendix \ref{app:HPL}.} with transcendentality up to $2\ell$.

A suitable ansatz is 
\begin{equation} \label{eq:ansatz}
F_1^{X Y, \mathrm{n}}=\frac{\lambda^4}{\chi(1-\chi)}\left(\sum_{\vec{a}} \alpha_{\vec{a}} H_{\vec{a}}+\chi \sum_{\vec{a}} \beta_{\vec{a}} H_{\vec{a}}\right)\,,
\end{equation}
where the transcendentality is then up to 3, as $\vec{a}=0,1,2,3$. In addition, the different OPE limits constrain $\vec{a}$ to contain only combinations of 0 and 1.

We then plot the numerical data and fit it with the ansatz. We find that the sum of numerical integrals has to be
\begin{equation} \label{eq:SigmaXYNumerical}
\begin{aligned}
F_1^{X Y, \mathrm{n}}= & \frac{\lambda^4}{24576 \pi^8} \frac{1}{\chi(1-\chi)}\left(2\left(6-\pi^2\right) H_1+3\left(H_{0,1}+H_{1,0}-2 H_{1,1}\right)\right. \\
& +3\left(3 H_{0,0,1}-H_{0,1,0}-2 H_{1,0,0}\right)+\chi\left(2\left(\pi^2-6\right) H_0+2\left(\pi^2-3\right) H_1\right) \\
& +6\left(H_{0,0}-H_{1,1}\right)+3\left(3\left(H_{0,0,1}+H_{1,1,0}\right)-\left(H_{0,1,0}+H_{1,0,1}\right)\right. \\
& \left.\left.\left.-2\left(H_{0,1,1}+H_{1,0,0}\right)\right)-9 \zeta_3\right)\right)\,.
\end{aligned}
\end{equation}

We then compare the numerical data with the analytic expression in Figure \ref{fig:numericalXYgraph}. We observe an impressive agreement between the function (\ref{eq:SigmaXYNumerical}) and the numerical integration of the sum of the terms in the first line of (\ref{eq:TwoLoopsXYDiagramOnec})-(\ref{eq:TwoLoopsXYDiagramFour}).

\begin{figure}[h]
\centering 
\includegraphics[scale=0.35]{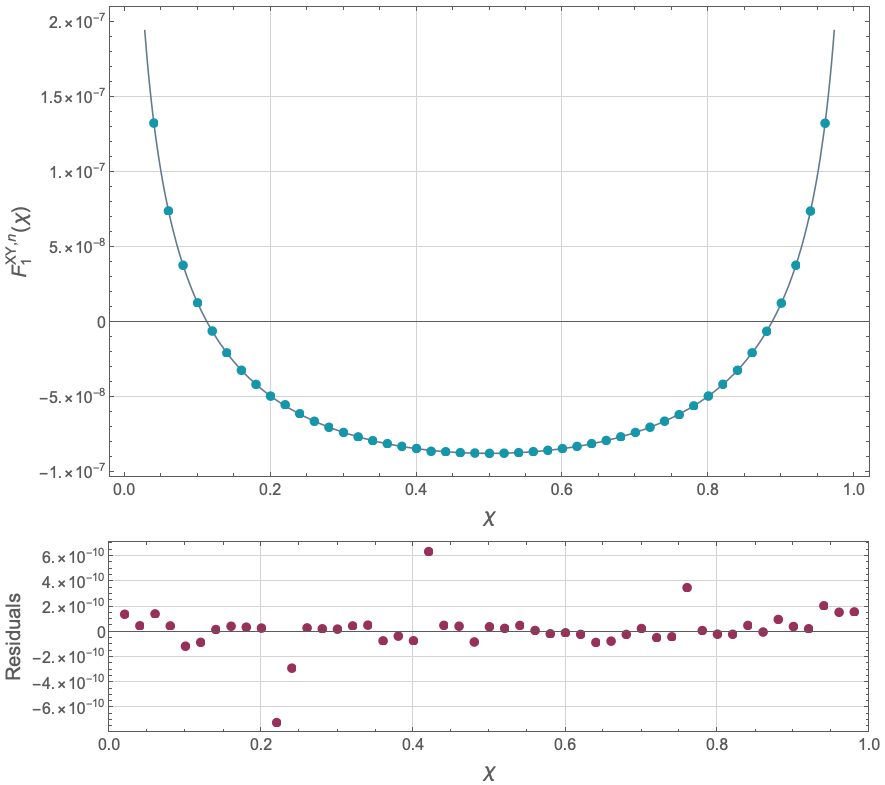}
\caption{\textit{The \textbf{\textcolor{gluoncolor}{light blue}} dots in this plot show the function $F_1^{X Y, n}(\chi)$, defined as the sum of the terms in the first line of (\ref{eq:TwoLoopsXYDiagramOnec})-(\ref{eq:TwoLoopsXYDiagramFour}), from which the $\log \tau_4^2$ dependence was removed as explained below (\ref{eq:SigmaXYAnalytical}). The grey line corresponds to the analytical expression (\ref{eq:SigmaXYNumerical}), obtained from comparing the ansatz (\ref{eq:ansatz}) to the numerical points. We observe a perfect agreement, and in particular, the residuals, indicated by the \textbf{\textcolor{vertexcolor}{magenta}} dots,  between the numerical data and the analytical curve are of order $10^{-10}$.}}
\label{fig:numericalXYgraph}
\end{figure}

\subsection{Final Result} \label{subsec:FinalResult}

We now combine all the elements in order to write down the full correlator at NNLO.  In particular, we sum equations (\ref{eq:SigmaAllAnalytical}) and (\ref{eq:SigmaXYNumerical}) that together make up the full expression of our $F_1$ channel. From here we extract $f(\chi)$, which we reinsert in (\ref{eq:F0FromLittlef}) and in (\ref{eq:F2FromLittlef}), allowing us to reconstruct the other channels, respectively $F_0$ and $F_2$, and to ultimately get the correlator through (\ref{eq:splittingR}). The last step is to unit-normalize the correlator as in (\ref{eq:unitnormalized}). This leads to
\begin{equation}
\begin{aligned}
F_1^{(2)}=&\frac{1}{192\pi^4\,\chi(1-\chi)}\bigl(\pi^2 H_1-3(H_{1,0,1}+H_{1,1,0}-2(H_{0,1,0}+H_{0,1,1}-H_{1,0,0})) \notag\\
&-3\chi(\frac{\pi^2}{3}H_0-(H_{0,0,1}+H_{1,1,0})+H_{0,1,0}+H_{1,0,1}+3\zeta_3)\bigr) \,,
\end{aligned}
\end{equation}
which agrees perfectly with \cite{Cavaglia:2022qpg}.

\section{Correlation Function of Unprotected Operators} \label{subsubsec:6666}

Since so far we explore four-point functions of protected operators only, in this section, we have a quick look at a correlator of unprotected operators.  Particularly, we insert the operator $\phi^6$ on the Wilson line.  Recall that $\phi^6$ is the only fundamental field that couples directly to the Wilson line. This introduces new types of diagrams as the one illustrated in Figure \ref{fig:Lambda} on the right.

As for the case of protected operators, we introduce a reduced correlator, which can be read from
\begin{equation} \label{eq:corr-6666}
\vvev{\phi^6 (\tau_1) \phi^6 (\tau_2) \phi^6 (\tau_3) \phi^6 (\tau_4)} = \frac{1} {\tau_{12}^{2\smash{\Delta_{\phi^6}}} \tau_{34}^{2\smash{\Delta_{\phi^6}}}} \Am^{6666} (\chi)\,.
\end{equation}
Similarly to \eqref{eq:perturbative}, the reduced correlator obeys the following perturbative expansion:
\begin{equation}
\Am^{6666} (\chi) = \sum_{k=0}^{\infty} \lambda^{k} \Am_{6666}^{(k)} (\chi)\,.
\label{eq:Am6666-pert}
\end{equation}
Note that, as opposed to the protected case presented above, this correlator consists of a \textit{single} $R$-symmetry channel.

At leading order, using (\ref{eq:recursioneventree}) and unit-normalizing we find that it agrees with $\Am_{1111}^{(0)}$:
\begin{align}
\Am_{6666}^{(0)} (\chi) &= \Am_{1111}^{(0)} (\chi) = \frac{1 - 2 \chi(1-\chi)}{(1-\chi)^2}\,.
\label{eq:4ptphi6LO}
\end{align}

At next-to-leading order, the conformal prefactor in \eqref{eq:corr-6666} produces logs when expanded around $\lambda \sim 0$ because of the anomalous dimension of $\phi^6$:
\begin{equation}
\frac{1}{\tau_{12}^{2\smash{\Delta_{\phi^6}}} \tau_{34}^{2\smash{\Delta_{\phi^6}}}} = \frac{1}{\tau_{12}^2 \tau_{34}^2} \left( 1 - \lambda\, \gamma^{(1)}_{\phi^6} \log \tau_{12}^2 \tau_{34}^2 + O(\lambda^2)\right)\,.
\label{eq:log6666}
\end{equation}
The log term must be taken into account in order to isolate the reduced correlator $\Am_{6666}^{(1)}$ at next-to-leading order. Moreover, as discussed in Section \ref{subsec:evencase}, the correlator can be expressed as a sum of building blocks and $U$-diagrams. Applying \eqref{eq:oldnew} and computing the integrals with the help of Appendix \ref{sec:integralsapp} results in the following elegant expression:
\begin{align}
\Am_{6666}^{(1)} (\chi) =& \Am_{1111}^{(1)} (\chi)  + \frac{\lambda}{12 (1-\chi)^2} \left( 1-2\chi(1-\chi) \phantom{\frac{3}{\pi^2}} \right. \notag \\
& \phantom{+ \frac{\lambda}{12 (1-\chi)^2} \left( \right.} \left. + \frac{3}{\pi^2} ( 3\chi(1-\chi) + \chi^2 \log \chi - (1-\chi(2-3\chi)) \log(1-\chi) ) \right)\,.
\label{eq:4ptphi6NLO}
\end{align}
The first line corresponds to the four-point correlator of protected scalars (\ref{eq:buildingblocksNLO}) that acts as a building block, while the extra contributions come from the computation of the $U$-diagrams.
 
We can also use our recursive algorithm to compute four-point functions involving the composite operators introduced in Section \ref{subsubsec:twopttwo}. Some examples can be found in \cite{Barrat:2022eim} and in the ancillary \textsc{Mathematica} notebook.

\section{Four-Point Data} \label{sec:blockexpansion4pt}
\begingroup
\allowdisplaybreaks

We conclude this chapter by expanding the four-point correlators in conformal blocks, as a consistency check and to extract CFT data. As described in Section \ref{sec:corrfunctionsandCFTdata}, the CFT data consists of the scaling dimensions and the OPE coefficients of the operators in the spectrum. 

For these four-point functions, we consider a particular OPE,  the comb channel, which we introduced in Section \ref{sec:conformalblockexpansion}, and we compare the simplest OPE coefficients (always involving either $\phi^i$ or $\phi^6$) to the results derived in Section \ref{subsec:threepoint}.  

Be aware that our analysis is purely bosonic. However, a superconformal analysis for the correlator $\vev{\phi_1 \phi_1 \phi_1 \phi_1}$ has been carried out in \cite{Liendo:2018ukf}, where the superconformal blocks were also derived.  It would be interesting to replicate this analysis also for the NNLO discussed in Section \ref{sec:2loop} in order to extract CFT.  We leave this interesting idea for the future.

Let us dive then into the expansion of the correlators $\vvev{\phi^i \phi^j \phi^k \phi^l}$ and $\vvev{\phi^6 \phi^6 \phi^6 \phi^6}$.
\begin{figure}[h]
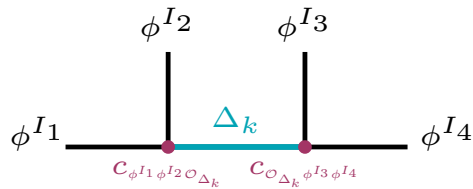

\centering
\fourcombblocks
\caption{\textit{Representation of four-point functions in the comb channel. One operator labeled \textbf{$\textcolor{gluoncolor}{\Op_{\Delta_k}}$} is being exchanged and the \textbf{\textcolor{vertexcolor}{OPE coefficients}} consist of three-point functions squared, when all the external operators are identical.}}
\label{fig:fourcombch}
\end{figure}

\subsubsection{Protected Operators} 

It can be seen from Figure \ref{fig:fourcombch} that the four-point function of protected operators $\vvev{\phi^i \phi^j \phi^k \phi^l}$ should contain the three-point function $\vvev{\phi^i \phi^j \phi^6}$,when the exchanged operator is $\Op_{\Delta_k}= \phi^6$. This correlator has been computed in \eqref{eq:ij6} at leading order.

%
This coefficient can be easily extracted from\footnote{In principle one needs the superconformal blocks of \cite{Liendo:2018ukf} in order to check that coefficient, but in practice, it turns out that the highest-weight channel $F_0$ can be expanded in the bosonic conformal blocks given in \eqref{eq:combblocks} if the goal is only to check the coefficient corresponding to the operator being exchanged with the lowest $\Delta$.}
\begin{align}
F_0 (\chi) &= 1 + c_{\phi^i \phi^j \phi^6} c_{\phi^6 \phi^k \phi^l} g_{\Delta_k=1}(\chi) + \ldots \notag \\
&= 1 + c_{\phi^i \phi^j \phi^6} c_{\phi^6 \phi^k \phi^l} \chi + \ldots\,,
\label{eq:expblock}
\end{align}
where the first term corresponds to the exchange of the identity operator $\mathds{1}$ and is $1$ due to the unit-normalization of the two-point function. According to equation \eqref{eq:ij6}, we have
\begin{equation}
c_{\phi^i \phi^j \phi^6} c_{\phi^6 \phi^k \phi^l} = \frac{\lambda}{8 \pi^2} + O(\lambda^2)\,.
\label{eq:tocheck}
\end{equation}

We now expand the correlator at $\chi \sim 0$ for the leading and next-to-leading orders and compare the order $O(\chi)$ to \eqref{eq:expblock} and \eqref{eq:tocheck}. From \eqref{eq:buildingblocksLO}, we see that
\begin{equation}
F^{(0)}_0 (\chi) = 1\,,
\end{equation}
and this implies that $c_{\phi^i \phi^j \phi^6} c_{\phi^6 \phi^k \phi^l}$ vanishes at $O(\lambda^0)$ as predicted by \eqref{eq:tocheck}. For the next order, we expand \eqref{eq:buildingblocksNLO} to find that
\begin{equation}
F^{(1)}_0 (\chi) = \frac{1}{8\pi^2} \chi + \ldots\,,
\end{equation}
which is in perfect agreement with \eqref{eq:expblock} and \eqref{eq:tocheck}.

\subsubsection{Unprotected Operators}

We focus now our attention on the four-point function of \textit{unprotected} operators $\phi^6$. In this case, it is clear from $R$-charge conservation that the only operator with (bare) scaling dimension $\Delta_k = 1$ that can appear in the exchange is the unprotected scalar $\phi^6$ itself.  Thus the correlator $\vvev{\phi^6 \phi^6 \phi^6 \phi^6}$ is expected to contain the three-point function $\vvev{\phi^6 \phi^6 \phi^6}$ in its expansion. This coefficient has been computed in \eqref{eq:3ptphi6} and can be compared to the four-point function obtained in Section \ref{subsubsec:6666}.
%
Expanding the correlator in blocks following \eqref{eq:expansionblocks}, we find
\begin{align}
\mathcal{A}^{6666} (\chi) &= 1 + c_{\phi^6 \phi^6 \phi^6}^2 g_{1} (\chi) + \ldots \notag \\
&= 1 + c_{\phi^6 \phi^6 \phi^6}^2 \chi + \ldots\,,
\end{align}
which we compare to the results listed in Section \ref{subsubsec:6666}.

From \eqref{eq:4ptphi6LO}, we find that the correlator at leading order can be expanded as
\begin{equation}
\mathcal{A}^{(0)}_{6666} (\chi) = 1+\chi^2+\ldots \,,
\end{equation}
and thus we observe that $c_{\phi^6 \phi^6 \phi^6}^2$ vanishes at $O(\lambda^0)$ as expected from \eqref{eq:3ptphi6} since there is no term of order $O(\chi)$. Again the first term corresponds to the exchange of the identity operator $\mathds{1}$ and it is $1$ due to the unit-normalization of the two-point function.

We are also able to derive a closed-form expression for the OPE coefficients with arbitrary $\Delta_k$:
\begin{equation}
c_{\phi^6 \phi^6 \Op_{\Delta_k}} c_{\Op_{\Delta_k} \phi^6 \phi^6} \rvert_{O(\lambda^0)} = \frac{4 \sqrt{\pi}\, (\Delta_k-1)\, \Gamma(\Delta_k+1)}{4^{\Delta_k}\,\Gamma(\Delta_k-\frac{1}{2})}\,.
\end{equation}
%
We expect that there exist several operators corresponding to the bare scaling dimensions $\Delta_k>1$, and thus that these coefficients are in fact \textit{averages} of three-point functions.

We now move to the next order and we expand the expression of the correlator in (\ref{eq:4ptphi6NLO}) to find
\begin{equation}
\mathcal{A}^{(1)}_{6666} (\chi) = \frac{9}{8 \pi^2} \chi + \ldots \,.
\end{equation}
This coefficient should be compared to \eqref{eq:3ptphi6}, which predicts
\begin{equation}
c_{\phi^6 \phi^6 \phi^6}^2 = \frac{9 \lambda}{8 \pi^2} + O(\lambda^2)\,.
\end{equation}
Thus, we observe a perfect match between the OPE coefficient obtained from the four-point function and the three-point function computed using the recursion relation.

\chapter{Higher-Point Functions} \label{ch:higherpoint}

It is now time to witness the real power of the recursion relations presented in Chapter~\ref{sec:recursionrelations}; we finally compute multipoint correlators. In particular, we focus on five- and six-point functions.  We write down some examples of these correlators, after discussing their kinematics. As for the four-point case, we also expand these correlators in conformal blocks, to perform consistency checks and to extract novel CFT data.  

The big pool of perturbative results derived thanks to the recursion relations allow us to conjecture multipoint Ward identities. We introduce them and discuss their properties. We even put them immediately to use in the bootstrap of a five-point function at strong coupling, to which is dedicated the conclusive section.

This chapter is based on \cite{Barrat:2021tpn, Barrat:2022eim} and some work in progress \cite{Barrat:2023pev}.

Before jumping straight into the five-point functions, it is worth mentioning the case of \textit{extremal} correlators, i.e. correlators for which the length of one operator is equal to the sum of the lengths of all the other operators:
\begin{equation*}
\vev{\phi_{\Delta_1} (\tau_1) \ldots \phi_{\Delta_{n-1}} (\tau_{n-1}) \phi_{\Delta_n} (\tau_n)}\,,
\end{equation*}
with $\Delta_n = \Delta_1 + \ldots + \Delta_{n-1}$.
We discussed in detail this case in \cite{Barrat:2021tpn}, where through the recursion relations we find a closed form of the correlators up to NLO.

\section{Five-Point Functions} \label{subsec:fivepoint}

Of course, we start from five-point functions, particularly from their kinematics. The discussion that follows on $R$-symmetry channels is limited to protected operators, so $\phi^i, \,i=1,\dots,5$.

\subsection{Kinematics} \label{subsubsec:5kinematics}

In this section, we review explicitly the kinematics introduced at the end of Section \ref{sec:wilsonlinecorrelators}. Here there are \textit{two} spacetime cross-ratios, which are defined as in (\ref{eq:crossratios}):
\begin{equation} \label{eq:crossratios5pt}
\chi_1 = \frac{\tau_{12} \tau_{45}}{\tau_{14} \tau_{25}}\,, \qquad \chi_2 = \frac{\tau_{13} \tau_{45}}{\tau_{14} \tau_{35}}\,,
\end{equation}
and \textit{five} $R$-symmetry cross-ratios defined following (\ref{eq:rRsymmecross}), (\ref{eq:sRsymmecross}) and (\ref{eq:tRsymmecross})\footnote{Note that we drop the subscript of $t_{ij}$,  since in this case there is only one $R$-symmetry cross-ratio of this kind.}:
\begin{gather} \label{eq:Rsymmcrossratios5pt}
r_1 = \frac{(u_1 \cdot u_2)(u_4 \cdot u_5)}{(u_1 \cdot u_4)(u_2 \cdot u_5)}\,, \qquad s_1 = \frac{(u_1 \cdot u_5)(u_2 \cdot u_4)}{(u_1 \cdot u_4)(u_2 \cdot u_5)}\,, \notag \\
r_2 = \frac{(u_1 \cdot u_3)(u_4 \cdot u_5)}{(u_1 \cdot u_4)(u_3 \cdot u_5)}\,, \qquad s_2 = \frac{(u_1 \cdot u_5)(u_3 \cdot u_4)}{(u_1 \cdot u_4)(u_3 \cdot u_5)}\,, \notag \\
t = \frac{(u_1 \cdot u_5)(u_2 \cdot u_3)(u_4 \cdot u_5)}{(u_1 \cdot u_4)(u_2 \cdot u_5)(u_3 \cdot u_5)}\,.
\end{gather}
Using these cross-ratios the correlators can be expressed in terms of $R$-symmetry channels. The number of channels depends on the scaling dimensions of the external operators (see Table \ref{table:Rsymmchannels} for some examples).  Understanding this number is only a combinatorial matter.

 Let us take as an example the simplest five-point: $\vvev{\phi_1 \phi_1 \phi_1 \phi_1 \phi_2}$\footnote{Note that the correlator $\vvev{\phi_1 \phi_1 \phi_1 \phi_1 \phi_1}$ is zero due to $R$-symmetry.}. We have to consider all the possible combinations of the $R$-symmetry vectors $u$: $u_1$, $u_2$, $u_3$, $u_4$, $u_5$ and again $u_5$, since the last operator is $\phi_2$. We have to remember that we cannot ``pair'' the $R$-symmetry vectors associated with the same operator,  in this case, the $u_5$ vectors. This is ensured by the properties of the $u$ vectors introduced in (\ref{eq:uvector}). Therefore we can make in total six different combinations that we write below:
\begin{equation}
\begin{aligned}
&(u_1\cdot u_2)(u_3\cdot u_5)(u_4\cdot u_5)\,, \qquad \qquad (u_1\cdot u_3)(u_2\cdot u_5)(u_4\cdot u_5)\,, \\
&(u_1\cdot u_4)(u_2\cdot u_5)(u_3\cdot u_5)\,,  \qquad \qquad (u_1\cdot u_5)(u_2\cdot u_3)(u_4\cdot u_5)\,, \\
&(u_1\cdot u_5)(u_2\cdot u_4)(u_3\cdot u_5)\,, \qquad \qquad (u_1\cdot u_5)(u_2\cdot u_5)(u_3\cdot u_4)\,.
\end{aligned}
\end{equation}
This strategy can be easily implemented to compute the $R$-symmetry channels of all the correlators. However, it is not straightforward to obtain a formula for the number of channels in the most generic case.  We thus determined it for the special case in which all external dimensions are equal to one,  where this number is reproduced by the double factorial: $(n-1)!! = 1 \cdot 3 \cdot 5 \cdot \ldots \cdot (n-1)$. Here $n$ is the (even) number of operators in the correlation function, and for example, the six-point function $\vvev{\phi_1 \phi_1 \phi_1 \phi_1 \phi_1 \phi_1}$ consists of $5!!=1 \cdot 3 \cdot 5=15$ $R$-symmetry channels.

In general,  at finite $N$,  all the channels would be present at any loop order but in the planar limit $N \to \infty$ many channels do not contribute, at least for the NLO computations.

In the following, we are presenting only two examples of five-point correlators: the simplest case with protected operators $\vvev{\phi_1 \phi_1 \phi_1 \phi_1 \phi_2}$ and the correlator $\vvev{\phi^6 \phi^6 \phi^6 \phi^6 \phi^6}$ constituted entirely by unprotected operators. However, many more correlators\footnote{Examples of correlators computed in \cite{Barrat:2021tpn,Barrat:2022eim} are $\vev{\phi_1 \phi_1 \phi_2 \phi_2 \phi_4}$, $\vev{\phi_1 \phi_1 \phi_2 \phi_2 \phi_2}$ and $\vvev{\phi^i \phi^j \phi^6 \phi^6 \phi^6}$. Note that the different brackets are not typos. In fact, in \cite{Barrat:2021tpn} we did not compute unit-normalized correlators, while in \cite{Barrat:2022eim} we did.} have been computed in \cite{Barrat:2021tpn,Barrat:2022eim} and in their respective ancillary \textsc{Mathematica} notebook using the recursion relations in combination with the pinching technique.

\begin{table}[h]
\centering
\caption{\textit{Number of $R$-symmetry channels for different five-point functions, obtained by pinching $n$-point functions $\vvev{\phi_1 \ldots \phi_1}$ from $n=6$ to $n=12$. }}
\begin{subtable}[c]{0.5\textwidth}
\centering
\begin{tabular}{cc}
\hline
 $\Delta_1$, $\Delta_2$, $\Delta_3$, $\Delta_4$, $\Delta_5$ & channels \\
\hline
\vspace{0.267cm}
$1,1,1,1,2$ & 6  \\
$1,1,1,1,4$ & 1  \\
$1,1,1,2,3$ & 6  \\
\vspace{0.26cm}
$1,1,2,2,2$ & 10\\
$1,1,1,2,5$ & 1  \\
$1,1,1,3,4$ & 6\\
$1,1,2,2,4$ & 6 \\
$1,1,2,3,3$ & 10  \\
$1,2,2,2,3$ & 15 \\
$2,2,2,2,2$ & 22 \\ \hline
\end{tabular}
\end{subtable}
\hspace{-2.2cm}
\begin{subtable}[c]{0.55\textwidth}
\centering
\begin{tabular}{cc}
\hline
$\Delta_1$, $\Delta_2$, $\Delta_3$, $\Delta_4$, $\Delta_5$ & channels \\
\hline
$1,1,1,3,6$ & 1 \\
$1,1,2,2,6$ & 1 \\
$1,1,1,4,5$ & 6 \\
$1,1,2,3,5$ & 6\\
$1,2,2,2,5$ & 6 \\
$1,1,2,4,4$ & 10 \\
$1,1,3,3,4$ & 10 \\
$1,2,2,3,4$ & 15 \\
$2,2,2,2,4$ & 21 \\
$1,2,3,3,3$ & 21\\
$2,2,2,3,3$ & 29 \\ \hline
\end{tabular}
\end{subtable}
\label{table:Rsymmchannels}
\end{table}

\subsection{$\vvev{\phi_1 \phi_1 \phi_1 \phi_1 \phi_2}$} \label{subsubsec:11112}

We start by considering the simplest pinching case, which is obtained by bringing the last two operators of the six-point function $\vev{\phi_1 \phi_1 \phi_1 \phi_1 \phi_1 \phi_1}$ together by taking the limits
\begin{equation}
(u_6,\tau_6) \to (u_5, \tau_5)\,,
\end{equation}
and it takes the following form:
\begin{equation}
\vvev{\phi_1 \phi_1 \phi_1 \phi_1 \phi_2} = \Km_{11112} (u, \tau) \Am_{11112}(\chi_1, \chi_2, r_1, r_2, s_1,s_2,t)\,.
\end{equation}
The spacetime cross-ratios have just been introduced in (\ref{eq:crossratios5pt}) as the $R$-symmetry ones in (\ref{eq:Rsymmcrossratios5pt}). It is particularly convenient to choose the superconformal prefactor to be:
\begin{equation}
\Km_{11112} := \frac{(u_1 \cdot u_4)(u_2 \cdot u_5)(u_3 \cdot u_5)}{\tau_{14}^2 \tau_{25}^2 \tau_{35}^2}\,.
\end{equation}

As for the four-point function,  this correlator satisfies crossing symmetry:
\begin{equation}
\vvev{\phi_1(\tau_1) \phi_1(\tau_2) \phi_1(\tau_3) \phi_1(\tau_4) \phi_2(\tau_5)} = \vvev{\phi_1(\tau_4) \phi_1(\tau_3) \phi_1(\tau_2) \phi_1(\tau_1) \phi_2(\tau_5)}\,,
\end{equation}
from which we identify:
\begin{equation} \label{eq:FivePoint_CrossingSymmetry}
\Am ( \chi_1, \chi_2 ; r_1, r_2, s_1, s_2, t ) = \Am ( 1 - \chi_2, 1 - \chi_1 ; s_2, s_1, r_2, r_1, t )\,.
\end{equation}
On the line of \cite{Liendo:2016ymz}, in Figure \ref{fig:fivepointcrossing} we give a graphical intuition for the crossing symmetry of this five-point.
\begin{figure}[h]
\centering 
\includegraphics[scale=0.45]{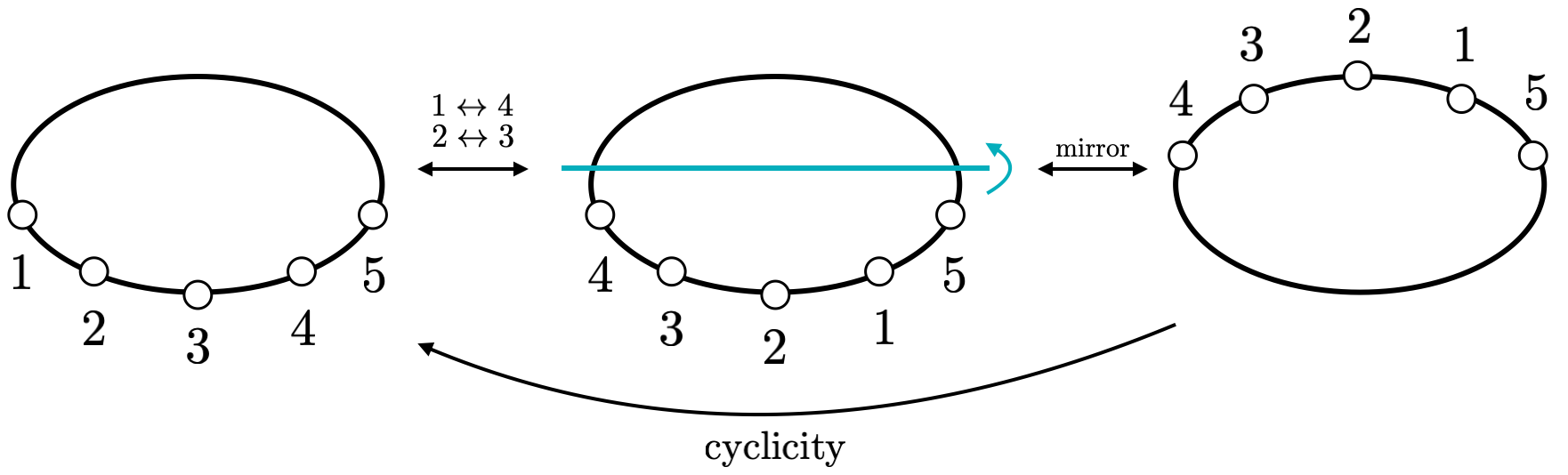}
\caption{\textit{In this figure we give a graphical illustration of how to interpret the crossing symmetry for the five-point function.}}
\label{fig:fivepointcrossing}
\end{figure}

This correlator consists of \textit{six} $R$-symmetry channels:
\begin{equation} \label{eq:5ptchannelsrs}
\mathcal{A}_{11112}=F_0+\frac{r_1}{\chi_1^2} F_1+\frac{s_1}{\left(1-\chi_1\right)^2} F_2+\frac{r_2}{\chi_2^2} F_3+\frac{s_2}{\left(1-\chi_2\right)^2} F_4+\frac{t}{\chi_{12}^2} F_5\,,
\end{equation}
where we suppress the dependency on the spacetime cross-ratios for compactness, i.e. $F_j := F_j (\chi_1, \chi_2)$. The $R$-symmetry channels have the following perturbative expansion
\begin{equation}
F_{j} = \sum_{k=1}^{\infty} \lambda^{\frac{k}{2}} F^{(k)}_{j}\, .
\end{equation}
The prefactor for each channel is chosen such that it becomes $1$ in the topological limit. 

In particular, the topological sector, which again corresponds to setting equal spacetime and $R$-symmetry cross-ratios,  takes the form for this five-point at large $N$:
\begin{equation}
\Fds = \frac{6\, \Ids_2^2}{\lambda\, \mathbb{I}_1^2} \frac{2 (\mathbb{I}_1 - 2) (\mathbb{I}_1 + 28) + \lambda ( 2 \mathbb{I}_1 - 19 ) }{ \sqrt{ 3 \lambda - (\mathbb{I}_1 - 2)(\mathbb{I}_1 + 10) } }\,,
\label{eq:FivePointTopologicalSector}
\end{equation}
where we define for compactness
\begin{equation}
\Ids_a := \frac{\sqrt{\lambda} I_0 (\sqrt{\lambda})}{ I_a (\sqrt{\lambda}) }\,.
\label{eq:TopologicalSector_HelpFunction}
\end{equation}
Expanded in the weak and strong coupling regimes, this function results in
\begin{align}
\Fds &\overset{\lambda \sim 0}{\sim} 3 + \frac{7 \lambda}{48} + O (\lambda^2)\,, \label{eq:FivePointTopologicalSector_WeakCoupling} \\
\Fds &\overset{\lambda \gg 1}{\sim} 6 \sqrt{2} - \frac{33}{\sqrt{2}} \frac{1}{\sqrt{\lambda}} + \frac{189}{8 \sqrt{2}} \frac{1}{\lambda} + O(\lambda^{-3/2})\,.
\label{eq:FivePointTopologicalSector_StrongCoupling}
\end{align}
These expansions have been obtained using matrix models in \cite{Giombi:2021zfb},  but they could also be derived by pinching the full correlator up to two- or three-point functions,  in order to compare it to the NLO results given in \eqref{eq:twopt} and \eqref{eq:threept}. 

We compute now the leading order to find that only three channels do not vanish:
\begin{equation} \label{eq:FivePoint_LO}
F_1^{(0)} = F_4^{(0)} = F_5^{(0)} = 1\,, \qquad F_0^{(0)} = F_2^{(0)} = F_3^{(0)} = 0\,.
\end{equation}

At next-to-leading order, almost all the channels are present and we obtain the following contributions:
\begin{subequations} \label{eq:FivePoint_NLOGroup2_3}
\begin{align}
F_1^{(1)} &= F_1^{(1), \text{4pt}} \left( \frac{\chi_1}{\chi_2} \right) + \frac{1}{48}\,,  \\
F_2^{(1)} &= F_0^{(1), \text{4pt}} \left( \frac{\chi_{21}}{1-\chi_1} \right)\,,  \\
F_3^{(1)} &= F_0^{(1), \text{4pt}} \left( \frac{\chi_1}{\chi_2} \right)\,,\\
F_4^{(1)} &= F_1^{(1), \text{4pt}} \left( \frac{1-\chi_2}{1 - \chi_1} \right) + \frac{1}{48}\,,  \\
F_5^{(1)} &= F_1^{(1), \text{4pt}} \left( \frac{\chi_{21}}{1 - \chi_1} \right) + F_1^{(1)} \left( \frac{\chi_{21}}{\chi_2} \right) + \frac{1}{48}\,.
\end{align}
\end{subequations}
The remaining channel $F_0$ is non-planar at this order. These results have been formulated in terms of $F^{\text{4pt}}_i$, being the $R$-symmetry channels of the four-point function $\vvev{\phi_1 \phi_1 \phi_1 \phi_1}$ written in (\ref{eq:buildingblocksNLO}).

Some checks can be performed on this result. First, the channels are individually finite, as expected for correlators of protected operators. In fact, the pinching of operators produces divergences, but they cancel again when summing up the different contributions. It is possible to further pinch the operators of the five-point function in order to produce e.g. four- and three-point functions. In particular we checked that pinching $\vvev{\phi_1 \phi_1 \phi_1 \phi_1 \phi_2}$ accordingly matches the known results for $\vvev{\phi_1 \phi_1 \phi_1 \phi_3}$, $\vvev{\phi_1 \phi_1 \phi_2 \phi_2}$ and $\vvev{\phi_1 \phi_2 \phi_3}$. 

We note also that the correlator given above seems to be the top component of a \textit{family} of correlators, namely $\vvev{\phi_1 \phi_1 \phi_1 \phi_k \phi_{k+1}}$, for which only the prefactor changes channel-wise.  The additional correlators of this family are in the ancillary \textsc{Mathematica} notebook of \cite{Barrat:2021tpn}. This classification into families of correlators is expected to hold only at next-to-leading order and in the planar limit.

\subsection{$\vvev{\phi^6 \phi^6 \phi^6 \phi^6 \phi^6}$} \label{subsubsec:66666}

We want to present a second example of five-point functions: the case of all unprotected operators. When the number of operators is odd, there are no building blocks, and the recursion relation provides the leading order of the correlators only. We therefore only compute the correlator of five unprotected scalars $\phi^6$ at leading order. It can be factorized in
\begin{equation}
\vvev{\phi^6 (\tau_1) \phi^6 (\tau_2) \phi^6 (\tau_3) \phi^6 (\tau_4) \phi^6 (\tau_5)} =\left( \frac{\tau_{42}}{\tau_{12}^{2}\tau_{32}^{\phantom{}}\tau_{43}^{\phantom{}}\tau_{45}^{2}}\right)^{\smash{\Delta_{\phi^6}}} \Am^{66666} (\tilde{\chi}_1, \tilde{\chi}_2)\,,
\end{equation}
where the cross-ratios are defined according to (\ref{eq:crossratio}) as follows:
\begin{equation}
\tilde{\chi}_1 = \frac{\tau_{12} \tau_{45}}{\tau_{14} \tau_{25}}\,, \qquad \tilde{\chi}_2 = \frac{\tau_{13} \tau_{45}}{\tau_{14} \tau_{35}}\,.
\end{equation}
and the prefactor is set as in \eqref{eq:prefactor}\footnote{In this section we are using a different convention compared to the previous section because one application of this result is the extraction of the CFT data. The bosonic conformal blocks were already present in the literature and to make use of them, we have to adapt our conventions to the one of those paper authors, as explained below (\ref{eq:crossratio}).}.

The reduced correlator obtained using the recursion relation \eqref{eq:recursionoddtree} obeys the following perturbative expansion:
\begin{equation}
\Am^{66666} (\tilde{\chi}_1,\tilde{\chi}_2) = \sum_{k=1}^{\infty} \lambda^{k/2} \Am_{66666}^{(k)} (\tilde{\chi}_1, \tilde{\chi}_2)\,.
\label{eq:Am1166-pert}
\end{equation}
The leading order is $O(\sqrt{\lambda})$ and we obtain the following expression:
\begin{align}
\Am_{66666}^{(1)} (\tilde{\chi}_1,\tilde{\chi}_2) = - \frac{3}{2\sqrt{2} \pi} &  \left( \frac{\tilde{\chi}_1 ( 2 (1-\tilde{\chi}_1) - \tilde{\chi}_1^2)}{1-\tilde{\chi}_1} + \frac{\tilde{\chi}_2 (1-\tilde{\chi}_2) - 1}{1-\tilde{\chi}_2} \right. \notag \\
& \quad \left. + \frac{\tilde{\chi}_1^2 (1-\tilde{\chi}_1)^2}{(1-\tilde{\chi}_1 - \tilde{\chi}_2)^2} + \frac{1 -2\tilde{\chi}_1 (1-\tilde{\chi}_1) (1+\tilde{\chi}_1)}{1-\tilde{\chi}_1 - \tilde{\chi}_2} \right)\,.
\label{eq:5ptphi6}
\end{align}

\subsection{Five-Point Data} \label{sec:blockexpansion5pt}

In this section, we expand in conformal blocks the five-point function $\vvev{\phi^6 \phi^6 \phi^6 \phi^6 \phi^6}$ just derived, as a consistency check and to extract CFT data.  We analyze in particular the comb channel, already introduced in Section \ref{sec:conformalblockexpansion},  performing analogous checks as for the four-point functions.

Note that we focus our attention on the five-point function of unprotected scalars, since we explore the expansion of the correlator $\vvev{\phi_1 \phi_1 \phi_1 \phi_1 \phi_2}$ involving protected operators at the end of this chapter, in the context of the strong coupling bootstrap of this correlator.

\begin{figure}[h]
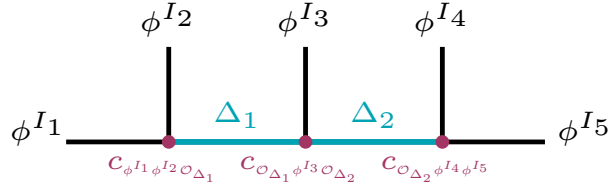

\centering
\fivecombblocks
\caption{\textit{Representation of five-point functions in the comb channel. Two operators are being exchanged, labeled in the diagram by the scaling dimensions \textbf{\textcolor{gluoncolor}{$\Delta_1$}} and \textbf{\textcolor{gluoncolor}{$\Delta_2$}}. The OPE coefficients consist of a product of three \textbf{\textcolor{vertexcolor}{three-point functions.}}}}
\label{fig:fivecombch}
\end{figure}

The comb channel for five-point functions is represented in Figure \ref{fig:fivecombch}. In this case, we are interested in checking the three-point function $\vvev{\phi^6 \phi^6 \phi^6}$, which can be accessed e.g. by setting $\Op_1 = \phi^6$, $\Op_2 = \mathds{1}$.
Thus, the expansion of the correlator in blocks up to this term reads
\begin{align}
\mathcal{A}^{66666} (\tilde{\chi}_1,\tilde{\chi}_2) &= c_{\phi^6 \phi^6 \phi^6} g_{1,0} (\tilde{\chi}_1,\tilde{\chi}_2) + \ldots \notag \\
&= c_{\phi^6 \phi^6 \phi^6} \tilde{\chi}_1 + \ldots\,,
\end{align}
where the OPE coefficient is just $c_{\phi^6 \phi^6 \phi^6}$ because $c_{\phi^6 \phi^6 \mathds{1}} = 1$. Note that, in this case, there is no term corresponding to the exchange of two identity operators, since the OPE coefficient $c_{\phi^6 \phi^6 \mathds{1}} c_{\mathds{1} \phi^6 \mathds{1}} c_{\mathds{1} \phi^6 \phi^6}$ vanishes due to the presence of a one-point function in the middle.

Here we only look at the leading order $O(\sqrt{\lambda})$ given in equation \eqref{eq:5ptphi6}, which upon expanding at $\tilde{\chi}_1, \tilde{\chi}_2 \sim 0$ reads
\begin{equation}
\Am_{66666}^{(0)} (\tilde{\chi}_1,\tilde{\chi}_2) = \frac{3}{2\sqrt{2} \pi} \tilde{\chi}_1 + \ldots\,,
\end{equation}
and thus we observe a perfect agreement of $c_{\phi^6 \phi^6 \phi^6}$ with equation \eqref{eq:3ptphi6}.

We can also derive a closed-form expression for the OPE coefficients at leading order:
\begin{align}
\left. c_{\phi^6 \phi^6 \Op_{\Delta_1}} c_{\Op_{\Delta_1} \phi^6 \Op_{\Delta_2}} c_{\Op_{\Delta_2} \phi^6 \phi^6} \right|_{O(\sqrt{\lambda})} =& \frac{12 \sqrt{2} \sqrt{\lambda}}{4^{\Delta_1 + \Delta_2}} \frac{\Gamma(\Delta_1 + \Delta_2)}{\Gamma(\Delta_1 - 1/2) \Gamma(\Delta_2 - 1/2)} \notag \\
& \times (\Delta_1 (\Delta_1 - 1) + \Delta_2 (\Delta_2 - 1) \delta_{\Delta_1,1})\,,
\end{align}
with $\Delta_1 < \Delta_2$.

\section{Six-Point Functions} \label{subsec:sixpoint}
\begingroup
\allowdisplaybreaks
We now turn our attention to six-point functions of operators of length $L=1$, involving both protected and unprotected scalars, that we compute using the recursion relations given in Section \ref{subsec:evencase}.  With the help of the \textsc{Mathematica} notebook coming with \cite{Barrat:2022eim}, it is just necessary to input the external operators to get the full result.

As before, the results can be extended to more complicated correlators by combining the formulae for length $L=1$ operators and the pinching technique.

%
%

\subsection{Building Blocks and $\vvev{\phi^i \phi^j \phi^k \phi^l \phi^m \phi^n}$} \label{subsubsec:sixpt-buildingblocks}

We start by analyzing the six-point function of protected operators and collecting the building blocks useful to express other correlators, including the ones involving the unprotected scalar $\phi^6$. As usual, we define a reduced correlator through
\begin{align}
\vvev{\phi^i (\tau_1) \phi^j (\tau_2) \phi^k (\tau_3) \phi^l (\tau_4) \phi^m (\tau_5) \phi^n (\tau_6)} &= \frac{\tau_{24}\tau_{35}}{\tau_{12}^2\tau_{23} \tau_{34}^2\tau_{45} \tau_{56}^2} \Am^{ijklmn} (\tilde{\chi}_1\,, \tilde{\chi}_2\,, \tilde{\chi}_3) \notag \\
&= \frac{1}{\tau_{12}^2 \tau_{34}^2 \tau_{56}^2} \frac{1}{\tilde{\chi}_2} \Am^{ijklmn} (\tilde{\chi}_1\,, \tilde{\chi}_2\,, \tilde{\chi}_3)\,,
\label{eq:6point}
\end{align}
where the notation on the second line turns out to be more convenient for expressing the correlator in terms of $R$-symmetry channels.  The prefactor is picked according to (\ref{eq:prefactor}).
We define \textit{three} independent spacetime cross-ratios from (\ref{eq:crossratio})\footnote{As anticipated, this is the convention we introduce to make contact with the existing literature, as stated below (\ref{eq:crossratio}).}:
\begin{equation}
\tilde{\chi}_1 = \frac{\tau_{12}\tau_{34}}{\tau_{13}\tau_{24}}\,, \qquad \tilde{\chi}_2=\frac{\tau_{23}\tau_{45}}{\tau_{24}\tau_{35}}\,, \qquad \tilde{\chi}_3= \frac{\tau_{34}\tau_{56}}{\tau_{35}\tau_{46}}\,.
\end{equation}
The reduced correlator consists of \textit{fifteen} $R$-symmetry channels, which we choose to define as follows\footnote{Note that here we decided to keep the $R$-symmetry indices open since this result is used as a building block for the correlator of unprotected operators, which does not present $R$-symmetry channels. }:
\begin{align}
\frac{1}{\tilde{\chi}_2} \Am^{ijklmn} =& \phantom{+ } \delta^{ij} \delta^{kl} \delta^{mn} F_0
+ \delta^{ik} \delta^{jl} \delta^{mn} \tilde{\chi}_1^2 F_1
+ \delta^{il} \delta^{jk} \delta^{mn} \frac{\tilde{\chi}_1^2}{(1-\tilde{\chi}_1)^2} F_{2}
+ \delta^{ij} \delta^{km} \delta^{ln} \tilde{\chi}_3^2 F_3 \notag \\
&+ \delta^{ij} \delta^{kn} \delta^{lm} \frac{\tilde{\chi}_3^2}{(1-\tilde{\chi}_3)^2} F_4
+ \delta^{ik} \delta^{jm} \delta^{ln} \frac{\tilde{\chi}_1^2 \tilde{\chi}_3^2}{(1 - \tilde{\chi}_2)^2} F_5
+ \delta^{im} \delta^{jk} \delta^{lm} \frac{\tilde{\chi}_1^2 \tilde{\chi}_3^2}{(1-\tilde{\chi}_1 - \tilde{\chi}_2)^2} F_6 \notag \\
&+ \delta^{ik} \delta^{jn} \delta^{lm} \frac{\tilde{\chi}_1^2 \tilde{\chi}_3^2}{(1 - \tilde{\chi}_2 - \tilde{\chi}_3)^2} F_7
+ \delta^{il} \delta^{jm} \delta^{kn} \frac{\tilde{\chi}_1^2 \tilde{\chi}_2^2 \tilde{\chi}_3^2}{(1-\tilde{\chi}_1)^2 (1-\tilde{\chi}_2)^2 (1-\tilde{\chi}_3)^2} F_{8} \notag \\
&+ \delta^{il} \delta^{jn} \delta^{km} \frac{\tilde{\chi}_1^2 \tilde{\chi}_2^2 \tilde{\chi}_3^2}{(1-\tilde{\chi}_1)^2 (1-\tilde{\chi}_2-\tilde{\chi}_3)^2} F_{9}
+ \delta^{im} \delta^{jl} \delta^{km} \frac{\tilde{\chi}_1^2 \tilde{\chi}_2^2 \tilde{\chi}_3^2}{(1-\tilde{\chi}_3)^2 (1 - \tilde{\chi}_1 - \tilde{\chi}_2)^2} F_{10} \notag \\
&+ \delta^{im} \delta^{jn} \delta^{kl} \frac{\tilde{\chi}_1^2 \tilde{\chi}_2^2 \tilde{\chi}_3^2}{(1 - \tilde{\chi}_1 - \tilde{\chi}_2)^2 (1 - \tilde{\chi}_2 - \tilde{\chi}_3)^2} F_{11} \notag \\
&+ \delta^{in} \delta^{jl} \delta^{km} \frac{\tilde{\chi}_1^2 \tilde{\chi}_2^2 \tilde{\chi}_3^2}{(1-\tilde{\chi}_1-\tilde{\chi}_2-\tilde{\chi}_3 + \tilde{\chi}_1 \tilde{\chi}_3)^2} F_{12} \notag \\
&+ \delta^{in} \delta^{jk} \delta^{lm} \frac{\tilde{\chi}_1^2 \tilde{\chi}_3^2}{(1-\tilde{\chi}_1-\tilde{\chi}_2-\tilde{\chi}_3 + \tilde{\chi}_1 \tilde{\chi}_3)^2} F_{13} \notag \\
&+ \delta^{in} \delta^{jm} \delta^{kl} \frac{\tilde{\chi}_1^2 \tilde{\chi}_2^2 \tilde{\chi}_3^2}{(1 - \tilde{\chi}_2)^2 (1 - \tilde{\chi}_1 - \tilde{\chi}_2 - \tilde{\chi}_3 + \tilde{\chi}_1 \tilde{\chi}_3)^2} F_{14}\,,
\label{eq:6Rchannels}
\end{align}
where we suppressed the dependency on the spacetime cross-ratios, i.e. $F_j := F_j(\tilde{\chi}_1, \tilde{\chi}_2, \tilde{\chi}_3)$.

As usual, these channels (the \textit{building blocks}) have the following perturbative expansion:
\begin{equation}
F_j (\tilde{\chi}_1,\tilde{\chi}_2,\tilde{\chi}_3) = \sum_{k=0}^{\infty} \lambda^{k} F^{(k)}_j (\tilde{\chi}_1,\tilde{\chi}_2,\tilde{\chi}_3)\,.
\end{equation}

At leading order, it is easy to determine the building blocks and they read
\begin{equation}
F_0^{(0)} = F_{2}^{(0)} = F_4^{(0)} = F_{13}^{(0)} = F_{14}^{(0)}= 1\,,  \qquad F_j^{(0)} = 0\,.
\label{eq:6ptbuildingblocksLO}
\end{equation}
At next-to-leading order the expressions are cumbersome and we gathered them in the ancillary notebook of \cite{Barrat:2022eim}. As an example, we give here the highest $R$-symmetry channel:
\begin{align}
8\pi^2 F_0^{(1)} =& \bar{L}_R \left( \frac{1}{\eta_1} \right) + \bar{L}_R \left( \frac{1-\eta_2}{\eta_{32}} \right) + \bar{L}_R \left( - \frac{\eta_2}{\eta_{32}} \right) + \bar{L}_R \left( \frac{\eta_{31}}{\eta_{32}} \right) \notag \\
& + 2 \left( L_R \left( - \frac{\eta_{21}}{\eta_1} \right)+ L_R \left( - \frac{\eta_{31}}{\eta_1} \right) \right) + \frac{\ell (\eta_1, \eta_2)}{\eta_{21}} - \left( \frac{\eta_2}{\eta_3 \eta_{21}} + \frac{i \pi}{\eta_{31}} \right) \ell (\eta_1, \eta_3) \notag \\
& + \left( \frac{1}{1-\eta_3} + \frac{\eta_1}{\eta_3 \eta_{21}} + \frac{i\pi}{\eta_{32}} \right) \ell(\eta_2,\eta_3) + \frac{\ell(\eta_1,1)}{1-\eta_1} - \left( \frac{1}{1-\eta_3} + \frac{i\pi}{1-\eta_2} \right) \ell(\eta_2,1) \notag \\
& + \frac{\ell(\eta_3,1)}{1-\eta_3} + \frac{i\pi \eta_3}{\eta_{31}} \log \eta_1 - \frac{i\pi (\eta_3 (1-\eta_2) - \eta_2 \eta_{32})}{(1-\eta_2) \eta_{32}} \log \eta_2 \notag \\
& - \frac{i\pi (2 \eta_1 \eta_{32} - \eta_3 (\eta_{31} + \eta_{32}))}{\eta_{31} \eta_{32}} \log \eta_3\,,
\label{eq:6ptF0NLO}
\end{align}
where we have defined the following help variables:
\begin{equation}
\eta_1 := \frac{\tilde{\chi}_1 \tilde{\chi}_2 \tilde{\chi}_3}{(1-\tilde{\chi}_1 - \tilde{\chi}_2)(1-\tilde{\chi}_2 - \tilde{\chi}_3)}\,, \quad \eta_2 := \frac{\tilde{\chi}_2 \tilde{\chi}_3}{(1-\tilde{\chi}_1 - \tilde{\chi}_2)(1- \tilde{\chi}_3)}\,, \quad \eta_3 := \frac{(1-\tilde{\chi}_1) \tilde{\chi}_3}{1-\tilde{\chi}_1 - \tilde{\chi}_2}\,,
\label{eq:newcrossratios}
\end{equation}
with $\eta_{ij} := \eta_i - \eta_j$. Note that with these definitions we have the ordering $0< \eta_1 < \eta_2 < \eta_3 <1$. We have used the functions $L_R(\tilde{\chi})$ and $\ell(\tilde{\chi}_1,\tilde{\chi}_2)$ defined respectively in \eqref{eq:Rogers} and \eqref{eq:ell}, while we introduced for compactness the new function
\begin{equation}
\bar{L}_R (\tilde{\chi}) :=  L_R (1-\tilde{\chi}) - L_R (\tilde{\chi})\,.
\end{equation}

Note that it is easy to derive the five-point function $\vvev{\phi_1 \phi_1 \phi_1 \phi_1 \phi_2}$ from this six-point, by pinching the last two operators:
\begin{equation}
\vvev{\phi_1 \phi_1 \phi_1 \phi_1 \phi_2} = \frac{n_1}{\sqrt n_2} \underset{6\rightarrow5}{\text{lim}} \vvev{\phi_1 \phi_1 \phi_1 \phi_1 \phi_1 \phi_1} \,,
\end{equation}
where the prefactor here takes into account the unit-normalization (\ref{eq:twopt}).

\subsection{$\vvev{\phi^6 \phi^6 \phi^6 \phi^6 \phi^6 \phi^6}$} \label{subsubsec:666666}

We give another example of a six-point function, namely the case where all the operators are the unprotected elementary scalar $\phi^6$. The reduced correlator is defined through
\begin{align}
\vvev{\phi^6 (\tau_1) \phi^6 (\tau_2) \phi^6 (\tau_3) \phi^6 (\tau_4) \phi^6 (\tau_5) \phi^6 (\tau_6)} &= \left( \frac{\tau_{24}\tau_{35}}{\tau_{12}^2\tau_{23} \tau_{34}^2\tau_{45} \tau_{56}^2} \right)^{\smash{\Delta_{\phi^6}}} \Am^{666666} (\tilde{\chi}_1\,, \tilde{\chi}_2\,, \tilde{\chi}_3) \notag \\
&=  \frac{1}{\tau_{12}^{2\smash{\Delta_{\phi^6}}}\tau_{34}^{2\smash{\Delta_{\phi^6}}}\tau_{56}^{2\smash{\Delta_{\phi^6}}}} \frac{1}{\tilde{\chi}_2^{\smash{\Delta_{\phi^6}}}} \Am^{666666} (\tilde{\chi}_1\,, \tilde{\chi}_2\,, \tilde{\chi}_3)\,.
\end{align}

At leading order, the correlator $\vvev{\phi^6 \phi^6 \phi^6 \phi^6 \phi^6 \phi^6}$ coincides with $\vvev{\phi^i \phi^j \phi^k \phi^l \phi^m \phi^n}$ with $i=j=k=l=m=n$, i.e.
\begin{align}
\frac{1}{\tilde{\chi}_2^{\smash{\Delta_{\phi^6}}}} \Am^{(0)}_{666666} &= \frac{1}{\tilde{\chi}_2} \left. \Am^{(0)}_{ijklmn} \right|_{i=j=k=l=m=n} \notag \\
& = 1 + \frac{\eta_1^2}{(1-\eta_1)^2} + \frac{\eta_{23}^2}{(1-\eta_3)^2} + \frac{\eta_1^2 \eta_{23}^2 (1-2 \eta_3 (1-\eta_3))}{\eta_3^2 (1-\eta_3)^2 \eta_{12}^2} \,,
  \label{eq:666666LO}
\end{align}
where in the second equality we have used again the cross-ratios defined in \eqref{eq:newcrossratios} for compactness.

The next order includes $U$-diagrams as well as the next-to-leading order building blocks $F^{(1)}_j$ and it is significantly more involved. These additional terms, as well as the full correlator, can be found in the ancillary notebook of \cite{Barrat:2022eim}.  Several other examples\footnote{Examples of six-point correlators in the ancillary notebook of \cite{Barrat:2022eim} are $\vvev{\phi^i \phi^j \phi^k \phi^l \phi^6 \phi^6}$ and $\vvev{\phi^i \phi^j \phi^6 \phi^6 \phi^6 \phi^6}$.} of six-point functions can be found in the ancillary notebook, and it is straightforward to extend these computations to correlators involving composite operators made of fundamental scalar fields.
\endgroup

\subsection{Six-Point Data} \label{sec:blockexpansion6pt}
\begingroup
\allowdisplaybreaks

We now expand in conformal blocks the six-point functions just derived, as we did with four- and five-point functions.  Interestingly, for $n \geq 6$, different OPE limits lead to decompositions with different topologies, and therefore there exist multiple $n$-point blocks. We expand our six-point correlators in two topologies: the comb and the snowflake channels, already introduced in Section \ref{sec:conformalblockexpansion}.

\subsubsection{Comb Channel}

We replicate the same analysis of Section \ref{sec:blockexpansion4pt} to the comb channel of the six-point functions of protected fundamental scalars $\vvev{\phi^i \phi^j \phi^k \phi^l \phi^m \phi^n}$ and of unprotected ones $\vvev{\phi^6 \phi^6 \phi^6 \phi^6 \phi^6 \phi^6}$.

\begin{figure}[h]
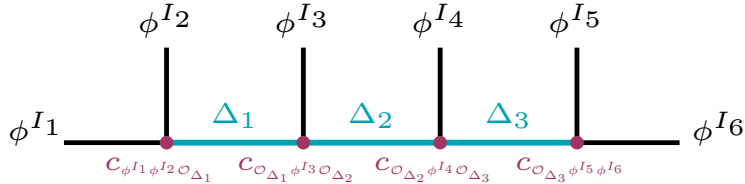

\centering
\sixcombblocks
\caption{\textit{Representation of six-point functions in the comb channel. Three operators are being exchanged, and the OPE coefficients consist of products of four \textbf{\textcolor{vertexcolor}{three-point functions}}.}}
\label{fig:sixcombch}
\end{figure}

\paragraph{$\vvev{\phi^i \phi^j \phi^k \phi^l \phi^m \phi^n}$.} We can start by expanding in conformal blocks the six-point function of protected operators studied in Section \ref{subsubsec:sixpt-buildingblocks}, and compare the three-point function $\vvev{\phi^i \phi^j \phi^6}$ computed in \eqref{eq:ij6} with the prediction obtained from the correlator. 

The comb channel for this correlator is represented in Figure \ref{fig:sixcombch}, and it is easy to see that the lowest coefficient we can check corresponds to setting $\Delta_1 = \Delta_2 = 1$, $\Delta_3 = 0$, for which the exchanged operators can only be $\Op_1 = \phi^6$, $\Op_2 = \phi^h\, (h=1,\ldots,5)$, $\Op_3 = \mathds{1}$, due to conservation of the $R$-charge. Noticing that the OPE coefficient vanishes when one $\Delta$ is equal to $1$ and the two other $\Delta$'s are $0$, we can expand the highest-weight channel $F_0$ in blocks in order to compare to that coefficient. We find that
\begin{equation}
F_0 (\tilde{\chi}_1, \tilde{\chi}_2, \tilde{\chi}_3) = 1 + c_{\phi^i \phi^j \phi^6} c_{\phi^6 \phi^k \phi^h} c_{\phi^h \phi^l \mathds{1}} c_{\mathds{1} \phi^m \phi^n} \tilde{\chi}_1 \tilde{\chi}_2 + \ldots\,,
\label{eq:exp6ptF0}
\end{equation}
where we note that $c_{\phi^h \phi^l \mathds{1}} c_{\mathds{1} \phi^m \phi^n} = 1$, due to the unit-normalization of the two-point function. This is also the reason why the leading term is $1$, in perfect analogy with the case of the four-point function. 



We now expand the $R$-symmetry channel $F_0$ of the six-point function studied in Section~\ref{subsubsec:sixpt-buildingblocks} in order to check whether we find a match for the OPE coefficient mentioned above. At leading order, we see from equation \eqref{eq:6ptbuildingblocksLO} that
\begin{equation}
F^{(0)}_0 (\tilde{\chi}_1, \tilde{\chi}_2, \tilde{\chi}_3) = 1\,,
\end{equation}
which matches the expectation that $c_{\phi^i \phi^j \phi^6} c_{\phi^6 \phi^k \phi^h}$ vanishes at order $O(\lambda^0)$, as it was the case for the four-point function as well.

At next-to-leading order, expanding equation \eqref{eq:6ptF0NLO} at $\tilde{\chi}_1, \tilde{\chi}_2, \tilde{\chi}_3 \sim 0$ results in
\begin{equation}
F^{(1)}_0 (\tilde{\chi}_1, \tilde{\chi}_2, \tilde{\chi}_3) = \frac{1}{8 \pi^2} \tilde{\chi}_1 \tilde{\chi}_2 + \ldots\,,
\end{equation}
where the coefficient of $\tilde{\chi}_1 \tilde{\chi}_2$ is to be identified with $c_{\phi^i \phi^j \phi^6} c_{\phi^6 \phi^k \phi^h}$ according to \eqref{eq:exp6ptF0}. We observe a perfect match with equation \eqref{eq:tocheck}.

\paragraph{$\vvev{\phi^6 \phi^6 \phi^6 \phi^6 \phi^6 \phi^6}$.} We can now focus on the expansion of the six-point function of unprotected operators discussed in Section \ref{subsubsec:666666}. 
This correlator is expected to contain the three-point function $\vvev{\phi^6 \phi^6 \phi^6}$, which can be checked against \eqref{eq:3ptphi6}. This coefficient can be accessed e.g. by setting as (bare) scaling dimensions $\Delta_1 = \Delta_2 = 1$ and $\Delta_3 = 0$. As for the previous cases, 
the exact correlator can be expanded in blocks and reads
\begin{align}
\mathcal{A}^{666666} (\tilde{\chi}_1,\tilde{\chi}_2,\tilde{\chi}_3) &= 1 + c_{\phi^6 \phi^6 \phi^6}^2 g_{1,1,0} (\tilde{\chi}_1\,, \tilde{\chi}_2\,, \tilde{\chi}_3) + \ldots \notag \\
&= 1 + c_{\phi^6 \phi^6 \phi^6}^2 \tilde{\chi}_1 \tilde{\chi}_2 + \ldots\,,
\end{align}
where the $1$ comes from the exchange of identity operators as always, and the OPE coefficient is just $c_{\phi^6 \phi^6 \phi^6}^2$ because of $c_{\phi^6 \phi^6 \mathds{1}}^2=1$. Other lower combinations such as $\Delta_1 = 1$, $\Delta_2 = \Delta_3 = 0$ vanish because one-point functions are zero in CFT.

We now extract this coefficient at leading and next-to-leading orders and compare it to the direct computation. At leading order, expanding \eqref{eq:666666LO} at $\tilde{\chi}_1\,, \tilde{\chi}_2\,, \tilde{\chi}_3 \sim 0$ gives
\begin{equation}
\Am^{(0)}_{666666} (\tilde{\chi}_1,\tilde{\chi}_2,\tilde{\chi}_3) = 1 + \tilde{\chi}_1^2 + \ldots\,,
\end{equation}
and thus $c_{\phi^6 \phi^6 \phi^6}^2$ vanishes at order $O(\lambda^0)$ as predicted by \eqref{eq:3ptphi6}.

We are also able to determine a closed form for the OPE coefficients at leading order:
\begin{align}
\left. c_{\phi^6 \phi^6 \Op_{\Delta_1}} c_{\Op_{\Delta_1} \phi^6 \Op_{\Delta_2}} c_{\Op_{\Delta_2} \phi^6 \Op_{\Delta_3}} c_{\Op_{\Delta_3} \phi^6 \phi^6} \right|_{O(\lambda^0)} =& - \frac{64 \pi^{3/2}}{4^{\Delta_1 + \Delta_2 + \Delta_3}} \frac{\Delta_1 (\Delta_1 -1) \Delta_{12}}{(2\Delta_1 - 1)(\Delta_1 + \Delta_2 - 1) } \notag \\
& \times \frac{\Gamma(\Delta_1 + \Delta_2)^2}{\Gamma(\Delta_2) \Gamma(\Delta_1 - 1/2)^2 \Gamma(\Delta_2 - 1/2)} \delta_{\Delta_1, \Delta_3}\,,
\end{align}
with $\Delta_{ij} := \Delta_i - \Delta_j$, valid when $\Delta_1<\Delta_2$.

At next-to-leading order, we expand the correlator given in the ancillary notebook and find
\begin{equation}
\Am^{(1)}_{666666} (\tilde{\chi}_1,\tilde{\chi}_2,\tilde{\chi}_3) = \frac{9}{8\pi^2} \tilde{\chi}_1 \tilde{\chi}_2 + \ldots\,,
\end{equation}
which is in full agreement with \eqref{eq:3ptphi6}.

\subsubsection{Snowflake Channel} \label{subsec:snowflake}

We now consider the snowflake channel for the same two correlators.  Note that here we use yet another convention for the cross-ratios. We commented on this choice in Section \ref{sec:snowflake} where the cross-ratios have also been introduced.

\paragraph{$\vvev{\phi^i \phi^j \phi^k \phi^l \phi^m \phi^n}$.} We start with the correlator of six protected fundamental scalars. As usual, let us focus on the highest-weight channel $F_0$, the expansion sketched in \eqref{eq:snowexp} then becomes
\begin{equation}
F_0 (z_1\,, z_2\,, z_3) = 1 + c_{\phi^i \phi^j \phi^6} c_{\phi^6 \phi^k \phi^l} z_1 z_2 + \ldots\,,
\label{eq:snowexp1}
\end{equation}
where the exchanged operator can only be $\phi^6$ because of $R$-charge conservation. At leading order, we have seen in equation \eqref{eq:6ptbuildingblocksLO} that
\begin{equation}
F^{(0)}_0 (z_1\,, z_2\,, z_3) = 1\,,
\end{equation}
and thus $c_{\phi^i \phi^j \phi^6} c_{\phi^6 \phi^k \phi^l} = 0$ at order $O(\lambda^0)$, in perfect agreement with \eqref{eq:ij6}.

At next-to-leading order, we can expand \eqref{eq:6ptF0NLO} at $z_1\,, z_2\,, z_3 \sim 0$ to obtain
\begin{equation}
F^{(1)}_0 (z_1\,, z_2\,, z_3) = \frac{1}{8 \pi^2} z_1 z_2 + \ldots\,,
\end{equation}
which perfectly matches the order $O(\lambda)$ of \eqref{eq:ij6} squared.

\paragraph{$\vvev{\phi^6 \phi^6 \phi^6 \phi^6 \phi^6 \phi^6}$.} Let us now perform checks on our result for the six-point function of unprotected scalars $\phi^6$ in the snowflake channel. In this case, equation \eqref{eq:snowexp} turns out to be
\begin{equation}
\Am^{666666} (z_1\,, z_2\,, z_3) = 1 + c_{\phi^6 \phi^6 \phi^6}^2 z_1 z_2 + \ldots\,,
\label{eq:snowexp2}
\end{equation}
where again the exchanged operator can only be $\phi^6$ because of conservation of the $R$-charge. At leading order, we can expand \eqref{eq:666666LO} and we find
\begin{equation}
\Am^{(0)}_{666666} (z_1\,, z_2\,, z_3) = 1 + z_1^2 z_2^2 + \ldots\,,
\end{equation}
from which we can read that $c_{\phi^6 \phi^6 \phi^6}^2 = 0$ at $O(\lambda^0)$, since there is no term of order $\Op(z_1 z_2)$. This is fully consistent with \eqref{eq:3ptphi6}.

At next-to-leading order, we can expand the correlator of the ancillary notebook in order to obtain
\begin{equation}
\Am^{(1)}_{666666} (z_1\,, z_2\,, z_3) = \frac{9}{8 \pi^2} z_1 z_2 + \ldots\,,
\end{equation}
where the prefactor perfectly matches the $c_{\phi^6 \phi^6 \phi^6}^2$ predicted by equation \eqref{eq:3ptphi6} at order $O(\lambda)$.

\endgroup

\section{Multipoint Ward Identities} \label{subsec:conjectureWI}

While it is undoubtedly true that the perturbative results we have just introduced are interesting in their own rights, it is especially true that what makes them even more interesting are the \textit{non-perturbative} constraints that we can derive from them. In this section, we introduce a conjecture for multipoint Ward identities that annihilate protected correlators of generic $n$.  We, then, discuss the properties of these differential equations and we solve them in a specific case. Basically,  we prepare the ground for an application of the Ward identities coming in Section \ref{sec:5ptbootstrap} that will prove how powerful these constraints are.

\subsection{The Conjecture}

From the pool of protected correlators that we obtained through the recursion relations up to $n=8$,  we find experimentally that \textit{all} our correlators are
annihilated by the following family of differential operators: 
\begin{equation}\label{eq:WIn}
\sum_{k=1}^{n-3} \left( \frac{1}{2} \partial_{\tilde{\chi}_k} + \alpha_k \partial_{r_k} - (1-\alpha_k) \partial_{s_k} \right) \, \Am_{\Delta_1 \ldots \Delta_n} \raisebox{-2ex}{$\Biggr 
|$}_{\raisebox{1.5ex}{$\begin{subarray}{l} r_i \to \alpha_i \tilde{\chi}_i \\
 s_i \to (1-\alpha_i)(1-\tilde{\chi}_i)
 \\ t_{ij} \to (\alpha_i-\alpha_j)(\tilde{\chi}_i-\tilde{\chi}_j)\end{subarray}$}} = 0\,,
\end{equation}
with $\alpha_k$ being \emph{arbitrary} real numbers. Notice that these operators are a natural generalization of the differential operator that captures the Ward identity \eqref{eq:WI4} for four-point functions of half-BPS operators.  We, then, conjecture that these equations are a multipoint extension of the superconformal Ward identity satisfied by the four-point functions. 

Even though we obtained these equations from NLO correlators, we expect these identities to be also satisfied in the strong-coupling expansion. This regime is captured by a well-understood AdS dual \cite{Giombi:2017cqn}, which has been used to calculate planar correlators in the $\lambda \to \infty$ limit. These are given by simple Wick contractions of the fluctuations of the dual fundamental string, i.e. the leading, disconnected order corresponds to the generalized free-field expression, e.g. (4.5) in \cite{Giombi:2017cqn}. It is then easy to check that all our $n$-point functions also satisfy \eqref{eq:WIn} in the extreme strong-coupling limit.  Moreover, a formula was recently derived \cite{Bliard:2022xsm} for computing arbitrary contact diagrams involving external scalar operators in AdS$_2$.  There it was independently checked that these next-to-leading order correlators also satisfy our conjectured Ward identities.

We have therefore \textit{four} non-trivial data points: the first two orders at weak and at strong coupling. It is then reasonable to assume that the constraints \eqref{eq:WIn} are non-perturbative and valid at \emph{all} loop orders.

\subsection{Considerations}

Let us start by saying that superconformal constraints are insensitive to gauge-theory quantum numbers, which means our identities (\ref{eq:WIn}) should go beyond the planar limit.  Indeed, we checked that they also hold for the first \textit{non-planar} corrections in the $N$ expansion, up to $n=8$ and for certain cases, we have checked that it also holds at the next-to-leading order.

Secondly, we expect the Ward identities we have just derived to be powerful constraints, as they encode the symmetries of the theory. As we have seen in Section \ref{sec:wardidentity4pt}, Ward identities can help in reducing the number of independent functions of a correlator.  In addition, they can be used to obtain superconformal blocks, as we show in Section \ref{sec:5ptbootstrap}, where we apply the Ward identities to derive the superblocks of the five-point function $\vvev{\phi_1 \phi_1 \phi_1 \phi_1 \phi_2}$.

We conclude these considerations with an important remark: our conjecture (\ref{eq:WIn}) cannot represent the full set of superconformal constraints on the correlators. That is because our analysis of protected operators only focuses on the highest-weight component, and we are ignoring possible fermionic descendants. Working in a suitable superspace, it is known that for four-points the full superconformal correlator can be reconstructed from the highest weight, and so it is safe to set the fermions to zero. Starting with five-point and up, one expects nilpotent invariants\footnote{We thank Paul Heslop for discussions on this topic.}. In general, for generic $n$-point functions the Ward identities should be a collection of partial PDEs relating the components associated with each fermionic structure. The fact that we obtained a differential operator that only acts on the highest weight and still annihilates the correlator is unexpected. It would be advisable to do a proper superspace analysis (similar to what was done in \cite{Chicherin:2015bza}) and prove that our ``experimental'' observation is indeed one of the constraints imposed by superconformal invariance.

\subsection{Solving the Ward Identities} \label{sec:solvingWI}

In Section \ref{sec:wardidentity4pt}, we also discussed the solution of the Ward identities for the four-point that was found in \cite{Liendo:2016ymz}.  We now want to extend this analysis to multipoint functions.  However, we focus on a particular example: the five-point function $\vvev{\phi_1 \phi_1 \phi_1 \phi_1 \phi_2}$. Not only the WI solution is useful to re-express the weak-coupling result in a compact way, but it is also fundamental for the strong-coupling analysis of Section \ref{sec:5ptbootstrap}.

We have already discussed the correlator $\vvev{\phi_1 \phi_1 \phi_1 \phi_1 \phi_2}$ at weak coupling in Section~\ref{subsubsec:11112}. The only difference is that here we use new $R$-symmetry cross-ratios $\zeta$ and $\eta$ instead of $r$ and $s$:
\begin{align}
r_1 = \zeta_1 \zeta_2\,, \quad s_1 = (1-\zeta_1) (1-\zeta_2)\,, \\
r_2 = \eta_1 \eta_2\,, \quad s_2 = (1-\eta_1) (1-\eta_2)\,.
\label{eq:FivePoint_CrossRatios2}
\end{align}
The $R$-symmetry channel decomposition \eqref{eq:5ptchannelsrs} then is rewritten as
\begin{align}
\Am(\chi_1;r_i,s_i,t)=\ & F_0+ \frac{ \zeta_1 \zeta_2 }{\chi_1^2} F_1 + \frac{ (1-\zeta_1) (1-\zeta_2) }{(1-\chi_1)^2} F_2 + \frac{ \eta_1 \eta_2 }{\chi_2^2} F_3 + \frac{ (1-\eta_1) (1-\eta_2) }{(1-\chi_2)^2} F_4+ \frac{t}{\chi_{12}^2} F_5\,.
\label{eq:FivePoint_RSymmetryChannels2}
\end{align}

This formulation is convenient, as the topological limit is reached for
\begin{align}
\zeta_i = \chi_1\,, \quad \eta_i = \chi_2\,, \quad t = \chi_{12}^2\,.
\label{eq:FivePoint_TopologicalLimit2}
\end{align}

In order to solve \eqref{eq:WIn}, we start by formulating an ansatz, keeping in mind that first all the $R$-symmetry channels of \eqref{eq:FivePoint_RSymmetryChannels2} have to appear and second we want to isolate the constant contribution corresponding to the topological sector, such that $\Am = \Fds$ when we set $R$-symmetry and spacetime variables equal to each other.

To make this clearer, let us look at the channel decomposition \eqref{eq:FivePoint_RSymmetryChannels2} and focus on the terms that have $\zeta_1$ and $\zeta_2$ as prefactors.
There are two degrees of freedom ($F_1$ and $F_2$), and the sum of these two functions is a polynomial of the form

\begin{equation}
(\zeta_1 + \zeta_2) g_1 (\chi_1, \chi_2) + \zeta_1 \zeta_2\, g_2 (\chi_1, \chi_2)\,.
\label{eq:FivePoint_ansatz1}
\end{equation}
Notice that the term without $\zeta$'s is left out, as it will be covered in the ansatz later on. Moreover, we have used the symmetry $\zeta_1 \leftrightarrow \zeta_2$ for reducing the ansatz to two functions $g_1$ and $g_2$ only. This preserves the number of degrees of freedom as expected.

To have vanishing prefactors in front of $g_1$ and $g_2$ in \eqref{eq:FivePoint_ansatz1} in the topological limit, we simply write:
\begin{equation}
\zeta_i \rightarrow \chi_1 - \zeta_i\,,
\label{eq:FivePoint_ansatz2}
\end{equation}
such that \eqref{eq:FivePoint_ansatz1} becomes
\begin{equation}
(\chi_1 - \zeta_1 + \chi_1 - \zeta_2) g_1 (\chi_1, \chi_2) + (\chi_1 - \zeta_1) (\chi_1 - \zeta_2)\, g_2 (\chi_1, \chi_2)\,.
\label{eq:FivePoint_ansatz3}
\end{equation}
Repeating this process for the $\eta$'s and for $t$ leads to the following ansatz:
\begin{align}
\Am(\chi_i;r_i,s_i,t)=\ & g_0 (\chi_1, \chi_2)+ (v_1 + v_2)\, g_1 (\chi_1, \chi_2)+ v_1 v_2\, g_2 (\chi_1, \chi_2) \notag \\
&+ (w_1 + w_2)\, g_3 (\chi_1, \chi_2)+ w_1 w_2\, g_4 (\chi_1, \chi_2) \notag \\
&+ z\, g_5 (\chi_1, \chi_2)\,,
\label{eq:FivePoint_ansatz4}
\end{align}
where we have defined the shorthand notation
\begin{equation}
v_i := \chi_1 - \zeta_i\,, \quad w_i := \chi_2 - \eta_i\,, \quad z := \chi_{12}^2 - t\,.
\label{eq:FivePoint_ansatz5}
\end{equation}
In this formulation, the topological sector fixes one function to be constant:
\begin{equation}
\Am(\chi_i;\chi_i,1-\chi_i,\chi_{ij})= g_0 (\chi_1, \chi_2) = \Fds\,,
\label{eq:FivePoint_ansatz6}
\end{equation}
as we requested before.  We discussed the topological sector of this correlator in Section~\ref{subsubsec:11112}.

We now apply the Ward identities on \eqref{eq:FivePoint_ansatz4}. This leads to three constraints, which can then be used for writing the general solution
\begin{equation}
\Am(\chi_i;r_i,s_i,t) = \Fds + \sum_{i=1}^3 \Dds_i f_i (\chi_1, \chi_2)\,,
\label{eq:FivePoint_SolutionWI}
\end{equation}
where we have defined the differential operators
\begin{subequations}
\begin{align}
\Dds_1 &:= (v_1 + v_2) + v_1 v_2\, (\pd_{\chi_1} + \pd_{\chi_2})\,, \label{eq:FivePoint_DifferentialOperator1} \\
\Dds_2 &:= (w_1 + w_2) + w_1 w_2\, (\pd_{\chi_1} + \pd_{\chi_2})\,, \label{eq:FivePoint_DifferentialOperator2} \\
\Dds_3 &:= z + \chi_{12} ( v_1 v_2 - w_1 w_2 )\, (\pd_{\chi_1} + \pd_{\chi_2})\,,
\label{eq:FivePoint_DifferentialOperator3}
\end{align}
\end{subequations}
and the functions $g$ have been renamed into $f$ as follows:
\begin{equation}
g_1 \to f_1\,, \quad g_3 \to f_2\,, \quad g_5 \to f_3\,.
\label{eq:RelabelgTof}
\end{equation}
In terms of the $R$-symmetry channels given in \eqref{eq:5ptchannelsrs} and \eqref{eq:FivePoint_RSymmetryChannels2}, the $f$-functions read
\begin{subequations}
\begin{align} 
f_1 (\chi_1, \chi_2) &= - \frac{1}{\chi_1} F_1 (\chi_1, \chi_2) + \frac{1}{1-\chi_1} F_2 (\chi_1, \chi_2)\,,  \\
f_2 (\chi_1, \chi_2) &= - \frac{1}{\chi_2} F_3 (\chi_1, \chi_2) + \frac{1}{1-\chi_2} F_4 (\chi_1, \chi_2)\,,  \\
f_3 (\chi_1, \chi_2) &= - \frac{1}{\chi_{12}^2} F_5 (\chi_1, \chi_2)\,. 
\end{align} \label{eq:FivePoint_LittlefFromF}
\end{subequations}
Note that $F_0$ does not appear in this decomposition. This is because the topological sector \eqref{eq:FivePoint_ansatz6} fully eliminates one channel.

To conclude, it might be possible to give a general solution of the Ward identities for a given number $n$ of external operators. For four-point functions, the general solution can be found in \cite{Liendo:2016ymz}.  Although,  it seems that for $n \geq 6$ is not possible to find a decomposition of the $R$-symmetry channels similar to \eqref{eq:FivePoint_RSymmetryChannels2}.  We can realize this by looking at e.g.  the six-point function $\vvev{\phi_1 \phi_1 \phi_1 \phi_1 \phi_1 \phi_1}$, whose $R$-symmetry channels are given in \eqref{eq:6Rchannels}. The fact that the $R$-symmetry cross-ratios appear in products makes it more difficult to formulate an ansatz on the line of \eqref{eq:FivePoint_ansatz4}. In any case, such a general analysis of the Ward identities and their solutions is left for future work.

\subsubsection{Re-Expressing the Weak Coupling Result} \label{subsubsec:RevisitingTheWeakCoupling}

The solution \eqref{eq:FivePoint_SolutionWI} to the Ward identity reduces the number of functions from \textit{six} $R$-symmetry channels to just \textit{three} functions $f_i$ and one constant $\Fds$, which can be computed from supersymmetric localization.

Our weak-coupling results can be expressed in terms of these functions. In particular,  we start from the $R$-symmetry channels decomposition given in (\ref{eq:FivePoint_RSymmetryChannels2}) and we insert the various $F_i^{(0)}$ from (\ref{eq:FivePoint_LO}) and $F_i^{(1)}$ from (\ref{eq:FivePoint_NLOGroup2_3}) in (\ref{eq:FivePoint_LittlefFromF}), to get the $f_i$, which at leading order read
\begin{equation}
\begin{alignedat}{2}
f_1^{(0)} (\chi_1, \chi_2) &= - \frac{1}{\chi_1}\,, \qquad && f_2^{(0)} (\chi_1, \chi_2) = \frac{1}{1-\chi_2}\,, \\
f_3^{(0)} (\chi_1, \chi_2) &= - \frac{1}{\chi_{12}^2}\,,\qquad && \Fds^{(0)} = 3\,.
\end{alignedat}
\label{eq:FivePoint_LittlefLO}
\end{equation}
Notice the following relations:
\begin{equation}
\begin{split}
f_1^{(0)} (\chi_1, \chi_2) & = - f_2^{(0)} (1-\chi_2, 1-\chi_1)\,, \\
f_3^{(0)} (\chi_1, \chi_2) & = f_3^{(0)} (1-\chi_2, 1-\chi_1)\,.
\end{split}
\label{eq:FivePoint_Crossing}
\end{equation}
These equations hold non-perturbatively, and follow from the crossing symmetry of the correlator, given by (\ref{eq:FivePoint_CrossingSymmetry}).

At next-to-leading order, the solution of the Ward identity reads
\begin{equation}
\begin{split}
f_1^{(1)} (\chi_1, \chi_2) =\ &- \frac{1}{8\pi^2} \biggl( \log\left(1-\frac{\chi_1}{\chi_2}\right)+\frac{\chi_1}{1-\chi_2} \log\left(1-\frac{\chi_1-\chi_2}{\chi_1-1}\right) \\
&-\frac{\chi_1}{\chi_1-\chi_2}\left( \log\left(\frac{\chi_2-1}{\chi_1-1}\right)+\log\left(\frac{\chi_1}{\chi_2}\right)\right)+ \frac{\pi^2}{6} \biggr)\,, \\
f_2^{(1)} (\chi_1, \chi_2) =\ & - f_1^{(1)} (1-\chi_2, 1-\chi_1)\,, \\
f_3^{(1)} (\chi_1, \chi_2) =\ & - \frac{1}{8 \pi^2 \chi_{12}^2 }\biggl( (\chi_2-1) \left( \log\left(\frac{\chi_2-1}{\chi_1-1}\right)+(\chi_1-\chi_2) \log\left(\frac{\chi_1-1\chi_2}{\chi_1-1}\right)\right)  \\
&+2 L_R\left(\frac{\chi_1-\chi_2}{\chi_1-1}\right) + (\chi_1 \leftrightarrow 1-\chi_2) + \frac{\pi^2}{6}
\biggr)\,, \\
\Fds^{(1)} =\ & \frac{7}{48}\,.
\end{split}
\label{eq:FivePoint_LittlefNLO}
\end{equation}
with $L_R(\chi)$ the Rogers dilogarithm defined in (\ref{eq:Rogers}).
The analysis done in this section is used for bootstrapping this correlator in the strong-coupling regime.

\section{Towards Multipoint Correlators at Strong Coupling} \label{sec:5ptbootstrap}

We now basically have (almost) all the ingredients we need to attempt the bootstrap of a correlator at strong coupling. 

In this chapter, we started from the weak coupling results of five- and six-point functions, that we used to conjecture multipoint Ward identities. These are fundamental to reducing the degrees of freedom of the correlation functions. In particular, we have just solved them for the case of the five-point function $\vvev{\phi_1 \phi_1 \phi_1 \phi_1 \phi_2}$, which is precisely our guinea pig for this bootstrap experiment.  In the following, we realize that the Ward identities are also central in deriving the superconformal blocks, which are the last essential element we need.

Just a quick comment, before starting. We are not bootstrapping a four-point function, even if that would be the obvious starting point because the correlator $\vvev{\phi_1 \phi_1 \phi_1 \phi_1}$ has already been derived using bootstrap at strong coupling up to NNNLO in \cite{Ferrero:2021bsb}. Indeed, we keep this remarkable result in mind, as its derivation is our guide in this attempt.

\subsection{Superconformal Blocks} \label{subsec:SuperconformalBlocks}

As promised, our first stop is the derivation of the superconformal blocks, which play a crucial role in the bootstrap strategy. In the case of the four-point functions \cite{Liendo:2018ukf}, the superconformal blocks were obtained with the help of the Ward identities \eqref{WIpedro}, bypassing the use of Casimir equations. We perform a similar analysis here and show that the five-point superblocks can also be entirely fixed thanks to the Ward identities.

\begin{figure}[h]
\centering 
\includegraphics[scale=0.35]{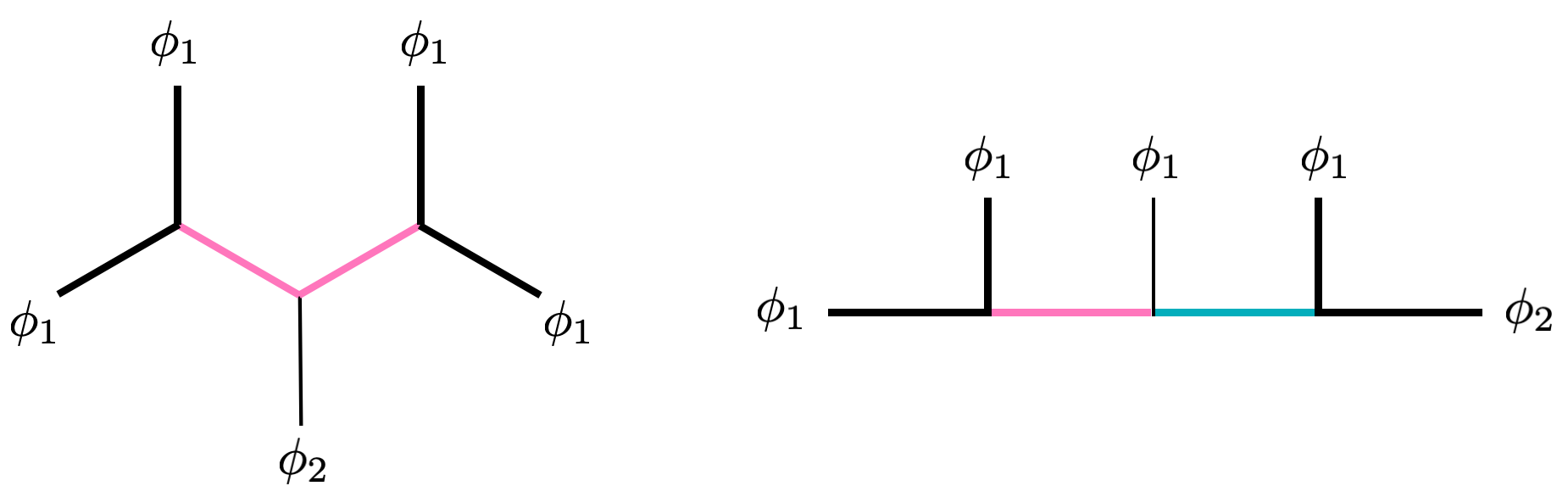}
\caption{\textit{We sketch the two possible ways of performing the OPEs for the correlator $\vvev{\phi_1 \phi_1 \phi_1 \phi_1 \phi_2}$. The figure on the left shows the symmetric channel \eqref{eq:FivePoint_BlockExpansionSymmetric}, while the one on the right presents the asymmetric channel given in \eqref{eq:FivePoint_BlockExpansionAsymmetric}. The \textbf{\textcolor{black}{black bold}} lines refer to the way the OPEs are being performed. \textbf{\textcolor{SEcolor}{Pink}} lines refer to the OPE \eqref{eq:OPEB1B1}, while the \textbf{\textcolor{gluoncolor}{light blue}} line corresponds to \eqref{eq:OPEB1B2}.}}
\label{fig:OPESymmetricAsymmetric}
\end{figure}

We consider here two ways, represented in Figure \ref{fig:OPESymmetricAsymmetric}, of performing the OPE. Indeed, one can either consider two pairs of $\phi_1 \times \phi_1$ and leave $\phi_2$ untouched, giving rise to what we call \textit{symmetric} channel,  or first perform $\phi_1 \times \phi_2$, which leads to the \textit{asymmetric} channel. These names refer to the fact that in the symmetric channel, the two OPEs are the same ($\phi_1 \times \phi_1$), while in the asymmetric one, they are different. Having two different OPE channels imply the derivation of two different block expansions, which we now go through in detail.

\subsubsection{The Symmetric Channel} \label{subsubsec:TheSymmetricChannel}

We have already introduced explicitly the relevant OPE in Section \ref{sec:1ddefectCFT}, particularly in equation (\ref{eq:OPEB1B1}). This OPE leads to the following expansion in superblocks
\begin{equation}
\begin{split}
\Am_{11112} (\chi_i;r_i, s_i, t) =\ & c_{112} ( \Gm_{\mathds{1}, \Bm_2} + \Gm_{\Bm_2, \mathds{1}} )
+ c_{112}^2 c_{222} \Gm_{\Bm_2, \Bm_2} \\
&+ \sum_{\Delta} c_{112} c_{11\Delta} c_{22\Delta} ( \Gm_{\Bm_2, \Lm_{\Delta}} + \Gm_{\Lm_{\Delta}, \Bm_2} ) \\
&+ \sum_{\Delta_1, \Delta_2} c_{11\Delta_1} c_{11\Delta_2} c_{2 \Delta_1 \Delta_2} \Gm_{\Lm_{\Delta_1},\Lm_{\Delta_2}}\,,
\end{split}
\label{eq:FivePoint_BlockExpansionSymmetric}
\end{equation}
where $\Delta$ indicates the dimension of the exchanged operators\footnote{Note that we write $\Delta$ without specifying if it is the operator exchanged on the left or the right because in this case, it does not matter. The formula is the same for both of them. Whenever it matters, we write it explicitly.} and $\Lm_{\Delta} := \Lm_{0,[0,0]}^{\Delta}$. Moreover, we omitted the dependence on the spacetime and $R$-symmetry variables on the RHS for compactness.

The superblocks themselves take the form
\begin{equation}
\Gm_{\mathcal{X},\mathcal{Y}} ( \chi_i; r_i,s_i,t) = \sum \alpha_{g} (\Delta_1,\Delta_2, [a,b][c,d])\, h_{[a,b][c,d]}(r_i,s_i,t)\, g_{\Delta_1,\Delta_2} (\chi_i)\,,
\label{eq:FivePoint_FormSuperblocks}
\end{equation}
where $h_{[a,b][c,d]}$ correspond to the $R$-symmetry blocks\footnote{$[a,b]$ are the Dynkin labels of the representation corresponding to the first multiplet exchanged on the left $\mathcal{X}$, while $[c,d]$ refers to the multiplet $\mathcal{Y}$ on the right.} and $g_{\Delta_1,\Delta_2}$ are the  \textit{bosonic} blocks\footnote{Here $\Delta_1$ and $\Delta_2$ correspond to the \textit{bosonic} scaling dimension of the operators being exchanged respectively on the left and on the right.}.
These blocks are known and they read:
\begin{equation}
g_{\Delta_1,\Delta_2} (\chi_i)=\chi_1^{\Delta_1-2} (1-\chi_2)^{\Delta_2-2} F_2(\Delta_1+\Delta_2-2,\Delta_1,\Delta_2,2\Delta_1,2\Delta_2;\chi_1,1-\chi_2)\,,
\end{equation}
where $F_2$ is the Appell Hypergeometric function. Unfortunately the relative coefficients $\alpha_{g} (\Delta_1,\Delta_2, [a,b][c,d])$ still have to be determined. For the most complicated block $\Gm_{\Lm_{\Delta_1},\Lm_{\Delta_2}}$, there are about $50$ coefficients to fix. Nevertheless,  remarkably \textit{all} the coefficients are fixed by \eqref{eq:WIn}, up to an overall normalization constant. This illustrates the great power of the conjectured Ward identities. 

We leave the full derivation, which is quite long and intricate, and the explicit expression of all the derived superblocks to the future paper \cite{Barrat:2023pev}. Although, we illustrate how to derive some specific cases for the asymmetric channel in Appendix \ref{app:superblocks}. 

Before moving to the other OPE channel, note that the expansion in superblocks largely truncates in the topological sector.
In particular, we have that 
\begin{equation}
\Fds = 2c_{112} \left( 1 - \frac{5}{2} c_{112} c_{222} \right)\,,
\label{eq:FivePoint_TopologicalSectorSuperblocks1}
\end{equation}
thanks to the fact that all the contributions from the longs drop out, which is a non-trivial check of the validity of these expressions.

\subsubsection{The Asymmetric Channel} \label{subsubsec:TheAsymmetricChannel}

We now turn our attention to the asymmetric channel. In this case, we need both OPEs in (\ref{eq:OPEB}). This results in an expansion in superblocks of the following form:
\begin{equation}
\begin{split}
\Am_{11112} (\chi_i;r_i, s_i, t) =\ & c_{112} \Gm_{\mathds{1}, \Bm_1}
+ c_{112}^3 \Gm_{\Bm_2, \Bm_1}
+ c_{112}^2 c_{123} \Gm_{\Bm_2, \Bm_1} \\
&+ \sum_{\Delta_1} \left(
c_{11\Delta_1}^2 \Gm_{\Lm_{\Delta_1}, \Bm_1}
+ c_{11\Delta_1} c_{13\Delta_1} c_{123} \Gm_{\Lm_{\Delta_1}, \Bm_3}
\right) \\
&+ \sum_{\Delta_2} c_{112} c_{12\Delta_2}^2 \Gm_{\Bm_2, \Lm_{0,[0,1]}^{\Delta_2}} \\
&+ \sum_{\Delta_1, \Delta_2} c_{11\Delta_1} c_{12\Delta_2} c_{1\Delta_1\Delta_2} \Gm_{\Lm_{\Delta_1} \Lm_{\Delta_2}}\,.
\end{split}
\label{eq:FivePoint_BlockExpansionAsymmetric}
\end{equation}
Here the subscripts in $\Delta_1$ and $\Delta_2$ are more than a label; they are a shorthand notation to refer to the long operators $\Lm_{0,[0,0]}^{\Delta_1}$ and $\Lm_{0,[0,1]}^{\Delta_2}$, which have different $R$-charges.

The topological sector in this channel reads
\begin{equation}
\Fds = c_{112} \left( 1 + 2 c_{112} - \frac{21}{2} c_{112} c_{123} \right)\,.
\label{eq:FivePoint_TopologicalSectorSuperblocks2}
\end{equation}

We remind that in Appendix \ref{app:superblocks} we derive some blocks in this channel and we give their expressions too.

\subsection{Bootstrapping a Five-Point Correlator} \label{subsec:BootstrappingTheCorrelator}

With this derivation of the superconformal blocks, we collected another key ingredient for our attempt to bootstrap $\vvev{\phi_1 \phi_1 \phi_1 \phi_1 \phi_2}$ at strong coupling. What we still need is the input from the Witten diagrams. In this section, we give the results for the leading and next-to-leading orders, while the higher-loop orders will be presented in the upcoming work \cite{Barrat:2023pev}.

\subsubsection{Generalized Free Field Theory} \label{subsubsec:GeneralizedFreeFieldTheory}

It is really important for the bootstrap of the higher orders to extract the CFT data of the leading order, which corresponds to the \textit{generalized free field} (GFF) theory\footnote{Generalized free-field theory, often called mean field theory, is a non-local theory of operators with generic dimension $\Delta$ and whose correlators are computed by Wick contractions.  GFF has no conserved stress tensor and its AdS$_2$ dual is the theory of a free massive scalar.}.

It is easy to compute the corresponding Witten diagrams, which are depicted in Figure~\ref{fig:LO5ptwitten}.
\begin{figure}[h]
\centering
\includegraphics[scale=0.37]{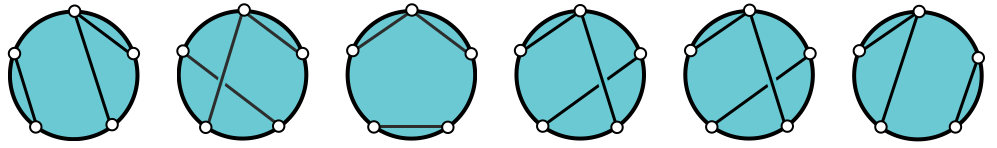}
\caption{\textit{Witten diagrams contributing to the leading order of the five-point function. }}
 \label{fig:LO5ptwitten}
\end{figure}
The AdS propagators correspond to \cite{Drukker:2011za,Correa:2012at}
\begin{equation}
\ScalarPropagatorStrong = 2\mathbb{B}(\lambda) \frac{\delta^{IJ}}{\tau_{12}^2}\,,
\end{equation}
where $\mathbb{B}(\lambda)$ is the Bremsstrahlung function, already encountered in equation (\ref{eq:normalization1}), whose expansion at strong coupling is
\begin{equation}
\mathbb{B}(\lambda)=\frac{\sqrt{\lambda}}{4\pi^2}-\frac{3}{8\pi^2}+O\left(\frac{1}{\sqrt{\lambda}}\right)\,.
\end{equation}

If we now take the leading order propagator, we get the $R$-symmetry channel decomposition (see \eqref{eq:5ptchannelsrs})
\begin{equation}
F_j^{(0)} = \sqrt{2}\,,
\end{equation}
which in terms of the solution of the Ward identity \eqref{eq:FivePoint_SolutionWI} translates into
\begin{equation}
\begin{alignedat}{2}
& f_1^{(0)} = - \sqrt{2} \frac{1-2\chi_1}{\chi_1 (1-\chi_1)}\,, \quad &&
f_2^{(0)} = - \sqrt{2} \frac{1-2\chi_2}{\chi_2 (1-\chi_2)}\,, \\
& f_3^{(0)} = - \frac{\sqrt{2}}{\chi_{12}^2}\,, && \mathds{F} = 6 \sqrt{2}\,.
\end{alignedat}
\end{equation}

This result can be expanded in the superconformal blocks introduced in Section \ref{subsec:SuperconformalBlocks}.
For the symmetric channel, we find the following OPE coefficients:
\begin{equation}
\begin{split}
&\left. c_{112} \right. = \sqrt{2}\,, \qquad\qquad c_{112}^2 c_{222} = - \frac{4 \sqrt{2}}{5}\,, \\
&\left. c_{112} c_{11 \Delta} c_{22 \Delta} \right|_{\Delta \text{ even}} = \frac{\sqrt{\pi} (\Delta - 1) \Gamma(\Delta + 3)}{ 2^{\frac{4 \Delta + 1}{2}} \Gamma \left(\Delta + \frac{3}{2} \right)}\,, \\
&\left.  c_{11\Delta_1} c_{11\Delta_2} c_{2 \Delta_1 \Delta_2} \right|_{\Delta_{1,2} \text{ even}} = \frac{\pi (\Delta_1)_3 (\Delta_2)_3 \Gamma(\Delta_1 + \Delta_2)}{ 2^{\frac{4 (\Delta_1 + \Delta_2) + 7}{2}} \Gamma\left( \Delta_1 + \frac{3}{2} \right) \Gamma\left( \Delta_2 + \frac{3}{2} \right)}\,.
\end{split}
\end{equation}
Note that the OPE coefficients with odd $\Delta$ vanish.

From the asymmetric channel, we obtain
\begin{equation}
\begin{split}
&\left. c_{112} \right. = \sqrt{2}\,, \qquad c_{112}^3 = \sqrt{2}\,, \qquad c_{112}^2 c_{123} = - \frac{2\sqrt{2}}{7}\,, \\
&\left. c_{11\Delta_1}^2 \right|_{\Delta_1 \text{ even}} = \frac{\sqrt{\pi} (\Delta_1 - 1) \Gamma(\Delta_1 + 3)}{ 2^{\frac{4 \Delta_1 + 3}{2}} \Gamma \left(\Delta_1 + \frac{3}{2} \right)}\,, \\
&\left. c_{112}c_{12\Delta_2}^2 \right|_{\Delta_2 \text{ even}} = \frac{ \sqrt{\pi} (\Delta_2 - 2)^2 ( \Delta_2 )_6 \Gamma (\Delta_2) }{2^{\frac{4 \Delta_2 + 7}{2}} 3 (\Delta_2 + 1) (\Delta_2 + 4) \Gamma \left( \Delta_2 + \frac{3}{2} \right) }\,, \\
&\left. c_{11\Delta_1}c_{12\Delta_2} c_{1 \Delta_1 \Delta_2} \right|_{\substack{ \Delta_1 \text{ even} \\ \Delta_2 > \Delta_1}} = \frac{ 2\sqrt{2} \Delta_{21} (\Delta_1 - 1)_{3} \Gamma (\Delta_1 + 3) \Gamma (\Delta_2 - 1) \Gamma (\Delta_1 + \Delta_2 + 4) }{ \Gamma ( 2 \Delta_1 +3 ) \Gamma ( 2 \Delta_2 +3 ) }\,,
\end{split}
\end{equation}
where all the other OPE coefficients are zero.

\subsubsection{First Correction} \label{subsubsec:FirstCorrection}

We now bootstrap the next order.  Note that deriving this order via Witten diagrams is still doable.  However,  the next-to-next-to-leading order is extremely involved. Therefore bootstrapping this order is fundamental to testing our bootstrap routine and the newly derived superblocks since we can double-check our result with the Witten diagrams. We leave the bootstrap of the NNLO for future publication \cite{Barrat:2023pev}.

Our starting point is the crossing symmetry given in \eqref{eq:FivePoint_CrossingSymmetry} that relates the different functions $f_j (\chi_1, \chi_2)$.
Additional \textit{braiding} constraints can be obtained by considering different orderings of the operators in $\vev{\phi_1 \phi_1 \phi_1 \phi_1 \phi_2}$. In this case, the functions are equal up to a phase.

We begin by writing an ansatz for $f_1$ and $f_3$\footnote{Recall that $f_2$ is fixed by \eqref{eq:FivePoint_CrossingSymmetry}.} based on the OPE limits. They take the form
\begin{equation}
\begin{split}
f_1 (\chi_1, \chi_2) =\ & p_1 (\chi_1, \chi_2)
+ r_1 (\chi_1, \chi_2) \log \chi_1
+ r_2 (\chi_1, \chi_2) \log \chi_2 \\
&+ r_3 (\chi_1, \chi_2) \log (1-\chi_1)
+ r_4 (\chi_1, \chi_2) \log (1-\chi_2) \\
&+ r_5 (\chi_1, \chi_2) \log \chi_{21}\,,
\end{split}
\label{eq:FivePoint_ansatzf1}
\end{equation}
and
\begin{equation}
\begin{split}
f_3 (\chi_1, \chi_2) =\ & p_3 (\chi_1, \chi_2)
+ s_1 (\chi_1, \chi_2) \log \chi_1
+ s_2 (\chi_1, \chi_2) \log \chi_2 \\
&+ s_3 (\chi_1, \chi_2) \log (1-\chi_1)
+ s_4 (\chi_1, \chi_2) \log (1-\chi_2) \\
&+ s_5 (\chi_1, \chi_2) \log \chi_{21}\,,
\end{split}
\label{eq:FivePoint_ansatzf3}
\end{equation}
where the functions $p_i$ and $r_j$ are \textit{rational} functions.

The crossing and braiding symmetries result in the following constraints on the rational functions present in \eqref{eq:FivePoint_ansatzf1} and \eqref{eq:FivePoint_ansatzf3}:
\begin{equation}
\begin{split}
p_1 (\chi_1, \chi_2) &= - p_1 (1-\chi_1, 1-\chi_2)\,, \\
r_1 (\chi_1, \chi_2) &= - r_3 (1-\chi_1, 1-\chi_2)\,, \\
r_2 (\chi_1, \chi_2) &= - r_4 (1-\chi_1, 1-\chi_2)\,, \\
r_5 (\chi_1, \chi_2) &= - r_5 (1-\chi_1, 1-\chi_2)\,,
\end{split}
\label{eq:FivePoint_Constraints1}
\end{equation}
and
\begin{equation}
\begin{split}
p_3 (\chi_1, \chi_2) &= p_3 (\chi_2, \chi_1)\,, \\
p_3 (\chi_1, \chi_2) &= p_3 (1-\chi_2, 1-\chi_1)\,, \\
s_2 (\chi_1, \chi_2) &= - s_1 (\chi_2, \chi_1)\,, \\
s_3 (\chi_1, \chi_2) &= - s_1 (1-\chi_1, 1-\chi_2)\,, \\
s_4 (\chi_1, \chi_2) &= - s_1 (1-\chi_2, 1-\chi_1)\,, \\
s_5 (\chi_1, \chi_2) &= - s_5 (\chi_2, \chi_1)\,, \\
s_5 (\chi_1, \chi_2) &= - s_5 (1-\chi_2, 1-\chi_1)\,.
\end{split}
\label{eq:FivePoint_Constraints2}
\end{equation}
This already eliminates five functions. Moreover, we also have relations that connect $f_1$ to $f_3$:
\begin{equation}
\begin{split}
p_1 (\chi_1, \chi_2) =\ & - \frac{1- \chi_1}{\chi_2^2} p_3 \left( \frac{\chi_1}{\chi_2}, \frac{1}{\chi_2} \right) + \frac{\chi_1}{(1-\chi_2)^2} p_3 \left( \frac{\chi_{12}}{1-\chi_2}, - \frac{\chi_2}{1-\chi_2} \right)\,, \\
r_1 (\chi_1, \chi_2) =\ & - \frac{1- \chi_1}{\chi_2^2} s_1 \left( \frac{\chi_1}{\chi_2}, \frac{1}{\chi_2} \right) + \frac{\chi_1}{(1-\chi_2)^2} s_5 \left( \frac{\chi_{12}}{1-\chi_2}, - \frac{\chi_2}{1-\chi_2} \right)\,, \\
r_2 (\chi_1, \chi_2) =\ & \frac{1-\chi_1}{\chi_2^2} \biggl(
s_1 \left( \frac{\chi_{21}}{\chi_2}, - \frac{1-\chi_2}{\chi_2} \right)
+  s_1 \left( - \frac{1-\chi_2}{\chi_2}, \frac{\chi_{21}}{\chi_2} \right) \\
& +  s_1 \left( \frac{\chi_1}{\chi_2}, \frac{1}{\chi_2} \right) 
+  s_1 \left( \frac{1}{\chi_2}, \frac{\chi_1}{\chi_2} \right)
+  s_5 \left( \frac{\chi_1}{\chi_2}, \frac{1}{\chi_2} \right)
\biggr) \\
&+ \frac{\chi_1}{(1-\chi_2)^2} s_1 \left( - \frac{\chi_2}{1-\chi_2}, \frac{\chi_{12}}{1-\chi_2} \right)\,, \\
r_5 (\chi_1, \chi_2) =\ & - \frac{1- \chi_1}{\chi_2^2} s_1 \left( \frac{\chi_{21}}{\chi_2}, - \frac{1-\chi_2}{\chi_2} \right) + \frac{\chi_1}{(1-\chi_2)^2} s_1 \left( \frac{\chi_{12}}{1-\chi_2}, - \frac{\chi_2}{1-\chi_2} \right)\,,
\end{split}
\label{eq:FivePoint_Constraints3}
\end{equation}
from which we can eliminate four more functions.

Thus, we are left with \textit{three} rational functions to fix, which we choose to be $p_3$, $s_1$, and $s_5$.
We now assume a Regge behavior for the anomalous dimensions, as well as the relations \eqref{eq:FivePoint_Constraints2} and boundary conditions following from the OPE expansion, which allow us to fix these functions up to \textit{two} constants, corresponding to $\Fds$ and $c_{112}$.
These are known from supersymmetric localization, and we use these results as input in our bootstrap routine.
$\Fds$ is given in \eqref{eq:FivePointTopologicalSector}, while $c_{112}$ can be computed from \cite{Giombi:2018qox} and reads
\begin{equation}
c_{112} = \sqrt{2} - \frac{3}{2\sqrt{2} \sqrt{\lambda}} - \frac{9}{16 \sqrt{2} \lambda} + O(\lambda^{-3/2})\,.
\label{eq:Lambda112_Localization}
\end{equation}
All in all, we find the remaining rational functions to be
\begin{equation}
\begin{split}
p_3 (\chi_1, \chi_2) &= \frac{1}{2\sqrt{2} \sqrt{\lambda}} \left( \frac{4}{\chi_1 \chi_2} + \frac{4}{(1-\chi_1)(1-\chi_2)} + \frac{19}{\chi_{12}^2} \right)\,, \\
s_1 (\chi_1, \chi_2) &= - \frac{\sqrt{2}}{\sqrt{\lambda}} \frac{\chi_1 ( \chi_1 (1 - 3 \chi_2) + 4 \chi_2^2 )}{\chi_2^2 \chi_{12}^3}\,, \\
s_5 (\chi_1, \chi_2) &= \frac{\sqrt{2}}{\sqrt{\lambda}} \left( \frac{1}{\chi_1^2} + \frac{1}{\chi_2^2} + \frac{1}{(1-\chi_1)^2} + \frac{1}{(1-\chi_2)^2}  \right)\,,
\end{split}
\label{eq:FivePoint_FixedFunctions}
\end{equation}
which gives a degenerate anomalous dimension, independent from the multiplying OPE coefficient
\begin{equation}
\gamma^{(1)} = \frac{\Delta (\Delta+2)}{2 \sqrt{\lambda}}\,,
\label{eq:FivePoint_gamma}
\end{equation}
solving the mixing between all the operators appearing in the OPE.

With this, we conclude our computation of the next-to-leading order. We are able to reconstruct all the $f$-functions from \eqref{eq:FivePoint_FixedFunctions} and the constraints listed above it. This however was just a warm-up for the real challenge, which is the next order.  As anticipated, computing it only via Witten diagram seems too complicated. Therefore,  our idea is to apply the same bootstrap algorithm to bootstrap this next order.  We are optimistic that this is doable.  We will (hopefully) report about this in \cite{Barrat:2023pev}. In any case, the next step in this direction consists of extracting the CFT data from the correlator presented in this section, which then can be used as input in the same way as it was done here.

\chapter{Mellin Amplitudes for 1$d$ Correlators} \label{ch:Mellinres}
  
In this chapter, we move a bit away from multipoint correlators and the Wilson line specifically, to meet a broader view on 1$d$ CFTs in general. In particular, we explore an alternative way of expressing correlators, which could help simplify the expression of correlation functions and also expose their interesting features. We are talking about the Mellin formalism.  In higher-dimensional theories, it has been proven already quite effective, especially in holographic setups.  It is therefore very appealing to apply this framework to one-dimensional CFTs.

Since there is a parallelism between Mellin and Mandelstam variables,  one would naturally expect a Mellin amplitude with the typical properties of a two-dimensional $S$-matrix for CFT$_1$ correlators.  One option to proceed in the definition of such a Mellin amplitude would be to start from the higher-dimensional definition and enforce the relation $u=v$\footnote{We have in mind a four-point function in $d>1$, which depends on two cross-ratios, $u$ and $v$ introduced in (\ref{eq:crossratios4pt}).} (so-called diagonal limit) among cross-ratios. This constraint does not entail a relation in the Mellin variables and thus does not provide an inherently one-dimensional Mellin definition. Nevertheless, given a certain Mellin representation of the correlator, one can integrate out one of the Mandelstam variables thus obtaining a one-dimensional Mellin transform. A similar approach was followed in~\cite{Ferrero:2019luz}, leading to a successful,  though technically involved, implementation of the Mellin-Polyakov bootstrap~\cite{Polyakov:1974gs,Gopakumar:2016wkt,Gopakumar:2016cpb}. 

Therefore, we follow a different route and we propose a new definition of the Mellin transform, inherently one-dimensional and inspired by the guiding principles outlined above. The general strategy is to infer the analytic properties of the Mellin amplitude $M(s)$ (where $s$ is the complex Mellin variable) from physical, nonperturbative requirements on the correlator, which we take to be a four-point of identical scalars. As we explain in Section \ref{sec:Mellindefinition},  more than one choice is possible and throughout most of our analysis we use the one which displays a transparent correspondence between the dimensions of the operators exchanged in the correlator OPE and the simple poles of the Mellin amplitude. 

Crucially,  building a non-perturbative definition requires a finite number of subtractions and analytic continuations, which we perform along the lines of~\cite{Penedones:2019tng}. This leads us to a Mellin counterpart of the conformal block expansion, which provides a clear picture of how to extract the CFT data in Mellin space, while the Regge behavior of the correlator imposes some powerful bounds on the growth of the Mellin amplitude at large $s$.

In addition, we derive an infinite set of non-perturbative sum rules from our Mellin definition. These are~\emph{not} dispersive sum rules, because they have only \emph{single zeros} at $\Delta_n=2\Delta_{\phi}+n$. Such single zeros prevent the presence of positivity properties that are typical of dispersive sum rules.  Even if such an absence of positivity limits their powerfulness and makes them harder to use within the standard toolkit of the modern conformal bootstrap, we test these sum rules on some known examples and discuss their applicability in a perturbative setting. 

The efficiency of the 1$d$ Mellin formalism that we propose is manifest in the perturbative setup. Below, in Section~\ref{sec:pert}, we consider first-order deformations from generalized free-field (GFF) theory produced by quartic interactions in a bulk AdS$_2$ field theory.  We limit our analysis to leading order contact diagrams, for which we find the perturbative Mellin amplitude in closed form. 

This result can be then efficiently used to extract CFT data. In particular, we find a closed-form expression for the first correction to the classical dimension of ``two-particle'' operators\footnote{In the literature these operators are also known as double-twist operators or double-trace operators. Since here there is no trace and no twist we opted for the label ``two-particle'' operators. To avoid any confusion, they are unambiguously defined as the conformal primary operators that are exchanged in the GFF correlator. Their scaling dimensions and OPE coefficients are corrected by the perturbation we are considering.}, reproducing existing bootstrap results \cite{Ferrero:2019luz}.

Finally, we also present an alternative definition of the Mellin transform for which even simpler results for Mellin amplitudes of contact interactions may be obtained. In that context, we attempt to extend this Mellin representation beyond correlators of identical scalars.
   
This chapter is mostly based on \cite{Bianchi:2021piu} but it also contains some unpublished results.   

\section{Correlation Functions in $1d$ CFT}  \label{sec:1dcorr}

We have already discussed 1$d$ correlators in Section \ref{sec:wilsonlinecorrelators} so the purpose of this section is mostly to set the additional notation we need.

In this chapter, we mainly focus on correlators of identical scalars of scaling dimension $\Delta_\phi$, even if in the end we consider other cases too.  We restrict our analysis to four-point functions only, which we write in the usual factorized form, stripping off a prefactor such that the reduced correlator depends only on the cross-ratio:
\begin{equation}\label{corr-id-chi}
\!\!\!\!\!\!
 \braket{\phi(\tau_1)\phi(\tau_2)\phi(\tau_3)\phi(\tau_4)}=\frac{1}{(\tau_{12}\,\tau_{34})^{2\Delta_\phi}} \, \mathcal{A}(\chi)\,,
\end{equation}
where $\chi$ is defined as in (\ref{eq:crossratios}).
We can also write the correlation function in terms of an alternative cross-ratio:
\begin{equation}\label{eta}
\eta=\frac{\chi}{1-\chi}=\frac{\tau_{12}\,\tau_{34}}{\tau_{14}\,\tau_{23}}>0\,,\qquad  \braket{\phi(\tau_1)\phi(\tau_2)\phi(\tau_3)\phi(\tau_4)}=\frac{1}{(\tau_{12}\,\tau_{34})^{2\Delta_\phi}} \, \mathcal{A}(\eta)\,.
\end{equation}
It is more convenient to use this variable in the following since normally the Mellin transform is defined between zero and infinity, while $\chi \in (0, 1)$.  Indeed, we recall that in $1d$ the order of the operators in the correlation function matters. Considering the ordering  $\tau_1\!< \tau_2 \!< \tau_3\! < \tau_4\!\,$, one can use conformal symmetry to fix $\tau_1=0$, $\tau_3 = 1$, $\tau_4 = \infty$, finding that $\tau_2 \equiv \chi \in (0, 1)$. Different orderings would generate different ranges for $\chi$.  Unlike the higher-dimensional case,  correlators obtained by exchanging $\tau_1\leftrightarrow \tau_2$ and $\tau_1\leftrightarrow \tau_4$ are not related to \eqref{corr-id-chi} by crossing.  For this reason, we keep the ordering of the operators fixed, so that our main focus is the function $\mathcal{A}(\chi)$, but keep an eye on its analytic continuation to complex values too.  It may seem unphysical to think about line correlators in these terms, but from the perspective of the diagonal limit of higher-dimensional correlators, it would correspond to Lorentzian regimes for which $\bar \chi\neq \chi^*$, but $\chi=\bar \chi$.  Further comments on this can be found in \cite{Bianchi:2021piu}.\\
To understand the analytical structure of the function $\mathcal{A}(\chi)$, we can consider the s-channel conformal block expansion\footnote{This corresponds to taking the OPE as illustrated in the LHS of (\ref{eq:OPEassociativity}).} in Table \ref{Table 1}, written both in terms of the cross-ratio $\chi$ and in terms of the alternative cross-ratio (\ref{eta}).
\begin{center}
{\renewcommand\arraystretch{1.3} 
\begin{tabular}{ c | c | c }
  & $\chi$ cross-ratio & $\eta$ cross-ratio \\
\hline 
OPE &  $\mathcal{A}(\chi)=\sum_{\Delta_k} \,c_{\Delta_\phi\Delta_\phi \Delta_k}^2 \,g_{\Delta_k} (\chi) $ &	$\mathcal{A}(\eta)= \sum_{\Delta_k} \,c_{\Delta_\phi\Delta_\phi \Delta_k}^2 \,g_{\D_k} (\eta)	$\\		
Blocks &  $g_{\Delta_k}(\chi)=\chi^{\Delta_k}\, {}_2F_1(\Delta_k,\Delta_k;2\Delta_k;\chi) $ &	$g_{\Delta_k}(\eta)=\eta^{\Delta_k}\, {}_2F_1(\Delta_k,\Delta_k;2\Delta_k;-\eta) $	\\					Crossing sym & $\left(1-\chi\right)^{2\Delta_\phi}\mathcal{A}(\chi)=\left(\chi\right)^{2\Delta_\phi}\,\mathcal{A}(1-\chi)$ & $ \mathcal{A}(\eta)=\eta^{2\Delta_\phi} \, \mathcal{A}\left(\tfrac{1}{\eta}\right)$	\\																				
\end{tabular}      }   
\captionof{table}{\textit{Properties of $\mathcal{A}$.}}\label{Table 1}
\end{center}

In Table~\ref{Table 1} we identify $\Delta_k$ with the dimension of the primary operators exchanged in the $\phi\times \phi$ OPE and $c_{\Delta_\phi\Delta_\phi \Delta_k}$ are the corresponding OPE coefficients.   $g_{\Delta_k}$  are standard $\mathfrak{sl}(2)$ blocks resumming the contribution of conformal descendants \cite{Dolan:2011dv}, as introduced in Section~\ref{sec:conformalblocks}.
This expansion, accordingly to physical expectations, shows the presence of three branch points at $\chi=0,1,\infty$. Furthermore, it can be shown that the OPE valid around $\chi=0$ converges everywhere but on the branch cuts  $(-\infty, 0]$ and $[1,\infty)$~\cite{Hogervorst:2013sma,Rychkov:2017tpc,Kravchuk:2020scc}. \\
The t-channel OPE expansion\footnote{This OPE corresponds to the RHS of (\ref{eq:OPEassociativity}).} for $\mathcal{A}(\chi)$ can be conveniently obtained from the crossing relation in Table \ref{Table 1},  derived from the symmetry of the correlator under the exchange $\tau_1\leftrightarrow \tau_3$, corresponding to $\chi\to 1-\chi$ or $\eta\to 1/\eta$~\footnote{Unlike the $\tau_1\leftrightarrow \tau_2$ and $\tau_1\leftrightarrow \tau_4$ exchanges, the $\tau_1\leftrightarrow \tau_3$ is an actual symmetry of the correlator as one can easily see by picturing the four points on a circle. Consistently, this exchange maps the interval $(0,1)$ for $\chi$ to itself.}.
For some applications, it is useful to introduce the crossing symmetric function $\tilde{\mathcal{A}}$, whose properties are summarized in Table \ref{Table 2}.
\begin{center}
{\renewcommand\arraystretch{1.3} 
\begin{tabular}{ c | c | c}
  & $\chi$ cross-ratio & $\eta$ cross-ratio \\
\hline 																						
		
Function & $\tilde{\mathcal{A}}(\chi)=\chi^{-2\Delta_\phi} \mathcal{A}(\chi)$ & $ \tilde{\mathcal{A}}(\eta)=\mathcal{A} \left(\tfrac{\eta}{1+\eta}\right)$	\\		

Crossing symmetry & $\tilde{\mathcal{A}}(\chi)=\tilde{\mathcal{A}}(1-\chi)$ & $ \tilde{\mathcal{A}}(\eta)=  \tilde{\mathcal{A}}(\tfrac{1}{\eta})$	\\																						
\end{tabular}      }   
\captionof{table}{\textit{Properties of $\tilde{\mathcal{A}}$.}}\label{Table 2}
\end{center}

There is another interesting limit we consider in the following, i.e. the~$\chi\to \frac12+ i \infty$ limit\footnote{We could take this limit along any direction excluding the real line to avoid the branch cuts, but for definiteness, we take it along the imaginary axis.}. This limit can be understood in terms of the higher-dimensional correlator in the diagonal limit, where it corresponds to the u-channel Regge limit\footnote{In $1d$ there is no u-channel OPE expansion as it is impossible to bring $\tau_1$ close to $\tau_3$ without $\tau_2$ in between. However, one can resort to the higher-dimensional picture to understand that while the u-channel OPE would correspond to $\chi\to i\infty$ and $\bar \chi\to -i\infty$, the \textit{u-channel Regge limit} is $\chi,\bar \chi\to i\infty$.}. In particular,  in a unitary CFT four-point functions are bounded in the Regge limit \cite{Maldacena:2015waa,Caron-Huot:2017vep} and we have \cite{Mazac:2018ycv}
\begin{equation}\label{Reggebound}
\left|\tilde{\mathcal{A}}\left(\textstyle{\frac{1}{2}}+iT\right)\right|\, \text{is bounded as } T\rightarrow \infty.
\end{equation}
Translating into the $\eta$ cross-ratio \eqref{eta}, the line parametrized by $\chi=\frac12+i \,\xi$ is mapped into the unit circle $\eta=e^{i \theta}$ for $\theta\in(-\pi,\pi)$ and the Regge limit occurs when $\theta\to \pi$. The Regge boundedness condition \eqref{Reggebound} for the function $\mathcal{A}(\eta)$ in~\eqref{eta} then reads
\begin{equation}\label{reggetheta}
 \mathcal{A}(e^{i\theta})=O\left((\pi-\theta)^{-2\Delta_\phi}\right) \qquad \theta\to \pi .
\end{equation}
Further details on the implication of the Regge boundedness condition on the Mellin amplitude can be found in Section \ref{sec:reggeandbound}.

 \section{Defining and Characterizing the Mellin Transform }  \label{sec:Mellindefinition}
  
Our goal is to find a well-defined Mellin transform for 1$d$ conformal correlators.  Since they depend on a single cross-ratio, the first natural step is to look at the textbook Mellin transform for a function $F(\eta)$ defined on the positive real axis
\begin{align}\label{textbook}
 \mathcal{M}[F](s)=\int_0^{\infty} \! d \eta\,  F(\eta)\, \eta^{-1-s} \,  ,
\end{align}
where it is natural to use $\eta$ in~\eqref{eta} to define the cross-ratio since it spans the correct range. Furthermore, more importantly, one has to choose which function of the cross-ratio should be identified with $F$.

As we described in Section~\ref{sec:1dcorr} different choices of the prefactor in \eqref{corr-id-chi} lead to different functions of the cross-ratio, related to each other by powers of $\eta$ and $(1+\eta)$. In contrast to the higher-dimensional case, where a rescaling by powers of the cross-ratios has the effect of shifting the corresponding Mellin variables, in 1$d$ a rescaling by powers of $\eta$ leads to a shift in $s$ in (\ref{textbook}), whereas a rescaling by powers of $(1+\eta)$ leads to different Mellin amplitudes. That is why the first issue to address is which criterion one should use to define the Mellin amplitude. 

Up to shifts in the $s$ variable, we define a one-parameter family of Mellin amplitudes:
\begin{align}\label{Mellin1param}
 \mathcal{M}_a(s)=\int_0^{\infty}\!\! d\eta \, \mathcal{A}(\eta) \Big(\frac{\eta}{1+\eta}\Big)^a\,\eta^{-1-s} \, ,
\end{align}
where the function $\mathcal{A}(\eta)$ is given in \eqref{eta}. Using the crossing relation in Table \ref{Table 1}, one immediately finds the functional relation for the Mellin amplitude
\begin{equation}\label{crossingmellin}
\mathcal{M}_a(s)=\mathcal{M}_a(2\Delta_\phi+a-s)\,,
\end{equation}
which is reminiscent of the crossing for $S$-matrix elements in two dimensions\footnote{Conservation of energy and momentum for the $2 \rightarrow 2$ scattering process in 2$d$ leads to the Mandelstam variables $\mathtt{s}=\left(p_1+p_2\right)^2, \mathtt{t}=\left(p_2-p_3\right)^2=4 m^2-\mathrm{s}, \mathtt{u}=\left(p_3-p_1\right)^2=0$, and the crossing symmetry is written as $S(\mathtt{s})=S\left(4 m^2-\mathtt{s}\right)$.}. However, the precise relation between $s$ and the ordinary flat space Mandelstam variable $\mathtt{s}$ requires a careful analysis of the flat space limit, which we do not address here\footnote{Here we just notice that the large $s$ regime is the relevant one for the flat space limit considered in~\cite{Penedones:2010ue}, where AdS scattering reduces to the scattering of massless excitations for large AdS radius. In that case, one would have the flat space relation $\mathcal{M}_a(s)=\mathcal{M}_a(-s)$ for any finite value of $a$. This relation would be consistent with 2$d$ massless scattering, where $\mathtt{s}=-\mathtt{t}$. There is however more than one approach to the flat space limit, see~\cite{Paulos:2016fap}.}. 

Up to shifts in the $s$ variable, the definition \eqref{Mellin1param} allows for different choices of prefactors in the correlator \eqref{eta}. For instance, the choice $a=0$ clearly corresponds to taking the Mellin transform of $\mathcal{A}(\eta)$, while the choice $a=-2\Delta_\phi$ effectively corresponds to taking the Mellin transform of the function $\tilde{\mathcal{A}}(\eta)$ in Table \ref{Table 2}. In \cite{Bianchi:2021piu}, we mostly study the case of $a=0$, which emerges naturally when considering the s-channel conformal block expansion. Though, we also introduce a possible alternative, with $a=-2\Delta_\phi+1$,  that leads to simple results in a perturbative expansion around GFF.

\subsection{A Non-Perturbative Definition}

To discuss the properties of the Mellin transform, we then focus on $\mathcal{M}_0(s)$ defined in \eqref{Mellin1param}, which we multiply by an overall factor for future convenience
\begin{equation}\label{Mellin}
M(s) =\frac{1}{\Gamma(s)\Gamma(2\Delta_\phi-s)}\int_0^\infty d\eta\, \mathcal{A}(\eta)\, \eta^{-1-s}\,.
\end{equation} 
In this case, the crossing relation \eqref{crossingmellin} reads
\begin{equation}\label{crossing1def}
M(s)=M(2\Delta_\phi -s)\, .
\end{equation}

We now want to infer the analytic properties of the Mellin amplitude $M(s)$ from the physical requirements on the correlator $\mathcal{A}(\eta)$. 
First of all, in \cite{Bianchi:2021piu}, following \cite{Penedones:2019tng}, we remind a general theorem for the one-dimensional Mellin transform \eqref{Mellin} about the exponential suppression of the Mellin when $|Im(s)|\rightarrow\infty$. The physical $1d$ correlator though violates the hypothesis of this theorem, but this problem can be overcome by carefully studying the Regge limit of $\mathcal{A}(\eta)$ and by relating it with the large $s$ asymptotics of $M(s)$. Indeed, in \cite{Bianchi:2021piu} we find a bound on the large $s$ behaviour of the Mellin amplitude $M(s)$ using the Regge behaviour of the function $\mathcal{A}(\eta)$, i.e. the limit $\eta\to e^{i\pi}$ described in \eqref{reggetheta}:
\begin{equation}
 M(c+i\rho)=O(|\rho|^0) \qquad |\eta| \to \infty \, .
\end{equation}
To obtain this result, we have considered the inverse Mellin transform 
\begin{align}\label{inverseMellin}
\mathcal{A}(\eta) = \int_\mathcal{C}\frac{ds}{2 \pi i}\,\Gamma(s)\Gamma(2\Delta_\phi-s)\,M(s)\,\eta^{s}\, ,
\end{align}
where the contour $\mathcal{C}$ is a straight line parametrized by $s=c+i\rho$ for some constant $2\D-\tilde \Delta_0<c<\tilde \Delta_0$ and $\rho \in\mathbb{R}$.\\
Assuming that no Stokes phenomenon occurs for physical correlators this behavior can be extended for any $\text{arg}(s)$ such that
\begin{equation}\label{resultasymM1}
 M(s)=O(|s|^0) \qquad |s|\to\infty \, .
\end{equation}
The result \eqref{resultasymM1} is valid for the full non-perturbative Mellin amplitude (\ref{Mellin}). The factorization of the product of the $\Gamma$-functions in (\ref{Mellin}) is crucial to obtain this result because it precisely accounts for the exponential behavior of the Mellin predicted by the theorem. We come back to this remarkable fact in Section \ref{sec:reggeandbound}.

\subsection{Convergence and Subtractions} \label{sec:convergenceandsubtractions}

Another crucial property that we need to address is the convergence of the Mellin.
Let $\mathcal{A}(\eta)$ be well-behaved for $\eta\in \mathbb{R}^+$,  i.e. with no divergences in $\eta$ other than at $\eta=0$ and $\eta\rightarrow \infty$. This behavior coincides with that of the CFT$_1$ correlators we are interested in. If we consider the behavior of $\mathcal{A}(\eta)$  close to $\eta=0$, we find that the leading power is $\mathcal{A}(\eta)\sim \eta^{\D_0}$ with $\Delta_0$ the dimension of the lightest exchanged operator. Analogously, using the crossing symmetry relation in Table \ref{Table 1}, we find that the large $\eta$ behaviour of $\mathcal{A}(\eta)$ is $\mathcal{A}(\eta)\sim \eta^{2\Delta_\phi-\D_0}$. Therefore the integral converges in the strip
\be\label{convergence}
2\Delta_\phi-\D_0<\text{Re}(s)<\D_0\,, 
\ee
which is a well-defined interval \emph{only for $\D_0>\Delta_\phi$}.  
In order to give a non-perturbative definition of the Mellin transform, which allows for lighter operators to be exchanged,  in \cite{Bianchi:2021piu} we perform some subtractions along the lines of~\cite{Penedones:2019tng}. One obvious example is the GFF case, where the identity operator is exchanged. We consider it explicitly in Section 3.1.3 of \cite{Bianchi:2021piu}. 

For the moment, we focus on the Mellin transform of the connected part of the correlator. Let us consider the following subtractions
\begin{subequations}
\begin{align}
 \Am_0(\eta)&=\Am_{\text{conn}}(\eta)-\sum_{\D_{0}\leq \Delta_k \leq \Delta_\phi}\sum_{j=0}^{[\Delta_\phi-\Delta_k]} c_{\Delta_k} \frac{(-1)^{j}}{j!} \frac{\Gamma(\Delta_k+j)^2\Gamma(2\Delta_k)}{\Gamma(\Delta_k)^2\Gamma(2\Delta_k+j)} \eta^{\Delta_k+j}\,, \label{subtr0} \\
 \Am_{\infty}(\eta)&=\Am_{\text{conn}}(\eta)-\sum_{\D_{0}\leq \D_k \leq\Delta_\phi}\sum_{j=0}^{[\D_\phi-\D_k]} c_{\D_k} \frac{(-1)^j}{j!} \frac{\Gamma(\D_k+j)^2\Gamma(2\D_k)}{\Gamma(\D_k)^2\Gamma(2\D_k+j)} \eta^{2\Delta_\phi-\D_k-j}\,,\label{subtrinf}
\end{align}
\end{subequations}
where, for convenience, we write  $c_{\Delta_\phi \Delta_\phi \Delta_k}^2\equiv c_{\Delta_k}$. What we are doing is subtracting from $\Am_0(\eta)$ the s-channel contribution of all the operators (primaries and descendants) with scaling dimension below the threshold $\D_k=\Delta_\phi$, making use of the series expansion of the hypergeometric function in the OPE appearing in Table \ref{Table 1}. This improves the behavior of the function at $\eta=0$. On the other hand, to improve the behavior at $\eta=\infty$ we subtract from $\Am_{\infty}(\eta)$ all the $s$-channel operators below threshold. The idea is to split the integral~\eqref{Mellin} into two parts, which are defined on (possibly non-overlapping) semi-infinite regions of the complex $s$ plane
\begin{subequations}
\begin{align}
 \psi_0(s)&=\int_0^1 dt\, \Am_{\text{conn}}(\eta) \,\eta^{-1-s} & \text{Re}(s)&< \D_0 \,  , \label{psi0def} \\
  \psi_{\infty}(s)&=\int_1^{\infty} dt \,\Am_{\text{conn}}(\eta) \,\eta^{-1-s} & \text{Re}(s)&>2\Delta_\phi- \D_0.\label{psiinfdef}
\end{align}
\end{subequations}
When the two regions do not overlap, we analytically continue $\psi_0(s)$ and $\psi_{\infty}(s)$ by considering the integrals of the functions  \eqref{subtr0} and \eqref{subtrinf} and adding a finite number of poles
\small
\begin{subequations}
\begin{align}
\!\!\!\! \psi_0(s)&=\int_0^1 \!\!\!dt \, \Am_0(\eta)\, \eta^{-1-s}+\!\!\!\!\!\!\sum_{\D_{0}\leq \D_k \leq\Delta_\phi}\!\!\! \! \sum_{j=0}^{[\Delta_\phi-\D_k]} c_{\D_k} \tfrac{(-1)^{j}}{j!} \tfrac{\Gamma(\D_k+j)^2\Gamma(2\D_k)}{\Gamma(\D_k)^2\Gamma(2\D_k+j)} \tfrac{1}{s-\D_k-j}\,, \qquad \small{\text{Re}(s)< \tilde \D_0} \, ,  \label{psi0ancont} \\
\!\!\!\! \psi_{\infty}(s)&=\int_1^{\infty} \!\!\!\!\!\!dt \, \Am_{\infty}(\eta)\, \eta^{-1-s}+\!\!\!\!\!\!\sum_{\D_{0}\leq \D \leq\Delta_\phi}\!\!\! \!\sum_{j=0}^{[\Delta_\phi-\D_k]} c_{\D_k} \tfrac{(-1)^{j}}{j!} \tfrac{\Gamma(\D_k+j)^2\Gamma(2\D_k)}{\Gamma(\D_k)^2\Gamma(2\D_k+j)}  \tfrac{1}{s-2\Delta_\phi+\D_k+j}\,,\small{\text{Re}(s)>2\Delta_\phi- \tilde \D_0} \, ,\label{psiinfancont}
\end{align}
\end{subequations}
\normalsize
where $\tilde \D_0>\Delta_\phi$ is the lightest exchanged operator above the threshold (notice that this operator could be either a primary or a descendant). Both these functions are now well defined on the non-vanishing strip $2\Delta_\phi- \tilde \D_0<\text{Re(s)}<\tilde \D_0$ and therefore their sum yields a well-defined Mellin transform
\begin{align}\label{Mstrip}
 M(s)&=\frac{\psi_0(s)+\psi_{\infty}(s)}{\Gamma(s)\Gamma(2\Delta_\phi-s)} \, ,& 2\Delta_\phi- \tilde \D_0&<\text{Re(s)}<\tilde \D_0 \, .
\end{align}
The price to pay is a deformation of the integration contour in the inverse Mellin transform (\ref{inverseMellin}).

To understand the form of the contour $\mathcal{C}$, we need to discuss the analytic structure of $M(s)$. In order to do this, one can follow the strategy described above to extend the definition \eqref{Mstrip} to the whole complex $s$ plane. To analytically continue $\psi_0(s)$ from the region $\text{Re}(s)<\D_0$ to the region $\text{Re}(s)<\tilde \D_0$ we subtracted a few exchanged operators in $\mathcal{A}(\eta)$ and added a finite number of poles in \eqref{psi0ancont}. By adding more and more poles, we can further extend the area of analyticity. We then conclude that the \emph{Mellin block expansion} defined by (\ref{Mstrip})
with
\begin{subequations} \label{eq:psi}
\begin{align} 
 \psi_0(s)&=\sum_{\D_k}\sum_{j=0}^{\infty} c_{\D_k }\frac{(-1)^{j+1}\,\Gamma(\D_k+j)^2\,\Gamma(2\D_k)}{j!\,\Gamma(\D_k)^2\,\Gamma(2\D_k+j)}\frac{1}{s-\D_k-j} \, ,\label{psi0sum}\\
 \psi_{\infty}(s)&=\sum_{\D_k}\sum_{j=0}^{\infty} c_{\D_k} \frac{(-1)^{j}\,\Gamma(\D_k+j)^2\,\Gamma(2\D_k)}{j!\,\Gamma(\D_k)^2\,\Gamma(2\Delta_k+j)}\frac{1}{s-2\Delta_\phi+\D_k+j} \label{psiinfsum}
\end{align}
\end{subequations}
provides a representation of $M(s)$ which is valid on the \emph{whole complex $s$ plane}, excluding the point at infinity which we mention in Section \ref{sec:reggeandbound}.

We can also perform the sum over $j$ in~\eqref{Mstrip}, resuming all the conformal descendants in a crossing-symmetric Mellin block expansion 
\begin{subequations}
\begin{align} \label{hypergeom}
 M(s) &= \frac{1}{\Gamma(s)\Gamma(2\Delta_\phi-s)} \sum_{\D_k}\,c_{\D_k} [\mathcal{F}_{\D_k}(s)+\mathcal{F}_{\D_k}(2\Delta_\phi-s)]\,,\\
\mathcal{F}_{\D_k}(s)&= \frac{{}_3F_2({\D_k, \D_k, \D _k- s}; {2 \D_k,1+ \D_k-s}; -1)}{\D_k-s} \, .
\end{align}
\end{subequations}
Interestingly the representation~\eqref{Mstrip} immediately allows us to read off the position of the poles of $M(s)$. For any exchanged primary operator of dimension $\D_k$ there are two infinite sequences of poles running to the right of $s=\D_k$ and the left of $s=2\Delta_\phi-\D_k$. Following the common nomenclature, we denote them as
\begin{subequations}
\begin{align}
\text{\emph{right} poles}: s_R&=\D_k+j\,, ~~\qquad\,\, \qquad j=0,1,2,\dots\,, \label{leftpoles}\\
\text{\emph{left} poles}: s_L&=2\Delta_\phi-\D_k-j\,, \qquad j=0,1,2,\dots\,, \label{rightpoles}\\
\text{Res}[M(s)]|_{s_L}&\equiv-\text{Res}[M(s)]|_{s_R}=\frac{(-1)^{j} \,\Gamma (2 \Delta_k)\, \Gamma (\D_k+j)}{j! \,\Gamma (\D_k)^2 \,\Gamma (2 \D_k+j) \,\Gamma (2\Delta_\phi-\D_k-j )}\,. \label{residues}
\end{align}
\end{subequations}
Notice that the precise identification of the sum over $j$ in \eqref{psi0sum} with the sum over descendants in the block expansion is a consequence of the choice $a=0$ in \eqref{Mellin1param}. Different choices of $a$ in \eqref{Mellin1param} would lead to a less transparent correspondence between poles and conformal descendants.

\begin{figure}[h]
\centering
\includegraphics[scale=0.45]{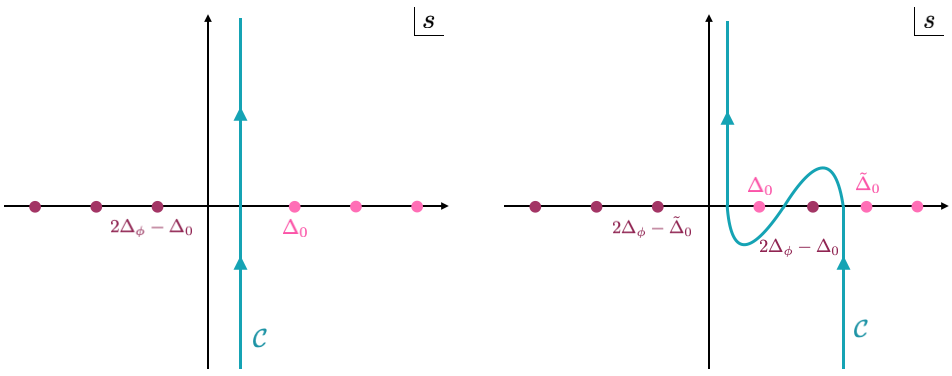}
\caption{\textbf{Left}: \textit{The \textbf{\textcolor{gluoncolor}{contour}} for the inverse Mellin transform when $\D_0>\Delta_\phi$. Left poles are marked in \textbf{\textcolor{vertexcolor}{magenta}} and right poles in \textbf{\textcolor{SEcolor}{pink}}. }\textbf{Right}: \textit{When $\D_0<\Delta_\phi$ left and right poles intersect and the contour needs to be deformed.}}
\label{fig:C}
\end{figure}

With this structure of poles, we can now give a precise definition of the contour $\mathcal{C}$ in \eqref{inverseMellin}. The contour $\mathcal{C}$ is chosen in such a way as to leave all the \emph{right} poles of $M(s)$ on its right and all the \emph{left} poles on its left. If the lightest exchanged operator has dimension $\D_0>\Delta_\phi$, no analytic continuation is required in \eqref{psi0def} and \eqref{psiinfdef} (in other words the set of left and right poles do not overlap) and any contour within the interval \eqref{convergence} suffices, see e.g. the straight one on the left in Figure \ref{fig:C}. 
When lighter operators are exchanged, the contour needs to be deformed because the set of right poles intersects with the set of left poles. In Figure~\ref{fig:C} we show an example with a single operator below threshold. It is clear from the picture that a more complicated situation arises when a left and a right pole coincide. This happens, for instance, for the GFF case, which we address in Section 3.1.3 of \cite{Bianchi:2021piu}. More generally, this happens whenever there is an exchanged operator with dimension $\D_k=\Delta_\phi+\frac{\mathbb{Z}}{2}$. In a generic spectrum, we do not expect this to be the case.

\subsection{Regge Limit and Mellin Boundedness} \label{sec:reggeandbound}  

We now want to motivate the choice of the $\Gamma$ prefactor in (\ref{inverseMellin}). We, therefore, take (\ref{inverseMellin}), where the contour $\mathcal{C}$ is a straight line parametrized by $s=c+i\rho$ for some constant $2\D-\tilde \Delta_0<c<\tilde \Delta_0$ (the additional poles that are included in \eqref{psi0ancont} and \eqref{psiinfancont} for the analytic continuation does not affect this argument) and $\rho \in\mathbb{R}$. We take $\eta=e^{i\theta}$ and we integrate over~$\rho$  
\begin{equation}\label{eq:inverseMellineta}
 \Am(e^{i \theta})=e^{i c \theta} \int_{-\infty}^{\infty} d\rho\, \Gamma(c+i \rho) \Gamma(2\Delta_{\phi}-c-i \rho)\, M(c+i\rho) \, e^{- \theta \rho} \, .
\end{equation}
We are interested in the behavior of the integrand for $|\rho|\to \infty$. In this limit 
\begin{equation}\label{Gammaasyn}
 \Gamma(c+i \rho) \Gamma(2\Delta_{\phi}-c-i \rho) \sim e^{-\pi|\rho|}\, \rho^{2\Delta_{\phi}-1} \qquad |\rho|\to\infty \, .
\end{equation}
This means that the $\Gamma$-function prefactor accounts for the exponential behavior of $\hat{M}(s)$ for $|\text{Im}(s)|\to\infty$, which is predicted by the theorem in Section 3.1.1 of \cite{Bianchi:2021piu}. This essentially motivates our choice of prefactor in \eqref{Mellin}. In particular, the exponential in \eqref{Gammaasyn} combined with that in \eqref{eq:inverseMellineta} shows that the regime $\theta\to \pm\pi$ is controlled by the region $\rho\sim \mp\infty$. 

We can make a more precise statement by defining
\begin{align}\label{Hdef}
 H(\rho)&\equiv\Gamma(c+i \rho) \Gamma(2\Delta_{\phi}-c-i \rho)\, M(c+ i \rho) \,e^{\pi|\rho|} 
\end{align}
and using it to rewrite (\ref{eq:inverseMellineta}), to show that the Regge behaviour \eqref{reggetheta} is reproduced by asking that
\begin{equation}
 H(\rho)\sim |\rho|^{2\Delta_{\phi}-1} \qquad |\rho|\to \infty \, .
\end{equation}
Combining this with \eqref{Gammaasyn} and \eqref{Hdef}, we conclude that 
\begin{equation}
 M(c+i\rho)=O(|\rho|^0) \qquad |\rho| \to \infty \, .
\end{equation}
Assuming that no Stokes phenomenon occurs\footnote{The absence of Stokes phenomenon is an assumption for which we do not have proof. This assumption however is verified in all our examples and it was made also in the higher-dimensional case \cite{Penedones:2019tng}.  } for physical correlators, we can extend this behavior for any $\text{arg}(s)$ such that
\begin{equation}\label{resultasymM}
 M(s)=O(|s|^0) \qquad |s|\to\infty \, .
\end{equation}
We have, therefore, found a bound on the large $s$ behavior of the Mellin amplitude $M(s)$, using the Regge behavior of the function $\Am(\rho)$, i.e. the limit $\rho\to e^{i\pi}$ described in \eqref{reggetheta}. The full derivation can be found in Section 3.2 of \cite{Bianchi:2021piu}.
In particular, notice that the function $M(s)$ has infinitely many poles that accumulate at $s=\infty$, where we would expect an essential singularity. Nevertheless, our prefactor in \eqref{Mellin} removes this singularity and leaves us with a bounded function \eqref{resultasymM}.

We conclude this section with an important remark about the perturbative regime. The result \eqref{resultasymM} is valid for the full non-perturbative Mellin amplitude. If the correlator contains a small parameter, it is often the case that order by order in the perturbative expansion the Regge behavior is worse than in the full non-perturbative correlator\footnote{A typical example of this phenomenon is the function $\frac{1}{1-g\chi}$, which is regular for $\chi\to\infty$ but its expansion at small $g$ is more and more divergent.}. In Appendix B.1 of \cite{Bianchi:2021piu}, this aspect is illustrated in detail in the context of the analytic sum rules, which we discuss in the next section.  Since these are subtleties, we decided not to include this appendix in this thesis.

In view of this aspect, it is then useful to formulate our result in a more general form. Let us consider a correlator $\Am(\chi)$ with a Regge behaviour
\begin{equation}
\Am(\chi)=O( \chi^{2\Delta_{\phi}+n}) \qquad \chi\to \frac12+i\infty
\end{equation}
for some positive integer $n$, then the associated Mellin amplitude has a large $s$ asymptotics
\begin{equation}\label{Mlarges}
 M(s)=O( |s|^n) \qquad |s|\to \infty \, .
\end{equation}

\section{Sum Rules}  \label{sumrules}

A common way to express the well-known fact that an arbitrary set of CFT data does not necessarily lead to a consistent CFT is through a set of sum rules for the CFT data. In the following, we start from our definition of Mellin amplitude and derive an infinite set of sum rules. These sum rules though are not dispersive, according to the definition of \cite{Caron-Huot:2020adz}. This is essentially related to the behavior at infinity obtained using our one-dimensional definition. In Section 3.2 of \cite{Bianchi:2021piu} and briefly in Section \ref{sec:reggeandbound} here, we described how the product of $\Gamma$-functions in our definition \eqref{Mellin} leads to a nice behavior for the Mellin amplitude $M(s)$ at $s=\infty$. However, the introduction of that prefactor leads also to the appearance of spurious poles in the integral \eqref{inverseMellin}. In a generic CFT, it is not expected that operators with the exact dimension $s=2\Delta_\phi+n$ are present in the spectrum. Thus, the poles of the $\Gamma$-functions must be compensated by zeros in the Mellin amplitude. This strategy was used in \cite{Penedones:2019tng,Carmi:2020ekr} to derive dispersive sum rules for the higher-dimensional case, where the Mellin amplitude needs to have double zeros. Here, we use the same idea to derive a new set of sum rules, which are characterized by single zeros of the Mellin amplitude. This makes these sum rules different and less powerful than the dispersive ones, but we believe that their derivation and their validity on a set of known example provides an important consistency check of our results.

One may be concerned because the presence of single or double zeroes for the Mellin amplitude seems to be related to the choice of the prefactor in \eqref{Mellin}. This is actually not the case. The choice to factor out a prefactor in \eqref{Mellin} is related to having a nice polynomial behavior for the function $M(s)$ at $s\to\infty$. 

Finally, let us emphasize some important differences compared to the higher-dimensional strategy of the Mellin Polyakov bootstrap \cite{Polyakov:1974gs,Gopakumar:2016wkt,Gopakumar:2016cpb}. The derivation of the non-perturbative Polyakov consistency conditions used in \cite{Penedones:2019tng,Carmi:2020ekr} is quite subtle due to the presence of accumulation points in the twist spectrum of higher-dimensional CFTs. In our case, however, the situation is simpler. The twist accumulation points are related to the presence of a spin or, equivalently, to the need of introducing two Mandelstam variables. For us, there is no spin and the only quantum number is the scaling dimension of the operators. Therefore, we do not expect any accumulation point in the spectrum and we can impose the conditions \eqref{zeroes} without recurring to any analytic continuations.

\subsection{Properties of $M(s)$ and Derivation of the Sum Rules}\label{derivationsumrules}

We start by summarizing the main properties of the Mellin amplitude $M(s)$ in~\eqref{Mellin}: 
\begin{itemize}
	\item $M$ is \textbf{crossing symmetric} 
	\begin{equation}
		M(s) = M(2\Delta_\phi-s) \, .
	\end{equation}
	\item $M$ has \textbf{poles} at the location of the physical exchanged operators in the two channels, i.e. $s=\D_k+j$ and $s=2\Delta_\phi-\D_k-j$ for $j\in \mathbb{N}  \, .$
	\item Generically, $M$ has single \textbf{zeros} compensating the poles of the prefactor 
	\begin{align}\label{zeroes}
		M(2\Delta_\phi+j) =0 \quad\text{and} \quad M(-j)=0 \quad \text{for} \quad j \in \mathbb{N} \,.
	\end{align}
	If the spectrum contains protected operators, some of these zeros might be absent.
	\item $M$ is \textbf{bounded} for $|s|\to \infty$, see \eqref{resultasymM} .

	\item $M$ admits a crossing-symmetric \textbf{Mellin block expansion} 
	\begin{align}
		M(s) = \sum_{\D_k} \, {c_{\D_k}}\, M_{\D_k}(s) \,,
	\end{align}
	with $M_{\D_k}(s)$ given by the comparison with \eqref{hypergeom}.
\end{itemize}

The properties above allow us to define a set of sum rules along the lines of \cite{Penedones:2019tng, Carmi:2020ekr}. 
Let $\omega_p$ be the functional 
\begin{align}\label{functional}
	\omega_{p_i} = \oint_{\mathbb{C}|_{\infty}} \frac{ds}{2\pi i}M(s)F_{p_i}(s)\, ,
\end{align}
where the contour here is a very large circle around infinity. When $F_{p_i}(s)$ is a sufficiently suppressed function at $s\to \infty$, we can take the limit of infinite radius for the circle and we get
\begin{align}
		\omega_{p_i} [M] =0	 \, .
\end{align}
For a non-perturbative Mellin amplitude characterized by the asymptotic behaviour \eqref{resultasymM} it is sufficient to ask that $F_{p_i}(s)\sim s^{-1-\e}$ for $\e>0$ as $|s|\to \infty$. 

The strategy to derive the sum rules simply consists in deforming the integration contour in \eqref{functional} to include all the poles of the integrand such that
 	\begin{align}\label{functionalwithM}
		\omega_{p_i} = \sum_{s^*} \text{Res}_{s=s^*}\left[M(s)\right]F_{p_i}(s^*)+\sum_{s^{**}} M(s^{**})\text{Res}_{s=s^{**}} \left[F_{p_i}(s)\right]=0 \, .
	\end{align}
	This equation already looks like a sum rule, but it depends on the value $M(s^{**})$ of the Mellin amplitude at the poles of $F_{p_i}(s)$. To avoid this issue one can simply choose $F_{p_i}(s)$ to have simple poles at the position of the zeros of $M(s)$. Therefore, we need a function $F_{p_i}(s)$ with poles at $s=-j$ or at $s=2\Delta_{\phi} +j$. Furthermore, the function $F_{p_i}(s)$ must not be crossing symmetric. Indeed, using the position of the poles in \eqref{leftpoles} and \eqref{rightpoles} and crossing symmetry for the residues \eqref{residues} we get
	\begin{align}\label{sumrulerightpoles}
		\omega_{p_i} = \sum_{s_R} \text{Res}_{s=s_R}(M(s))(F_{p_i}(s_R)-F_{p_i}(2\Delta_\phi-s_R)) \, ,
	\end{align}
so that any crossing symmetric function $F$ would lead to a trivial vanishing of $\omega_{p_i}$. Using the explicit expression for the residues \eqref{residues} we find the set of sum rules
\small
\begin{align}\label{sumrulesgenericF}
	\sum_{\D_k,j} c_{\D_k} \frac{(-1)^{j+1}\Gamma(2\D_k)\Gamma(\D_k+j)}{\Gamma(\D_k)^2\Gamma(2\D_k+j)\Gamma(2\Delta_\phi-\D_k-j)\Gamma(j+1)}(F_{p_i}(\Delta_k+j)-F_{p_i}(2\Delta_\phi-\Delta_k-j))=0 \, .
\end{align}
\normalsize
A natural choice for the function $F$ is 
\begin{align}\label{Fp1p2}
 F_{p_1,p_2}(s)=\frac{1}{(s+p_1)(s+p_2)} \, ,\qquad p_1,p_2\in \mathbb{N} \, .
\end{align}
Notice that, despite the function $F_{p_1,p_2}(s)\sim \frac{1}{s^2}$ for $s\to \infty$, thanks to \eqref{sumrulerightpoles} only the crossing antisymmetric part of it matters, i.e. $F_{p_1,p_2}(s)-F_{p_1,p_2}(2\Delta-s)$ and one can easily check that this combination decays as $\frac{1}{s^3}$ for $s\to\infty$. Using this function we can derive the non-perturbative sum rules
\begin{align}\label{sumrulesnonpert}
	\sum_{\D_k,j} c_{\D_k} \tfrac{(-1)^{j+1}\Gamma(2\D_k)\Gamma(\D_k+j)}{\Gamma(\D_k)^2\Gamma(2\D_k+j)\Gamma(2\Delta_\phi-\D_k-j)\Gamma(j+1)}\tfrac{2(\Delta_k+j-\Delta_{\phi})(p_1+p_2+2\Delta_{\phi})}{(\Delta_k+j+p_1)(\Delta_k+j+p_2)(2 \Delta_{\phi}-\Delta_k-j+p_1)(2 \Delta_{\phi}-\Delta_k-j+p_2)}=0 \, .
\end{align}
Performing the sum over $j$ one obtains sum rules of the form 
\begin{align}
 \sum_{\D_k} c_{\D_k}  \a_{\Delta_k}=0\,,
\end{align} \label{formsumrule}
with 
\footnotesize
\begin{subequations}
\begin{align}\label{alphasumrule}
\a_{\Delta_k}&=\frac{\Gamma(\Delta_k)}{\Gamma(2\Delta_k)\Gamma(2\Delta_{\phi}-\Delta_k)} \left(\mathcal{F}_{p_1,p_2}(\Delta_k)+\mathcal{F}_{-2\Delta_{\phi}-p_1,-2\Delta_k-p_2}(\Delta) \right) \, ,\\
\mathcal{F}_{p_1,p_2}(\Delta_k)&=\tfrac{1}{p_1-p_2}\left(\tfrac{{}_3F_2(\Delta_k,p_1+\Delta_k,1+\Delta_k-2\Delta_{\phi};2\Delta_k,1+p_1+\Delta_k;1)}{(p_1+\Delta_k)}-\tfrac{{}_3F_2(\Delta_k,p_2+\Delta_k,1+\Delta_k-2\Delta_{\phi};2\Delta_k,1+p_2+\Delta_k;1)}{(p_2+\Delta_k)}\right)\, .
 \end{align}
 \end{subequations}
 \normalsize
The absence of the positivity property we mentioned before makes these sum rules less powerful and harder to use with the standard method of the modern conformal bootstrap. That is why testing the sum rules \eqref{sumrulesnonpert} or  \eqref{sumrulesgenericF} on a fully non-perturbative spectrum is a task that is momentarily out of reach.  For this reason, we restrict our test to some known examples (we apply them to the GFF case in Section 4.2.1 of \cite{Bianchi:2021piu}) and to the perturbative setting.

\subsection{Perturbative Sum Rules}

 We invite the reader to find more details about the GFF case in \cite{Bianchi:2021piu}, while we focus here on the perturbations around GFF, constructed by introducing an effective field theory in AdS$_2$ background and considering the $1d$ boundary conformal field theory through the usual holographic dictionary. These quartic interactions with derivatives are classified in e.g. \cite{Mazac:2018ycv}, where the authors find there is a one-parameter family labeled by~$L$, where~$4L$ is the number of derivatives in the schematic interaction~$(\partial^L \Phi)^4$. \\
Let us consider a perturbed correlator 
\begin{equation}\label{pertcorr}
 \mathcal{A}(\eta)=\Am^{\text{GFF}}(\eta)+g_{L} \Am_L^{(1)}(\eta)+O(g_L^2)\,,
\end{equation}
where $L$ labels the maximum number of derivatives in the quartic interaction (i.e. the interaction term may involve a combination of terms with $\ell\leq L$ derivatives) and $g_L$ is the associated coupling. The Regge behavior of this correlator is determined by the term with the maximum number of derivatives and it reads \cite{Mazac:2018ycv}
\begin{equation}
 \Am_L^{(1)}(\chi)\sim \chi^{2\Delta+2L-1} \qquad \chi \to \frac12+ i \infty\,.
\end{equation}
According to our discussion in Section \ref{derivationsumrules}, the associated Mellin amplitude behaves as
\begin{equation}
 M^{(1)}_L(s)\sim |s|^{2L-1} \qquad |s|\to \infty\,,
\end{equation}
and we need to choose a function $F_p(s)$ which vanishes at infinity faster than $|s|^{-2L}$. \\
The strategy we implement is the following. We use equation \eqref{sumrulesgenericF} to write down non-perturbative sum rules with a specific function $F_p(s)$ which is chosen to decay sufficiently fast at $|s|\to \infty$ at a given value of $L$. We then expand the CFT data as
\begin{subequations}
\begin{align}\label{expansionDelta}
 \Delta_k&=2 \Delta_{\phi}+ 2n+g_L \gamma_{L,n}^{(1)}+O(g_L^2) \, ,\\
 c_{\Delta_k}&= c^{(0)}_n+ g_L c^{(1)}_{L,n} +O(g_L^2)\,,\label{expansionc}
\end{align}
\end{subequations}
and derive perturbative sum rules for $\gamma^{(1)}_{L,n}$ and $c^{(1)}_{L,n}$. We then check that these sum rules are satisfied by the $L=0,1$ results obtained in \cite{Mazac:2018ycv}, which read
\begin{subequations}
\begin{align}\label{andimL0}
	\gamma_{0,n}^{(1)} &= \frac{\left(\frac{1}{2}\right)_n \left((\Delta_{\phi} )_n\right){}^2 \left(2 \Delta_{\phi} -\frac{1}{2}\right)_n}{(1)_n \,(2 \Delta_{\phi} )_n \left(\left(\Delta_{\phi} +\frac{1}{2}\right)_n\right){}^2}\, ,\\ \label{andimL1}
	\gamma_{1,n}^{(1)}&=A_{\Delta_{\phi}}^{-1} \gamma_{0,n}^{(1)}  \frac{2 n (4 \Delta_{\phi} +2 n-1)}{(\Delta_{\phi} +n-1) (2 \Delta_{\phi} +2 n+1)}( 16 \Delta_{\phi} ^5-13 \Delta_{\phi} ^3-3 \Delta_{\phi} ^2+16 \Delta_{\phi}  n^4+8 n^4+64 \Delta_{\phi} ^2 n^3\nonumber \\
&+16 \Delta_{\phi}  n^3-8 n^3+96 \Delta_{\phi} ^3 n^2+8 \Delta_{\phi} ^2 n^2-24 \Delta_{\phi}  n^2-2 n^2+64 \Delta_{\phi} ^4 n-28 \Delta_{\phi} ^2 n-2 \Delta_{\phi}  n+2 n)\,,
\end{align}
\end{subequations}
where the result for $\gamma^{(1)}_{1,n}$ differs from \cite{Mazac:2018ycv} by an overall factor 
\begin{equation}
A_{\Delta_{\phi}} \equiv \frac{\Delta_{\phi}  (\Delta_{\phi} +1) (\Delta_{\phi} +2) (4 \Delta_{\phi}-1) (4 \Delta_{\phi} +1)^2 (4 \Delta_{\phi} +3)}{(2 \Delta_{\phi} +1)^2 (2 \Delta_{\phi} +3)}  \,  ,
\end{equation}
which we introduced to normalize the anomalous dimension as $\gamma_{1,1}^{(1)}=1$. Notice that $\gamma_{1,0}^{(1)}=0$. This is equivalent to a choice of basis for the set of independent interactions that can be built with up to one derivative. We discuss this issue in detail in Section \ref{sec:pert}. The OPE coefficients $c^{(1)}_n$ are given by the relation
\begin{equation}\label{OPEcoeff1}
 c^{(1)}_{L,n}=\partial_n(\gamma^{(1)}_{L,n} c^{(0)}_{n}) \, .
\end{equation}
In \cite{Bianchi:2021piu} we explicitly checked the case of $L=0$ and $L=1$.

\section{Perturbative Results} \label{sec:pert}

In this section, we consider deformations from generalized free field theory produced by effective interactions in a bulk AdS$_2$ field theory. In this holographic AdS$_2$/CFT$_1$ setup the background AdS$_2$ metric is not dynamical, corresponding to the absence of a stress tensor in the boundary CFT$_1$. According to the usual dictionary, a massive free scalar field $\Phi$ in AdS$_2$ is dual to a boundary $1d$ generalized free field $\phi$. We deform this theory by quartic self-interactions with an arbitrary number $L$ of derivatives 
\begin{equation}\label{phi4lagrangian}
S=\int dx dz\,\sqrt{g}\,\big[ \,g^{\mu\nu}\,\partial_\mu\Phi\,\partial_\nu\Phi+m^2_{\Delta_\phi}\Phi^2+g_L\,(\partial^L\Phi)^4\,\big]\,,\qquad L=0,1,\dots\,,
\end{equation}
where we use the AdS$_2$ metric in Poincar\'e coordinates $ds^ 2=\frac{1}{z^2}(dx^ 2+dz^ 2)$. The mass $m^2_{\Delta_\phi}=\Delta_\phi(\Delta_\phi-1)$ is fixed in units of the AdS radius so that $\Delta_\phi$ is the dimension, independent of $g_L$, of the field $\Phi$ evaluated at the boundary, $\phi(x)$\footnote{When we introduce an interaction, such as \eqref{phi4lagrangian}, there are Witten diagrams contributing to the mass renormalization of $\Phi$. We can always choose the bare mass in such a way that the dictionary is preserved and $\Delta_{\phi}$ is not modified.}.  We limit our analysis to leading order correlators and thus consider only contact diagrams, which we have already introduced in Section \ref{sec:holographiccfts} and whose building blocks are the $D$-functions~\cite{Liu:1998ty,DHoker:1999kzh, Dolan:2003hv} defined for the general case of AdS$_{d+1}$ as
\small
\begin{equation} \label{D-function}
\!\!\!\!\!\!\!\!
D_{\Delta_1 \Delta_2 \Delta_3 \Delta_4}\left(x^\prime_1, x^\prime_2, x^\prime_3, x^\prime_4\right)=\int \frac{d z d^d x}{z^{d+1}} \tilde{K}_{\Delta_1}\left(z, x ; x^\prime_1\right) \tilde{K}_{\Delta_2}\left(z, x ; x^\prime_2\right) \tilde{K}_{\Delta_3}\left(z, x ; x^\prime_3\right) \tilde{K}_{\Delta_4}\left(z, x ; x^\prime_4\right)\,,
\end{equation}
\normalsize
via the bulk-to-boundary propagator in $d$ dimensions
\begin{equation} \label{bulktoboundary}
K_{\Delta}(z,x;x') = {\cal C}_{\Delta} \Big[\frac{z}{z^2+(x-x')^2}\Big]^{\Delta} \equiv  {\cal C}_{\Delta}\,  \tilde{K}_{\Delta}(z,x;x')\,,
\,\qquad {\cal C}_{\Delta_{\phi}} =\frac{\Gamma\left(\Delta_{\phi}\right)}{2\,\sqrt{\pi}\,\Gamma\left(\Delta_{\phi}+{1\ov 2}\right)} \,.
\end{equation}
The writing $(\partial^L\Phi)^4$ in (\ref{phi4lagrangian}) is symbolic, denoting a complete and independent set of quartic vertices with four fields and up to $4L$ derivatives\footnote{The fact that a complete and independent basis of vertices is labeled by 1/4 the number of derivatives can be seen using integration by parts and the equations of motion, or noticing that the counting of physically distinct four-point interactions is equivalent to the counting of crossing-symmetric polynomial $S$-matrices in 2D Minkowski space, see discussion in~\cite{Mazac:2018ycv, Ferrero:2019luz}.}. In the following, we present a particularly convenient basis for these interactions, which allow us to derive a closed-form expression for the leading order correlator in Mellin space. Consider the interaction Lagrangian
\begin{equation}\label{interactionlag}
 \mathcal{L}_{L}=g_{L}\left[\prod_{j=0}^{L-1}\left(\tfrac{1}{2}\partial_\mu \partial^\mu-(\Delta_{\phi}+j)(2(\Delta_{\phi}+j)-1)\right)\Phi^2 \right]^2 \, .
\end{equation}
This looks like a very complicated term, but it contains four fields $\Phi$ and $4L$ derivatives, so by the argument above it must be effectively a linear combination of operators like $(\partial^{\ell}\Phi)^4$ for $\ell\leq L$. The advantage of this interaction is that the corresponding correlator computed via Witten diagrams reads\footnote{In \cite{Bianchi:2021piu} we showed how the interaction \eqref{interactionlag} leads to the correlator~\eqref{ansatzcorrelator} through explicit Witten diagrammatics. Moreover, for the cases $L=0,1,2$, we indicate how other interaction terms lead to results that can be rearranged as linear combinations of the eigenfunctions \eqref{ansatzcorrelator}.}:
\begin{equation}\label{ansatzcorrelator}
\Am^{(1)}_L(\chi) = \frac{4^{L-1}\pi^{-\frac32}\Gamma(2\Delta_{\phi}-\frac{1}{2}+2L)}{\Gamma(\Delta_{\phi}+\frac12)^4}  \chi^{2\Delta_\phi}\,(1+\chi^{2L}+(1-\chi)^{2L})\bar{D}_{\Delta_\phi+L, \Delta_\phi+L, \Delta_\phi+L, \Delta_\phi+L}(\chi) \, ,
\end{equation}
with the $\bar{D}$-functions \cite{Dolan:2003hv} defined as  ($\Sigma \equiv \frac{1}{2}\sum_i \Delta_i$) 
\begin{equation} \label{Dbar}
\!\!\!\!
D_{\Delta_{1}\Delta_{2}\Delta_{3}\Delta_{4}}= 
\frac{\pi^{d\ov 2}\Gamma\left(\Sigma-{d\ov2}\right)}{2\, \Gamma\left(\Delta_1\right)\Gamma\left(\Delta_2\right)\Gamma\left(\Delta_3\right)\Gamma\left(\Delta_4\right)}
\frac{\tau_{14}^{2(\Sigma-\Delta_1-\Delta_4)} \tau_{34}^{2(\Sigma-\Delta_3-\Delta_4)}}
{\tau_{13}^{2(\Sigma-\Delta_4)} \tau_{24}^{2\Delta_2}}\bar{D}_{\Delta_1\Delta_2\Delta_3\Delta_4}(u,v)\,,
\end{equation}
and depending only on the cross-ratios defined in (\ref{eq:crossratios4pt}).  Their explicit expression in terms of a Feynman 
parameter integral reads in the general case
\begin{equation}
	\bar{D}_{\Delta_1\Delta_2\Delta_3\Delta_4}(u,v)=
	\int d\alpha d\beta d\gamma\  \delta(\alpha+\beta+\gamma-1)\ 
	\alpha ^{\Delta _1-1} \beta ^{\Delta _2-1} \gamma ^{\Delta _3-1} 
	\frac{\Gamma \left(\Sigma-\Delta _4\right) \Gamma\left(\Delta_4\right)}
	{\big(\alpha  \gamma + \alpha  \beta\, u  + \beta  \gamma\, v\big)^{\Sigma-\Delta_4}}\,.
	\label{Dbar-integral}
\end{equation}
More details on the $D$-functions can be found in Appendix \ref{app:Dfunctions}.

If one then starts with some specific $4L$-derivative interaction, such as $(\partial^{L}\Phi)^4$, the explicit computation through Witten diagrams shows the appearance of several other combinations of $D$-functions with different weights. Nevertheless, by the argument above these results cannot be independent of those obtained using $\mathcal{L}_{L}$ and therefore the result must be expressible as a linear combination $\sum_\ell a_{\ell}\Am^{(1)}_\ell(\chi)$. This requires a series of non-trivial identities among $\bar D$ functions, some of which have been derived in \cite{Bianchi:2021piu}. 
Using \eqref{ansatzcorrelator} as a basis for $4L$-derivative results, we can take its Mellin transform. 

The first step is to compute the Mellin transform of the function $\bar{D}_{\Delta_\phi\Delta_\phi\Delta_\phi\Delta_\phi}(\eta)$.  In this section, we consider the reduced Mellin amplitude $\hat{M}(s)\equiv M(s) \Gamma(s)\Gamma(2\Delta_{\phi}-s)$ and we need to compute
\be\label{MellinDbar}
\hat{M}_{\Delta_\phi}(s) = \int_0^\infty d\eta \,\bar D_{\Delta_\phi\Delta_\phi\Delta_\phi\Delta_\phi}(\eta)  \, \Big(\frac{\eta}{1+\eta}\Big)^{2\Delta_\phi}\,\eta^{-1-s}   \, .
\ee
A closed-form expression for the $\bar D$ functions is not available and dealing with integral representations is quite hard. Therefore, we considered the case of integer $\Delta_{\phi}$, where simple explicit expressions for the $\bar D$ functions are known and we inferred the general form
\begin{subequations}
\begin{align}\label{MellinL0}
	\hat{M}_{\Delta_\phi}(s) &=\pi \csc(\pi s)\,\Big(  \pi \cot(\pi s)P_{\Delta_\phi}(s) -\sum_{j=1}^{2\Delta_\phi-1}\frac{P_{\Delta_\phi}(j)}{s-j} \Big)\, ,\\
		P_{\Delta_\phi}(s) &= 	2 \sum_{n=0}^{\Delta_{\phi}-1}   (-1)^{n} \frac{\Gamma(2 n + 1) \Gamma^4(\Delta_{\phi}) \Gamma(\Delta_{\phi}+n)}{\Gamma^4(n + 1)\Gamma(\Delta_{\phi}-n)\Gamma(2(\Delta_{\phi}+n))} (2\Delta_{\phi}-s)_n(s)_n\,.
\end{align}
\end{subequations}
The functions $P_{\Delta_\phi}(s)$ are effectively just polynomials of order $2\Delta_\phi-2$. Defining \be
Q_{\Delta_{\phi}} (s(s-2\Delta_{\phi}))\equiv P_{\Delta_{\phi}}(s) \, ,
\ee
we have, for the first few cases
\begin{align} \label{DfunctMellintable}
{\renewcommand\arraystretch{1.3} 
\begin{tabular}{ c | c }
$\Delta_{\phi}$ &  Q$_{\Delta_{\phi}} (x)$ \\
\hline 
1  &  2 		\\																										
2  & $\frac{1}{15} (5 + x) $ 	\\																					
3  & $\frac{1}{315} (84 + 17 \,x + x^2) $		\\											
4  & $\frac{1}{30030} (15444 + 2889 \,x + 206\, x^2 + 5\, x^3)$	\\
5  & $\frac{1}{765765} (1400256 + 239640\, x + 17387\, x^2 + 570\, x^3 + 7\, x^4)$	\\
\end{tabular}      }   
\end{align}

The functions $P_{\Delta_\phi}(s)$ can also be rewritten as
\begin{subequations}
\begin{align}\label{PDelta}
		P_{\Delta_\phi}(s) &= 	2 \frac{\Gamma(\Delta_\phi)^4}{\Gamma(2\Delta_\phi)}	{}_4{F}_3(\{\textstyle\frac{1}{2},s,1-\Delta_\phi,2\Delta_\phi-s\};\{1,1,\Delta_\phi+\frac{1}{2}\};1)\, .
\end{align}
\end{subequations}

We remark on the important fact that in cross-ratio space a closed-form expression for the $\bar D$ functions is not known, while in Mellin space it looks reasonably simple, at least for integer $\Delta_{\phi}$. This is similar to what happens in the higher-dimensional case, where this occurrence is even more striking as the reduced Mellin transform of the $\bar D$ functions is simply a product of $\Gamma$-functions (see Section \ref{sec:holographiccfts}). In the one-dimensional case, we could not find such a simple representation for the contact interactions, but the fact we could write the result in a closed form is already a notable improvement compared to cross-ratio space and as we will see, it allows us to successfully extract new CFT data. Furthermore, in Section \ref{sec:alternativeMellin} we present an alternative definition of the Mellin transform which leads to simpler results for the contact interaction.  

Knowing the Mellin transform for the $\bar D$ functions, it is simple to compute the Mellin transform of \eqref{ansatzcorrelator}
\begin{align}\label{ML}
	\hat{M}^{(1)}_{L}(s)&=\int_0^\infty dt \, \Am^{(1)}_L(\eta)\, \Big(\frac{\eta}{1+\eta}\Big)^{2\Delta_\phi}\,\eta^{-1-s}=  \,\sum_{j=0}^{2L}	c_{j,L} \, \hat M_{\Delta_\phi+L}(s+j)\, ,\\
2 c_{j,L} &= \frac{\Gamma(2L+1)}{\Gamma(j+1)\Gamma(2L-j+1)}+\delta_{j,0}+\delta_{j,2L}\, .
\end{align}
Notice that the presence of double poles for integer values of $s$ in~\eqref{MellinL0} is not in contradiction with the general single-pole structure of the non-perturbative Mellin amplitude~\eqref{Mstrip}-\eqref{eq:psi}, but just a consequence of the perturbative expansion of those single poles at this first order of perturbation theory. Moreover, the structure of~\eqref{MellinL0}  is such that both single and double poles cancel (as they should) within the region of convergence~\eqref{convergence} of the integral, which in this case ($\Delta_0=2\Delta_\phi$) is $0<\text{Re}(s)<2\Delta_\phi$.  

We stress that equation \eqref{ML} is a closed-form expression for the first-order perturbation around GFF generated by a quartic interaction with any number of derivatives. Under the assumption that the deformation from GFF described by these interactions only modifies two-particle data, one can extract these data.

\subsection{CFT Data}

In this section, we briefly sketch how to extract the anomalous dimension of two-particle operators, following the line of Section 5.2 of \cite{Bianchi:2021piu}.
Given the closed-form expression \eqref{ML} for the perturbative Mellin amplitude, we can use it to extract the CFT data
\begin{subequations} \label{perturbdeltaOPE}
\begin{align}\label{pertdelta}
\Delta&\equiv \D_{n,L}=2\D_\phi+2n+g_L\, \hat{\gamma}_{ L, n}^{(1)}(\Delta_{\phi})+\dots\, ,\\\label{pertOPE}
c_{\D}&\equiv c_{n,L}=c_n^{(0)}(\Delta_{\phi})+g_L\,c_{L,n}^{(1)}(\Delta_\phi)+\dots \,,
\end{align}
\end{subequations}
where $n\in \mathbb{N}$ and
\begin{equation}
c_n^{(0)}= \frac{2\Gamma(2n+2\Delta_\phi)^2\Gamma(2n+4\Delta_\phi-1)}{\Gamma(2\Delta_\phi)^2\Gamma(2n+1)\Gamma(4n+4\Delta_\phi-1)}\,.
\end{equation}
In~\eqref{pertdelta}-\eqref{pertOPE} we are assuming that the AdS interaction only modifies the CFT data of the two-particle operators exchanged in GFF. If we insert~\eqref{pertdelta}-\eqref{pertOPE} in the general Mellin OPE expansion~\eqref{Mstrip}, at first order in $g_L$  one obtains double and single poles at $s=2\Delta_{\phi}+p$,  $p\in \mathbb{N}$. One can then compare the corresponding residues with the ones in the tree-level Mellin amplitude~\eqref{ML}, which amounts to solving the equations formally written as
\begin{subequations}
\begin{align}\label{doublepoles}
 &\sum_{n}  c_n^{(0)}(\Delta_{\phi}) \hat{\gamma}_{ L, n}^{(1)}(\Delta_{\phi})\,{\textstyle{\frac{(-1)^p \Gamma(4\Delta_{\phi}+4n)\Gamma(2\Delta_{\phi}+p)^2}{\Gamma(2\Delta_{\phi}+2n)^2\Gamma(4\Delta_{\phi}+2n+p)\Gamma(p-2n+1)}}}=\lim_{s\rightarrow 2\Delta_{\phi}+p}(s-2\Delta_{\phi}-p)^2 \hat{M}^{(1)}_L(s) \,,
\\\label{singlepoles}
&\sum_{n}(c_{L,n}^{(1)}(\Delta_{\phi})\!+c_{n}^{(0)}(\Delta_{\phi}) \hat{\gamma}_{ L, n}^{(1)}(\Delta_{\phi})\partial_n)\,{\textstyle{\frac{(-1)^p \Gamma(4\Delta_{\phi}+4n)\Gamma(2\Delta_{\phi}+p)^2}{\Gamma(2\Delta_{\phi}+2n)^2\Gamma(4\Delta_{\phi}+2n+p)\Gamma(p-2n+1)}}} =\text{Res}_{s= 2\Delta_{\phi}+p} \hat{M}^{(1)}_L(s)\,,
\end{align}
\end{subequations}
for the leading-order corrections $\hat{\gamma}_{ L, n}^{(1)}(\Delta_{\phi})$ and $c_{L,n}^{(1)}(\Delta_{\phi})$.  Since the function $\hat{M}^{(1)}_L(s)$ is known explicitly, it is possible to write down a linear system for the anomalous dimensions $\hat{\gamma}_{ L, n}^{(1)}(\Delta_{\phi})$, which we find to be:
\begin{align}\label{greatresult}
 \hat{\gamma}_{ L, n}^{(1)}(\Delta_{\phi})&=\frac{\Gamma(L+\Delta_{\phi})^4}{\Gamma(2L+2\Delta_{\phi})}\sum_{p=0}^{2n} \sum_{j=2L-p}^{2L}\sum_{l=0}^{j+p-2L}(-1)^j c_{j,L} \times\\   &\frac{(4\Delta_{\phi}+2n-1)_p (-2n)_p (2L-j-p)_l (1-\Delta_{\phi}-L)_l (2\Delta_{\phi}+j+p)_l (\tfrac12)_l}{(l!)^3 (2\Delta_{\phi})_p (\tfrac12+\Delta+L)_l} \nonumber  \, .
\end{align}
In order to compare these results with those computed with bootstrap methods in~\cite{Mazac:2018ycv, Ferrero:2019luz} for $L=0,1,2,3$, we have to change the basis in the space of couplings. Since the bootstrap approach is blind to the specific values of the couplings $g_L$ in \eqref{perturbdeltaOPE}, one needs to establish a criterion to organize the set of independent data. The criterion that is used in~\cite{Mazac:2018ycv, Ferrero:2019luz} consists in setting
\begin{equation}\label{condvanandim}
 \gamma_{L,n}(\Delta_{\phi})= 0\,, \qquad n<L \, .
\end{equation}
In our approach, this is implemented by taking a linear combination
\begin{equation}\label{lincombandim}
\gamma_{L,n} =\sum_{\ell=0}^{L} a_{\ell} \hat\gamma_{\ell,n}
\end{equation}
and fixing the $L+1$ $a_\ell$ coefficients in~\eqref{lincombandim}, using the $L$ conditions \eqref{condvanandim} and the normalization $\gamma_{L,L}(\Delta_{\phi})=1$. Following this strategy in Section 5.2 of \cite{Bianchi:2021piu}, we reproduce the known results for $L\leq 3$ and present new results for $L\leq 8$.
We stress however that equation \eqref{greatresult} is valid for any $L$, so, up to the algorithmic procedure of fixing the $a_{\ell}$ coefficients, one can easily extract the result for any given $L$.\\
In particular, it turns out that expression \eqref{greatresult} can be rewritten as
\begin{align}\label{gammaLB2}
	\gamma_{ L, n}^{(1)}(\Delta_{\phi})&=\mathcal{G}_{L,n}(\Delta_{\phi}) \mathcal{P}_{L,n}(\Delta_{\phi}) \,,
\end{align}
where
\begin{align}\label{gammaLB3}
	\mathcal{G}_{L,n}(\Delta_{\phi})=\frac{4^{-L} \left(L+\frac{1}{2}\right)_{\Delta _{\phi }} \left(L+\Delta _{\phi }\right)_{\Delta _{\phi }} (-L+n+1)_{\Delta _{\phi }-1} \left(L+n+\Delta _{\phi }+\frac{1}{2}\right)_{\Delta _{\phi }-1}}{\Gamma \left(\Delta _{\phi }\right) \left(\Delta _{\phi }\right)_{3 L} \left(2 L+\Delta _{\phi }+\frac{1}{2}\right)_{\Delta _{\phi }-1} \left(L+2 \Delta _{\phi }-\frac{1}{2}\right)_{2 L} \left(n+\frac{1}{2}\right)_{\Delta _{\phi }} \left(n+\Delta _{\phi }\right)_{\Delta _{\phi }}} \,,
\end{align}
while $\mathcal{P}_{L,n}(\Delta_\phi)$ is a polynomial of degree $4L$ in $n$ and $5L$ in $\Delta_\phi$. The explicit polynomials for the first few values of $L$ are detailed in Appendix \ref{Ap: anomalous dimension} and, up to $L=3$, they perfectly agree with the result of \cite{Ferrero:2019luz}. The \textsc{Mathematica} notebook attached to \cite{Bianchi:2021piu} has values of $L$ ranging from $L=0$ to $L=8$ as well as a function~\texttt{FindBootstrapPolynomial[L,$\Delta$,n]} to compute $\mathcal{P}_{L,n}(\Delta_\phi)$ for arbitrary $L$.

\section{Alternative Formulation of the Mellin Transform}  \label{sec:alternativeMellin}
 
We conclude this chapter by pointing out that there is another noteworthy definition of Mellin transform,  identified by taking $a=-2\Delta_{\phi}+1$ in equation (\ref{Mellin1param}):
\be\label{Mellin2}
\mathcal{M}_{-2\Delta_{\phi}+1}(s)\equiv\tilde M_{\Delta_{\phi}}(s)=\int_0^\infty d\eta\, \Am(\eta)\,\left(\frac{\eta}{1+\eta}\right)^{-2\Delta_{\phi}+1}\,\eta^{-1-s} .
\ee 
The interesting feature of this particular choice is to provide the simplest representation for Mellin amplitudes of $\bar D$-functions.  We can motivate this claim by looking at the more general expression (\ref{Mellin1param}), where $a$ is a free parameter. We rewrite this in terms of the $\bar D$-functions using the identity in Table \ref{Table 2} between $\Am(\eta)$ and $\tilde{\Am}(\eta)$:
\begin{equation} \label{altMellina}
\begin{aligned}
\mathcal{M}_a(s) =\int_{0}^{\infty} d \eta \, \bar{D}_{\Delta_{\phi}\Delta_{\phi}\Delta_{\phi}\Delta_{\phi}}(\eta) \left(\frac{\eta}{1+\eta}\right)^{2\Delta_{\phi}}\left(\frac{\eta}{1+\eta}\right)^{a} \eta^{-1-s}
\end{aligned}.
\end{equation}
In particular, if we consider for example $\bar D_{1111}$, we obtain
\begin{equation}\label{resaltMellina}
\begin{aligned}
\mathcal{M}_a(s) =\tfrac{2 \Gamma(s-1) \Gamma(a+2\Delta_{\phi}-s)(\psi(s-1)-\psi(a+2\Delta_{\phi}-s))}{\Gamma(a+2\Delta_{\phi}-1)} +\tfrac{2 \Gamma(s-1) \Gamma(a+2\Delta_{\phi}-s-1)(\psi(a+2\Delta_{\phi}-2)-\psi(s-1))}{\Gamma(a+2\Delta_{\phi}-2)}
\end{aligned}.
\end{equation}
This expression simplifies for an integer value of the parameter $a$ below a certain threshold, namely $a \leq -2\Delta_{\phi}+1$. Considering the region of convergence of (\ref{altMellina}), we need 
\begin{equation} 
\begin{aligned}
\Delta_0 > \Delta_{\phi}-\frac{a}{2} ,
\end{aligned}
\end{equation}
and given that $\Delta_0=2\Delta_{\phi}$,  the only simple convergent integral has $a=2\Delta_{\phi}+1$,  yielding (\ref{Mellin2}) and the simple Mellin amplitude for $\bar{D}_{1111}$
\begin{equation} 
\begin{aligned}
\tilde M_{1}(s)= 2\, \Gamma(-s)\Gamma(s-1).
\end{aligned}
\end{equation}
See equations (\ref{MellinL0}) and (\ref{altMellinpat}) to appreciate the difference between the two representations of the $\bar{D}$-functions, obtained respectively with (\ref{MellinDbar}) and (\ref{Mellin2}).\\
Apart from this property,  (\ref{Mellin2}) satisfies \eqref{crossingmellin}, which reads
\be\label{crossing2gen}
\tilde{M}(s)=\tilde{M}(1-s)\,.
\ee
Using the properties listed in Table \ref{Table 1}, we can then derive the strip of convergence of this Mellin definition
\be\label{convergence2gen}
2\Delta_\phi-\Delta_0<\text{Re}(s)<1+\Delta_0-2\Delta_\phi\,,
\ee
which translates, perturbatively ($\Delta_0=2\Delta_\phi$), in 
\be\label{convergence2}
0<\text{Re}(s)<1~.
\ee
The inverse Mellin transform reads 
\be
\Am(\eta)=\int_{c-i\infty}^{c+i\infty}\frac{ds}{2\pi i}\,\tilde{M}(s)\left(\frac{\eta}{1+\eta}\right)^{2\Delta_{\phi}-1}\,\eta^s\,,
\ee
where the range of the real constant $c$ is the same of $\text{Re}(s)$ in \eqref{convergence2gen}, and therefore in perturbation theory the contour of the integral in the complex $s$-plane is any straight line within the interval \eqref{convergence2}. \\
We can finally report a general structure for the Mellin transform of the $\bar{D}$-functions
\be \label{altMellinpat}
\tilde M_{\Delta_\phi}(s) = P_{\Delta_{\phi}} (s)\, \Gamma(-s-2\Delta_{\phi}+2) \Gamma(s-2\Delta_{\phi}+1)\,,
\ee
where 
\begin{equation}
\resizebox{.97\hsize}{!}{$P_{\Delta_{\phi}} (s) = 2\, \Gamma(2\Delta_{\phi}-1) {}_{4}F_{3} \left(\{-s-2\Delta_{\phi}+2,s-2\Delta_{\phi}+1,1-\Delta_{\phi},1-\Delta_{\phi}\};\{2-2\Delta_{\phi},2-2\Delta_{\phi},2-2\Delta_{\phi}\};1\right)$}.
\end{equation}
Note that $P_{\Delta_{\phi}}$ is a polynomial for integer $\Delta_{\phi}$, that we now tabulate for the first few cases, using a more convenient rewriting, $Q_{\Delta_{\phi}} (s(s-1))= P_{\Delta_{\phi}}(s)$:
\begin{align} \label{DfunctMellintable2}
{\renewcommand\arraystretch{1.3} 
\begin{tabular}{ c | c }
$\Delta_{\phi}$ & Q$_{\Delta_{\phi}} (x)$ \\
\hline 
1  &  2 		\\																										
2  & $2\,(2+x)	$ 	\\																									
3  & $32 \,(24 + 22 \,x + x^2)$		\\											
4  & $2592\,(720 + 876  \,x + 100 \, x^2 + x^3)$	\\
5  & $663552 \,(40320 + 58416  \,x + 10508 \, x^2 + 300 \, x^3 + x^4)$	\\
\end{tabular}      }   
\end{align}
We report then an alternative closed-form expression valid for integer values of $\Delta_{\phi}$
\be \label{MellinD}
\tilde{M}_{\Delta_\phi}(s)=\! \sum^{\Delta_\phi-1}_{n=0}\frac{2\,(-1)^n\,\Gamma(\Delta_\phi)^2\Gamma(2\Delta_\phi-1-n)^3}{\Gamma(n+1)\Gamma(\Delta_\phi-n)^2}  
\Gamma(-s-2\Delta_{\phi}+2+n)\, \Gamma(s-2\Delta_\phi+1+n)\,,
\ee
which is a linear combination of squared $\Gamma$-functions. \\
Despite this nice representation of the $\bar{D}$-functions, the correspondence between the poles and the physical exchanged operators is more obscure, in contrast with (\ref{leftpoles}) and (\ref{rightpoles}) for the Mellin transform (\ref{Mellin}).  We, therefore, reckoned that the Mellin transform defined in (\ref{Mellin}) is the most suitable for the applications we presented in the previous sections, which have as a main goal the extraction of CFT data.

\subsection{Extension to Non-Identical Operators}

We conclude this chapter with some unpublished results. In particular, we extend this Mellin representation to $\bar{D}$-functions of non-identical operators.

For $\bar{D}$-functions of the form $\bar{D}_{\Delta_1\Delta_2\Delta_1\Delta_2}$ which are invariant under crossing symmetry ($1\leftrightarrow3,2\leftrightarrow4$) we can easily extend (\ref{MellinD}) to
\begin{equation}
\begin{split}
\tilde{M}_{\Delta_1 \Delta_2\Delta_1 \Delta_2}=2 \sum^{\Delta_1-1}_{n=0}&\frac{(-1)^n\Gamma(\Delta_1)\Gamma(\Delta_2)}{\Gamma(\Delta_1-n)\Gamma(\Delta_2-n)\Gamma(n+1)}  \Gamma(\Delta_1+\Delta_2-1-n)^3\\
&\Gamma(s-(\Delta_1+\Delta_2-1-n))\Gamma(1-s-(\Delta_1+\Delta_2-1-n)) \,.
\end{split}
\end{equation}
There are however other types of crossing-invariant $\bar{D}$-functions, e.g.  $\bar{D}_{2123}$ and $\bar{D}_{2321}$. These cases are more complex as for the first type we get a Mellin of the form that we encountered in (\ref{MellinL0}), while in the second case, something similar to the non-crossing-invariant case, which we are exploring in the following, happens. 

Generalizing this Mellin representation to generic $\bar{D}$-functions is still out of reach but we manage to write down a closed-form expression for non-crossing-invariant functions obtained by setting $\Delta_1 = \Delta_4, \Delta_2 =\Delta_3$ and $\Delta_1=\Delta_2, \Delta_3 =\Delta_4$,  which are connected by the relation
\begin{equation} \label{4.13}
\bar{D}_{\Delta_1 \Delta_2 \Delta_2 \Delta_1}\left(\frac{\eta}{1+\eta}\right)= \bar{D}_{\Delta_2 \Delta_2 \Delta_1 \Delta_1}\left(\frac{1}{1+\eta}\right).
\end{equation}
However, it is not immediate to get their Mellin representation since the integral does not converge.  Therefore we first need to regularize it by isolating the ``singular'' terms.
We start by looking at the series expansion near the extremes of the integral. We consider e.g.  $\bar{D}_{1221}$ and $\bar{D}_{2211}$:
\begin{equation}
\begin{split}
&\bar{D}_{1221}(\eta) \overset{\eta\rightarrow0}{\sim} \frac{1}{2} - \text{log}\left(\frac{\eta}{\eta+1}\right) \,,\\
&\bar{D}_{1221}(\eta) \overset{\eta\rightarrow\infty}{\sim} (1+\eta)^2 +(1+\eta)+\frac{1}{18} \left(13 + 6\,\text{log}\left(\frac{1}{1+\eta}\right)\right)\,,\\
&\bar{D}_{2211}(\eta) \overset{\eta\rightarrow0}{\sim} \left(\frac{1+\eta}{\eta}\right)^2 +\left(\frac{1+\eta}{\eta}\right)+\frac{1}{18} \left(13 + 6\,\text{log}\left(\frac{\eta}{\eta+1}\right)\right)\,,\\
&\bar{D}_{2211}(\eta) \overset{\eta\rightarrow\infty}{\sim} \frac{1}{2} - \text{log}\left(\frac{1}{1+\eta}\right)\,.
\end{split}
\end{equation}

We see that the behavior of $\bar{D}_{1221}(0)$ is related by crossing to the behavior of $\bar{D}_{2211}(\infty)$ and vice versa, as expected. Moreover, the terms
\be
(1+\eta)^2 +(1+\eta)\,, \qquad \qquad \left(\frac{1+\eta}{\eta}\right)^2 +\left(\frac{1+\eta}{\eta}\right)
\ee
are precisely the ``singular'' terms of $\bar{D}_{1221}$ and $\bar{D}_{2211}$ respectively.  Of course, we cannot compute the Mellin transform of these terms because it diverges. We, therefore, use the technique introduced in \cite{Penedones:2019tng} and described in Section \ref{sec:convergenceandsubtractions}, which is based on the deformation of the integration contour. We then split the integral
\begin{equation}\label{14.6}
\tilde{M}(s)=\int^{1}_{0} d\eta\, \Am(\eta)\frac{\eta^{s-1}}{1+\eta}\,, \qquad \qquad \tilde{M}(s)=\int_{1}^{\infty} d\eta\, \Am(\eta)\frac{\eta^{s-1}}{1+\eta} \,,
\end{equation}
such that these two integrals converge separately and give us precisely the Mellin transform of the divergent terms we are looking for. We can then find a formula for these singular contributions in cross-ratios and Mellin space terms:
\begin{subequations}
\begin{align}
& ST(\eta)= 2 \frac{\Gamma(\Delta_1)^3 \Gamma(\Delta_3)}{\Gamma(\Delta_1+\Delta_3)}\, (1+\eta)\left(\Delta_3(1+\eta)-\Delta_1\,\eta\right)\,, \\ 
& ST(s) =2\,\frac{\Gamma(\Delta_1)^3\, \Gamma(\Delta_3)}{\Gamma(\Delta_1+\Delta_3)}\,\frac{(\Delta_1-s\Delta_1-3\Delta_3+2s\Delta_3)}{(2-3s+s^2)}\,. 
\end{align}
\end{subequations}
Thus we can add $ST(\eta)$ to the $\Am(\eta)$ that we want to Mellin transform to obtain a convergent Mellin amplitude. Vice versa we can add $ST(s)$ to $\tilde{M}(s)$ to reproduce the correct $\Am(\eta)$.
Recall that this analysis is only restricted to the non-crossing-invariant $\bar{D}$-functions of the form considered in (\ref{4.13}).

Now that we know how to obtain a convergent Mellin transform, let's find the closed-form expression for this type of non-crossing-invariant $\bar{D}$-functions.
Below we list some of the Mellin we compute using (\ref{Mellin2}):
\begin{center}
{\renewcommand\arraystretch{1.3} 
\begin{tabular}{  c | c | c }
$\Delta_1\Delta_2\Delta_3\Delta_4$ & Polynomial part & $\Gamma$s \\
\hline 
  1\,2\,2\,1   &  2 (1 - 3 s)																												& $\Gamma(s - 3) \Gamma(-s - 1)$\\
 2\,2\,1\,1  &  2 (-2 + 3 s) 																											& $\Gamma(s - 2) \Gamma(-s - 2)$\\
 2\,3\,3\,2 &  $8 (24 + s (-55 + (12 - 5 s) s)) 	$ 																						& $\Gamma(s - 5) \Gamma(-s - 3)$\\
 3\,3\,2\,2 &  $8 (-24 + s (46 + (-3 + 5 s) s))$																					& $\Gamma(s - 4) \Gamma(-s - 4)$\\
 4\,3\,3\,4 &  $288 (1080 + s (-2454 + s (1121 + s (-491 + s (31 - 7 s)))))$													& $\Gamma(
  s - 7) \Gamma(-s - 5)$\\
 4\,4\,3\,3 &  $288 (-720 + s (1596 + s (-236 + s (437 + s (-4 + 7 s)))))$												& $\Gamma(
   s - 6) \Gamma(-s - 6)$\\
\end{tabular}}
\end{center}

Finding a closed form becomes more complicated. However, we can always use the property $s\,\Gamma(s)=\Gamma(s+1)$ to change the $\Gamma$ part of the Mellin and consequently the polynomial in front.  It turns out that, if we play a bit, we find a nice representation of this $\bar{D}$-functions. To illustrate it, let us take a specific example:
\be
\begin{split}
&\tilde{M}_{1221}(s)= -16 \Gamma(s - 3) \Gamma(-s - 1) - 6 \Gamma(s - 2) \Gamma(-s - 1)\,,\\
&\tilde{M}_{2211}(s) = -16 \Gamma(s - 2) \Gamma(-s - 2) - 6 \Gamma(s - 2) \Gamma(-s - 1)
\end{split}
\ee
for $\bar{D}_{1221}$ and $\bar{D}_{2211}$.

These Mellin representations consist of two terms: the second ones are identical and crossing invariant, while the first ones display the crossing-relation that connects the respective $\bar{D}$-functions:
\begin{equation}
16 \Gamma(s - 3) \Gamma(-s - 1) \overset{s\rightarrow 1-s} {=} 16 \Gamma(s - 2) \Gamma(-s - 2) \,.
\end{equation}
We can then rewrite all the other $\bar{D}$-functions of the type (\ref{4.13}) with this criterion, which makes finding a pattern relatively easy:
\be
\begin{split}
\tilde{M}_{\Delta_1 \Delta_2 \Delta_3 \Delta_4} &= \sum^{\Delta_1-\Delta_3}_{m=0} \sum^{\Delta_3-1}_{n=0}
  (-1)^{(\Delta_1 - \Delta_3) - n} 2^{1 + (\Delta_1 - \Delta_3) - m} \frac{1}{\Gamma(n+1)}\\
 &\frac{\Gamma(\Delta_3)\Gamma(\Delta_4)}{\Gamma(\Delta_3-n)\Gamma(\Delta_4-n )} 
  \frac{ \Gamma(\Delta_1+\Delta_2+\Delta_3+\Delta_4-2-2n)} {\Gamma(\Delta_1+\Delta_2+\Delta_3+
  \Delta_4-2-2n-m)} \\
&  \Gamma(\Delta_3 +\Delta_4 -1 -n) \Gamma(\Delta_1 + \Delta_2 - 1 - n - m)^2\\
  &\Gamma(s-(\Delta_2+\Delta_3-1-n))\Gamma(-s-(\Delta_1+\Delta_2-2-n-m)) .
\end{split}
\ee

\chapter{Summary and Outlook} \label{sec:conclusion}

This thesis is an exploration of the realm of 1$d$ CFTs, which we believe are and are going to be of extreme importance in the developments of the various non-perturbative techniques central in higher-dimensional CFTs.  Our focus has been particularly on the Wilson line defect CFT. We wandered particularly in the world of multipoint correlators of operators inserted on this defect, but we also investigated alternative ways of representing correlators through the Mellin formalism.  A lot of exciting directions stream from this work.  In this chapter, some of these prospects are discussed after a summary of the main results.

\section{Main Results}

In this section, we dedicate our attention to a summary of the main results obtained in this work. Particularly,  we opened Part II with a detailed discussion of an efficient algorithm for computing multipoint correlation functions of scalar fields inserted on the $\Nm=4$ SYM Wilson line defect.  This algorithm consists of recursion relations that encode the possible interactions between protected and unprotected operators of elementary scalars up to next-to-leading order at weak coupling. 

The outcome of the application of these recursive formulae has been shown in Chapters~\ref{sec:23point},\ref{ch:4point}, and \ref{ch:higherpoint}.  In Chapter~\ref{sec:23point}, we computed two-point functions of fundamental scalars, which are necessary in order to normalize higher-point correlators properly. Moreover, we calculated two-point functions of composite operators of length two. This was an important consistency check for the recursion relations, but it also allowed us to access the anomalous dimensions of these operators.  Lastly, we explored three-point functions. These are relevant because they give us access to some OPE coefficients, whose knowledge is fundamental to performing other consistency checks using the expansion of the correlators in terms of conformal blocks.

In Chapter \ref{ch:4point}, we moved then to four-point functions.  Our first focus was on the simplest four-point function of scalar operators of protected dimension $\Delta=1$ $\vev{\phi_1 \phi_1 \phi_1 \phi_1}$, which we computed up to next-to-next-to-leading order thanks to the constraining power of the Ward identities.  We also computed another four-point function made of unprotected operators only. Then, we extracted the bosonic CFT data from both these correlators.

Chapter \ref{ch:higherpoint} finally focused on the investigation of multipoint correlators.  The recursion relations allowed us to gather a huge pool of perturbative results. In particular, in this thesis, we focused on five- and six-point functions, which we analyze in detail, also by extracting some bosonic CFT data.  All these new perturbative results are interesting in their own right. However, they remarkably gave us access to non-perturbative constraints, which we conjectured to be an extension of the Ward identities satisfied by four-point functions. Multipoint Ward identities constituted a fundamental piece of the puzzle in the bootstrap of a multipoint correlator at strong coupling, which has been scrutinized at the end of this chapter.

Finally, in Chapter \ref{ch:Mellinres}, we explored another tool to represent and study 1$d$ correlators. We developed an inherently one-dimensional Mellin transform, which can be defined at the non-perturbative level with appropriate subtractions and analytical continuations.  This definition allowed us to derive an infinite set of non-perturbative sum rules whose characteristics have been discussed in detail and applications sketched.  The efficiency of this 1$d$ Mellin formalism is manifest at the perturbative level.  We found a closed-form expression for the Mellin transform of $\bar{D}$-functions, which has been used to extract CFT data.

\section{The Way Forward} \label{sec:outlook}

The research presented in this thesis is situated at the intersection of two ambitious programs: the study of conformal defects and analytic bootstrap.  The numerous connections between CFTs and physical phenomena grant these fields wide applications \cite{Hartman:2022zik}.  In the following, however, we sketch only the future directions streaming directly from the results we have just summarized. Some possible outlooks have already been mentioned along the way.

\subsubsection{More General Recursion Relations?}

A first natural question is whether extending the recursion relations beyond scalar fields, considering then fields transforming non-trivially under transverse rotations, is possible.  An example would be fermionic fields.  At first and quick thought, there do not seem to be conceptual walls preventing this extension. Of course, one needs to be careful with the ordering of the operators, which plays a non-trivial role. It is also possible that the inclusion of these extra fields could be technically challenging. Nevertheless, we have already approached a computation of a fermionic loop diagram computing the next-to-next-to-leading order of $\vev{\phi_1 \phi_1 \phi_1 \phi_1}$ in Section \ref{subsec:BulkDiagrams}. There, the integral associated with this diagram turned out to be solvable analytically in the conformal frame and using the star-triangle relation (\ref{eq:startriangle}). Therefore, we are hopeful that setting up a recursion relation, including fermionic fields, is possible.

An alternative idea for developing a recursive algorithm including both scalars and fermions would be to promote the scalar fields to superfields.  These would require an upgrade of the Feynman rules we introduced in Section \ref{sec:N=4Feynmanrules} to Feynman rules in terms of superfields, and of course, the same would have to be done for the defect Feynman rules outlined in Section \ref{sec:insertionrulesdefect}. It might be helpful then to consider the superfield formulation of the supersymmetric Wilson line presented in \cite{Beisert:2015jxa}.

\subsubsection{Push the Recursion Relations to the Next Order?}

A second natural research direction could be to push the recursion relations to NNLO.  Particularly, we mostly refer to the recursive algorithm for protected scalars since the inclusion of the unprotected scalar $\phi^6$ already greatly increases the complexity of the recursion.  In Section \ref{sec:recursionrelations}, we discussed that the input for the recursion relation at leading order (\ref{eq:recursiontreediagrams}) is the two-point function, while for the next-to-leading order (\ref{eq:recursionloop}) the starting value is the four-point function at NLO.  In the case of the NNLO, we would have to input the six-point function $\vev{\phi_1 \phi_1 \phi_1 \phi_1 \phi_1 \phi_1}$ at that order.  

In Section \ref{sec:2loop}, we showed how to compute a four-point function at NNLO. The computations of the associated Feynman diagrams proved doable only thanks to the constraining power of the Ward identity,  reducing the unknowns from three independent functions to a single one.  Nevertheless, we had to integrate numerically the integrals representing the boundary diagrams since we have not find a way to solve them analytically yet. A similar challenge would for sure arise when computing the building blocks of the six-point functions, which are what we really need in order to set up the NNLO recursive formula.

\subsubsection{Other Four-Point Functions at NNLO?}

After having computed the simplest four-point function of protected operators at NNLO in Section \ref{sec:2loop}, it seems natural to compute other four-point functions. In particular, one may start by generalizing the dimension of the last two operators, therefore computing $\vev{\phi_1 \phi_1 \phi_k \phi_k}$.  Having operators of higher length means, in principle, an increase in the complexity of the Feynman diagrams we have to compute. However, we are only looking at the simplest $R$-symmetry channel where, in practice, we do not expect to obtain new types of diagrams. Of course, in any case, a careful analysis is required. In Figure \ref{fig:diagrams11kk}, we draw some examples of diagrams that will appear in the computation.

\begin{figure}[h]
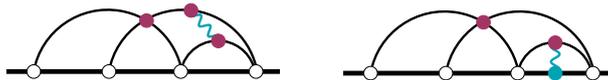

\centering
\XHookk \qquad \XYookk
\caption{\textit{Example of NNLO order diagrams of $\vev{\phi_1 \phi_1 \phi_2 \phi_2}$. It should be relatively straightforward to compute these diagrams from the calculations already done for $\vev{\phi_1 \phi_1 \phi_1 \phi_1}$. Moreover, it should be quite easy to generalize to $\vev{\phi_1 \phi_1 \phi_k \phi_k}$. }}
\label{fig:diagrams11kk}
\end{figure}

\subsubsection{Proof of the Conjectured Ward Identities?}

Of course, it would be very interesting to prove the multipoint Ward identities we conjectured with the usual superspace techniques, where a careful analysis of possible nilpotent invariants is also necessary. Some relevant techniques were developed recently in \cite{Chicherin:2015bza} for $\mathcal{N}=4$ without defects, and they could possibly be adapted to a one-dimensional setup.  This analysis would be fundamental also to retrieving the full set of superconformal Ward identities. As Section \ref{subsec:conjectureWI} points out,  our analysis of protected operators only focuses on the highest-weight component,  ignoring possible fermionic descendants.  While this is not a problem for the four-point function, since the full superconformal correlator can be reconstructed from the highest weight, for higher-point functions the Ward identities should be a collection of partial PDEs relating bosonic and fermionic components. Therefore it is unexpected to find this set of superconformal Ward identities only acting on the highest weight.  A superspace analysis could shed some light on why this happens and indicate if this is a feature we expect only in 1$d$ theories or if it is also valid in higher $d$.

\subsubsection{More Ward identities?} \label{sec:morewardidentities}

Multipoint superconformal Ward identities are, in general, an underexplored subject.  In principle, one could try to repeat the strategy presented here in more general setups, starting perhaps with the parent theory of the Wilson line defect CFT: $4d$ $\Nm=4$ SYM. For this theory, a recursion relation has already been worked out for $n$-point functions of half-BPS operators at next-to-leading order \cite{Drukker:2008pi}. Therefore we would already have some samples on which to base our conjecture. However, in the $4d$ case, there are more spacetime cross-ratios, $n(n-3)/2$ in particular, the same number as the $R$-symmetry cross-ratios. This would possibly lead to a slightly more complicated ansatz to test. In any case, with the current results, it should be possible to derive the WI.

It is fascinating to think that these conjectured Ward identities could also open the door to other setups. Particularly it was noted in \cite{Liendo:2016ymz} that the superconformal Ward identity for four-point functions, that we introduced in Section \ref{sec:wardidentity4pt}, can be analytically continued between two-point functions of bulk half-BPS operators in the presence of a half-BPS line and four-point functions of defect half-BPS operators along the same line. An analogous observation can also be made for the superconformal blocks appearing in the OPE associated with this setup. Similarly, the Ward identities for four-point functions also capture the superconformal symmetry of two-point functions of half-BPS operators in the presence of a half-BPS \textit{boundary}, as well as four-point functions of half-BPS operators on the boundary, which can be interpreted as a $3d$ $\Nm=4$ gauge theory. Moving from one setup to the other is possible with the appropriate analytic continuations. Therefore, we wonder if the same can be done with the Ward identities. We leave this intriguing question to further exploration.

\subsubsection{Bootstrap of a Six-Point Function?}

A clear follow-up of the bootstrap of the five-point function discussed in Section \ref{sec:5ptbootstrap} is the bootstrap of the six-point function $\vev{\phi_1 \phi_1 \phi_1 \phi_1 \phi_1 \phi_1}$. This correlator presents \textit{fifteen} independent $R$-symmetry channels. Of course, we expect the Ward identity to reduce this number, but we wonder if they would be constraining enough to succeed in determining the full correlator.  Possibly the fact that this correlator is more symmetric compared to the five-point will provide sufficiently many additional constraints.

\subsubsection{Transfer to Other Setups?}

The analysis developed in this thesis revolving around a recursive algorithm to compute correlators could also be applied to other setups.  One example could be the $\epsilon-$expansion applied to a localized magnetic field line defect, for which a perturbative analysis was initiated recently \cite{Gimenez-Grau:2022czc}. There has been some exciting development also on scalar-fermion models, such as the Gross-Neveu-Yukawa model, which admits natural 1$d$ defects given by the exponential of a scalar field integrated along a straight line \cite{Barrat:2023ivo}. Finally, models such as the fermionic Wilson line defect in ABJM \cite{Drukker:2009hy} would be an interesting target both from a perturbative side and a bootstrap perspective.

\subsubsection{A Bridge Between Multipoint Correlators and the Mellin Formalism?}

It would be ideal to establish a direct contact between the work on multipoint correlators and the inherently one-dimensional Mellin transform for four-point functions. Since Mellin space offers an alternative, sometimes simpler, formulation for correlation functions, it would be interesting to generalize this 1$d$ Mellin representation to higher-point correlators. The first step in this direction would be a generalization to non-identical scalars. Some attempts have been presented in Section \ref{sec:alternativeMellin}. Finding this representation for higher-point could allow us to get a closed-form expression for them and hopefully discover some hidden features of these functions.  

Moreover, given the progress in understanding the analytic structure in cross-ratio space~\cite{Ferrero:2019luz, Ferrero:2021bsb, Alday:2021odx}, it would be interesting to map this knowledge into Mellin space and see whether, as in the higher-dimensional counterpart, implementing the bootstrap directly in Mellin space leads to significant simplifications.

\subsubsection{What About the Flat-Space Limit?}

It would be worth analyzing the connections of our Mellin formalism with the ambitious program of the $S$-matrix bootstrap~\cite{Paulos:2016fap} by carefully studying a proper flat space limit. In this context, it is interesting to notice that, unlike higher-dimensional examples, accumulation points of poles do not seem to appear in this picture. It is, therefore, important to understand how the singularities of $2d$ $S$-matrices appear in the flat space limit of our Mellin amplitude.

\subsubsection{Can We Expose Integrability?}

To conclude,  the $1d$ CFT defined by the supersymmetric Wilson line in $\mathcal{N}=4$ SYM is also the most important application motivating our work on a 1$d$ Mellin formalism.  In this case, the holographic dual is described through an effective string worldsheet theory in AdS$_2$.  Little is known about integrability in curved spaces, and how the power of integrability can be exploited in this setting is still an open question. This could be done by identifying, in Mellin space,  the analog of the $S$-matrix factorization, the latter being one of the highlights of integrability in flat space.

\appendix

\part{Appendices}

\chapter{Four-Dimensional Superconformal Algebra} \label{app:4dsuperconformalalgebra}

This appendix is dedicated to the four-dimensional superconformal algebra $\mathfrak{su}(2,2|\mathcal{N})$, where for physical theories $\mathcal{N}=1, \ldots, 4$. We have already introduced in detail the superconformal algebra in Section \ref{sec:superconformalalgebra}. We give here the rich list of (anti)commutation relations, which we did not include in the main text.

We do not use vector indices, since it is more natural to write $P, K$ and $L$ with spinor indices, such that the conformal algebra reads
\begin{equation}
\begin{aligned}
& {\left[D, P_{\alpha \dot{\alpha}}\right]=P_{\alpha \dot{\alpha}}} \,,\\
& {\left[D, K^{\dot{\alpha} \alpha}\right]=-K^{\dot{\alpha} \alpha}}\,, \\
& {\left[K^{\dot{\alpha} \alpha}, P_{\beta \dot{\beta}}\right]=\delta^{\dot{\alpha}}{ }_{\dot{\beta}} L_\beta {}^\alpha+\delta^\alpha {}_\beta \bar{L}_{\dot{\beta}} {}^{\dot{\alpha}}+\delta_\beta{ }^\alpha \delta^{\dot{\alpha}}{ }_{\dot{\beta}} D} \,, \\
& {\left[L_\alpha {}^\beta, L_\gamma {}^\delta\right]=\delta_\gamma {}^\beta L_\alpha {}^\delta-\delta_\alpha {}^\delta L_\gamma {}^\beta} \,, \\
& {\left[\bar{L}^{\dot{\alpha}} {}_{\dot{\beta}}, \bar{L}^{\dot{\gamma}} {}_{\dot{\delta}} \right]=-\delta^{\dot{\gamma}} {}_{\dot{\beta}} \bar{L}^{\dot{\alpha}} {}_{\dot{\delta}} +\delta^{\dot{\alpha}} {}_{\dot{\delta}} \overline{L}^{\dot{\gamma}}}{}_{\dot{\beta}} \,, \\
& {\left[L_\alpha {}^\beta, P_{\gamma \dot{\gamma}}\right]=\delta_\gamma {}^\beta P_{\alpha \dot{\gamma}}-\frac{1}{2} \delta_\alpha {}^\beta P_{\gamma \dot{\gamma}}}\,,  \\
& {\left[\bar{L}^{\dot{\alpha}} {}_{\dot{\beta}}, P_{\gamma \dot{\gamma}}\right]=\delta^{\dot{\alpha}} {}_{\dot{\gamma}} P_{\gamma \dot{\beta}}-\frac{1}{2} \delta^{\dot{\alpha}} {}_{\dot{\beta}} P_{\gamma \dot{\gamma}}} \,, \\
& {\left[L_\alpha {}^\beta, K^{\dot{\gamma} \gamma}\right]=-\delta_\alpha {}^\gamma K^{\dot{\gamma} \beta}+\frac{1}{2} \delta_\alpha {}^\beta K^{\dot{\gamma} \gamma}} \,, \\
& {\left[\bar{L}^{\dot{\alpha}} {}_{\dot{\beta}}, K^{\dot{\gamma} \gamma}\right]=-\delta^{\dot{\gamma}} {}_{\dot{\beta}} K^{\dot{\alpha} \gamma}+\frac{1}{2} \delta^{\dot{\alpha}} {}_{\dot{\beta}} K^{\dot{\gamma} \gamma} \,.}
\end{aligned}
\end{equation}

The bosonic subgroup also includes an $SU(\mathcal{N})$ $R$-symmetry
\begin{equation}
\left[R^i {}_j, R^k {}_l\right]=\delta^k {}_j R^i {}_l-\delta^i {}_l R^k {}_j\,,
\end{equation}
and the $U(1)_R$ charge $R$. There are $4 \mathcal{N}$ Poincaré supercharges $Q_\alpha^A$ and $\bar{Q}_{B \alpha}$, with the indices running over $A,B=1, \ldots, \mathcal{N}$ and $\alpha, \dot{\alpha}=1,2$. 

Under the bosonic subalgebra, these Poincaré and superconformal supercharges transform as
\begin{equation}
\begin{aligned}
& {\left[D, S_i^\alpha\right]=-\frac{1}{2} S_i^\alpha}\,, \\
& {\left[D, \bar{S}^{i \dot{\alpha}}\right]=-\frac{1}{2} \bar{S}^{i \dot{\alpha}}}\,, \\
& {\left[R, S_i^\alpha\right]=+S_i^\alpha}\,, \\
& {\left[R, \bar{S}^{i \dot{\alpha}}\right]=-S^{i \alpha}}\,, \\
& {\left[P_{\alpha \alpha}, S_i^\beta\right]=-\delta_\alpha^\beta \bar{Q}_{i \alpha}}\,, \\
& {\left[P_{\alpha \alpha}, \bar{S}^{i \beta}\right]=-\delta_\alpha^\beta Q_\alpha^i}\,, \\
& {\left[L_\alpha^\beta, S_i^\gamma\right]=-\delta_\alpha^\gamma S_i^\beta+\frac{1}{2} \delta_\alpha^\beta S_i^\gamma}\,, \\
& {\left[\vec{L}_\beta^\alpha, \vec{S}^{i \hat{\gamma}}\right]=-\delta_\beta^\psi \bar{S}^{i \dot{\alpha}}+\frac{1}{2} \delta^{\dot{\alpha}}{ }_\beta \bar{S}^{i \dot{\gamma}}}\,, \\
& {\left[R^i, Q_a^k\right]=\delta^k{ }_j Q_\alpha^i-\frac{1}{N} \delta^i{ }_j Q_a^k}\,, \\
& {\left[R_j^i, S_k^\alpha\right]=-\delta_k^i S_\alpha^j+\frac{1}{\mathcal{N}} \delta_j^i S_k^\alpha}\,, \\
& {\left[R_j^i, \bar{Q}_{k \dot{\alpha}}\right]=-\delta_k^i \bar{Q}_{j \dot{\alpha}}+\frac{1}{N} \delta_j^i \bar{Q}_{k \dot{\alpha}}}\,, \\
\end{aligned}
\qquad \quad 
\begin{aligned}
& {\left[D, S_i^\alpha\right]=-\frac{1}{2} S_i^\alpha} \,,\\
& {\left[D, \bar{S}^{i \dot{\alpha}}\right]=-\frac{1}{2} \bar{S}^{i \dot{\alpha}}} \,,\\
& {\left[R, S_i^\alpha\right]=+S_i^\alpha}\,, \\
& {\left[R, \bar{S}^{i \dot{\alpha}}\right]=-\bar{S}^{i \dot{\alpha}}}\,, \\
& {\left[P_{\alpha \dot{\alpha}}, S_i^\beta\right]=-\delta^\beta{ }_\alpha \bar{Q}_{i \dot{\alpha}}}\,, \\
& {\left[P_{\alpha \dot{\alpha}}, \bar{S}^{i \dot{\beta}}\right]=-\delta^{\dot{\beta}}{ }_{\dot{\alpha}} Q_\alpha^i}\,, \\
& {\left[L_\alpha{ }^\beta, S_i^\gamma\right]=-\delta_\alpha{ }^\gamma S_i^\beta+\frac{1}{2} \delta_\alpha{ }^\beta S_i^\gamma}\,, \\
& {\left[\bar{L}^{\dot{\alpha}}{ }_{\dot{\beta}}, \bar{S}^{i \dot{\gamma}}\right]=-\delta^{\dot{\gamma}}{ }_{\dot{\beta}} \bar{S}^{i \dot{\alpha}}+\frac{1}{2} \delta^{\dot{\alpha}}{ }_{\dot{\beta}} \bar{S}^{i \dot{\gamma}}}\,, \\
& {\left[R^i{ }_j, S_k^\alpha\right]=-\delta^i{ }_k S_\alpha^j+\frac{1}{\mathcal{N}} \delta^i{ }_j S_k^\alpha}\,, \\
& {\left[R^i{ }_j, \bar{S}^{k \dot{\alpha}}\right]=\delta^k{ }_j \bar{S}^{i \dot{\alpha}}-\frac{1}{\mathcal{N}} \delta^i{ }_j \bar{S}^{k \dot{\alpha}} \,.}
\end{aligned}
\end{equation}
Finally, the supercharges anticommute as follows:
\begin{equation}
\begin{aligned}
& \left\{Q_\alpha^A, \bar{Q}_{B \dot{\alpha}}\right\}=\delta^A{ }_B P_{\alpha \dot{\alpha}}, \\
& \left\{S_A^\alpha, \bar{S}^{B \dot{\alpha}}\right\}=\delta^B{ }_A K^{\dot{\alpha} \alpha}\,, \\
& \left\{Q_\alpha^A, S_B^\beta\right\}=\delta_B {}^A \delta_\alpha{ }^\beta\left(\frac{D}{2}+R \frac{4-\mathcal{N}}{4 \mathcal{N}}\right)+\delta^A{ }_B L_\alpha{ }^\beta-\delta_\alpha{ }^\beta R^A {}_B, \\
& \left\{\bar{S}^{A \dot{\alpha}}, \bar{Q}_{B \dot{\beta}}\right\}=\delta^A {}_B \delta^{\dot{\alpha}}{ }_{\dot{\beta}}\left(\frac{D}{2}-R \frac{4-\mathcal{N}}{4 \mathcal{N}}\right)+\delta^A {}_B \bar{L}^{\dot{\alpha}}{ }_{\dot{\beta}}+\delta^{\dot{\alpha}} { }_{\dot{\beta}} R^A {}_B\,.
\end{aligned}
\end{equation}
For $\mathcal{N}=4$, the $U(1)_R$ charge can be quotiented out from the algebra, leading to the $\mathfrak{psu}(2,2 \mid 4)$ superalgebra. Under Hermitian conjugation in radial quantization, the generators transform as
\begin{equation}
\begin{aligned}
& D^{\dagger}=D\,, \quad P_{\alpha \dot{\alpha}}^{\dagger}=K^{\dot{\alpha} \alpha}\,, \quad\left(R^A {}_B\right)^{\dagger}=R^B {}_A\,, \quad R^{\dagger}=R\,, \\
& \left(L_\alpha{}^\beta\right)^{\dagger}=L_\beta{}^\alpha\,, \quad\left(\bar{L}^{\dot{\alpha}} {}_{\dot{\beta}}\right)^{\dagger}=\bar{L}^{\dot{\beta}}{}_{\dot{\alpha}}\,, \quad\left(Q_\alpha^A\right)^{\dagger}=S_A^\alpha\,, \quad\left(\bar{Q}_{\dot{\alpha} A}\right)^{\dagger}=\bar{S}^{\dot{\alpha} A} \,. \\
\end{aligned}
\end{equation}

\chapter{Conventions and $\mathfrak{u}(N)$ Identities} \label{sec:convidentities}

In this appendix, we fix some conventions and state some identities that are useful to compute Feynman diagrams in $\mathcal{N}=4$ SYM and consequently in the Wilson line defect CFT.

In particular, we recall that we consider $SU(N)$ as the gauge group of the Yang-Mills theory, while $SO(6)_R$ is the $R$-symmetry group.  However, there is little difference between $SU(N)$ and $U(N)$ in the large $N$ limit we are interested in. In any case, the \textit{unitary group} is defined as the group of unitary complex $N \times N$ matrices:
\begin{equation}
U^\dag U=\mathbb{1}\,,
\end{equation}
and it has dimension $\operatorname{dim} U(N)=N^2$. The \textit{special unitary group} $SU(N)$ is defined by the matrices $U \in U(N)$ that also satisfy the requirement det $U=+1$. This implies that the $SU(N)$ group has one degree of freedom less than the $U(N)$ group, and thus its dimension is $\operatorname{dim} S U(N)=N^2-1$

Both $U(N)$ and $S U(N)$ are connected to the identity and hence are Lie groups. Their elements can be expressed infinitesimally as:
\begin{equation}
U=\mathbb{1}+i \alpha_a T^a+\mathcal{O}(\alpha^2)\,,
\end{equation}
with $T^a$ the generators of the groups and $\alpha$ an infinitesimal real parameter. As a consequence, the generators must satisfy:
\begin{equation}
T^a \stackrel{!}{=}\left(T^a\right)^{\dagger}\,.
\end{equation}
$U(N)$ and $S U(N)$ can be formulated such that they share $N^2-1$ traceless generators, i.e. $U(N)$ has an extra generator, which we define to be $T^0$ and which is not traceless. 

We can now have a look at the Lie algebras $\mathfrak{u}(N)$ and $\mathfrak{su}(N)$. We normalize the generators such that:
\begin{equation} \label{eq:tracegen}
\tr T^a T^b=\frac{\delta^{a b}}{2}\,,
\end{equation}
while the commutator takes the form
\begin{equation}
\left[T^a, T^b\right]=i\,f^{ab}{}_c\, T^c\,.
\end{equation}

We now collect a bunch of identities for both traces and structure constants. We start with trace identities. First, it is useful to derive the following \textit{completeness relation}:
\begin{equation}
T_{i j}^a T_{a, l k}=\frac{1}{2} \delta_{i k} \delta_{j l}\,,
\end{equation}
which allows us to obtain the two following useful identities:
\begin{subequations}
\begin{align}
\tr T^a A\, \tr T_a B&=T_{i j}^a T_{a, l k} A^{j i} B^{k l}=\frac{1}{2} \tr A B\,, \label{eq:idone}\\
\tr T^a A\, T_a B&=T_{i j}^a T_{a, l k} A^{j l} B^{i k}=\frac{1}{2} \tr A\, \tr B\,. \label{eq:idtwo}
\end{align}
\end{subequations}
It is straightforward to find the expression for the trace of one generator. The trace of $T^a$ is non-zero only for $a=0$, and we have:
\begin{equation}
\left(\tr T^a\right)^2=\tr T^a \mathbb{1}\, \tr T_a \mathbb{1}=\frac{N}{2} .
\end{equation}
Thus we find:
\begin{equation}
\tr T^a=\sqrt{\frac{N}{2}} \delta^{a 0} .
\end{equation}
Note that, as mentioned before, the generators of the $\mathfrak{su}(N)$ algebra are traceless and hence are zero for all $a$.

The trace of two generators has been defined by the normalization condition \eqref{eq:tracegen}. When the indices are contracted, we have then the following equality:
\begin{equation}
\tr T^a T_a=\frac{N^2}{2} \,.
\end{equation}
For the trace of four generators, we have the following:
\begin{equation}
T^a T^b T^c T^d=\frac{1}{4}\left(\left[T^a, T^b\right]+\left\{T^a ,T^b\right\}\right)\left(\left[T^c, T^d\right]+\left\{T^c ,T^d\right\}\right) \,.
\end{equation}

Moving now to structure constants, we start by noting that:
\begin{equation}
\tr\left[T^a, T^b\right]=\tr T^a T^b-\tr T^b T^a=0 \stackrel{!}{=} i f^{a b}{ }_c \tr T^c \propto f^{a b 0}\,,
\end{equation}
and thus:
\begin{equation}
f^{a b 0}=0 \,.
\end{equation}
The product of two structure constants with two indices free out of six can be obtained as follows:
\begin{equation}
\begin{aligned}
f^{a c d} f^b{}_{c d} & =-4 \tr([T^a, T^c] T^d) \tr([T^b, T_c] T_d) \\
& =-2 \tr([T^a, T^c][T^b, T_c]) \\
& =-2 \tr T^a \tr T^b+2 N \tr T^a T^b\,,
\end{aligned}
\end{equation}
where in the second line we have used \eqref{eq:idone}, and in the third line equation \eqref{eq:idtwo}. The remaining traces give:
\begin{equation}
f^{a c d} f_{c d}^b=N\left(\delta^{a b}-\delta^{a 0} \delta^{b 0}\right)=N \delta^{\tilde{a} \tilde{b}}\,,
\end{equation}
where $\tilde{a}, \tilde{b} \equiv 1, \ldots, N$ are $\mathfrak{s u}(N)$ indices (one generator less). By abuse of notation, we often drop the tilde in this thesis. This is harmless in the large $N$ limit.
From the previous result, we immediately obtain the case in which all indices are contracted:
\begin{equation}
f^{a b c} f_{a b c}=N\left(N^2-1\right) \sim N^3\,,
\end{equation}
where the last equality holds in the large $N$ limit.

We conclude with an important identity which relates the generators and the structure constants:
\begin{equation}
f^{abc}=-2 i \,\tr([T^a, T^b]T^c)\,.
\end{equation}

\chapter{Correlators on the Wilson Line}

\section{Integrals and Regularization} \label{sec:integralsapp}

In this appendix, we compute the \textit{bulk} and \textit{boundary} integrals we encounter in this work. We add interesting limits and identities of these integrals whenever they are known or we find ones. 

\subsection{Bulk Integrals} \label{subsec:bulkintegrals}

First, we focus on the bulk integrals defined in Section \ref{sec:N=4Feynmanrules}, starting from the $Y$-integrals.

The $Y$-integral can be easily obtained from this expression by taking the following limit:
\begin{align} \label{eq:Y123}
Y_{123} &= \lim_{x_4 \to \infty} (2\pi)^2 x_4^2\ X_{1234} \notag \\
&= \frac{I_{12}}{8\pi^2} \left( \frac{\tau_{12}}{\tau_{23}} \log |\tau_{13}| + \frac{\tau_{12}}{\tau_{31}} \log |\tau_{23}| + \frac{\tau_{12}^2}{\tau_{23}\tau_{31}} \log |\tau_{12}| \right)\,.
\end{align}

In $1d$ the $X$-integral is given by
\begin{equation} \label{eq:X1234}
\frac{X_{1234}}{I_{13} I_{24}} = - \frac{1}{8 \pi^2} \frac{\ell(\chi, 1)}{\chi(1-\chi)}\,,  \qquad\qquad \chi^2 := \frac{\tau_{12}^2 \tau_{34}^2}{\tau_{13}^2 \tau_{24}^2}\,,
\end{equation}
with $\ell(\chi_1,\chi_2)$ defined in \eqref{eq:ell}.

The $H$-integral seems to have no known closed form so far, but $F_{12,34}$ can, fortunately, be reduced to a sum of $Y$- and $X$-integrals in the following way \cite{Beisert:2002bb}:
\begin{align} \label{eq:FXYidentity}
F_{12,34} &= \frac{X_{1234}}{I_{13}I_{24}} - \frac{X_{1234}}{I_{14}I_{23}} + \left( \frac{1}{I_{14}} - \frac{1}{I_{24}} \right) Y_{124} + \left( \frac{1}{I_{23}} - \frac{1}{I_{24}} \right) Y_{234} \notag \\
& \qquad \qquad \qquad \qquad \qquad + \left( \frac{1}{I_{23}} - \frac{1}{I_{13}} \right) Y_{123} + \left( \frac{1}{I_{14}} - \frac{1}{I_{13}} \right) Y_{134}\,.
\end{align}

The integrals above appear in their respective pinching limits, i.e. when two external points are brought close to each other. The integrals simplify greatly in this limit, but they exhibit a logarithmic divergence that is tamed using point-splitting regularization. For the $Y$-integral, we define
\begin{equation*}
Y_{122} := \lim_{x_3 \to x_2} Y_{123}\,, \qquad \qquad \qquad \lim_{x_3 \to x_2} I_{23} := \frac{1}{(2\pi)^2 \epsilon^2}\,.
\end{equation*}
Inserting this in (\ref{eq:Y123}) and expanding up to order $O (\log \epsilon^2)$, we obtain
\begin{equation}
\diagYonetwotwo\ := Y_{112} = Y_{122} = - \frac{I_{12}}{16 \pi^2} \left( \log \frac{I_{12}}{I_{11}} - 2 \right)\,.
\label{eq:Y112}
\end{equation}
This result coincides with the expression in e.g. \cite{Drukker:2008pi}.

For completion, we also give the pinching limit of the $X$- and $F$-integrals. The first one reads
\begin{equation}
\diagXoneonetwothree\ := X_{1123} = - \frac{I_{12} I_{13}}{16 \pi^2} \left( \log \frac{I_{12} I_{13}}{I_{11} I_{23}} - 2 \right)\,,
\label{eq:X1123}
\end{equation}
which is again the same as in \cite{Drukker:2008pi}.

The pinching limit $\tau_2 \to \tau_1$ of the $F$-integral gives
\begin{align}
F_{13,14} &= F_{14,13} = - F_{13,41} \notag \\
& = -\frac{X_{1134}}{I_{13} I_{14}} + \frac{Y_{113}}{I_{13}} + \frac{Y_{114}}{I_{14}} + \left(\frac{1}{I_{13}} + \frac{1}{I_{14}} - \frac{2}{I_{34}} \right) Y_{134}\,.
\label{eq:F1314}
\end{align}

Moving to the integrals appearing for the first time at NNLO,  the $K_{ij}$ has been defined in (\ref{eq:Kij}), and in particular, the integrals $K_{12}$ and $K_{23}$  can be calculated analytically by taking the $\tau_4 \to \infty$ limit of the well-known two-loop \textit{kite} integral \cite{Usyukina:1992wz}:
\begin{align}
K_{13,24} :&= \frac{1}{I_{13}} \int d^4 x_5\, I_{15} I_{25} I_{35}\, X_{1345} \notag \\
&= \frac{I_{13} I_{24}}{128 \pi^4 \chi (1-\chi)}
\bigl(
5 ( H_{0,0,1} + H_{0,1,0} + 2 H_{0,1,1} ) \notag \\
&\phantom{=\ }+ 2 ( H_{1,0,0} + 2 H_{1,1,0} )
+ 7 H_{1,0,1} + 6 H_{1,1,1}
\bigr)\,,
\label{eq:KiteIntegralDefinition}
\end{align}
whose evaluation we give compactly in terms of HPL's. This integral can be represented diagrammatically as
\begin{equation*}
\KiteIntegral\ \,.
\end{equation*}
In this case, the conformal limit can be considered as cutting one leg from the diagram. Indeed for $K_{12}$ and $K_{23}$ we have
\begin{equation}
K_{ij} =\ \Kij\ \,,
\end{equation}
with $k \neq i,j, 4$.  We can also easily compute the integral $K_{i4}$ in the limit $\tau_4 \to \infty$,  which can be reduced to $Y$-integrals
\begin{equation} \label{eq:Ki4}
K_{i4} =\ \KiFour\ = \frac{I_{24}}{32 \pi^2} \left( \log \frac{I_{12} I_{13}}{I_{34}^2} + 4 \right) Y_{123} + O\left(\frac{1}{\tau_4^3}\right)\,,
\end{equation}
where $i,j,k \neq 4$.
Finally a special case of $K_{ij}$ is $K_{44}$ which reads
\begin{align} \label{eq:K44}
K_{44} &:= \int d^4x_5\, Y_{445}\, I_{15} I_{25} I_{35}  \notag \\
& = - \frac{I_{24}}{16\pi^2} \left( \log \veps^2 - \log \tau_4^2 + 2 \right) Y_{123} + O\left(\frac{1}{\tau_4^3}\right)\,.
\end{align}

If we restrict to the full conformal frame,i.e. $(\tau_1, \tau_2, \tau_3, \tau_4) \to (0,\chi,1,\infty)$, we can also find interesting identities relating the some $K$-integrals to some $H$-integrals: 
\begin{equation} \label{eq:KHidentity}
\frac{K_{12}}{I_{12}} = H_{13,23}\,, \quad \frac{K_{23}}{I_{23}} = H_{12,23}\,,
\end{equation}
which, however, is only valid in the conformal frame.

We can derive similar identities also for the $A$-integrals and the $Y$-integrals, if, again, we restrict to the conformal frame:
\begin{equation} \label{eq:AkIdentity}
A_1 + A_2 + A_3 = ( \log x_{12}^2 + \log x_{23}^2 ) Y_{123}\,.
\end{equation}

In the conformal frame, there is also a beautiful identity relating the $H$-integrals together:
\begin{equation} \label{eq:Hidentity}
\chi H_{12,13} + (1-\chi) H_{13,23} - \chi(1-\chi) H_{12,23} = \frac{3\, \zeta_3}{512 \pi^6}\,.
\end{equation}

We finally collect some identities that allow us to solve the sixteen-dimensional integral of the spider diagram: first, some identities between $\Gamma$-functions
\begin{equation} \label{eq:TwoLoopsTraceOfGammas}
\tr \Gamma^i \gamma_\mu
\Gamma^j \gamma_\nu
\Gamma^k \gamma_\rho
\Gamma^l \gamma_\sigma =
(\delta^{ij} \delta^{kl}
+ \delta^{il} \delta^{jk}
- \delta^{ik} \delta^{jl})
\tr \gamma_\mu \gamma_\nu \gamma_\rho \gamma_\sigma\,,
\end{equation}
and recalling that the $\gamma$-matrices are sixteen-dimensional, we have
\begin{equation} \label{eq:TwoLoopsTraceOfGammasb}
\tr \sx_1 \sx_2 \sx_3 \sx_4 = 16\, [ (x_1 \cdot x_2)(x_3 \cdot x_4) + (x_1 \cdot x_4)(x_2 \cdot x_3) - (x_1 \cdot x_3)(x_2 \cdot x_4) ]\,.
\end{equation}

We can now write down the fermionic star-triangle identity 
\begin{equation} \label{eq:startriangle}
\int d^d x_4 \spd_4 I_{14} I_{24} \spd_4 I_{34}=-\pi^2 \sx_{12} \sx_{23} I_{12} I_{13} I_{23}\,,
\end{equation}
and, to conclude, a useful algebraic identity
\begin{equation} \label{eq:TwoLoopsAlgebraicIdentity}
x_{12} \cdot x_{34} = - \frac{1}{8\pi^2} \left( \frac{1}{I_{13}} +  \frac{1}{I_{24}} -  \frac{1}{I_{23}} -  \frac{1}{I_{14}} \right)\,.
\end{equation}

\subsection{Boundary Integrals} \label{subsec:boundaryintegrals}

In the computations, we deal with two types of boundary integrals: the $T$- and $U$-integrals. In the following two sections, we gather their explicit expressions and interesting properties.

\subsubsection{$T$-Integrals} \label{subsubsec:Tintegrals}
 
Before entering into details about the various integrals, we should define a function that keeps track of the change of sign in all the boundary diagrams, characterized by a gluon coupling to the Wilson line, e.g in Section \ref{subsec:BoundaryDiagrams}:
\begin{equation} \label{eq:epsilon}
\veps (\tau_i\, \tau_j\, \tau_k)=\text{sign}(\tau_i-\tau_j)\text{sign}(\tau_j-\tau_k)\text{sign}(\tau_i-\tau_k)\,,
\end{equation}
with $\tau_i<\tau_j<\tau_k$.

Moving to the $T$-integrals, the following identity can be used in order to ``swap'' the limits of integration:
\begin{equation} \label{eq:T3}
\left. T_{jk;il} \right|_{i<j<k<l} = -\frac{I_{jk}}{12} - T_{jk;li}\,,
\end{equation}
where the integration range $(li)$ on the right-hand side has to be understood as the union of segments $(l,+\infty) \cup (-\infty,i)$.

There also exists another relevant combination for the computations at one loop relating the $T$- and $Y$-integrals:
\begin{equation} \label{eq:T5}
I_{ik} T_{jk;ki} + I_{jk} T_{ik;jk} = - \frac{I_{ik} I_{jk}}{12} +I_{ik} I_{jk}\left(\frac{1}{I_{ik}} + \frac{1}{I_{jk}}-\frac{2}{I_{ij}} \right)Y_{ijk}\,.
\end{equation}

In general, the integrals can be performed explicitly for the different possible orderings of the $\tau$'s, and here we give the results assuming $\tau_1 < \tau_2 < \tau_3 < \tau_4$:
\begin{subequations} \label{eq:Tint} 
\begin{gather}
T_{12;34} = \frac{1}{32 \pi^4 \tau_{12}^2} \left( 4 L_R \left( \frac{\tau_{12}}{\tau_{14}} \right) - 4 L_R \left( \frac{\tau_{12}}{\tau_{13}} \right) - C_{123} + C_{124} \right)\,, \\
T_{34;12} = \frac{1}{32 \pi^4 \tau_{34}^2} \left( 4 L_R \left( \frac{\tau_{34}}{\tau_{14}} \right) - 4 L_R \left( \frac{\tau_{34}}{\tau_{24}} \right) - C_{341} + C_{342} \right)\,, \\
T_{14;23} = \frac{1}{32 \pi^4 \tau_{14}^2} \left( 4 L_R \left( \frac{\tau_{24}}{\tau_{14}} \right) - 4 L_R \left( \frac{\tau_{34}}{\tau_{14}} \right) - C_{412} - C_{143} \right)\,, \\
T_{23;41} = \frac{1}{32 \pi^4 \tau_{23}^2} \left( -4 L_R \left( \frac{\tau_{23}}{\tau_{13}} \right) - 4 L_R \left( \frac{\tau_{23}}{\tau_{24}} \right) - C_{234} - C_{123} \right)\,,
\end{gather}
\end{subequations}
where we have defined the following help function:
\begin{equation}
C_{ijk} := - 32 \pi^4 \tau_{ij} (\tau_{ik} + \tau_{jk}) Y_{ijk}\,,
\end{equation}
and where the Rogers dilogarithm $L_R(x)$ is defined in \eqref{eq:Rogers}.

It is easy to take pinching limits of the integrals given above. For example, we can have
\begin{equation}
T_{12;23} = \frac{1}{32 \pi^4 \tau_{12}^2} \left( 4 L_R \left( \frac{\tau_{12}}{\tau_{13}} \right) - \frac{2 \pi^2}{3} + C_{123} \right) + Y_{112}\,,
\end{equation}
using the fact that $L_R(1) = \frac{\pi^2}{6}$. All the other pinching limits can be performed in the same way.

\subsubsection{$U$-Integrals} \label{subsubsec:Uintegrals}

The integrals $U_{a;ij}$ introduced in Section \ref{sec:mainU} can be explicitly performed and here is the expression: 
\begin{equation} \label{eq:defU}
U_{a;ij}= \frac{1}{4\pi^2} \left( \frac{1}{\tau_i - \tau_a} +\frac{1}{\tau_a -\tau_j} \right)\,,
\end{equation}
which is valid both when $\tau_a < \tau_i <\tau_j$ and $\tau_i < \tau_j <\tau_a$. Though,  variations of this $U$-integral appear in the recursion, for example, when $\tau_a = \tau_i$. In these cases, we just take the appropriate limit, regularizing the divergences in the integral with point-splitting regularization
as for the integrals of the previous sections. 
If $i,j=\pm\infty$, we again take the limit of the expression above. 

We move now to the second type of $U$-integrals: $U^{(2)}_{ab;ij}$.  The explicit expressions for the three different configurations are
\begin{subequations} \label{eq:Utwoexpl}
\begin{align}
 \Utwoint  =\frac{1}{16 \pi^4 } \Bigg( \frac{\tau_{ij}} {\tau_{ai}\tau_{aj}\tau_{jb}} +& \frac{1}{\tau_{ab}^2\tau_{ai}\tau_{aj}} \left( (\tau_a^2+ \tau_i \tau_j)  \log \frac{\tau_{ai}\tau_{bj}}{\tau_{bi}\tau_{aj}} \right.  \notag \\
&  \left. +\tau_a(\tau_i +\tau_j) \log \frac{\tau_{bi}\tau_{aj}}{\tau_{ai}\tau_{bj}} + \tau_{ba} \tau_{ij} \right) \Bigg)  \,,\\
\Utwointtwo= \frac{1}{16 \pi^4 } \Bigg( \frac{\tau_{ji}} {\tau_{ai}\tau_{ib}\tau_{aj}} +& \frac{1}{\tau_{ab}^2\tau_{ai}\tau_{aj}} \left( (\tau_a^2+ \tau_i \tau_j) \log \frac{\tau_{bi}\tau_{aj}}{\tau_{ai}\tau_{bj}}  \right.  \notag \\
& \left. +\tau_a(\tau_i +\tau_j) \log \frac{\tau_{ai}\tau_{bj}}{\tau_{bi}\tau_{aj}} + \tau_{ab} \tau_{ij} \right) \Bigg) \,,\\
\Utwointthree  = \frac{1}{16 \pi^4 } \Bigg( \frac{\tau_{ij}} {\tau_{bi}\tau_{bj}\tau_{ja}} +& \frac{1}{\tau_{ab}^2\tau_{bi}\tau_{bj}} \left((\tau_b^2+ \tau_i \tau_j)  \log \frac{\tau_{bi}\tau_{aj}}{\tau_{ai}\tau_{bj}} \right.  \notag \\
&  \left. +\tau_b(\tau_i +\tau_j)\log \frac{\tau_{ai}\tau_{bj}}{\tau_{bi}\tau_{aj}} + \tau_{ab} \tau_{ij} \right) \Bigg) \,, 
\end{align}
\end{subequations}
with $\tau_{ij}:= \tau_i -\tau_j$ and assuming $\tau_1 < \tau_2 < \tau_3 < \tau_4$.

\section{NNLO Recursion Relation for Even $\phi^6$} \label{sec:recursionapp}
\begingroup
\allowdisplaybreaks

In this appendix, we give the formal expression for the recursion relation given in equation \eqref{eq:recursionevenNLO} in a diagrammatic way. A close look at \eqref{eq:recursionevenNLO} reveals that there are \textit{two types} of $U$-integrals that one can encounter. These two types are represented in Figure \ref{fig:Udiagrams}, and correspond to whether the integration limits are ``connected'' or not. This distinction is made clear in the definitions of the integrals, which can be found in the appendix Section~\ref{subsubsec:Uintegrals}. Equation \eqref{eq:recursionevenNLO} contains both types of integrals, and in order to write an effectively usable formula, we must extract the $U^{(2)}$ contributions. It is easy to check visually which terms contain a $U^{(2)}$:
\begin{align*}
\recursionphione\, \overset{m=l+2}{\supset}\,  \recursionphioneb\, \\
\recursionphithree\, \overset{m=l+2}{\supset}\, \recursionphithreeb\, \\
\recursionphisix\ \overset{m=l+2}{\supset}\, \recursionphisixb\,.
\end{align*}

Hence we only have to change the summation range for these three terms and add them by hand with an explicit mention of $U^{(2)}$. This gives the following expression:
\begin{align}
\eqref{eq:recursionevenNLO} =& \frac{\lambda^2}{4} \sum_{j=1}^{n-1} \sum_{k=j+1}^n \delta^{I_j 6} \delta^{I_k 6} \notag \\
& \times \Biggl( \sum_{l=k}^{n-2} \sum_{m=l+4}^n U_{j;m(m+1)} U_{k;l(l+1)} A_\text{\tiny{LO}}^{I_1 \ldots I_{j-1}} A_\text{\tiny{LO}}^{I_{j+1} \ldots I_{k-1}} A_\text{\tiny{LO}}^{I_{k+1} \ldots I_l} A_\text{\tiny{LO}}^{I_{l+1} \ldots I_m} A_\text{\tiny{LO}}^{I_{m+1} \ldots I_n} \notag \\
& \phantom{\Biggl(} + \sum_{l=k}^n \sum_{m=j}^{k-1} U_{j;l(l+1)} U_{k;m(m+1)} A_\text{\tiny{LO}}^{I_1 \ldots I_{j-1}} A_\text{\tiny{LO}}^{I_{j+1} \ldots I_{m}} A_\text{\tiny{LO}}^{I_{m+1} \ldots k_l} A_\text{\tiny{LO}}^{I_{k+1} \ldots I_l} A_\text{\tiny{LO}}^{I_{l+1} \ldots I_n} \notag \\
& \phantom{\Biggl(} + \sum_{l=0}^{j-3} \sum_{m=l+4}^{j-1} U_{j;m(m+1)} U_{k;l(l+1)} A_\text{\tiny{LO}}^{I_1 \ldots I_{l}} A_\text{\tiny{LO}}^{I_{l+1} \ldots I_{m}} A_\text{\tiny{LO}}^{I_{m+1} \ldots j_l} A_\text{\tiny{LO}}^{I_{j+1} \ldots I_{k-1}} A_\text{\tiny{LO}}^{I_{k+1} \ldots I_n} \notag \\
& \phantom{\Biggl(} + \sum_{l=j}^{k-1} \sum_{m=0}^{j-1} U_{j;l(l+1)} U_{k;m(m+1)} A_\text{\tiny{LO}}^{I_1 \ldots I_{m}} A_\text{\tiny{LO}}^{I_{m+1} \ldots I_{j-1}} A_\text{\tiny{LO}}^{I_{j+1} \ldots I_l} A_\text{\tiny{LO}}^{I_{l+1} \ldots I_{k-1}} A_\text{\tiny{LO}}^{I_{k+1} \ldots I_n} \notag \\
& \phantom{\Biggl(} + \sum_{l=j}^{k-1} \sum_{m=k}^n U_{j;l(l+1)} U_{k;m(m+1)} A_\text{\tiny{LO}}^{I_1 \ldots I_{j-1}} A_\text{\tiny{LO}}^{I_{j+1} \ldots I_{l}} A_\text{\tiny{LO}}^{I_{l+1} \ldots k_l} A_\text{\tiny{LO}}^{I_{k+1} \ldots I_m} A_\text{\tiny{LO}}^{I_{m+1} \ldots I_n} \notag \\
& \phantom{\Biggl(}  + \sum_{l=j}^{k-3} \sum_{m=l+4}^{k-1} U_{j;l(l+1)} U_{k;m(m+1)} A_\text{\tiny{LO}}^{I_1 \ldots I_{j-1}} A_\text{\tiny{LO}}^{I_{j+1} \ldots I_{l}} A_\text{\tiny{LO}}^{I_{l+1} \ldots m} A_\text{\tiny{LO}}^{I_{m+1} \ldots I_{k-1}} A_\text{\tiny{LO}}^{I_{k+1} \ldots I_n} \notag \\
& \phantom{\Biggl(} + \sum_{l=0}^{j-1} \sum_{m=k}^n U_{j;l(l+1)} U_{k;m(m+1)} A_\text{\tiny{LO}}^{I_1 \ldots I_{l}} A_\text{\tiny{LO}}^{I_{l+1} \ldots I_{j-1}} A_\text{\tiny{LO}}^{I_{j+1} \ldots k_l} A_\text{\tiny{LO}}^{I_{k+1} \ldots I_m} A_\text{\tiny{LO}}^{I_{m+1} \ldots I_n} \notag \\
& \phantom{\Biggl(} + \sum_{l=0}^{j-1} \sum_{m=j}^{k-1} U_{j;l(l+1)} U_{k;m(m+1)} A_\text{\tiny{LO}}^{I_1 \ldots I_{l}} A_\text{\tiny{LO}}^{I_{l+1} \ldots I_{j-1}} A_\text{\tiny{LO}}^{I_{j+1} \ldots I_m} A_\text{\tiny{LO}}^{I_{m+1} \ldots I_{k-1}} A_\text{\tiny{LO}}^{I_{k+1} \ldots I_n}
\Biggr) \notag \\
& + \frac{\lambda^2}{4} \sum_{j=1}^{n-3} \sum_{k=j+2}^{n-1} \delta^{I_j I_k} I_{jk} \notag \\
& \times \Biggl( \sum_{l=k+1}^n \sum_{m=l}^n \delta^{I_l 6} U_{l;m(m+1)} A_\text{\tiny{LO}}^{I_1 \ldots I_{j-1}} A_\text{\tiny{LO}}^{I_{j+1} \ldots I_{k-1}} A_\text{\tiny{LO}}^{I_{k+1} \ldots I_{l-1}} A_\text{\tiny{LO}}^{I_{l+1} \ldots I_{m}} A_\text{\tiny{LO}}^{I_{m+1} \ldots I_n} \notag \\
& \phantom{\Biggl(} + \sum_{l=k+1}^n \sum_{m=k}^{l-1} \delta^{I_l 6} U_{l;m(m+1)} A_\text{\tiny{LO}}^{I_1 \ldots I_{j-1}} A_\text{\tiny{LO}}^{I_{j+1} \ldots I_{k-1}} A_\text{\tiny{LO}}^{I_{k+1} \ldots I_{m}} A_\text{\tiny{LO}}^{I_{m+1} \ldots I_{l-1}} A_\text{\tiny{LO}}^{I_{l+1} \ldots I_n} \Biggr) \notag \\
& + \frac{\lambda^2}{4} \sum_{j=2}^{n-2} \sum_{k=j+2}^n \delta^{I_j I_k} I_{jk} \notag \\
& \times \Biggl( \sum_{l=1}^{j-1} \sum_{m=l}^{j-1} \delta^{I_l 6} U_{l;m(m+1)} A_\text{\tiny{LO}}^{I_1 \ldots I_{l-1}} A_\text{\tiny{LO}}^{I_{l+1} \ldots I_{m}} A_\text{\tiny{LO}}^{I_{m+1} \ldots I_{j-1}} A_\text{\tiny{LO}}^{I_{j+1} \ldots I_{k-1}} A_\text{\tiny{LO}}^{I_{k+1} \ldots I_n} \notag \\
& \phantom{\Biggl(} + \sum_{l=1}^{j-1} \sum_{m=0}^{l-1} \delta^{I_l 6} U_{l;m(m+1)} A_\text{\tiny{LO}}^{I_1 \ldots I_{m}} A_\text{\tiny{LO}}^{I_{m+1} \ldots I_{l-1}} A_\text{\tiny{LO}}^{I_{l+1} \ldots I_{j-1}} A_\text{\tiny{LO}}^{I_{j+1} \ldots I_{k-1}} A_\text{\tiny{LO}}^{I_{k+1} \ldots I_n} \Biggr) \notag \\
& + \frac{\lambda^2}{4} \sum_{j=1}^{n-3} \sum_{k=j+2}^{n-1} \sum_{l=k+1}^n \sum_{m=0}^{j-1} \delta^{I_j I_k} \delta^{I_l 6} \notag \\
& \qquad\qquad \times I_{jk} U_{l;m(m+1)} A_\text{\tiny{LO}}^{I_1 \ldots I_{m}} A_\text{\tiny{LO}}^{I_{m+1} \ldots I_{j-1}} A_\text{\tiny{LO}}^{I_{j+1} \ldots I_{k-1}} A_\text{\tiny{LO}}^{I_{k+1} \ldots I_{l-1}} A_\text{\tiny{LO}}^{I_{l+1} \ldots I_{n}} \notag \\
& + \frac{\lambda^2}{4}  \sum_{j=3}^{n-2} \sum_{k=j+2}^{n} \sum_{l=1}^{j-1} \sum_{m=k}^{n} \delta^{I_j I_k} \delta^{I_l 6} \notag \\
& \qquad\qquad \times  I_{jk} U_{l;m(m+1)} A_\text{\tiny{LO}}^{I_1 \ldots I_{l-1}} A_\text{\tiny{LO}}^{I_{l+1} \ldots I_{j-1}} A_\text{\tiny{LO}}^{I_{j+1} \ldots I_{k-1}} A_\text{\tiny{LO}}^{I_{k+1} \ldots I_{m}} A_\text{\tiny{LO}}^{I_{m+1} \ldots I_{n}} \notag \\
& + \frac{\lambda^2}{4} \sum_{j=1}^{n-5} \sum_{k=j+2}^{n-3} \sum_{l=k+1}^{n-2} \sum_{m=l+2}^n \delta^{I_j I_k} \delta^{I_l I_m} \notag \\
& \qquad\qquad \times  I_{jk} I_{lm} A_\text{\tiny{LO}}^{I_1 \ldots I_{j-1}} A_\text{\tiny{LO}}^{I_{j+1} \ldots I_{k-1}} A_\text{\tiny{LO}}^{I_{k+1} \ldots I_{l-1}} A_\text{\tiny{LO}}^{I_{l+1} \ldots I_{m-1}} A_\text{\tiny{LO}}^{I_{m+1} \ldots I_{n}} \notag \\
& + \frac{\lambda^2}{4} \sum_{j=1}^{n-1} \sum_{k=j+1}^n \delta^{I_j 6} \delta^{I_k 6} \notag \\
& \qquad\qquad \times \Biggl( \sum_{l=k}^n U^{(2)}_{j;k;l(l+1)} A_\text{\tiny{LO}}^{I_1 \ldots I_{j-1}} A_\text{\tiny{LO}}^{I_{j+1} \ldots I_{k-1}} A_\text{\tiny{LO}}^{I_{k+1} \ldots I_{l}} A_\text{\tiny{LO}}^{I_{l+1} \ldots I_{n}} \notag \\
& \qquad\qquad \phantom{\Biggl(} + \sum_{l=j}^{k-1} U^{(2)}_{j;k;l(l+1)} A_\text{\tiny{LO}}^{I_1 \ldots I_{j-1}} A_\text{\tiny{LO}}^{I_{j+1} \ldots I_{l}} A_\text{\tiny{LO}}^{I_{l+1} \ldots I_{k-1}} A_\text{\tiny{LO}}^{I_{k+1} \ldots I_{n}} \notag \\
& \qquad\qquad \phantom{\Biggl(} + \sum_{l=0}^{j-1} U^{(2)}_{j;k;l(l+1)} A_\text{\tiny{LO}}^{I_1 \ldots I_{l}} A_\text{\tiny{LO}}^{I_{l+1} \ldots I_{j-1}} A_\text{\tiny{LO}}^{I_{j+1} \ldots I_{k-1}} A_\text{\tiny{LO}}^{I_{k+1} \ldots I_{n}} \Biggr) \,,
\label{eq:recursionevenNLOformal}
\end{align}
where the $U^{(2)}$-integrals are contained in the three last terms.

This formula is the one that was effectively implemented in the ancillary \textsc{Mathematica} notebook of \cite{Barrat:2022eim}, and that has been used for producing the results of Sections \ref{sec:23point}, \ref{ch:4point} and \ref{ch:higherpoint}.

\endgroup

\chapter{Harmonic Polylogarithms} \label{app:HPL}

Harmonic polylogarithms (HPL's) are special functions of weight $w$ and argument $x$ identified by a set of $w$ indices, grouped into a $w$-dimensional vector $\vec{m}_w$. There are indicated by $H(\vec{M}_w;x)$.
In particular, for the case $w=1$,  there are three HPL's:
\begin{subequations}
\begin{align}
H(0;x)&=\text{ln} x\,,\\
H(1;x)&=\int^x_0 \frac{dx^\prime}{1-x^\prime}= -\text{ln}(1-x)\,,\\
H(-1;x)&=\int^x_0 \frac{dx^\prime}{1+x^\prime}= \text{ln}(1+x)\,.
\end{align}
\end{subequations}
Taking their derivatives, one has 
\begin{equation}
\frac{d}{dx}H(a;x)=f(a;x)\,,
\end{equation}
where the index $a$ takes three values $0,+1,-1$ and the three rational functions $f(a;x)$ are given by
\begin{equation}
\begin{aligned}
f(0 ; x) & =\frac{1}{x}\,, \\
f(1 ; x) & =\frac{1}{1-x}\,, \\
f(-1 ; x) & =\frac{1}{1+x}\,.
\end{aligned}
\end{equation}

For $w>1$, we write more in general
\begin{equation}
\vec{m}_w=\left(a, \vec{m}_{w-1}\right)\,,
\end{equation}
where $a=m_w$ is the leftmost index, and $\vec{m}_{w-1}$ is a vector of the remaining $w-1$ components. Moreover,  $\vec{0}_w$ is the vector whose $w$ components are all equal to the index 0. The harmonic polylogarithms of weight $w$ are then defined as follows:
\begin{equation}
\mathrm{H}\left(\vec{0}_w ; x\right)=\frac{1}{w !} \ln ^w x\,,
\end{equation}
and similarly for $\vec{1}_w,(-\vec{1})_w$, which are the vectors whose components are all equal to 1 or -1. By applying recursively the definitions we obtain 
\begin{subequations}
\begin{align}
& \mathrm{H}\left(\vec{1}_w ; x\right)=\frac{1}{w !}(-\ln (1-x))^w\,,\\
& \mathrm{H}\left((\vec{-1})_w ; x\right)=\frac{1}{w !} \ln ^w(1+x) \,.
\end{align}
\end{subequations}

On the other hand, if $\vec{m}_w \neq \vec{0}_w$, one defines
\begin{equation}
\mathrm{H}\left(\vec{m}_w ; x\right)=\int_0^x d x^{\prime} f\left(a ; x^{\prime}\right) \mathrm{H}\left(\vec{m}_{w-1} ; x^{\prime}\right)\,.
\end{equation}
The derivatives can be written in the compact form
\begin{equation}
\frac{d}{d x} \mathrm{H}\left(\vec{m}_w ; x\right)=f(a ; x) \mathrm{H}\left(\vec{m}_{w-1} ; x\right)\,.
\end{equation}

We can now look at the indices' first few values. For $w=2$ there are nine combinations
\begin{subequations}
\begin{align}
\mathrm{H}(0,0 ; x) & =\frac{1}{2 !} \ln ^2 x\,, \\
\mathrm{H}(0,1 ; x) & =\int_0^x \frac{d x^{\prime}}{x^{\prime}} \mathrm{H}\left(1 ; x^{\prime}\right)=-\int_0^x \frac{d x^{\prime}}{x^{\prime}} \ln \left(1-x^{\prime}\right), \\
\mathrm{H}(0,-1 ; x) & =\int_0^x \frac{d x^{\prime}}{x^{\prime}} \mathrm{H}\left(-1 ; x^{\prime}\right)=\int_0^x \frac{d x^{\prime}}{x^{\prime}} \ln \left(1+x^{\prime}\right), \\
\mathrm{H}(1,0 ; x) & =\int_0^x \frac{d x^{\prime}}{1-x^{\prime}} \mathrm{H}\left(0 ; x^{\prime}\right)=\int_0^x \frac{d x^{\prime}}{1-x^{\prime}} \ln x^{\prime}, \\
\mathrm{H}(1,1 ; x) & =\int_0^x \frac{d x^{\prime}}{1-x^{\prime}} \mathrm{H}\left(1 ; x^{\prime}\right)=-\int_0^x \frac{d x^{\prime}}{1-x^{\prime}} \ln \left(1-x^{\prime}\right), \\
\mathrm{H}(1,-1 ; x) & =\int_0^x \frac{d x^{\prime}}{1-x^{\prime}} \mathrm{H}\left(-1 ; x^{\prime}\right)=\int_0^x \frac{d x^{\prime}}{1-x^{\prime}} \ln \left(1+x^{\prime}\right), \\
\mathrm{H}(-1,0 ; x) & =\int_0^x \frac{d x^{\prime}}{1+x^{\prime}} \mathrm{H}\left(0 ; x^{\prime}\right)=\int_0^x \frac{d x^{\prime}}{1+x^{\prime}} \ln x^{\prime}, \\
\mathrm{H}(-1,1 ; x) & =\int_0^x \frac{d x^{\prime}}{1+x^{\prime}} \mathrm{H}\left(1 ; x^{\prime}\right)=-\int_0^x \frac{d x^{\prime}}{1+x^{\prime}} \ln \left(1-x^{\prime}\right), \\
\mathrm{H}(-1,-1 ; x) & =\int_0^x \frac{d x^{\prime}}{1+x^{\prime}} \mathrm{H}\left(-1 ; x^{\prime}\right)=\int_0^x \frac{d x}{1+x^{\prime}} \ln \left(1+x^{\prime}\right) \,,
\end{align}
\end{subequations}
which can all be expressed in terms of logarithmic and dilogarithmic functions; indeed, if
\begin{equation}
\mathrm{Li}_2(x)=-\int_0^x \frac{d x^{\prime}}{x^{\prime}} \ln \left(1-x^{\prime}\right)
\end{equation}
is the usual Euler's dilogarithm, one finds
\begin{subequations}
\begin{align}
\mathrm{H}(0,1 ; x) & =\mathrm{Li}_2(x)\,,\\
\mathrm{H}(0,-1 ; x) & =-\operatorname{Li}_2(-x)\,, \\
\mathrm{H}(1,0 ; x) & =-\ln x \ln (1-x)+\mathrm{Li}_2(x)\,, \\
\mathrm{H}(1,1 ; x) & =\frac{1}{2 !} \ln ^2(1-x)\,, \\
\mathrm{H}(1,-1 ; x) & =\operatorname{Li}_2\left(\frac{1-x}{2}\right)-\ln 2 \ln (1-x)-\operatorname{Li}_2\left(\frac{1}{2}\right)\,, \\
\mathrm{H}(-1,0 ; x) & =\ln x \ln (1+x)+\operatorname{Li}_2(-x)\,, \\
\mathrm{H}(-1,1 ; x) & =\operatorname{Li}_2\left(\frac{1+x}{2}\right)-\ln 2 \ln (1+x)-\mathrm{Li}_2\left(\frac{1}{2}\right) \,,\\
\mathrm{H}(-1,-1 ; x) & =\frac{1}{2 !} \ln ^2(1+x) \,.
\end{align}
\end{subequations}
Something similar happens for polylogarithms of weight three, but it does not hold further. 

Note that it is convenient to rewrite HPL's using
\begin{equation}
H(w_1, \dots, \underbrace{0, \dots, 0}_{\text{k times}},w_i, \dots;x)= H(w_1, \dots, w_i+k, \dots;x)\,.
\end{equation}
To conclude, we highlight a useful identity for HPL's:
\begin{equation}
\begin{aligned}
H(a;x)H(w_p\dots, w_1;x)& =\mathrm{H}\left(a, w_p \cdots, w_1 ; x\right) \\
& +\mathrm{H}\left(w_p, a, w_{p-1} \cdots, w_1 ; x\right) \\
& +\mathrm{H}\left(w_p, w_{p-1}, a, w_{p-2} \cdots w_1 ; x\right) \\
& +\cdots \\
& +\mathrm{H}\left(w_p, \cdots, w_1, a ; x\right) \,.
\end{aligned}
\end{equation}

For further details, we invite the reader to check out \cite{Remiddi:1999ew}.

\chapter{Snowflake Casimir} \label{app:casimir}

This short appendix explains how we calculated the snowflake-channel conformal blocks, which we introduce in Section \ref{sec:snowflake}. Explicit expressions for these blocks have already appeared in the literature \cite{Fortin:2020zxw}; however here we use different cross-ratios that make the blocks symmetric in all its arguments.

Defining
\begin{equation} 
a_1 = \frac{1}{2}(\Delta_2 - \Delta_1)	\,, 
\quad a_2 = \frac{1}{2}(\Delta_4 - \Delta_3)\,,
\quad a_3 = \frac{1}{2}(\Delta_6 - \Delta_5),
\end{equation}
we can write the following Casimir\footnote{ We introduced the notion of Casimir operator in Section \ref{sec:conformalblocks}.} operators:
\begin{subequations}
\begin{align}
\begin{split}
\mathcal{C}_{2}^{(12)} & =
-(z_1-1)z_1^2 \partial_{z_1}^2 + (z_2-1) z_2 z_1^2 \partial_{z_1} \partial_{z_2} 
+ z_1^2 (-2 a_2 z_2 + 2 a_1 -1) \partial_{z_1}
\\
& -2a_1(z_2-1)z_2 z_1 \partial_{z_2} - 2 a_1 z_3 z_1 \partial_{z_3} + 4 a_1 a_2 z_2 z_1 +z_3 z_1^2 \partial_{z_1} \partial_{z_3} \,,
\end{split}
\\
\begin{split}
\mathcal{C}_{2}^{(34)} & =
-(z_2-1)z_2^2 \partial_{z_2}^2 + (z_3-1) z_3 z_2^2 \partial_{z_2} \partial_{z_3} 
+ z_2^2 (-2 a_3 z_3 + 2 a_2 -1) \partial_{z_2}
\\
& -2a_2(z_3-1)z_3 z_2 \partial_{z_3} - 2 a_2 z_1 z_2 \partial_{z_1} + 4 a_2 a_3 z_3 z_2 +z_1 z_2^2 \partial_{z_2} \partial_{z_1} \,,
\end{split}
\\
\begin{split}
\mathcal{C}_{2}^{(56)} & =
-(z_3-1)z_3^2 \partial_{z_3}^2 + (z_1-1) z_1 z_3^2 \partial_{z_3} \partial_{z_1} 
+ z_3^2 (-2 a_1 z_1 + 2 a_3 -1) \partial_{z_3}
\\
& -2a_3(z_1-1)z_1 z_3 \partial_{z_1} - 2 a_3 z_2 z_3 \partial_{z_2} + 4 a_3 a_1 z_1 z_3 +z_2 z_3^2 \partial_{z_3} \partial_{z_2}\,.
\end{split}
\end{align}
\end{subequations}
The conformal blocks are then eigenfunctions of the following Casimir equations:
\begin{subequations}
\begin{align}
\mathcal{C}^{(12)}_2 g_{\Delta_a,\Delta_b,\Delta_c}(z_1,z_2,z_3) & = \Delta_a (\Delta_a-1) g_{\Delta_a,\Delta_b,\Delta_c}(z_1,z_2,z_3)\,,
\\
\mathcal{C}^{(34)}_2 g_{\Delta_a,\Delta_b,\Delta_c}(z_1,z_2,z_3) & = \Delta_b (\Delta_b-1) g_{\Delta_a,\Delta_b,\Delta_c}(z_1,z_2,z_3)\,,
\\
\mathcal{C}^{(56)}_2 g_{\Delta_a,\Delta_b,\Delta_c}(z_1,z_2,z_3) & = \Delta_c (\Delta_c-1) g_{\Delta_a,\Delta_b,\Delta_c}(z_1,z_2,z_3)\,.
\end{align}
\end{subequations}
In order to solve these equations, we give the ansatz
\begin{equation}
g_{\Delta_{a}, \Delta_b, \Delta_c}\left(z_1\,, z_2\,,z_3 \right) = z_1 ^{\Delta_a} z_2^{\Delta_b} z_3^{\Delta_c} \sum_{n_1, n_2, n_3} \bar{c}_{n_1, n_2, n_3} z_1^{n_1} z_2^{n_2} z_3^{n_3} \,,
\end{equation}
and since we are only interested in extracting low-lying CFT data, we content ourselves with a handful of low-lying coefficients:
\begin{align} 
\bar{c}_{0,0,0} & = 1\,,
\\
\bar{c}_{1,0,0} & = \frac{(-2 a_1 + \Delta_a ) (\Delta_a + \Delta_b - \Delta_c)}{2 \Delta_a}\,,
\\
\bar{c}_{0,1,0} & = \frac{(-2 a_2 + \Delta_b ) (\Delta_b + \Delta_c - \Delta_a)}{2 \Delta_b}\,,
\\
\bar{c}_{0,0,1} & = \frac{(-2 a_3 + \Delta_c ) (\Delta_c + \Delta_a - \Delta_b)}{2 \Delta_c}\,,
\\
\bar{c}_{1,1,0} & = -\frac{(-2 a_1 + \Delta_a)(-2 a_2 + \Delta_b)(1 + \Delta_a - \Delta_b - \Delta_c)(\Delta_a + \Delta_b - \Delta_c)}{4 \Delta_a \Delta_b}\,,
\\
\bar{c}_{0,1,1} & = -\frac{(-2 a_2 + \Delta_b)(-2 a_3 + \Delta_c)(1 + \Delta_b - \Delta_c - \Delta_a)(\Delta_b + \Delta_c - \Delta_a)}{4 \Delta_b \Delta_c}\,,
\\
\bar{c}_{1,0,1} & = -\frac{(-2 a_3 + \Delta_c)(-2 a_1 + \Delta_a)(1 + \Delta_c - \Delta_a - \Delta_b)(\Delta_c + \Delta_a - \Delta_b)}{4 \Delta_c \Delta_a}\,.
\end{align}
We refer the reader to \cite{Fortin:2020zxw} for a more detailed analysis of the snowflake channel and for a closed-form expression for the $\bar{c}_{n_1,n_2,n_3}$ coefficients (albeit in a different convention).

\chapter{Superconformal Blocks for the Five-Point Function} \label{app:superblocks}

\begingroup
\allowdisplaybreaks

In this appendix, we detail the derivation of the superconformal blocks for the asymmetric OPE channel discussed in Section \ref{subsubsec:TheAsymmetricChannel}.

We consider the first block appearing in (\ref{eq:FivePoint_BlockExpansionAsymmetric}).
\paragraph{$\mathbb{1},\mathcal{B}_1$:} the first superblock we want to determine is $\mathcal{G}_{\mathbb{1},\mathcal{B}_1}$ which is a trivial but still instructive case. Let us start by having a closer look at the multiplet $\mathcal{B}_1$:
\begin{equation}
[0,1]^{\Delta=1}_{s=0}\,\, \longrightarrow [1,0]^{\Delta=\frac{3}{2}}_{s=1}\,\,\longrightarrow [0,0]^{\Delta=2}_{s=2}\,,
\end{equation}
where the numbers $[a,b]$ refer to the $R$-symmetry group $Sp(4)_R \sim SO(5)$. Only scalars can appear in the superblocks; hence the only relevant component is the head of the multiplet $[0,1]^1_0$.

On the other hand, the identity is only composed of $[0,0]^0_0$.  We must then identify whether $[0,1]$ appears in the tensor product decomposition $[0,0] \otimes [0,1]$. In fact, the contribution $\mathcal{B}_1$ arises from the OPE between an external operator $\mathcal{B}_1$\footnote{The head of this multiplet is our scalar $\phi_I$.} and the identity $\mathds{1}$. Of course, here the case is trivial since $[0,0]$ is the identity, but in other cases, it is important.

The superblock then reads
\begin{equation}
\mathcal{G}_{\mathds{1},\Bm_1} = \alpha \, h_{[0,0], [0,1]} g_{0,1}\,,
\end{equation}
with
\begin{align} \label{eq:1B1}
\mc{G}_{\mathbb{1}, \Bm_1}=\frac{r_1 s_2}{\chi_1^2 (1-\chi_2)^2}\,,
\end{align}
where $\alpha$ is only an overall normalization constant which we fix after determining all the blocks.

It is more useful to decompose (\ref{eq:1B1}) in terms of $f^i$, analogously to the rewriting of the correlator in (\ref{eq:FivePoint_LittlefFromF}):
\begin{equation}
f^1_{\mathbb{1}, \Bm_1} =1\,, \qquad f^1_{\mathbb{1}, \Bm_1} =0\,, \qquad f^2_{\mathbb{1}, \Bm_1} =0\,, \qquad f^3_{\mathbb{1}, \Bm_1} =-\frac{1}{(\nu_1 \nu_2)^2}\,,
\end{equation}
with
\begin{equation}
\nu_1:=\frac{\chi_1-\chi_2}{1-\chi_2}\,, \qquad \nu_2:=1-\chi_2\,.
\end{equation}

\paragraph{$\Bm_2,\Bm_1$:} We can now analyze a less trivial case.  To derive the superblock $\mathcal{G}_{\Bm_2,\Bm_1}$, we look at the multiplet of $\Bm_2$. In general, any $\Bm_k$ multiplet reads
\begin{equation*}
\includegraphics[scale = 0.3]{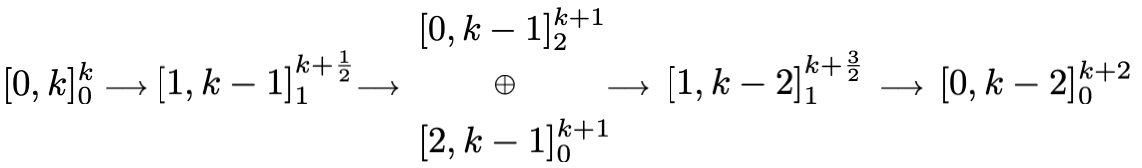}
\end{equation*}

We see that only three components can contribute at maximum, the ones with $s=0$:  $[0,2]^2_0, [2,0]^3_0, [0,0]^4_0$. 

We must now do the tensor product decompositions $[a,b] \otimes [c,d]$ of $SU(4)$ in order to check whether those also appear as a result of the OPE $\Bm_1$ (the external operator) and $\Bm_2$ (the exchanged operator on the right). Using any Lie algebra package, we find
\begin{equation}
[0,1] \otimes [0,1] \to [0,2] + [2,0] + [0,0]
\end{equation}
and thus, all the components are relevant for the superblock, which reads
\begin{equation}
\mathcal{G}_{\Bm_2,\Bm_1} = \alpha_0\, h_{[0,2],[0,1]} g_{2,1} + \alpha_1\, h_{[2,0],[0,1]} g_{3,1} + \alpha_2\, h_{[0,0],[0,1]} g_{4,1}\,. 
\end{equation}
The block in terms of $f^i$ reads
\begin{subequations}
\begin{align}
	f^0_{\Bm_2,\Bm_1} &=1\,,\\
	f^1_{\Bm_2,\Bm_1} &=\frac{1}{\nu_2}\sum _{k_2} -\frac{6 (k_2-2) \nu_1^{k_2}}{(k_2+2) (k_2+3) (k_2+4)}\,,\\
	f^2_{\Bm_2,\Bm_1} &=\frac{1}{\nu_2}\sum _{k_2} \frac{12 (1+k_2) \nu_1^{k_2}}{(k_2+2) (k_2+3) (k_2+4)}\,,\\
	f^3_{\Bm_2,\Bm_1} &=\frac{1}{\nu_2^2}\sum _{k_2} \frac{12 (k_2-2) \nu_1^{k_2}}{(k_2+3) (k_2+4) (k_2+5)}\,.
\end{align}
\end{subequations}
Note that the normalization has been chosen so that the OPE coefficient of the protected operator $c_{112}$ corresponds to the localization result. 

\paragraph{$\Lm_{0,[0,0]}^\Delta,\Bm_1$:} We conclude with an example of a block including the long multiplet. The logic remains the same, but the multiplet is richer:
\begin{equation*}
\includegraphics[scale = 0.3]{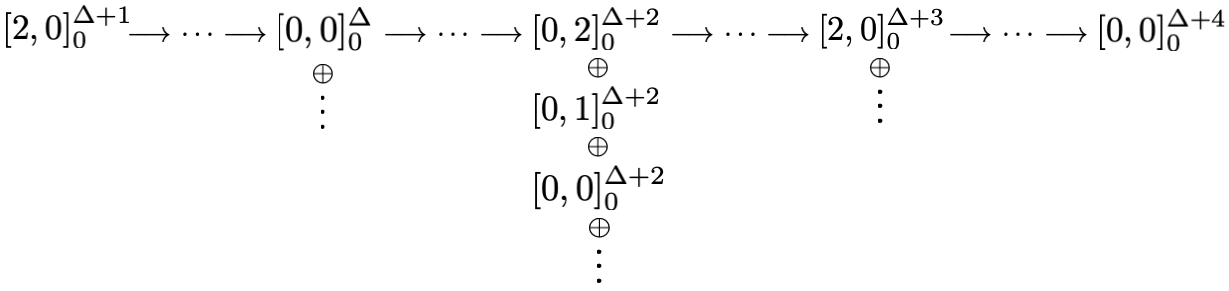}
\end{equation*}
where we have already kept only the contributions with $s=0$.

After performing the relevant tensor product decomposition, we obtain
\begin{align}
\mathcal{G}_{\Lm_{0,[0,0]}^\Delta,\Bm_1} =&
\,\alpha_0\, h_{[0,0],[0,1]} g_{\Delta,1} +
\alpha_1\, h_{[2,0],[0,1]} g_{\Delta+1,1} +
\alpha_2\, h_{[0,2],[0,1]} g_{\Delta+2,1} \notag \\ &+
\alpha_3\, h_{[0,0],[0,1]} g_{\Delta+2,1} +
\alpha_4\, h_{[2,0],[0,1]} g_{\Delta+3,1} +
\alpha_5\, h_{[0,0],[0,1]} g_{\Delta+4,1}\,.
\end{align}
Note that here there is no real need to perform the tensor product decomposition since cutting off the two rightmost external legs produces a four-point function, for which the superblocks are known. We can, therefore, just read the relevant contributions from there (and that is what we did).

The block, in this last case, reads
\begin{subequations}
\begin{align}
	f^0_{\mc{L}^\Delta_{0,[0,0]},\Bm_1} &=0\,,\\
	f^1_{\mc{L}^\Delta_{0,[0,0]},\Bm_1} &=\sum_{k_1} \Delta^3+\Delta^2 (k_1+1)+h_2 k_1-(k_1-3) k_1\,G_{\mc{L}^\Delta_{0,[0,0]},\Bm_1}^{k_1,k_2}\,,\\
	f^2_{\mc{L}^\Delta_{0,[0,0]},\Bm_1} &=\sum _{k_1}-((\Delta + \Delta^2 - 2 k1) (\Delta + k1))\,G_{\mc{L}^\Delta_{0,[0,0]},\Bm_1}^{k_1,k_2}\,, \\
	f^3_{\mc{L}^\Delta_{0,[0,0]},\Bm_1} &= \sum _{k_1}-\frac{\left(\Delta^2+\Delta-2 k_1\right) (\Delta+k_1-1)}{\nu_1 \nu_2}\,G_{\mc{L}^\Delta_{0,[0,0]},\Bm_1}^{k_1,k_2}\,,
\end{align}
\end{subequations}
where
\begin{align}
	G_{\mc{L}^\Delta_{0,[0,0]},\Bm_1}^{k_1,k_2}&=\frac{(-1)^{k_1+1} \Gamma (2 \Delta+4) \Gamma (\Delta+k_1) \Gamma (\Delta+k_1+1)}{\left(\Delta^2-1\right) \Gamma (\Delta+1)^2 \Gamma (k_1+1) \Gamma (2 \Delta+k_1+4)}(-\nu_1)^{\Delta+k_1-1}  \nu_2^{-1}\,.
\end{align}

Remarkably in the most complicated case, namely, when these two long multiplets are exchanged, $\mathcal{G}_{\Lm_{0,[0,0]}^{\Delta_1},\Lm_{0,[0,1]}^{\Delta_2}}$,  there are fifty coefficients to fix and the Ward identity (\ref{eq:WIn}) for the five-point function determines them all!

To conclude, in this appendix, we showed some examples of the derivation of the superconformal blocks. The same algorithm can be applied to the remaining blocks in this channel and the symmetric channel. We, therefore, leave the complete list of blocks to the future paper \cite{Barrat:2023pev}.

\endgroup

\chapter{1$d$ Mellin Transform}

\section{D-Functions} \label{app:Dfunctions}

\begingroup
\allowdisplaybreaks

The quartic contact diagrams with external conformal dimensions $\D_i$ are expressed in terms of $D$-functions \cite{Liu:1998ty,DHoker:1999kzh, Dolan:2003hv},  defined  for the general case of AdS$_{d+1}$ as
\be \label{D-function}
\!\!\!\!\!\!\!\!
D_{\Delta_{1}\Delta_{2}\Delta_{3}\Delta_{4}}(x_1,x_2,x_3,x_4) =\! \!\int \!\!\frac{dz d^dx}{z^{d+1}} 
\tilde{K}_{\Delta_{1}}\!(z,x;x_1) \tilde{K}_{\Delta_{2}}\!(z,x;x_2) \tilde{K}_{\Delta_{3}}\!(z,x;x_3) \tilde{K}_{\Delta_{4}}\!(z,x;x_4)\,,
\ee
via the bulk-to-boundary propagator in $d$ dimensions:
\be \label{bulktoboundary}
K_{\Delta}(z,x;x') = {\cal C}_{\Delta} \Big[\frac{z}{z^2+(x-x')^2}\Big]^{\Delta} \equiv  {\cal C}_{\Delta}\,  \tilde{K}_{\Delta}(z,x;x')\,,
\,\qquad {\cal C}_{\Delta_{\phi}} =\frac{\Gamma\left(\Delta_{\phi}\right)}{2\,\sqrt{\pi}\,\Gamma\left(\Delta_{\phi}+{1\ov 2}\right)} \,.
\ee
For vertices with derivatives, the following identity is useful 
\bal 
&g^{\m\n}\partial_\m \tilde{K}_{\Delta_1}(z,x;x_1)\ \partial_\n\tilde{K}_{\Delta_2}(z,x;x_2) 
\\   & \qquad = 
\Delta_1 \Delta_2
\left[\tilde{K}_{\Delta_1}(z,x;x_1)\tilde{K}_{\Delta_2}(z,x;x_2)-2x_{12}^2 \tilde{K}_{\Delta_1+1}(z,x;x_1)\tilde{K}_{\Delta_2+1}(z,x;x_2)\right]\\,,
\label{identityder}
\eal
where $g^{\m\n}= {z^2}\delta^{\m\n}$ and $\del_\m=(\del_z,\del_r)$,  $r=0,1, 2, ...,d-1$.  

To make explicit the covariant form of the correlator, it is useful to consider the ``reduced'' functions $\bar D$ that we introduced in (\ref{Dbar}). In $d=1$ as usual they only depend on the single variable $\chi$ ($u=\chi^2$, $v=(1-\chi)^2$):
\be\label{Dbar1d}
\bar D_{\Delta\Delta\Delta\Delta}(z)=\frac{\Gamma(\Delta)^4}{\Gamma(2\Delta)} (1-z)^{-2\Delta}\!\int_{-\infty}^{+\infty}\!d\tau\,e^{-\tau} {}_2F_1\big(\Delta,\Delta,2\Delta,\textstyle-\frac{4z}{(1-z^2)}\cosh^2\frac{\tau}{2}\big)\, .
\ee

Some explicit expressions for $\bar{D}$-functions read
\begin{subequations}
\begin{eqnarray}\label{Dbar-explicit}
	\bar D_{1111}&=&-\frac{2 \log (1-z )}{z }-\frac{2 \log (z )}{1-z }\, ,\\
	\bar{D}_{2222} &=&-\frac{2 \left(z ^2-z +1\right)}{15 (1-z )^2 z ^2}+\frac{\left(2 z ^2-5 z +5\right) \log (z )}{15 (z -1)^3}-\frac{\left(2 z ^2+z +2\right) \log (1-z )}{15 z ^3}\, ,\\
	\bar{D}_{3333}&=&\frac{\left(8 z ^4-36 z ^3+64 z ^2-56 z +28\right) \log(z)}{105 (z -1)^5}+\frac{\left(-8 z ^4-4 z ^3-4 z ^2-4 z -8\right) \log (1-z )}{105 z ^5}\nonumber\\
	&&+\frac{-24 z ^6+72 z ^5-74 z ^4+28 z ^3-74 z ^2+72 z -24}{315 (z -1)^4 z ^4}\, .\label{Dbar-explicit-end}
\end{eqnarray}
\end{subequations}

Other expressions can be found through the identities in~\cite{Dolan:2003hv}.  Useful relations between $\bar D$-function of consequent weight are ($\Sigma \equiv \frac{1}{2}\sum_i \Delta_i$)
%
\begin{subequations}
\begin{align}\label{Dbarconseq}
	\Delta\,\bar{D}_{\Delta \Delta \Delta \Delta }&= \bar{D}_{\Delta \Delta \Delta+1 \Delta+1}+\bar{D}_{\Delta \Delta+1 \Delta \Delta +1}+\bar{D}_{\Delta+1 \Delta \Delta \Delta +1} \, ,\\
\label{Dfunctioncrossing1}
	(\Delta_2+\Delta_4-\Sigma) \,\bar{D}_{\Delta_1 \Delta_2 \Delta_3 \Delta_4} &= \bar{D}_{\Delta_1 \Delta_2+1 \Delta_3 \Delta_4+1} - \bar{D}_{\Delta_1+1 \Delta_2 \Delta_3+1 \Delta_4}\, ,\\	
	(\Delta_1+\Delta_4-\Sigma) \,\bar{D}_{\Delta_1 \Delta_2 \Delta_3 \Delta_4}& =\bar{D}_{\Delta_1+1 \Delta_2 \Delta_3 \Delta_4+1}-(1-z)^2 \bar{D}_{\Delta_1 \Delta_2+1 \Delta_3+1 \Delta_4}\, ,\\
	(\Delta_3+\Delta_4-\Sigma)\, \bar{D}_{\Delta_1 \Delta_2 \Delta_3 \Delta_4}&= \bar{D}_{\Delta_1 \Delta_2 \Delta_3+1 \Delta_4+1}-z^2\bar{D}_{\Delta_1+1 \Delta_2+1 \Delta_3 \Delta_4}\, ,\\
\label{Dfunctioncrossing2}
	\bar{D}_{\Delta_1 \Delta_2 \Delta_3 \Delta_4}&=(1-z)^{2(\Delta_1+\Delta_4-\Sigma)}\bar{D}_{\Delta_2 \Delta_1 \Delta_4 \Delta_3}\\
	&=\bar{D}_{\Sigma-\Delta_3 \Sigma- \Delta_4 \Sigma-\Delta_1 \Sigma-\Delta_2}\\
	&=z^{2(\Delta_3+\Delta_4-\Sigma)}\bar{D}_{\Delta_4 \Delta_3 \Delta_2 \Delta_1}\, .
\end{align}
\end{subequations}

 \section{Anomalous Dimensions for Higher Derivative Interactions} \label{Ap: anomalous dimension}
 
This section lists some results for the polynomial part of the anomalous dimension in equation (\ref{gammaLB2}). The \textsc{Mathematica} notebook attached to \cite{Bianchi:2021piu} has values of $L$ ranging from $L=0$ to $L=8$ as well as a function~\texttt{FindBootstrapPolynomial[L,$\Delta$,n]} to compute $\mathcal{P}_{L,n}(\Delta_\phi)$ for arbitrary $L$ (the function gets slower and slower at higher $L$, but in principle it works for any $L$).

\scriptsize
\begin{align}
	\mathcal{P}_{0,n}(\Delta)&=1\,,\\
	\hspace{0.1cm}\nonumber \\
		\mathcal{P}_{1,n}(\Delta)&=8 (2 \Delta +1) n^4+8 \left(8 \Delta ^2+2 \Delta -1\right) n^3+2 (2 \Delta -1) (2 \Delta +1) (12 \Delta +1) n^2\nonumber \\
		&+\left(64 \Delta ^4-28 \Delta ^2-2 \Delta +2\right) n+\Delta ^2 \left(16 \Delta ^3-13 \Delta -3\right)\,,\\
		\hspace{0.1cm}\nonumber \\
	\mathcal{P}_{2,n}(\Delta)&= 64 (2 \Delta +3) (2 \Delta +5) n^8+128 (2 \Delta +3) (2 \Delta +5) (4 \Delta -1) n^7\nonumber \\
	&+32 (2 \Delta +3) (2 \Delta +5) \left(56 \Delta ^2-22 \Delta -1\right) n^6\nonumber \\
	&+32 (2 \Delta +3) (2 \Delta +5) (4 \Delta -1) (28 \Delta ^2-5 \Delta -5) n^5\nonumber \\
	&+4 (2 \Delta +3) \left(2240 \Delta ^5+4800 \Delta ^4-2924 \Delta ^3-2156
	\Delta ^2+246 \Delta -415\right) n^4\nonumber \\
	&+8 (2 \Delta +3) (4 \Delta -1) (224 \Delta ^5+576 \Delta ^4-158 \Delta ^3-572 \Delta
	^2-243 \Delta -160) n^3\nonumber \\
	&+4 (2 \Delta -1) (2 \Delta +3) (448 \Delta ^6+1392 \Delta ^5+84 \Delta ^4-2183
	\Delta ^3-2091 \Delta ^2-1134 \Delta -105) n^2\nonumber \\
	&+4 (2 \Delta +3) (4 \Delta -1) (64 \Delta ^7+208 \Delta ^6-36 \Delta ^5-605 \Delta
	^4-554 \Delta ^3-30 \Delta ^2+243 \Delta +90) n\nonumber \\
		&+(\Delta -2) (\Delta -1) \Delta ^2 (\Delta +1)^2 (4 \Delta +3) (4 \Delta +5) (4 \Delta +7) (4 \Delta +9)\,,\\
		\hspace{0.5cm}\nonumber \\
	\mathcal{P}_{3,n}(\Delta)&= 512 (2 \Delta +5) (2 \Delta +7) (2 \Delta +9) n^{12}\nonumber \\
	&+1536 (2 \Delta +5) (2 \Delta +7) (2 \Delta +9) (4 \Delta -1) n^{11}\nonumber \\
	&+128 (2 \Delta +5) (2 \Delta +7) (2 \Delta +9) (264 \Delta ^2-102 \Delta +5) n^{10}\nonumber \\
	&+640 (2 \Delta +5) (2 \Delta +7) (2 \Delta +9) (4 \Delta -1) (44 \Delta ^2-7 \Delta -3) n^9\nonumber \\
	&+96 (2 \Delta +5) (2 \Delta +7) (5280 \Delta ^5+22080 \Delta ^4-8610 \Delta ^3-4790 \Delta ^2-798 \Delta -2931) n^8\nonumber \\
	&+96 (2 \Delta +5) (2 \Delta +7) (4 \Delta -1) (2112 \Delta ^5+9792 \Delta ^4+268 \Delta ^3-5448 \Delta ^2-4628 \Delta -5493) n^7\nonumber \\
	&+8 (2 \Delta +5) (2 \Delta +7) \left(118272 \Delta ^7+556416 \Delta ^6-8736 \Delta ^5-656280 \Delta ^4-661308 \Delta ^3-560400 \Delta ^2 \right. \nonumber \\
	&\quad \left.+371392 \Delta +17415 \right) n^6\nonumber \\
	&+24 (2 \Delta +5) (2 \Delta +7) (4 \Delta -1) \left(8448 \Delta ^7+45504 \Delta ^6+22128 \Delta ^5-79708 \Delta ^4-143680 \Delta ^3-114082 \Delta ^2\right. \nonumber \\
	&\quad \left.+52985 \Delta +27645\right) n^5\nonumber \\
	&+4 (2 \Delta +5) \left( 253440 \Delta ^{10}+2327040 \Delta ^9+5816640 \Delta ^8-1506240 \Delta ^7-22985970 \Delta ^6-33151830 \Delta ^5 \right. \nonumber \\
	&\quad \left.-9079800 \Delta ^4+25792815 \Delta ^3+10370477 \Delta
	^2-446534 \Delta +2131794 \right) n^4\nonumber \\
	&+8 (2 \Delta +5) (4 \Delta -1)\left(14080 \Delta ^{10}+142080 \Delta ^9+423840 \Delta ^8+8160 \Delta ^7-2172753 \Delta ^6-4187481 \Delta ^5 \right. \nonumber \\
	&\quad \left.-1812050 \Delta ^4+3606930 \Delta ^3+3965596 \Delta ^2+1661325
	\Delta +791091 \right) n^3 \nonumber \\
	&+6 (2 \Delta -1) (2 \Delta +5) \left(11264 \Delta ^{11}+125184 \Delta ^{10}+437120 \Delta ^9+118880 \Delta ^8-2771604 \Delta ^7-6808095 \Delta ^6 \right. \nonumber \\
	&\quad \left.-4248981 \Delta ^5+6860955 \Delta ^4+13140919 \Delta
	^3+9496058 \Delta ^2+4002384 \Delta +360360\right) n^2\nonumber \\
	&+2 (2 \Delta +5) (4 \Delta -1) \left(3072 \Delta ^{12}+36096 \Delta ^{11}+132224 \Delta ^{10}-3360 \Delta ^9-1214676 \Delta ^8-2926395 \Delta ^7 \right. \nonumber \\
	&\quad \left.-970776 \Delta ^6+6196080 \Delta ^5+10143424 \Delta
	^4+5128059 \Delta ^3-1542528 \Delta ^2-3028860 \Delta -907200\right) n\nonumber \\
	&+(\Delta -3) (\Delta -2) (\Delta -1) \Delta ^2 (\Delta +1)^2 (\Delta +2)^2 (4 \Delta +5) (4 \Delta +7) (4 \Delta +9) (4 \Delta +11) (4 \Delta +13) (4 \Delta +15)\,.
\end{align}

\normalsize
\endgroup


\bibliography{Bibliography}


\end{document}